\documentclass[11pt, oneside]{article}   	
\usepackage{jheppub}
\usepackage{amsmath}
\usepackage{amssymb}
\usepackage{amsfonts}
\usepackage{amsthm}
\usepackage{amsbsy}
\usepackage{array}
\usepackage{mathtools}
\usepackage[usenames,dvipsnames]{xcolor}
\usepackage{subcaption}
\usepackage{dsdshorthand}
\usepackage{graphicx}
\usepackage[vcentermath]{youngtab}
\usepackage{cancel}

\usepackage{tikz}
\usetikzlibrary{decorations.markings}
\usetikzlibrary{decorations.pathmorphing}
\usetikzlibrary{positioning,decorations.pathreplacing}
\tikzset{snake it/.style={decorate, decoration=snake}}

\newtheorem*{assumption}{Assumption}

\makeatletter
\def\@fpheader{\ }
\makeatother

\title{Universal Asymptotics for High Energy CFT Data}
\author[a]{Nathan Benjamin,}
\author[a]{Jaeha Lee,}
\author[a,b]{Hirosi Ooguri,}
\author[a]{David Simmons-Duffin}
\affiliation[a]{Walter Burke Institute for Theoretical Physics, Caltech, Pasadena, California 91125, USA}
\affiliation[b]{Kavli Institute for the Physics and Mathematics of the Universe (WPI), \\ University of Tokyo, Kashiwa 277-8583, Japan}
\emailAdd{nbenjami@caltech.edu}
\emailAdd{jaeha@caltech.edu}
\emailAdd{ooguri@caltech.edu}
\emailAdd{dsd@caltech.edu}

\date{}
\abstract{Equilibrium finite temperature observables of a CFT can be described by a local effective action for background fields --- a ``thermal effective action." This effective action determines the asymptotic density of states of a CFT as a detailed function of dimension and spin. We discuss subleading perturbative and nonperturbative corrections to the density, comparing with free and holographic examples. We furthermore show how to use the thermal effective action on more complicated geometries at special locations called ``hot spots." The hot spot idea makes a prediction for a CFT partition function on a higher-dimensional version of a genus-2 Riemann surface, in a particular high temperature limit. By decomposing the partition function into a novel higher-dimensional version of genus-2 conformal blocks (which we compute at large scaling dimension), we extract the asymptotic density of heavy-heavy-heavy OPE coefficients in a higher-dimensional CFT. We also compute asymptotics of thermal 1-point functions using the same techniques.}

\preprint{CALT-TH 2023-014, IPMU 23-0020}

\renewcommand\beforetochook{\pagestyle{myplain}\pagenumbering{roman}\renewcommand{\baselinestretch}{0.947}}

\begin{document}

\maketitle
\renewcommand{\baselinestretch}{1}
\pagenumbering{roman}
\setcounter{page}{2}
\newpage
\pagenumbering{arabic}
\setcounter{page}{1}

\section{Introduction}
\label{sec:intro}

What is the behavior of conformal field theory (CFT) data at high energies? This question is well-studied in two dimensions. For instance, the density of states of any 2d CFT at high energies takes the following universal form, known as the Cardy formula \cite{Cardy:1986ie}:
\be
\rho^{d=2}(\Delta, J) \sim \exp\left[\sqrt{\frac{c}3}\pi \left(\sqrt{\Delta + J - \frac{c}{12}} + \sqrt{\Delta - J - \frac{c}{12}}\right)\right], ~~~\Delta - |J| \gg c.
\label{eq:Cardy2d}
\ee
Here, $\rho^{d=2}(\Delta, J)$ is the density of local operators (equivalently states on $S^1$) with scaling dimension $\Delta$ and spin $J$. The entropy at high energies is controlled by a single theory-dependent number: the central charge $c$. The Cardy formula follows from modular invariance of the genus one partition function.

Though (\ref{eq:Cardy2d}) is valid for all 2d CFTs, it has a particularly nice interpretation for CFTs dual to quantum gravity in weakly-curved AdS$_3$. In such theories, the entropy $\log \rho^{d=2}(\De,J)$ is interpreted as the area of a BTZ black hole with spin $J$ and mass $M$ given by \cite{Strominger:1997eq}
\begin{equation}
    M = \frac{1}{\ell_{\text{AdS}}}\left(\Delta - \frac{c}{12}\right),
\end{equation}
in an AdS$_3$ space with \cite{Brown:1986nw}
\begin{equation}
    c = \frac{3\ell_{\text{AdS}}}{2G_N}.
\end{equation}
The Cardy formula then becomes a statement of universality of black hole entropy, regardless of the microscopic details of the quantum gravity theory.

OPE coefficients of heavy operators in 2d CFTs obey similar, though perhaps less well-known, universal formulas. In \cite{Cardy:2017qhl}, a formula for average squared OPE coefficients of three heavy Virasoro primaries was derived using modular invariance of the genus two partition function. For example, when all three operators have roughly equal dimensions $\De_i=\Delta\gg c$, it takes the form
\begin{equation}
    (C^{d=2}_\textrm{HHH})^2 \sim \left(\frac{27}{16}\right)^{3\Delta} e^{-6\pi\sqrt{\frac{c-1}{24}\Delta}} \Delta^{\frac{5c-11}{36}}, ~~~ \Delta \gg c.
\end{equation}
Similar formulas were derived for OPE coefficients with one or two heavy operator(s) (see e.g. \cite{Kraus:2016nwo}). These formulas were subsequently unified in \cite{Collier:2019weq}, with interesting connections to the DOZZ formula. In holographic theories, the formula for $(C^{d=2}_\textrm{HHH})^2$ matches the contribution of a two-sided wormhole connecting a pair of boundary three-point functions \cite{Chandra:2022bqq}.

In this paper, we explore whether similar universal formulas exist for higher dimensional CFTs. We will use purely field-theoretic methods, so our results will be applicable to both holographic and non-holographic theories. An immediate puzzle is that there is no simple analog of modular invariance in higher dimensional geometries like $S^1 \x S^{d-1}$ ($d\geq 3$). (See \cite{Shaghoulian:2015kta,Belin:2016yll,Shaghoulian:2016gol,Horowitz:2017ifu, Belin:2018jtf, Luo:2022tqy} for some discussion and progress on modular invariance in higher dimensions.) However, we can instead use a beautiful idea from \cite{Bhattacharyya:2007vs,Jensen:2012jh,Banerjee:2012iz}, which was used to count the density of states in higher dimensional CFTs in \cite{Bhattacharyya:2007vs,Shaghoulian:2015lcn}. (Similar ideas were used for studying supersymmetric indices in \cite{DiPietro:2014bca}.) The key point is that finding  the leading asymptotics of CFT data doesn't require full modular invariance --- we just need a sufficiently powerful effective theory for a CFT dimensionally reduced on a circle.

The dimensional reduction of a $d$-dimensional CFT is generically a gapped theory in $d{-}1$ dimensions. Fortunately for our purposes, the exponential decay of correlations in a gapped theory makes it very {\it flexible}: we can place it on many different geometries, and in this way extract myriad predictions for the $d$-dimensional CFT. 

A gapped theory can be described by a local action for background fields, obtained by integrating out the gapped degrees of freedom. In the context of a dimensionally-reduced CFT (with thermal boundary conditions), we call this local action the ``thermal effective action." It describes hydrodynamic observables of the CFT in equilibrium.  The derivative expansion of the thermal effective action is an expansion in the inverse temperature $\b=1/T$. This construction was explained in \cite{Jensen:2012jh,Banerjee:2012iz}, and has been explored extensively in the hydrodynamics literature, see e.g.\ \cite{Jensen:2012kj,Bhattacharyya:2013lha,Crossley:2015evo,Jensen:2013kka,Jensen:2012jy}. We review it in section~\ref{sec:thermaleffective}, along the way discussing some subtleties related to the Weyl anomaly. 

Placing the thermal effective theory on $S^1 \x S^{d-1}$ leads to simple universal predictions for the density of CFT operators at large $\De$. For example, in 3d CFTs, one obtains  
\begin{equation}
    \log \rho^{d=3}(\Delta, J) = 3\pi^{1/3} f^{1/3} (\Delta^2-J^2)^{1/3} - \frac23 \log(\Delta^2-J^2) + \mathcal{O}(\Delta^{0}).
    \label{eq:3dcardylike}
\end{equation}
 Here, $f$ is a theory-dependent positive real number, equal to minus the free energy density of the CFT, as we review in section~\ref{sec:cosmoconst}.
 The leading term in the high temperature partition function for the canonical ensemble of a CFT was first written down using hydrodynamic techniques in \cite{Bhattacharyya:2007vs}. It was subsequently transformed to the microcanonical ensemble in \cite{Shaghoulian:2015lcn}.\footnote{The density of states was also studied \cite{Verlinde:2000wg, Kutasov:2000td}.}  The thermal effective theory approach in this work allows us to reproduce those results and systematically explore subleading corrections.

The quantity $f$ controls the leading density of states in both 2d (where $f=\pi c/6$) and higher dimensions. However, unlike in 2d, where the Cardy formula is valid up to nonperturbative corrections in $\De$, the entropy in higher dimensional CFTs receives perturbative corrections in $1/\De$, coming from higher-derivative terms in the thermal effective action. The derivation of (\ref{eq:3dcardylike}) using the thermal effective action is given in section~\ref{sec:density}. There, we also describe the leading higher-derivative corrections. (Furthermore in section~\ref{sec:casimir}, we clarify some subtleties related to the Casimir energy on $S^{d-1}$ in higher-dimensional CFTs.) We also briefly discuss nonperturbative corrections to the density of states in section~\ref{sec:nonpertcorrections}. Then, in section~\ref{sec:examplesdensityofstates}, we compare these general formulas to free theories and holographic theories, determining Wilson coefficients in those cases by matching their partition functions to the effective theory.

In addition to the density of states, we will also find universal formulas for OPE coefficients of three heavy operators in higher-$d$ CFTs.\footnote{Formulas for heavy OPE coefficients weighted by light OPE coefficients, e.g.\ $C_\textrm{HHH}C_\textrm{HLL}^3$, were derived in \cite{Anous:2021caj} using crossing symmetry of 6-point functions of local operators. By contrast, our focus will be on {\it un-weighted} heavy OPE coefficients $C_\textrm{HHH}$, which are controlled by different physics. For example, the leading behavior of $C_\textrm{HHH}$ is determined by the free energy density $f$, which does not (to our knowledge) appear in a simple way in a 6-point function of light local operators.} Our strategy will be to put the theory on a higher-dimensional version of a genus-2 Riemann surface, obtained by gluing a pair of three-punctured $S^d$'s with three cylinders $S^{d-1}\x I$ (where $I$ is an interval). We describe this ``genus-2" geometry in detail in section~\ref{sec:partitionfunction}.\footnote{A special case of this geometry with no angular fugacities was studied recently in \cite{Belin:2021ibv}.}

\begin{figure}
\centering
\begin{tikzpicture}
\begin{scope}[scale=1.2]
  \begin{scope}[yscale=0.3, xscale=1.3, line width=1pt]
      
      \draw [dotted] (-2,0) arc (180:0:2);
      \draw [] (2,0) arc (0:-180:2);
      
      \draw [dotted] (-1, 0) circle (1);
      
      \draw [dotted] (1, 0) circle (1);
  \end{scope}
  
  \begin{scope}[yscale=0.3, xscale=1.3, shift={(0,5)}, line width=1pt]
      
      \draw [] (-1, 0) circle (1);
      
      \draw [] (1, 0) circle (1);
  \end{scope}
  \draw [very thick,red,postaction={decorate,decoration={markings,mark=at position 0.3 with {\arrow[scale=0.8]{>}}}}] (-0.02,0) to[out=96,in=-96] (-0.02,1.5);
  \draw [very thick,red,postaction={decorate,decoration={markings,mark=at position 0.3 with {\arrow[scale=0.8]{>}}}}]  (0.02,1.5) to[out=-84,in=84] (0.02,0);
  \draw [very thick,red] (-0.02,0) to[out=-84,in=-96] (0.02,0);
  \draw [very thick,red] (-0.02,1.5) to[out=84,in=96] (0.02,1.5);
  \begin{scope}[shift={(-2.6,0)}]
    \draw [very thick,red,postaction={decorate,decoration={markings,mark=at position 0.3 with {\arrow[scale=0.8]{>}}}}] (-0.02,0) to[out=96,in=-96] (-0.02,1.5);
  \draw [very thick,red,postaction={decorate,decoration={markings,mark=at position 0.3 with {\arrow[scale=0.8]{>}}}}]  (0.02,1.5) to[out=-84,in=84] (0.02,0);
  \draw [very thick,red] (-0.02,0) to[out=-84,in=-96] (0.02,0);
  \draw [very thick,red] (-0.02,1.5) to[out=84,in=96] (0.02,1.5);
  \end{scope}
  \begin{scope}[shift={(2.6,0)}]
    \draw [very thick,red,postaction={decorate,decoration={markings,mark=at position 0.3 with {\arrow[scale=0.8]{>}}}}] (-0.02,0) to[out=96,in=-96] (-0.02,1.5);
  \draw [very thick,red,postaction={decorate,decoration={markings,mark=at position 0.3 with {\arrow[scale=0.8]{>}}}}]  (0.02,1.5) to[out=-84,in=84] (0.02,0);
  \draw [very thick,red] (-0.02,0) to[out=-84,in=-96] (0.02,0);
  \draw [very thick,red] (-0.02,1.5) to[out=84,in=96] (0.02,1.5);
  \end{scope}

  \end{scope}
  \begin{scope}[scale=1.2]
  \begin{scope}[yscale=0.3, xscale=1.3, shift={(0,5)}, line width=1pt]
      \draw [] (0, 0) circle (2);
  \end{scope}
  \end{scope}
\end{tikzpicture}
\caption{The ``genus-2" geometry and its ``hot spots." The top is a ball $B^d$ with two balls removed. It is topologically equivalent to a three-punctured $S^d$. The bottom is the same. The top and bottom are glued together with three cylinders. In the limit that the cylinders get short, there are shrinking circles indicated in red that run down one cylinder and up another. The neighborhoods of each of these circles are ``hot spots," where the thermal effective action receives a large contribution.
\label{eq:hotspotfigure}}
\end{figure}
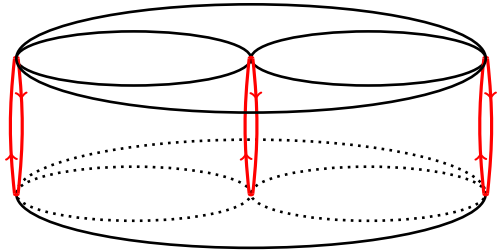

A glaring problem is that the ``genus-2" geometry is {\it not} a circle fibration, so it is not immediately obvious how to apply the thermal effective action. However, in a ``high temperature" limit where the cylinders get short, the geometry contains shrinking circles. We claim that these shrinking circles can be treated like thermal circles in local regions that we call ``hot spots," see figure~\ref{eq:hotspotfigure}. We furthermore conjecture that the effective action of the hot spots gives the singular part of the partition function on our ``genus-2" geometry. (The remaining parts of the geometry are not described by thermal EFT, but contribute non-singular corrections to the partition function at high temperature.) With the ``hot spot" conjecture, we can determine the partition function in the regime where it is dominated by heavy CFT data.

To extract heavy-heavy-heavy OPE coefficients, we must furthermore understand the decomposition of the partition function into a higher-dimensional version of genus-2 (global) conformal blocks. These are interesting special functions that to our knowledge have not previously appeared in the CFT literature. We explore them in section~\ref{sec:conformalblock}, determining their behavior at large $\De$ using the shadow formalism and saddle-point analysis. We then decompose the partition function into ``genus-2" blocks using an appropriate inverse Laplace transform on the moduli space of ``genus-2" conformal structures in higher dimensions. In the end, we obtain a universal formula for average squared heavy-heavy-heavy OPE coefficients in a $d$-dimensional CFT. For example, for three scalar operators with similar dimensions $\De$ in $d=3$, we find
\be
\label{eq:densityofstates}
\rho^{d=3}(\De,0)^3 (C^{d=3}_\textrm{HHH})^2 &\sim \p{\frac{3}{2}}^{6\De} e^{3\sqrt{2\pi f \De}} \x \dots,
\ee
where ``$\dots$" are subleading corrections in $\De$. We give a formula for OPE coefficients of three operators with arbitrary Lorentz representations (with spin held constant as $\De\to \oo$) in arbitrary $d$ below in (\ref{eq:generalope}).

In section~\ref{sec:thermalonept}, we apply similar (but simpler) methods to compute asymptotic thermal 1-point functions of heavy operators. This can be viewed as a particularly simple limit of heavy-heavy-heavy OPE coefficients.

In section~\ref{sec:discuss}, we discuss (\ref{eq:densityofstates}), its generalizations, and some implications and future directions. In holographic theories, we speculate that (\ref{eq:densityofstates}) describes a three-point function of three black holes surrounded by highly entangled matter. Three point functions of three ``pure" black holes are likely atypical from the point of view of (\ref{eq:densityofstates}), but perhaps could be determined from an appropriate holographic calculation. In appendix~\ref{sec:warmup}, we discuss a simple warmup example of the thermal effective action for a two-point function of momentum generators. In appendix~\ref{app:freebosonconstants}, we discuss some aspects of free theories, including novel formulas for nonperturbative corrections to density of states. Other appendices contain detailed calculations to supplement the main text.

\section{The thermal effective action}
\label{sec:thermaleffective}

Consider a $d$-dimensional CFT at finite temperature $T$. Generically, thermal fluctuations cause equilibrium correlators to decay exponentially with distance:
\be
\<\cO(\vec x_1)\cO(\vec x_2)\>_\beta &\sim e^{-|\vec x_1-\vec x_2|/\xi}.
\ee
By dimensional analysis, the correlation length $\xi$ must be inversely proportional to the temperature, $\xi\propto 1/T$. Exponentially-decaying correlators can be expanded in a series in $\de$-functions and their derivatives. (Equivalently, in momentum space, they can be expanded in a power series in momenta.) This expansion is summarized by a local effective action for background fields that we call the {\it thermal effective action}.

It is useful to adopt the geometric perspective on the thermal effective action explained in \cite{Banerjee:2012iz}. Equilibrium thermal correlators are computed by compactifying the Euclidean theory on a thermal circle of length $\beta=1/T$. Generically, when a $d$-dimensional CFT is compactified on a circle, the result is a {\it gapped\/} theory in $(d{-}1)$ dimensions. A rough argument is that compactification of a CFT does not involve tuning any parameters, since all $\beta$ are equivalent by $d$-dimensional scale invariance. Thus, it would be non-generic for the resulting $(d{-}1)$-dimensional theory to be at a critical point. Instead, it will typically have a nonzero mass gap $m_\mathrm{gap}\propto T$, and  a finite correlation length $\xi=1/m_\mathrm{gap}$.\footnote{By contrast, when a theory with an intrinsic scale is compactified, one generally obtains different dynamics at different compactification radii. By tuning $\beta$ it may be possible to reach a critical point. An example is 4d $\SU(2)$ pure Yang-Mills theory, which is expected to possess a critical point in the Ising universality class at a particular temperature, see e.g.\ \cite{Fukushima:2010bq} for a review.} We can think of this gapped theory as the modular transform of the $d$-dimensional CFT.

This argument fails when a symmetry protects gapless modes in the compactified theory, such as in free theories or supersymmetric compactifications (where we twist by $(-1)^F$ around the circle). It could also fail in theories that spontaneously break a continuous symmetry at finite temperature, such as those recently constructed (in fractional spacetime dimensions) in \cite{Chai:2020zgq}. It is currently unknown whether such theories exist in integer spacetime dimensions. See \cite{Chai:2020zgq,Halperin_2018} and references therein for more discussion. In this work, we will focus on theories that are gapped at finite temperature.

An efficient way to capture correlators of the CFT is to couple to classical background fields. For example, stress tensor correlators are captured by coupling to a $d$-dimensional background metric $G_{\mu\nu}$. If the metric possesses a circle isometry, then by a suitable choice of coordinates, we can put it in Kaluza-Klein form
\be
\label{eq:kkgeometry}
G_{\mu\nu} dx^\mu dx^\nu &= g_{ij}(\vec x) dx^i dx^j + e^{2\phi(\vec x)}(d\tau + A_i(\vec x))^2,\qquad \tau\in [0,1),
\ee
where the periodic direction is $x^0=\tau$. The $(d{-}1)$-dimensional fields are a metric $g_{ij}$, a gauge field $A_i$, and a dilaton $\f$. We choose conventions so that $\tau$ has periodicity $1$. Thus, for the thermal compactification $S_\beta^1 \x \R^{d-1}$ with the flat metric, we have $e^{\f}=\beta$. However, it will be interesting in what follows to allow the $(d{-}1)$-dimensional fields $g_{ij},A_i,\f$ to be spatially varying.

By our assumption above, the partition function of the $d$-dimensional CFT on the Kaluza-Klein geometry (\ref{eq:kkgeometry}) becomes the partition function of a gapped $(d{-}1)$-dimensional theory coupled to $(d{-}1)$-dimensional background fields:
\be
Z_\textrm{CFT}[G] &= Z_\textrm{gapped}[g,A,\f].
\ee
The partition function of a trivially gapped QFT at long distances can be expanded in a sum of local counterterms in the background fields. In this case, we have\footnote{A theory that spontaneously breaks a discrete symmetry at finite temperature can display mild violations of (\ref{eq:thermalactionintro}), see \cite{Chai:2020zgq}. In general, if the finite temperature theory is nontrivially gapped, then the thermal effective action must include a nontrivial TQFT. We focus on the trivially gapped case in this work, though most of our results are simple to adapt to a more general thermal TQFT.}
\be
\label{eq:thermalactionintro}
Z_\textrm{CFT}[G] &= Z_\textrm{gapped}[g,A,\f] \sim e^{-S_\textrm{th}[g,A,\f]}.
\ee
The thermal effective action $S_\textrm{th}$ is a sum of local terms in $g_{ij},A_i,\f$ that captures Euclidean correlators at length scales that are large compared to the correlation length $\xi=1/m_\textrm{gap}$ (equivalently, at momenta small compared to $m_\textrm{gap}$).\footnote{Note that the thermal effective action {\it does not\/} in general capture long-distance real time observables, even at small nonzero frequencies $0<\w\ll m_\textrm{gap}$, where dissipation is an important effect.} In (\ref{eq:thermalactionintro}), ``$\sim$" means agreement up to exponential corrections of the form $e^{-L/\xi}$, where $L$ is a characteristic length scale.

In QFT, we usually have the freedom to add arbitrary local counterterms in background fields. This is called a change of ``scheme." In the thermal effective action (\ref{eq:thermalactionintro}), it is important that we are only allowed to add local $d$-dimensional counterterms to the CFT (which enter $S_\textrm{th}$ via dimensional reduction). We are {\it not} allowed to add arbitrary local $d{-}1$-dimensional counterterms. Thus, $S_\textrm{th}$ can contain physical, scheme-independent information.

The thermal effective action is highly constrained by symmetries. Firstly, coordinate-invariance in $d$-dimensions implies that $S_\textrm{th}$ is invariant under $(d{-}1)$-dimensional coordinate transformations, as well as gauge transformations of the KK gauge field $A_i$. For simplicity, in this work we focus on CFT$_d$'s with vanishing gravitational anomaly.  When the gravitational anomaly is non-vanishing (for example in a 2d CFT with $c_L\neq c_R$), the anomaly must be matched by gravitational Chern-Simons terms in the thermal effective action, see \cite{Jensen:2012kj,Jensen:2013rga,Golkar:2015oxw,DiPietro:2014bca,Chang:2019uag} for more details. Such terms could be easily incorporated into the analysis that follows.\footnote{We thank Yifan Wang for discussion on these points.}

Secondly, $S_\textrm{th}$ is constrained by Weyl invariance of the $d$-dimensional theory. Under a Weyl transformation, the CFT partition function changes by
\be
\label{eq:weyltransf}
Z_\textrm{CFT}[e^{2\s} G] &= Z_\textrm{CFT}[G]e^{-S_\textrm{anom}[G,\s]},
\ee
where $S_\textrm{anom}[G,\s]$ is the contribution from the Weyl anomaly. Because $\f$ transforms with a shift $\f\to \f+\s$ under $\tau$-independent Weyl transformations, we can use (\ref{eq:weyltransf}) to completely determine the $\f$-dependence of $Z_\textrm{CFT}[G]$ \cite{Eling:2013bj}. Note that
\be
Z_\textrm{CFT}[G] &= Z_\textrm{CFT}[\hat G] e^{-S_\textrm{anom}[\hat G,\f]},
\ee
where $\hat G\equiv e^{-2\f}G$. Plugging in (\ref{eq:thermalactionintro}), this implies
\be
\label{eq:solveweyltrans}
S_\textrm{th}[g,A,\f] &= S_\textrm{th}[\hat g,A,0] + S_\textrm{anom}[\hat G,\f] \nn\\
&\equiv S[\hat g,A] + S_\textrm{anom}[\hat G,\f].
\ee

In equation (\ref{eq:solveweyltrans}), $S_\textrm{anom}[\hat G,\f]$ plays the role of matching the Weyl anomaly in the thermal effective action. We discuss this contribution later in section~\ref{sec:weyl}. The remaining term $S[\hat g,A]$ is invariant under $\tau$-independent Weyl transformations. It depends only on $A_i$, and the Weyl-invariant ``effective metric"
\be
\hat g_{ij}\equiv e^{-2\f}g_{ij}.
\ee
Note that $\hat g$ transforms as a metric under coordinate-transformations.
Thus, $S[\hat g,A]$ can be organized in a derivative expansion in coordinate invariants built out of $\hat g$ and $A$.
Classifying the terms in $S[\hat g,A]$ is similar to classifying local interactions in Einstein-Maxwell theory (without the freedom to perform field redefinitions). The first few terms are:
\be
\label{eq:theffact}
S[\hat g,A] &= \int d^{d-1} x \sqrt{\hat g}\p{-f + c_1 \hat R + c_2 F^2 + \dots}.
\ee
Here, $\hat R$ is the Riemann curvature built from the metric $\hat g$, and $F_{ij}=\ptl_i A_j - \ptl_j A_i$ is the field strength of the KK gauge boson. Indices are everywhere contracted using $\hat g$, for example
\be
F^2 &= \hat g ^{ik}\hat g^{jl}F_{ij}F_{kl}.
\ee
This ensures that the derivative expansion for $S_\textrm{th}$ becomes an expansion in $\vec p/T$, since $\hat g^{ij}$ contains a factor of $e^{2\f}$, where $e^{-\f}$ is a local temperature. Again, in (\ref{eq:theffact}), we have assumed gravitational anomalies are absent.

The construction of the thermal effective action (\ref{eq:theffact}) closely mimics the construction of the dilaton effective action in theories where scale invariance is spontaneously broken to $d$-dimensional Poincare invariance \cite{Komargodski:2011vj}. The key difference is that here we have an effective action in $(d{-}1)$-dimensions instead of $d$ dimensions.

\subsection{The cosmological constant term}
\label{sec:cosmoconst}

The leading term in the thermal effective action is a cosmological constant $-\int d^d x \sqrt{\hat g} f$ for the effective metric $\hat g$. The coefficient $f$ has at least three important interpretations. 
\begin{enumerate}
\item $-f$ is (the scheme-independent part of) the free energy density of the CFT at finite temperature in flat space.\footnote{Our convention for $f$ is opposite to the one in \cite{Iliesiu:2018fao}, $f_\textrm{here}=-f_\textrm{there}$.} To see why, we place the theory on $\R^{d-1}\x S^1_\b$, where the thermal circle has length $\b$. In this geometry, the effective metric is $\hat g_{ij} = \b^{-2}\de_{ij}$. Only the cosmological constant term in $S_\textrm{th}$ is nonzero, and it contributes
\be
F/T = -\log Z_\textrm{CFT}[\R^{d-1}\x S^1_\b] &= S_\textrm{th} = -\int d^d x f \b^{-(d-1)} = -f T^{d-1} \vol\, \R^{d-1}.
\ee
Thus, the free energy density is $-fT^{d}$. 
\item $f$ is proportional to the thermal one-point function of the stress tensor in flat space $\R^{d-1}$. Lorentz and scale invariance dictate that
\be
\label{eq:stressonept}
\<T^{\mu\nu}(t,\vec x)\>_\b &= b_T T^d \p{\de^\mu_0 \de^\nu_0 - \frac 1 d \de^{\mu\nu}},
\ee
for some dimensionless coefficient $b_T$. Meanwhile, the energy density can be computed from the derivative of the partition function\footnote{Note that we are using the conventions of Euclidean field theory, where the generator of time translations includes a minus sign $H=-\int d\vec x T^{00}$. The minus sign can be understood via Wick rotation from Lorentzian signature $T^{00}_E=(i)^2 T^{00}_L$.}
\be
\label{eq:energydens}
-\<T^{00}(0,\vec x)\>_\b &= -\frac{1}{\vol\,\R^{d-1}}\pdr{}{\b} \log Z_\textrm{CFT}[\R^{d-1}\x S^1_\b] = (d-1)f T^d.
\ee
Equating (\ref{eq:stressonept}) and (\ref{eq:energydens}), we find $b_T=-d f$. In particular, positivity and extensivity of the energy density  on $\R^{d-1}$ implies that $f$ is positive:
\be
\label{eq:negativityoff}
f &> 0.
\ee
\item $-f$ is the Casimir energy density of the CFT compactified on a circle. To see why, we choose $x^1$ as a time direction. The Hamiltonian density for the compactified theory is then
\be
\label{eq:casenergy}
\frac{E_\mathrm{cas}}{\vol(S^1_\beta \x \R^{d-2})} &= -\<T^{11}\>_\beta = -f T^d.
\ee
In particular, (\ref{eq:negativityoff}) gives a simple proof that the Casimir energy of a CFT compactified on a circle is always negative. Ultimately, this is a consequence of tracelessness of the stress tensor. The components $T^{00}$ compute the energy in the thermal ensemble (which is positive), while the components $T^{11}$ compute the energy in the compactified theory. The two have opposite signs by tracelessness of $T^{\mu\nu}$. This proof applies, for example, to electromagnetism.
\end{enumerate}

Thus, the coefficient $f$ is a simple and important observable of the CFT. We will see that it controls a huge amount of the physics at finite temperature. As a simple warmup example, in appendix~\ref{sec:warmup}, we show how $f$ determines a (regularized) two-point function of momentum operators at finite temperature. This result can be understood both from ``bootstrap" arguments using properties of stress tensor correlators, and from the thermal effective action.

\subsection{Weyl anomaly terms}
\label{sec:weyl}

The Weyl anomaly terms $S_\textrm{anom}[\hat G,\f]$ in the thermal effective action were given in \cite{Eling:2013bj}. Let us write them down in detail. Such terms are of course absent when $d$ is odd, so we focus on even $d$ in this subsection. 

As a review, the infinitesimal form of the Weyl anomaly in $d$ dimensions is
\be
\label{eq:sigvariation}
\de_\s(-\log Z_\textrm{CFT}[G]) &= \int d^d x \sqrt{G} \s \cA[G],
\ee
where $\de_\s$ is defined by rescaling the metric $G\to (1+2\s) G$ with $\s$ infinitesimal. Here, $\cA[G]$ is a local functional of $G$ such that (\ref{eq:sigvariation}) solves the Wess-Zumino consistency condition $[\de_{\s_1},\de_{\s_2}]\log Z=0$. The infinitesimal Weyl anomaly can be integrated by considering a family of metrics $e^{2t\s}G$ where $t\in [0,1]$, using (\ref{eq:sigvariation}) to write a differential equation in $t$, and solving the differential equation. The result is the finite Weyl transformation rule (\ref{eq:weyltransf}), where $S_\textrm{anom}$ is given by \cite{Schwimmer:2010za}
\be
\label{eq:sanomresult}
S_\textrm{anom}[G,\s] &= \int_0^1 dt \int d^d x \,\sqrt{\det(e^{2t\s}G)}\, \s \cA[e^{2t\s}G].
\ee

The general solution of the Wess-Zumino consistency condition in $d$-dimensions is \cite{LBonora1986,Boulanger:2007st}:
\be
\label{eq:wzsolution}
\int d^d x \sqrt{G} \s \cA[G] &= \frac{1}{(4\pi)^{d/2}}\int d^d x \sqrt{G}\, \s\p{(-1)^{d/2}a_d E_d - {\textstyle\sum_k} c_{dk} I^{(d)}_k} + \de_\s S_\textrm{ct}.
\ee
Here, $E_d$ is the Euler density,\footnote{In 2d, we have $a_2=c/6$, and in 4d we write $a_4=a$.} and $\sqrt{G} I^{(d)}_k$ are local Weyl-invariants of $G$. For example, in 4d there is one such Weyl-invariant, given by the square of the Weyl tensor:
\be
I^{(4)}_1 &= C^2 = C_{\mu\nu\rho\s}C^{\mu\nu\rho\s} \qquad (d=4).
\ee
In 6d there are three such Weyl invariants $I^{(6)}_{k=1,2,3}$, and in general the number grows with $d$. The factor $1/(4\pi)^{d/2}$ in (\ref{eq:wzsolution}) is a convention.

The remaining terms $\de_\s S_\textrm{ct}$ in (\ref{eq:wzsolution}) are Weyl variations of local counterterms. For instance, in 4d we can have 
\be
\label{eq:schemedeppart}
S_\textrm{ct} &=-\frac{b}{12(4\pi)^2}\int d^d x \sqrt{G} R^2 \qquad (d=4),
\ee
which leads to a contribution $\de_\s R^2 \sim b\Box R$ in the Weyl anomaly. We sometimes refer to $\de_\s S_\textrm{ct}$ as ``$b$-type" terms. Such terms trivially obey the Wess-Zumino consistency condition. They are scheme-dependent because they can be shifted by adding local counterterms to the action of the CFT. We comment more on this below in section~\ref{sec:casimir}.

Plugging these results into (\ref{eq:sanomresult}), we find $S_\textrm{anom}$ for example in $d=2$ and $d=4$:
\be
S^\textrm{2d}_\textrm{anom}[G,\s] &= -\frac{c}{24\pi} \int d^2 x \sqrt{G}(\s R + (\ptl \s)^2) \nn\\
S^\textrm{4d}_\textrm{anom}[G,\s] &= \frac{a}{(4\pi)^2} \int d^4 x \sqrt{G} \p{\s E_4 - 4\ptl_\mu \s \ptl_\nu \s(R^{\mu\nu}-\tfrac 1 2 G^{\mu\nu} R)-4 (\ptl \s)^2 \Box \s - 2 (\ptl \s)^4}\nn\\
&\quad - \frac{c}{(4\pi)^2}\int d^4 x \sqrt G \s C^2 + S_\textrm{ct}[e^{2\s}G] - S_\textrm{ct}[G].
\ee
Note that the scheme-dependent part of the Weyl anomaly $\de_\s S_\textrm{ct}$ integrates trivially to give $S_\textrm{ct}[e^{2\s}G] - S_\textrm{ct}[G]$. The Weyl-invariant terms are also simple to integrate because the integrand (\ref{eq:sanomresult}) is $t$-independent for those terms.

Putting everything together, the Weyl-anomaly contribution to the thermal effective action is
\be
\label{eq:sanomthermal}
S_\textrm{anom}[\hat G,\f] &= S_\textrm{Euler}-\frac{1}{(4\pi)^{d/2}}\sum_k\,c_{dk}\int d^{d-1} x \sqrt {\hat g} \f\, \mathrm{DR}[I_k^{(d)}[\hat G]] + \mathrm{DR}[S_\textrm{ct}[G]]-\mathrm{DR}[S_\textrm{ct}[\hat G]],
\ee
where
\be
\label{eq:seuler}
S_\textrm{Euler} &= \frac{(-1)^{d/2}a_d}{(4\pi)^{d/2}} \int_0^1 dt \int d^{d-1} x\, e^{dt\f}\sqrt{\hat g}\, \f\,\mathrm{DR}[ E_d[e^{2t\f}\hat G]].
\ee
Here, the dimensional reduction operation $\textrm{DR}[\cdots]$ means evaluating in the KK metric (\ref{eq:kkgeometry}) and integrating over $\tau$. 

Note that the $I_k^{(d)}$ terms in (\ref{eq:sanomthermal}) are linear in $\f$, and thus can lead to temperature dependence of the form $\log(\beta/\beta_0)$ in certain geometries. Here, we note that the coefficient of $\log \b$ is a genuine prediction of $S_\textrm{th}$, but the scale $\b_0$ is scheme-dependent. The reason is that $\b_0$ can be shifted by adding local Weyl-invariant counterterms to the action of the CFT:
\be
S_\textrm{CFT} &\to S_\textrm{CFT} + \frac{1}{(4\pi)^{d/2}}\int d^d x \sqrt G\sum_k r_k I_k^{(d)}.
\ee
This ambiguity shifts the coefficients of $\mathrm{DR}[I_k^{(d)}]$ in the thermal effective action:
\be
\label{eq:ambiguityincoefficients}
-c_{dk} \f &\to r_k - c_{dk} \f,
\ee
and consequently shifts $\b_0$. This ambiguity will not play a further role in this work, since we will always consider CFTs in conformally-flat\footnote{We call a manifold ``conformally-flat" if in a neighborhood of each point, the metric is Weyl-equivalent to a flat metric. This is sometimes called ``locally conformally-flat." A 3-manifold is conformally-flat if and only if the Cotton tensor vanishes, and a $d$-manifold with $d\geq 4$ is conformally-flat if and only if the Weyl tensor vanishes.} geometries where $I_k^{(d)}$ vanishes.

\subsection{2 dimensions}

In 2 dimensions, a very nice thing happens. The cosmological constant term is the {\it only\/} local gauge-invariant combination of $\hat g$ and $A_i$ that we can write down. Furthermore, there are no nontrivial local Weyl invariants. Thus, the Weyl-invariant part of the thermal effective action truncates to a single term!
\be
\label{eq:twodcase}
S[\hat g,A] &= -\int d^1 x \sqrt{\hat g} f \qquad (d=2).
\ee
 The action (\ref{eq:twodcase})  describes equilibrium thermal physics in 2d to all perturbative orders in $1/T$.
Using the connection between $f$ and the Casimir energy (\ref{eq:casenergy}), we find 
\be
f=\frac{2\pi c}{12},
\label{eq:fcrelation2d}
\ee
where $c$ is the central charge.

This result is not a surprise. A 2d CFT at high temperature can be described by performing a modular transformation, reinterpreting the thermal circle as a spatial circle. The states propagating in the modular-transformed theory have energies $E_i=\frac{2\pi}{\beta}(\De_i - \frac{c}{12})$, where $\De_i$ are scaling dimensions in the CFT. The effective action (\ref{eq:twodcase}) simply captures the contribution of the ground state in the modular transformed theory, with energy $E_0=-\frac{2\pi}{\beta}\frac c {12}$. The energy gap to the next state is the ``mass gap" of the thermal theory 
\be
m_\textrm{gap}&=E_1-E_0 = \frac{2\pi}{\beta} \De_1 \qquad (d=2).
\ee
States with energies at or above the mass gap $E_i-E_0\geq m_\textrm{gap}$ contribute nonperturbative corrections in $\beta$ of the form $e^{-2\pi\De_i/\beta}$, which are not captured by $S_\textrm{th}$.

\section{The density of high-dimension states}
\label{sec:density}

The spectrum of a $d$-dimensional CFT is captured by the partition function on $S_\b^1 \x S^{d-1}$. In this section, we compute this partition function using the thermal effective action, and decompose the result into conformal characters to extract the density of high dimension states. We will recover the leading-order formulas from \cite{Bhattacharyya:2007vs,Shaghoulian:2015lcn}, and also discuss subleading corrections. The precise expression for the partition function involves the Casimir energy of the CFT on $S^{d-1}$. To start, we review the Casimir energy and discuss some details of its relation to the thermal effective action.

\subsection{The Casimir energy on $S^{d-1}$}
\label{sec:casimir}

The partition function on $S_\b^1 \x S^{d-1}$ is a sum over states on $S^{d-1}$ weighted by Boltzmann factors $e^{-\beta E_i}$. By the state-operator correspondence, states on $S^{d-1}$ are in one-to-one correspondence with local CFT operators $\cO_i$. In even dimensions the energy $E_i$ of the state $|\cO_i\>$ is equal to the dimension $\De_i$ plus a contribution from the Casimir energy on the sphere:
\be
E_i &= \De_i + E_0,
\ee
where $\De_i$ is the scaling dimension of $\cO_i$.
For example, in 2d, the Casimir energy is $E_0=-\frac c {12}$ (in units where the $S^{d-1}$ has radius $1$). The Casimir energy $E_0$ will play an important role in higher dimensions as well, so let us recall how to derive it. 

We follow the discussion of \cite{Assel:2015nca}. Let $W[G]\equiv -\log Z_\textrm{CFT}[G]$, so that the Weyl anomaly is
$W[e^{2\s} G]-W[G] = S_\textrm{anom}[G,\s]$.
To compute the stress tensor on the cylinder $\R \x S^{d-1}$, we consider the Weyl rescaling from the plane to the cylinder
\be
dr^2 + r^2 d\Omega_{d-1}^2 &\to \frac{dr^2}{r^2} + d\Omega_{d-1}^2 = e^{-2\log r} \de_{\mu\nu} dx^\mu dx^\nu,
\ee
which corresponds to $\s=-\log r$. Plugging this into $S_\textrm{anom}[G,\s]$, we obtain the partition function on the cylinder as a function of the partition function on the plane. Taking a derivative with respect to $G_{\mu\nu}$ and using that the one-point function $\<T^{\mu\nu}\>$ on the plane vanishes, we obtain $\<T^{\mu\nu}\>$ on the cylinder, from which we can read off the Casimir energy.

There is a small shortcut that will be useful in what follows. We can simply compute $W[G]$ on the infinite cylinder $\R\x S^{d-1}$ using the Weyl anomaly. This has an infinite part of the form $E_0 \vol\, \R$, from which we can read off the Casimir energy $E_0$. As an example, in 2d, we have
\be
W[e^{-2\log r} \de] - W[\de] &= -\frac{c}{24\pi} \int r\, dr\, d\th \frac{1}{r^2} = -\frac{c}{12} \int d\tau\qquad (d=2),
\ee
where we defined $\tau=\log r$.
This gives the expected result $E_0=-\frac c {12}$. In 4d, we find
\be
W[e^{-2\log r} \de] - W[\de] &= \frac{a\, \vol\, S^3}{(4\pi)^2} \int r^3 dr \p{\frac{6}{r^4}} + S_\textrm{ct}[e^{2\s}\de] 
\nn\\
&= \p{\frac{3 a}{4} - \frac{3 b}{8}} \int d\tau,
\ee
where we used the form of $S_\textrm{ct}$ in (\ref{eq:schemedeppart}), together with the fact that the curvature of $S^3$ is $R=6$. Thus, the Casimir energy in 4d is
\be
\label{eq:fourdcasimir}
E_0 &= \frac{3 a}{4} - \frac{3 b}{8}\qquad (d=4).
\ee

\subsubsection{Choice of scheme}
\label{sec:cancelscheme}

As noted in \cite{Assel:2015nca}, the 4d Casimir energy (\ref{eq:fourdcasimir}) is scheme-dependent --- it can be shifted by redefining the local counterterm coefficient $b$. Similar statements hold in any even $d\geq 4$. However, CFT data is scheme-independent. To study it, we are free to choose whatever scheme is most convenient.

In what follows, we will choose a scheme where $S_\textrm{ct}=0$, so that $b$-type terms are absent from both the Casimir energy and the Weyl anomaly. To define such a scheme in practice, one must choose a regulator, compute the Weyl anomaly with that regulator, and then add appropriate local counterterms to cancel the $b$-type terms.\footnote{This requires a sufficiently ``flexible" regulator that we can compute the Weyl anomaly. For example it is not obvious how to do this with a lattice regulator. Furthermore, such a scheme choice might clash with other symmetries, e.g.\ SUSY \cite{Assel:2015nca}. We thank Zohar Komargodski for discussion on these points}

In this $S_\textrm{ct}=0$ scheme, the $b$-type terms $\mathrm{DR}[S_\textrm{ct}]$ are not present in the thermal effective action, and the partition function on $S_\b^1 \x S^{d-1}$ is given by
\be
\label{eq:theresultfirst}
\Tr\left[e^{-\b(D+\varepsilon_0)}\right] &\sim e^{-S_\textrm{th}} = e^{-S[\hat g,A]-S_\textrm{Euler}},
\ee
where $\varepsilon_0$ is the {\it $a_d$-type contribution to the Casimir energy alone}. Here, ``$\sim$" means equality up to exponentially suppressed corrections in $1/\b$. $S_\textrm{Euler}$ is given in (\ref{eq:seuler}). We have used that $S^1_\b\x S^{d-1}$ is conformally-flat to drop the Weyl-invariants $I_k^{(d)}$. In appendix~\ref{app:scheme}, we describe how (\ref{eq:theresultfirst}) comes about in a general scheme.

The value of $\varepsilon_0$ was computed in general $d$ in \cite{Herzog:2013ed}:
\be
\label{eq:unambiguouscasimirenergy}
\varepsilon_0 &\equiv \textrm{$a_d$-type contribution to Casimir energy on $S^{d-1}$}
\nn\\
 &= \frac{\sqrt{\pi}}{\G(\tfrac{1-d}{2})}a_d = \begin{cases} \frac{(d-1)!!}{(-2)^{d/2}} a_d, & \textrm{$d$ even}, \\
0 & \textrm{$d$ odd},
\end{cases}
\ee
where $(d-1)!!=(d-1)\cdots 3 \cdot 1$ for even $d$. Note that in 2d, we have $a_2=c/6$. 

\subsection{The partition function from the thermal effective action}
\label{sec:partitionthermaleff}

Let us now study the density of CFT operators with various dimensions and spins. We can obtain this from the partition function of the CFT on $S^1_\b \x S^{d-1}$ with a spin fugacity:
\be
Z(\b,\vec{\Omega})=\text{Tr}\left[ e^{-\b (D+\varepsilon_0)+i\beta\vec{\Omega} \cdot \vec{M}}\right] \sim e^{-S[\hat g,A]-S_\textrm{Euler}}.
\label{eq:pftn}
\ee
Here, $D$ is the dilatation operator, $\vec{M}$ are the $n \coloneqq \lfloor \tfrac d 2 \rfloor$ generators of the Cartan subalgebra of the rotation group $\SO(d)$, and $\vec{\Omega}$ are spin fugacities.

 Geometrically, (\ref{eq:pftn}) is computed by a path integral on $S_\b^1 \x S^{d-1}$, with a twist by $\beta \vec\Omega$ as we move around the thermal circle. The metric is 
\be
    ds_\text{cylinder}^2 = \beta^2 d\tau^2 + ds_\text{sphere}^2,
\ee
where $\tau\in [0,1]$ is a coordinate on $S^1$ and $ds_\text{sphere}^2$ is the metric on $S^{d-1}$. 

To write down the metric on the sphere, let us choose coordinates that make Cartan rotations manifest.
The Cartan generators are rotations in $n$  orthogonal 2-planes. We use radius-angle coordinates $\{ r_a,\theta_a\}$ for each plane $(a=1,\dots,n)$, so the Cartan generators are simply $i\partial_{\theta_a}$.
If $d$ is odd, we have an extra axis and we use the coordinate $r_{n+1}$ for it. The radii satisfy the constraint $\sum_{a=1}^{n+\epsilon}  r_a^2=1$, where $\epsilon=0$ for even $d$ and $\epsilon = 1$ for odd $d$.
In these coordinates, the metric of the sphere is 
\be
    ds_\text{sphere}^2
    =               
    \sum_{a=1}^{n+\epsilon}d r_a^2+\sum_{a=1}^n r_a^2 d\theta_a^2,
    \label{eq:themetricintwisted}
\ee
where we have the constraint $\sum_{a=1}^{n+\epsilon}  r_a d r_a = 0$. 

In the twisted geometry that computes $Z(\beta,\vec\Omega)$, we identify the points
\be
\label{eq:ourtwist}
(\tau,\theta_a) &\sim (\tau+1,\theta_a - \beta\Omega_a).
\ee
Because of this identification, shifts in $\tau$ with fixed $\theta_a$ are not periodic isometries, and thus the metric (\ref{eq:themetricintwisted}) is not in Kaluza-Klein (KK) form. To place it in KK form, we redefine $\theta_a\to \theta_a+\beta\Omega_a\tau$, which removes the twist (\ref{eq:ourtwist}), and produces the new metric 
\begin{align}
    ds^2 &= \beta^2d\tau^2+\sum_{a=1}^{n+\epsilon}d r_a^2+\sum_{a=1}^n r_a^2 (d\theta_a+\beta\Omega_a d\tau)^2\nn\\
    &= \beta^2\left(1+\sum_{a=1}^n r_a^2\Omega_a^2\right)\left(d\tau+\frac 1 \beta \sum_{a=1}^n\frac{ r_a^2\Omega_a}{1+\sum_{b=1}^n r_b^2\Omega_b^2}d\theta_a\right)^2\nn\\
    &\quad +\sum_{a=1}^{n+\epsilon}d r_a^2 + \sum_{a,b=1}^n \left( r_a r_b\delta_{ab}-\frac{ r_a^2 r_b^2\Omega_a\Omega_b}{1+\sum_{c=1}^n r_c^2\Omega_c^2}\right)d\theta_ad\theta_b.
    \label{eq:untwistedmetric}
\end{align}

Comparing \eqref{eq:untwistedmetric} with \eqref{eq:kkgeometry}, we identify a metric $g_{ij}$, a KK gauge field $A_i$, and a dilaton $\phi$ given by
\begin{align}
    e^{2\phi} &= \beta^2\left(1+\sum_{a=1}^{n} r_a^2\Omega_a^2\right),
    \nn\\
    A &= \frac 1 \beta \sum_{a=1}^{n}\frac{ r_a^2\Omega_a}{1+\sum_{b=1}^n r_b^2\Omega_b^2}d\theta_a,
    \nn\\
    g &= \sum_{a=1}^{n+\epsilon}d r_a^2+\sum_{a,b=1}^{n}\left( r_a r_b\delta_{ab}-\frac{ r_a^2 r_b^2\Omega_a\Omega_b}{1+\sum_{c=1}^n r_c^2\Omega_c^2}\right)d\theta_a d\theta_b.
    \label{eq:phiag}
\end{align}
The effective metric $\hat g = e^{-2\f}g$, together with $A$, then appears in the thermal effective action $S[\hat g,A]$.

Explicitly, the cosmological constant term in the effective Lagrangian is
\begin{align}
    \sqrt{\hat{g}}
    &=
    T^{d-1}\left(1+\sum_{i=1}^nr_i^2 \Omega_i^2 \right)^{-\frac{d}{2}}\prod_{i=1}^n  r_i,
\end{align}
while the Maxwell and Einstein densities are
\begin{align}
\label{eq:highercurvatures}
    F^2
    &=
    8T^{-2}
    \left(
    \sum_{i=1}^n\Omega_i^2-\frac{\sum_{i=1}^n r_i^2\Omega_i^2(1+\Omega_i^2)}{1+\sum_{i=1}^n r_i^2\Omega_i^2}
    \right),\nn\\
    \hat{R} 
    &=
    T^{-2}\left(d(d+1)\frac{1-\sum_{i=1}^n r_i^2\Omega_i^4}{1+\sum_{i=1}^n r_i^2\Omega_i^2}
    -2(2d-1)\p{1-\sum_{i=1}^n\Omega_i^2}\right).
\end{align}
As expected, at high temperature, the cosmological constant term gives the leading contribution, while the Einstein and Maxwell terms are subleading by $1/T^2$, since they are two-derivative terms.
Finally, the thermal effective action on our geometry is given by integrating over $S^{d-1}$: 
\be
S[\hat g,A]&=\int_{S^{d-1}} \sqrt{\hat{g}}\left(-f+c_1\hat{R}+c_2F^2+\dots\right)\nn\\ 
&=
\frac{\vol\, S^{d-1}}{\prod_{i=1}^n(1+\Omega_i^2)}
    \left[-fT^{d-1}
    +(d-2)\left((d-1)c_1+(2c_1+\tfrac{8}{d}c_2)\sum_{i=1}^n\Omega_i^2\right)T^{d-3}+\dots\right],
\label{eq:Zformula}
\ee
where $\vol\, S^{d-1}=\frac{2\pi^{d/2}}{\G(d/2)}$ is the volume of the $d{-}1$-sphere.\footnote{To evaluate (\ref{eq:Zformula}), we use the following explicit coordinates on $S^{d-1}$. For even $d=2n$, the integral is
\begin{equation*}\int_0^1 dr_1 \ldots \int_0^1 dr_n \int_0^{2\pi} d\theta_1 \ldots \int_0^{2\pi} d\theta_n \delta\left(\sqrt{ r_1^2+\dots+ r_n^2}-1\right) \sqrt{\hat g}\left(-f + c_1 \hat R + c_2 F^2+\ldots\right),\end{equation*} and for odd $d=2n+1$, the integral is \begin{equation*}\int_0^1 dr_1 \ldots \int_0^1 dr_{n} \int_{-1}^1 dr_{n+1}\int_0^{2\pi} d\theta_1 \ldots \int_0^{2\pi} d\theta_{n}\delta\left(\sqrt{ r_1^2+\dots+ r_{n+1}^2}-1\right) \sqrt{\hat g}\left(-f + c_1 \hat R + c_2 F^2+\ldots\right).\end{equation*}
To compute either case, we used the Feynman parametrization identity:
\begin{equation}
    \frac{1}{A_1^{\alpha_1}\dots A_k^{\alpha_k}}=\frac{\Gamma(\alpha_1+\dots+\alpha_k)}{\Gamma(\alpha_1)\dots\Gamma(\alpha_k)}\int_0^1 du_1\dots \int_0^1 du_k
   \frac{\delta(u_1+\dots+u_k-1)u_1^{\alpha_1-1}\dots u_k^{\alpha_k-1}}{(u_1A_1+\dots+u_kA_k)^{\alpha_1+\dots+\alpha_k}}.
\end{equation}
In order for (\ref{eq:Zformula}) to be valid, we require that the convex hull of $1+\Omega_i^2$ not contain $0$. Otherwise, the integral diverges (as expected from the unitarity bound of the CFT).
} 

Note that the cosmological constant predicts the entire leading term in (\ref{eq:Zformula}) as a detailed function of the spin fugacities $\Omega_i$. This leading term was first written down in \cite{Bhattacharyya:2007vs}.\footnote{Our $\vec \Omega$ is related to the one in \cite{Bhattacharyya:2007vs} by $i\vec \Omega_\textrm{here} = \vec\Omega_\textrm{there}$.}
Using the thermal effective action, it is straightforward to incorporate more terms in $S_\textrm{th}$ and characterize the form of subleading corrections. Overall, we obtain a formula for the thermal partition function of a CFT with a spin fugacity in a systematic expansion in $1/T$. 
          
The leading term in (\ref{eq:Zformula}) has poles at $\Omega_i = \pm i$. These poles are related to the unitarity bound because states close to the unitarity bound are not penalized by Boltzmann factors $e^{-\beta(\Delta - i\Omega J)}$ in this regime of angular fugacities. In our calculation, the poles come from locations on the sphere where $r_i=1$ (and the remaining $r_j$ vanish). Despite additional poles in the expressions (\ref{eq:highercurvatures}), the higher-order corrections $\hat R$ and $F^2$ do {\it not} lead to further enhanced poles in the partition function.  The reason is that $\hat R$ and $F^2$ are actually finite at $r_i=1$ when $\Omega_i=\pm i$. We expect that this remains true for all higher-order corrections in the thermal effective action, so that the pole structure of (\ref{eq:Zformula}) holds to arbitrary (perturbative) order in $1/T$. Specifically, we expect that the coefficient of $T^{d-2k-1}$ is a degree-$2k$ polynomial in the $\Omega_i$, times an overall factor $1/\prod_{i=1}^n (1+\Omega_i^2)$.

Finally, let us compute $S_\textrm{Euler}$ by plugging (\ref{eq:phiag}) into (\ref{eq:seuler}). In $d=2,4,6$, we find
\be
S^{d=2}_\textrm{Euler}&=0,
\nn\\
S^{d=4}_\textrm{Euler}
&= -\frac{a_4\beta}{12}  \frac{(\Omega _1^2-\Omega _2^2)^2}{(1+\Omega _1^2) (1+\Omega _2^2)},
\nn\\
S^{d=6}_\textrm{Euler}
&=
-\frac{3 a_6 \beta}{80}
\frac{1}{(1+\Omega _1^2) (1+\Omega _2^2) (1+\Omega _3^2)} \nn\\
&\quad \x \Big[
\Omega _1^6+\Omega _2^6+\Omega _3^6
 -4 (\Omega _1^4(\Omega_2^2+\Omega _3^2)+\Omega _2^4 (\Omega _3^2+\Omega _1^2)+\Omega _3^4 (\Omega _1^2+\Omega _2^2))
 +21 \Omega _2^2 \Omega _3^2 \Omega _1^2
 \nn\\
&\quad\quad+5 (\Omega _1^2 \Omega _2^2+\Omega _2^2 \Omega _3^2 +\Omega _3^2\Omega _1^2 )-5 (\Omega _1^4+\Omega _2^4+\Omega _3^4)
\Big].
\ee
In all these cases, $S_\textrm{Euler}$ has the same functional form as expected from the $O(T^{-1})$ terms in the Weyl-invariant part of the thermal effective action $S[\hat g,A]$ --- namely a polynomial of degree $d$ in the $\Omega_i$'s times $\b/\prod_{i=1}^n (1+\Omega_i^2)$. Thus, the effects of $S_\textrm{Euler}$ cannot be distinguished from $S[\hat g,A]$ in the CFT partition function on $S^1\x S^{d-1}$. It would be interesting to try to distinguish these terms in some example theories, perhaps by studying stress-tensor correlators in thermal flat space.

\subsection{Leading asymptotic formula}
\label{sec:laplacetransf}

From (\ref{eq:Zformula}) we can extract the high energy density of states for any CFT\footnote{A previous version of this paper had minor typos in the density of states that we have corrected. We thank Sasha Diatlyk and Yifan Wang for pointing them out to us.}.  Let us first consider the leading term of the high-temperature partition function:
\begin{equation}
    \log Z(T, \Omega_i) = \frac{\vol\, S^{d-1} f T^{d-1}}{\prod_{i=1}^{n} (1+\Omega_i^2)} + O(T^{d-3}),
    \label{eq:freeenergy}
\end{equation}
where $n = \left \lfloor{d/2}\right \rfloor$.  
To extract the density of states, we perform an inverse Laplace transform on the partition function, which can be done by saddle point approximation. Before we do the general $d$ case however, let us first do $d=2$ and $d=3$ explicitly.

We pause to note that we can compute either the asymptotic density of all operators of the CFT, including both primary and descendent operators, or the asymptotic density of only the conformal primary operators. We compute the latter by decomposing the partition function into the conformal characters. In $d$ dimensions the characters are given by\footnote{The characters (\ref{eq:charactersgend}) are for long representations of the conformal group. For special values of $\Delta, J_i$ (e.g.\ states at the unitarity bound), the representation may be shortened and the expression for the character will be modified. However, since we are interested in reading off the density of primary operators at large dimension, short representations will not play a role.}
\begin{equation}
\chi_{\Delta, J_i}(\beta, \Omega_i)  = \begin{cases}
\frac{e^{-\beta\Delta}\chi_{J_i}(\beta\Omega_i)}{\prod_{i=1}^n (1-e^{-\beta(1 + i \Omega_i)})(1-e^{-\beta(1 - i \Omega_i)})}, & d~\text{even},\\
\frac{e^{-\beta\Delta}\chi_{J_i}(\beta\Omega_i)}{(1-e^{-\beta})\prod_{i=1}^n (1-e^{-\beta(1 + i \Omega_i)})(1-e^{-\beta(1 - i \Omega_i)})}, & d~\text{odd},
\end{cases}
\label{eq:charactersgend}
\end{equation}
where $\chi_{J_i}(\theta_i)$ is the character of the $\SO(d)$ representation $\l=(J_1,\dots,J_n)$. The partition function is a sum over characters, with an additional inclusion of the Casimir energy $\varepsilon_0$ defined in (\ref{eq:unambiguouscasimirenergy}):
\be
Z(T,\Omega_i) &= e^{-\beta \varepsilon_0}\sum_{\De_i,J_i}\chi_{\De_i,J_i}(\beta,\Omega_i).
\ee

\subsubsection{$d=2$}

From (\ref{eq:freeenergy}), our high-temperature expression for the partition function in $d=2$ is
\begin{align}
Z(T,\Omega)_{d=2} &\approx \exp\left(\frac{2\pi f T}{1+\Omega^2}\right) \nonumber \\
&= \exp\left(\frac{4\pi^2 c T}{12(1+\Omega^2)}\right),
\label{eq:d2approx}
\end{align}
where we used the relation between $f$ and $c$ for 2d CFTs (\ref{eq:fcrelation2d}). We would like to take the inverse Laplace transform to extract the high-energy density of states. This calculation is precisely Cardy's calculation for the high-energy density of states \cite{Cardy:1986ie}, but we include it for completeness. It is convenient to first change variables:
\begin{align}
    \beta_L \coloneqq \frac1T + \frac{i\Omega}T,~~~~~\beta_R &\coloneqq \frac1T - \frac{i\Omega}T,
    \label{eq:changeofvariablesbeta}
\end{align}
which gives
\be
Z(\beta_L, \beta_R)_{d=2} \coloneqq \text{Tr}\left(e^{-\beta_L(\frac{\Delta - J}2-\frac{c}{24})}e^{-\beta_R(\frac{\Delta + J}2-\frac{c}{24})}\right) \approx e^{\frac{2\pi^2 c}{12} \left(\frac1{\beta_L}+\frac1{\beta_R}\right)},
\label{eq:z2dlowt}
\ee
where we include the Casimir shift described in Sec \ref{sec:cancelscheme}.
Taking the inverse Laplace transform then gives the following integral:
\begin{equation}
\rho^{\text{states}}_{d=2}(\Delta,J) \sim \frac12\left[\frac1{2\pi i}\int_{\gamma-i\infty}^{\gamma+i\infty} d\beta_L e^{\frac{2\pi^2 c}{12\beta_L} + \beta_L\left(\frac{\Delta-J}2-\frac{c}{24}\right)}\right]\left[\frac1{2\pi i}\int_{\gamma-i\infty}^{\gamma+i\infty} d\beta_R e^{\frac{2\pi^2 c}{12\beta_R} + \beta_R\left(\frac{\Delta+J}2-\frac{c}{24}\right)}\right].     \label{eq:d2saddleintegraltodo}
\end{equation}
Although we can do this integral by saddle, it actually can be done exactly. The tree level piece is
\begin{equation}
    \rho^{\text{states}}_{d=2}(\Delta, J) \sim \exp\left[\sqrt{\frac{2c}3}\pi \left(\sqrt{\frac{\Delta+J}2-\frac{c}{24}} + \sqrt{\frac{\Delta-J}2-\frac{c}{24}}\right)\right].
    \label{eq:finald2}
\end{equation}
We see this is none other than the Cardy formula \cite{Cardy:1986ie}.
If we do the integrals in (\ref{eq:d2saddleintegraltodo}) exactly\footnote{Strictly speaking the integral in (\ref{eq:d2saddleintegraltodo}) diverges. The precise statement is \begin{equation} \frac{1}{2\pi i} \int_{\gamma-i\infty}^{\gamma+i\infty} d\beta e^{\beta \Delta} \left(e^{\frac{\pi f}{\beta}}-1\right) = \sqrt{\frac{\pi f}{\Delta}} I_1\left(\sqrt{4\pi f \Delta}\right).\end{equation} This leads to an additional factor of $\delta(\Delta)$ in the inverse Laplace transform. However, since we are using this method to read off the large energy density of states, it does not affect our final expression (\ref{eq:cardylog}).}, we get
\begin{align}
&\rho^{\text{states}}_{d=2}(\Delta, J) 
\nn\\
&\sim \frac{\pi^2 c}{6\sqrt{(\Delta + J - \frac{c}{12})(\Delta - J-\frac{c}{12})}} I_1\left(\sqrt{\frac{2c}3}\pi\sqrt{\frac{\Delta + J}2-\frac{c}{24}}\right) I_1\left(\sqrt{\frac{2c}3}\pi\sqrt{\frac{\Delta - J}2-\frac{c}{24}}\right) \nonumber \\
&= \frac{\sqrt{c}}{\sqrt{48}(\Delta+J-\frac{c}{12})^{3/4}(\Delta-J-\frac{c}{12})^{3/4}}\exp\left[\sqrt{\frac{2c}3}\pi\left(\sqrt{\frac{\Delta+J}2-\frac{c}{24}} + \sqrt{\frac{\Delta-J}2-\frac{c}{24}}\right)\right]\nonumber \\ &~~~~~~~~~~~\times\left(1 + O(\Delta^{-1/2})\right),
\label{eq:cardylog}
\end{align}
where $I_1$ is a modified Bessel function of the first kind.
The expression (\ref{eq:cardylog}) indeed gives the known logarithmic corrections to Cardy's formula \cite{Carlip:2000nv}.

So far, (\ref{eq:cardylog}) is counting the density of all states rather than the density of global or Virasoro primaries. A more natural object from the CFT perspective may be to count the density of primary operators. In order to generalize more easily to higher dimensions, we will now compute the asymptotic density of global (not Virasoro) primary operators. The calculation is almost identical, except now instead of taking the inverse Laplace transform of (\ref{eq:z2dlowt}), we include the characters (\ref{eq:charactersgend}):
\begin{equation}
\int d\Delta dJ \rho^{\text{primaries}}_{d=2}(\Delta,J) e^{-\beta_L(\frac{\Delta - J}2-\frac{c}{24})}e^{-\beta_R(\frac{\Delta + J}2-\frac{c}{24})} \approx e^{\frac{2\pi^2 c}{12} \left(\frac1{\beta_L}+\frac1{\beta_R}\right)}(1-e^{-\beta_L})(1-e^{-\beta_R}).
\label{eq:d2primeq}
\end{equation}
Taking the inverse Laplace transform we then get
\begin{align}
\rho^{\text{primaries}}_{d=2}(\Delta, J) &= \frac{c^{3/2}\pi^2}{\sqrt{432}(\Delta+J-\frac{c}{12})^{5/4}(\Delta-J-\frac{c}{12})^{5/4}}
\nn\\
&\quad\x\exp\left[\sqrt{\frac{2c}3}\pi\left(\sqrt{\frac{\Delta+J}2-\frac{c}{24}} + \sqrt{\frac{\Delta-J}2-\frac{c}{24}}\right)\right]
\nonumber \\ 
&\quad\times\left(1 + O(\Delta^{-1/2})\right).
\label{eq:cardy2dglobalprim}
\end{align}

\subsubsection{$d=3$}

Now let us redo this analysis for $d=3$. The logic is the same except now the form of the partition function is different:
\be
Z(T,\Omega)_{d=3} \approx \exp\left(\frac{4\pi f T^2}{1+\Omega^2}\right).
\label{eq:d3approx}
\ee
Using the same change variables (\ref{eq:changeofvariablesbeta}) we get
\be
Z(\beta_L, \beta_R)_{d=3} \coloneqq \text{Tr}\left(e^{-\beta_L(\frac{\Delta - J}2)}e^{-\beta_R(\frac{\Delta + J}2)}\right) \approx \exp\left(\frac{4\pi f}{\beta_L \beta_R}\right).
\ee
To extract the density of states we again use an inverse Laplace transform:
\begin{equation}
    \rho^{\text{states}}_{d=3}(\Delta, J) \approx \frac12\left(\frac1{2\pi i}\right)^2\int_{\gamma-i\infty}^{\gamma+i\infty} d\beta_L d\beta_R \exp\left(\frac{4\pi f}{\beta_L \beta_R} + \beta_L\left(\frac{\Delta-J}2\right)+\beta_R\left(\frac{\Delta+J}2\right)\right).
\label{eq:3ddensity}
\end{equation}
The integral in (\ref{eq:3ddensity}) is more complicated so now we do it via saddle point analysis. The saddles in $\beta_L, \beta_R$ are located at
\begin{align}
\beta_L^* &= \left(\frac{8\pi f (\Delta+J)}{(\Delta-J)^2}\right)^{1/3},~~~~~\beta_R^* = \left(\frac{8\pi f (\Delta-J)}{(\Delta+J)^2}\right)^{1/3}.
\label{eq:saddled3beta}
\end{align}
This gives a remarkably simple expression for the tree-level density of states:
\begin{equation}
    \rho^{\text{states}}_{d=3}(\Delta, J) \sim 
    \exp\left[3 \pi^{1/3}f^{1/3}\left(\Delta+J\right)^{1/3}\left(\Delta-J\right)^{1/3}\right].
    \label{eq:finald3}
\end{equation}
Keeping the one-loop terms we get
\begin{align}
     \rho^{\text{states}}_{d=3}(\Delta, J) &\sim \frac{f^{1/3}}{\sqrt 3 \pi^{2/3}(\Delta+J)^{2/3}(\Delta-J)^{2/3}}    \exp\left[3 \pi^{1/3}f^{1/3}\left(\Delta+J\right)^{1/3}\left(\Delta-J\right)^{1/3}\right]. 
    \label{eq:finald3v2}
\end{align}
The expression (\ref{eq:finald3v2}) again is counting the asymptotic density of states rather than conformal primaries. We can read off the density of primaries from the character formula (\ref{eq:charactersgend}). We get:
\begin{align}
   \rho^{\text{primaries}}_{d=3}(\Delta,J) &\sim \frac{8 \pi^{2/3} f^{5/3} (2J+1)\Delta}{\sqrt3(\Delta+J+\frac12)^{7/3}(\Delta-J-\frac12)^{7/3}}\exp\left[3 \pi^{1/3}f^{1/3}\left(\Delta+J+\frac12\right)^{1/3}\left(\Delta-J-\frac12\right)^{1/3}\right].
\label{eq:d3prim}
\end{align}
Note that in (\ref{eq:d3prim}), $2J+1$ is the dimension of the spin $J$ representation of $SO(3)$. This is reminiscent of the formulas found in \cite{Harlow:2021trr, Kang:2022orq}, where the density of a global symmetry representation $\rho$ is proportional to its dimension $\dim \rho$. This comes about because the high temperature partition function can be approximated by a delta-function on a group (the rotation group in this case) centered at the identity, whose harmonic transform is the Plancherel measure $\dim \rho/\vol\,G$.

\subsubsection{General $d$}

For general $d>3$, there are several chemical potentials to turn on, so the final leading formula for the density of states is a little more cumbersome. For simplicity, we will simply compute the tree-level asymptotic density of states (i.e.\ the value at the saddle-point, not including the Gaussian determinant). Note that because in this section we are computing the tree-level contribution, the formula is identical for states and for primaries. 

First, let us consider the case where we only turn on information about one spin, for simplicity. The saddles in temperature and chemical potential are located at
\begin{align}
\label{eq:saddlesonechem}
T_* &= \left(\frac{(\Delta + \varepsilon_0-i J \Omega_*)(1+\Omega_*^2)}{(d-1)f \vol\, S^{d-1}}\right)^{1/d}, \nonumber \\
\Omega_* &= -i\frac{\sqrt{(\Delta+\varepsilon_0)^2 + (d-3)(d-1)J^2}-(\Delta+\varepsilon_0)}{J(d-3)},
\end{align}
which lead to a high energy density of states of
\begin{align}
\log \rho_d(\Delta, J) 
&\sim
\frac{d}{d-1}(\Delta+\varepsilon_0)^\frac{d-1}{d}
\left(\frac{\vol\, S^{d-1} f(d-1)}{2}
\left(1 + \sqrt{1 + \frac{(d-3)(d-1)J^2}{(\Delta+\varepsilon_0)^2}}\right)\right)^{\frac1d} \nonumber \\ &~~~~~~~~~~~~~
\times\left(\frac{d-2}{d-3}-\frac{1}{d-3}\sqrt{1 + \frac{(d-3)(d-1)J^2}{(\Delta+\varepsilon_0)^2}}\right)^{1-\frac2d}.
\label{eq:rhogenerald}
\end{align}
This leading order formula matches the result in Equation (49) of \cite{Shaghoulian:2015lcn}.

The expression (\ref{eq:rhogenerald}) reproduces (\ref{eq:finald2}) and (\ref{eq:finald3}) for $d=2$ and $d=3$ respectively. Note that to reproduce (\ref{eq:finald3}) we set the Casimir energy $\varepsilon_0 = 0$ and use the fact that
\be
\lim_{d\rightarrow 3}\frac{(d-2)\Delta - \sqrt{\Delta^2 + (d-3)(d-1)J^2}}{d-3} = \frac{\Delta^2 - J^2}{\Delta}.
\ee

Now let us consider all the chemical potentials $\Omega_i$ turned on. The leading term for the partition function is given in (\ref{eq:freeenergy}), which means
\begin{equation}
\rho_d(\Delta, J_i) \sim \int dT d\Omega_i \exp\left(\frac{\Delta+\varepsilon_0}T + \frac{\vol\, S^{d-1} f T^{d-1}}{\prod_{i=1}^n (1+\Omega_i^2)} - \sum_{i=1}^n \frac{i\Omega_i J_i}T\right),
\label{eq:gendintpresad}
\end{equation}
where for even dimensions, $\varepsilon_0$ is the Casimir energy as defined in (\ref{eq:unambiguouscasimirenergy}).
The saddle for the $T$ integral is easy to compute, and is located at
\begin{equation}
T_* = \left(\frac{(\Delta + \varepsilon_0 - i \sum_i J_i \Omega_i)\prod_i (1+\Omega_i^2)}{\vol\, S^{d-1} f(d-1)}\right)^{1/d}.
\label{eq:Tsaddle}
\end{equation}
Plugging this back into (\ref{eq:gendintpresad}), we get
\begin{equation}
\rho_d(\Delta, J_i) \sim \int d\Omega_i  \exp\left[d(d-1)^{-\frac{d-1}d}(\vol\, S^{d-1} f)^{\frac1d} \left(\Delta +\varepsilon_0 - i \sum_i J_i \Omega_i\right)^{\frac{d-1}d}\prod_i \left(1+\Omega_i^2\right)^{-\frac1d}\right].
\label{eq:genafterdoingt}
\end{equation}
We now want to find the saddles in the chemical potentials. Taking a derivative with respect to each $\Omega_j$ gives us the following equations to solve for the saddle $\Omega_{*,j}$:
\begin{equation}
    -\frac{i(d-1)J_j}2 = (\Delta + \varepsilon_0 - i \sum_i J_i \Omega_{*,i})\frac{\Omega_{*,j}}{1+\Omega_{*,j}^2}. 
\label{eq:omegasaddles}
\end{equation}
If we solve (\ref{eq:omegasaddles}) for $\Omega_{*,j}$ for $j=1, 2, \ldots, \lfloor \frac d2 \rfloor$, and plug into (\ref{eq:genafterdoingt}), we get the tree-level density of states (or primaries) in $d$ dimensions. To describe the solution for \eqref{eq:omegasaddles}, it is useful to define the quantity $a$ as the following:
\begin{equation}
    a \coloneqq \frac{(\Delta+\varepsilon_0)-i\sum_i J_i\Omega_{*,i}}{d-1}.
\label{eq:adefinition}
\end{equation} 
$a$ is a function of $\Delta+\varepsilon_0$ and the spins $J_i$, and it is a symmetric function of $J_i$. Each saddle point of $\Omega$ satisfies the following equations:
\begin{equation}
    J_j = \frac{2 i a \Omega_{*,j}^2}{1+\Omega_{*,j}^2}, \ \ \ \ \ \ \ \  
    \Omega_{*,j} = -i\,\frac{-a+\sqrt{a^2+J_j^2}}{J_j}.
\label{eq:omegasaddleina}
\end{equation}

When we substitute this relation into \eqref{eq:omegasaddles}, we get the density of states at leading order to be:
\begin{align}
    \log \rho_d(\Delta,J_1,\dots,J_n)&\sim
    d (\vol\, S^{d-1} f)^{\frac1d} a^\frac{d-1}{d}\prod_{i=1}^n\left(\frac{1+\sqrt{1+\frac{J_i^2}{a^2}}}{2}\right)^{\frac{1}{d}},
\end{align}
where $a$ is the positive real solution of:
\begin{equation}
    \Delta+\varepsilon_0 = \lfloor \tfrac{d-1}2\rfloor a + \sum_i\sqrt{a^2+J_i^2}.
\label{eq:aeqn}
\end{equation}
Because the right-hand side of \eqref{eq:aeqn} is a monotonically increasing function of $a$, there is only one positive real $a$ satisfying \eqref{eq:aeqn} for a given energy $\Delta+\varepsilon_0$ and spins $J_i$.

\subsection{Perturbative corrections to the density of states}

Besides corrections coming from higher-loop terms in the saddle point analysis, the first nontrivial correction for $d>2$ comes from higher derivative terms in the effective action of $S_{\text{th}}$ in (\ref{eq:Zformula}). Note that in $d=2$ these corrections are absent (see (\ref{eq:twodcase})). This is another way of understanding that the corrections to the first line of (\ref{eq:cardylog}) are non-perturbatively suppressed in $\Delta$ (and come from the modular $S$ transformation of the lightest non-vacuum operator).

For $d>2$, the first correction comes from the Maxwell and Einstein terms in the effective action. This will turn out to induce a correction to the entropy of the form $\Delta^{\frac{d-3}d}$. To see this, let us look at our expression for the thermal effective action (\ref{eq:Zformula}).

In $d=3$ we have
\begin{equation}
    \log Z_{d=3}(T, \Omega) = \frac{4\pi}{1+\Omega^2}\left(f T^2 - (2c_1 + (2c_1 + \frac 83 c_2)\Omega^2) + O(T^{-2})\right).
    \label{eq:d3correctedaction}
\end{equation}
From this we can extract a correction to the density of states from our saddle (\ref{eq:saddled3beta}). We get
\begin{align}
    \log \rho^{\text{states}}_{d=3}(\Delta, J) &= 3\pi^{1/3}f^{1/3} \left(\Delta + J\right)^{1/3}\left(\Delta-J\right)^{1/3} - \frac23 \log(\Delta^2-J^2) + \frac13 \log\left(\frac{f}{3\sqrt 3 \pi^2}\right) \nonumber \\ &- 8\pi c_1 + \frac{32 c_2 J^2 \pi}{3(\Delta^2-J^2)} + O(\Delta^{-1/3}), \nonumber\\
       \log \rho^{\text{primaries}}_{d=3}(\Delta, J) &= 3\pi^{1/3}f^{1/3} \left(\Delta + J + \frac12\right)^{1/3}\left(\Delta-J-\frac12\right)^{1/3}+\log\left(\frac{\Delta(2J+1)}{(\Delta^2-(J+\frac12)^2)^{7/3}}\right)\nonumber \\ &+\log\left(\frac{8\pi^{2/3}f^{5/3}}{\sqrt 3}\right) - 8\pi c_1 + \frac{32 c_2 (J+\frac12)^2 \pi}{3(\Delta^2-(J+\frac12)^2)} + O(\Delta^{-1/3}).
       \label{eq:d3correcteddensityofstates}
\end{align}

Note that the size of the Maxwell term ($c_2$) compared to higher-derivative terms depends on the order of limits in $\Delta, J$. If we take $\Delta \gg J \gg 1$, then the Maxwell term scales as $\Delta^{-2}$ (instead of $\Delta^0$), can be neglected at this order in the derivative expansion. However, if we instead take a limit where $\Delta/J$ is fixed and then take $\Delta$ to infinity, then it is important.

For $d>3$, we have the following correction to the density of states:
\begin{align}
    \log &~\rho_d(\Delta,J_1,\dots,J_n)\sim
    d (\vol\, S^{d-1} f)^{\frac1d} a^\frac{d-1}{d}\prod_{i=1}^n\left(\frac{1+\sqrt{1+\frac{J_i^2}{a^2}}}{2}\right)^{\frac{1}{d}}\nonumber\\
    &-\frac{d-2}{f}(\vol\, S^{d-1} f)^{\frac3d}a^\frac{d-3}{d}
    \left((d-1)c_1-\left(2c_1+\frac{8}{d}c_2\right)\sum_i\frac{\sqrt{1+J_i^2/a^2}-1}{\sqrt{1+J_i^2/a^2}+1}\right)
    \prod_{i=1}^n\left(\frac{1+\sqrt{1+\frac{J_i^2}{a^2}}}{2}\right)^{\frac{3}{d}}\nonumber\\
    &+ O\left(a^{\frac{d-5}d}\right),
\end{align}
where $a$ is defined in \eqref{eq:aeqn}. 
In order to trust the high-temperature expansion, we demand that the temperature at the saddle point be large. The saddle temperature \eqref{eq:Tsaddle} is proportional to $a^\frac1d$, with the remaining factors being $\mathcal{O}(1)$. 
Therefore, we can expand our formula in $a$.
We will discuss the regime of validity of our formulas in more detail in Sec \ref{sec:regime}.

If we keep the information from only one spin $J$, then we get a correction of the form
\begin{align}
\log \rho_d(\Delta, J) 
&\sim
\frac{d}{d-1}(\Delta+\varepsilon_0-\alpha J)^\frac{d-1}{d}\left((d-1)\vol\, S^{d-1} f\right)^\frac1d\left(1 + \frac{(d-3)\alpha J}{\Delta+\varepsilon_0}\right)^{\frac1d}
\left(1 - \frac{\alpha J}{\Delta+\varepsilon_0}\right)^{1-\frac2d} \nonumber \\
&-\frac{(d-2)(\vol\, S^{d-1}f)^{3/d}}{(d-1)^{\frac{d-3}d}f}
(\Delta + \varepsilon_0 -\alpha J)^{\frac{d-3}d}(1-\alpha^2)^{-3/d}\left((d-1)c_1-\frac{8c_2+2c_1 d}d\alpha^2\right)\nonumber\\
&+ O\left((\Delta+\varepsilon_0)^{\frac{d-5}d}\right),
\label{eq:onespincorrected}
\end{align} 
where $\alpha\coloneqq \frac{J}{\Delta+\varepsilon_0}\frac{d-1}{1+\sqrt{1+(d-1)(d-3)J^2/(\Delta+\varepsilon_0)^2}}$. Because $\Delta+\varepsilon_0$ is larger than $J$, $|\alpha|$ is always in between $0$ and $1$. 
The corrections from the Maxwell and Einstein terms are in fact more important than the Gaussian fluctuations about the saddle point.

Again, if $\Delta + \varepsilon_0 \gg |J|$, then $|\alpha| \ll 1$, so the Einstein ($c_1$) term dominates in (\ref{eq:onespincorrected}) and the Maxwell ($c_2$) term is subleading. However, if $(\Delta + \varepsilon_0)/J$ is fixed as $\Delta\rightarrow\infty$, then $\alpha\sim O(1)$, and the two terms are comparable.

\subsection{Regime of validity}
\label{sec:regime}

Because we are doing a high-temperature expansion, in order for our formulas to be valid, we need the saddle-point value of temperature, $T_*$, to be large. From \eqref{eq:Tsaddle}, we see that we need to take $\Delta \gg f$. However, large $\Delta$ is not a sufficient condition --- it is possible that the saddles in $\Omega$ are sufficiently close to $i$ to make the saddle in $T$ no longer large. This puts a condition on the twist $\Delta - \sum_i |J_i|$ of the operators. In particular, if $m$ of the $\lfloor \frac d2 \rfloor$ spins are large, meaning $|J_1|, \ldots |J_{m}| \gg \Delta - \sum_i |J_i|$, then our universal entropy formula is only valid when
\begin{equation}
    \Delta-\sum_i|J_i|\gg \left(f\prod_{i=1}^m |J_i|\right)^{\frac{1}{m+1}}.
\label{eq:generaltwistcond}
\end{equation}
For example, in CFT$_3$, this is equivalent to the entropy formula only being valid in the regime
\begin{equation}
    \Delta - |J| \gg \sqrt{f \Delta}.
\label{eq:twistcondition}
\end{equation}
(This can be seen easily by demanding the saddles in (\ref{eq:saddled3beta}) are very small.)
Operators outside of this window (for example operators along a Regge trajectory with $\Delta$ large but $\Delta - |J|$ growing slower than $\sqrt J$) will not obey our universal entropy formula.

In $d=2$, the regime of validity is larger. In order to trust the saddle point analysis, it is only necessary to take $\Delta \rightarrow \infty$ with 
\begin{equation}
\Delta - |J| \gg c
\end{equation}
(or equivalently take $h, \bar{h} \gg c$). 

\subsection{Non-perturbative corrections to the density of states}
\label{sec:nonpertcorrections}

So far, we have discussed an infinite set of perturbative corrections in $1/T$ to $\log Z(T, \Omega_i)$, parametrized by an infinite set of terms in the thermal effective action. In this section we briefly consider nonperturbative corrections to $\log Z(T, \Omega_i)$, namely corrections that scale as $e^{-\# T}$.

In general we expect the first nonperturbative correction to be proportional to
\be
e^{-2\pi m},
\label{eq:firstnonpert}
\ee
where $m$ is the mass of the lightest massive state in the dimensionally reduced theory. By dimensional analysis, $m\propto T$, so (\ref{eq:firstnonpert}) is indeed a nonperturbative correction. The reason for (\ref{eq:firstnonpert}) is the following. Consider the CFT on $S^1_\b \x S^{d-1}$ as a gapped theory on $S^{d-1}$. Corrections of the form (\ref{eq:firstnonpert}) will be generated by world-line instantons associated with a massive particle moving along a great circle of $S^{d-1}$ of length $2\pi$. Similar world line instantons were studied in \cite{Dondi:2021buw,Grassi:2019txd,Hellerman:2021yqz,Hellerman:2021duh,Caetano:2023zwe} in the context of the large-charge expansion.\footnote{We thank Yifan Wang for pointing out this interpretation and associated references.}

We can also understand (\ref{eq:firstnonpert}) from a Hamiltonian perspective. We can compute the partition function of the gapped theory on $S^{d-1}$ by slicing the path integral on $S^{d-2}$ spatial slices at various polar angles $\theta\in [0,\pi]$. These slices have varying radius $\sin \theta$, and therefore varying Hamiltonian $H(\theta)$. Overall, we time-evolve by Euclidean time $\pi$ as we move from the south pole to the north pole of the $S^{d-1}$. This gives an expression for the partition function of the form
\be
Z &= \<\psi_0|\mathcal T\exp\p{-\int_{0}^\pi d\theta\, H(\theta)}|\psi_0\>,
\ee
where $|\psi_0\>$ is the state in the $S^{d-2}$ Hilbert space created by the path integral near the south pole, and $\mathcal T$ denotes Euclidean time ordering. In general, the spectrum of $H(\tau)$ could be quite complicated. However, when $\b$ is small, most spatial slices are large compared to the mass gap, and we expect the low-lying spectrum of $H(\tau)$ to be close to the gapped spectrum in flat space $\R^{d-2}$. In particular, there is a contribution from a particle-antiparticle pair nucleated at the south pole, which propagate for Euclidean time $\pi$ before annihilating at the north pole. This leads to (\ref{eq:firstnonpert}).

 In general, we expect similar corrections of the form $e^{-2\pi m_i}$ for each massive state in the gapped theory on $\R^{d-2}$. There will also be L{\"u}scher corrections \cite{Luscher1986} that it would be interesting to study in more detail.

We can check the prediction (\ref{eq:firstnonpert}) explicitly in free theories. In Appendices \ref{app:nonpert} and  \ref{app:nonpertferm}, we write down the high-temperature expansions of the partition function for a $d$-dimensional free boson and free Dirac fermion respectively. In even dimensions, we write down the exact expression, including all non-perturbative corrections; in odd dimensions, the perturbative expansion is asymptotic rather than convergent, but we are still able to write down the first non-perturbative correction. In all cases, we show that the first non-perturbative correction at high temperature to the partition function take the form as predicted by (\ref{eq:firstnonpert}).

\section{Density of states: examples}
\label{sec:examplesdensityofstates}

In this section, we study partition functions of various CFTs to illustrate the general results of the previous sections. The examples we consider are: the free scalar, the free scalar with a $\mathbb{Z}_2$ twist, the free fermion, and holographic CFTs where the entropy is well-approximated by that of a Kerr-AdS black hole. In these examples, we check the partition function against our general formula \eqref{eq:Zformula} and determine the unknown coefficients $f$, $c_1$, and $c_2$ when the thermal effective action applies. Furthermore, in appendix~\ref{sec:3dising}, we compare the predictions of the thermal effective action to numerical bootstrap data for the 3d Ising model, obtaining an estimate for $f$.

\subsection{Free scalar}
\label{sec:scalarsfree}

Our first example is the free scalar in $d$ dimensions. When compactified, this theory contains a gapless sector corresponding to a free scalar in one lower dimension. Therefore, it violates the central assumption of the thermal effective action, and the predictions of the thermal EFT should be violated in some way.

The partition function of this theory can be computed exactly. For a review, see appendix~\ref{app:freebosonconstants}. Expanding the result at high temperature, we find for example
\begin{equation}
    \log Z(T, \Omega_i) = \frac{1}{\prod_{i=1}^{\lfloor \frac d2\rfloor}(1+\Omega_i^2)}\left(2\zeta(d) T^{d-1} - \frac{d-4 - 2\sum_{i=1}^{\lfloor \frac d2\rfloor} \Omega_i^2}{12}\zeta(d-2)T^{d-3} + O(T^{d-5})\right).
\label{eq:freefieldlogeven}
\end{equation}
Importantly, in even $d$, we find that the high temperature expansion contains a term proportional to $T^0$, while in odd dimensions there is a term proportional to $\log T$. (When $d=3$, the logarithm is visible in (\ref{eq:freefieldlogeven}) via the pole in the $\zeta$-function at $1$.) Such terms are inconsistent with the derivative expansion of the thermal effective action (which contains powers of the form $T^{d-2k-1}$ for integer $k$). They represent contributions from the gapless sector. We discuss these terms more explicitly in Appendix \ref{app:nonpert}.

\subsection{Free scalar with a $\mathbb Z_2$ twist}

To remove the gapless sector in the compactified free scalar, we can insert a $\mathbb Z_2$ twist on the $S^1$, where we identify $\phi(\tau=1)=-\phi(\tau=0)$ as we go around the thermal circle. Computing the partition function with this twist inserted (this can be treated using methods in e.g. \cite{Dowker:2021gqj}), we find
\begin{align}
    &\log Z(T, \Omega_i)=
    \frac{1}{\prod_{i=1}^{\lfloor \frac d2\rfloor}(1+\Omega_i^2)} \times \nonumber \\&
    \left(-2\left(1-\frac{1}{2^{d-1}}\right)\zeta(d)T^{d-1}+\left(1-\frac{1}{2^{d-3}}\right)\frac{(d-4)-2\sum_{i=1}^{\lfloor \frac d2\rfloor}\Omega_i^2}{12}\zeta(d-2)T^{d-3}+ O(T^{d-5})\right).
    \label{eq:freenergyfortwist}
\end{align}
This result is now consistent with the thermal effective action (even though the compactification is no longer ``thermal"). For example, when $d=3$, the extra factor $(1-1/2^{d-3})$ cancels the pole in $\zeta(d-2)$, so there is no $\log T$ term. Matching with (\ref{eq:Zformula}), we get
\begin{align}
    f &= -\frac{2\zeta(d)}{\vol\, S^{d-1}}\left(1-\frac{1}{2^{d-1}}\right),\nn\\
    c_1 &= -\frac{(d-4)\zeta(d-2)}{12(d-1)(d-2)\vol\, S^{d-1}}\left(1-\frac{1}{2^{d-3}}\right),\nn\\
    c_2 &= \frac{d(2d-5)\zeta(d-2)}{48(d-1)(d-2)\vol\, S^{d-1}}\left(1-\frac{1}{2^{d-3}}\right),
    \label{eq:fc1c2fortwist}
\end{align}
for a free scalar with a $\mathbb Z_2$ twist.

Note that $f<0$ in (\ref{eq:fc1c2fortwist}). This is not a contradiction because the partition function computed in (\ref{eq:freenergyfortwist}) is not a positive-definite sum of states (due to the insertion of the $\mathbb Z_2$ twist). In fact $f<0$ implies a strong cancellation between $\mathbb{Z}_2$-even and odd operators in the theory. 

\subsection{Free fermion}

Next, let us consider a free Dirac fermion in $d$ dimensions. We compactify the theory with thermal boundary conditions, where we do not insert $(-1)^F$. This leads to a massive $(d-1)$-dimensional theory. (With the $(-1)^F$ operator inserted, there would be a gapless sector.)

We compute the partition function explicitly in Appendix \ref{app:freebosonconstants}. The leading two terms in the free energy are given by
\begin{align}
    \log Z(T,&\Omega_i)
    =
    \frac{2^{\lfloor \frac{d}{2} \rfloor + 1}}{\prod_{i=1}^{\lfloor \frac{d}{2} \rfloor}(1+\Omega_i^2)} \times \nonumber \\ &\left[\left(1-\frac{1}{2^{d-1}}\right)\zeta(d)T^{d-1}-
    \left(1-\frac{1}{2^{d-3}}\right)\frac{(d-1)+\sum_{i=1}^{\lfloor \frac{d}{2} \rfloor}\Omega_i^2}{24}\zeta(d-2)T^{d-3}+O(T^{d-5})\right].
\label{eq:fermionresult}
\end{align}
Unlike in the free scalar case, the expression (\ref{eq:fermionresult}) has no $\log T$ terms or $T^0$ terms in even $d$, consistent with the thermal compactification being gapped.
Matching with (\ref{eq:Zformula}), we find
\begin{align}
    f &= \frac{2^{\lfloor \frac d2 \rfloor+1}\left(1-\frac{1}{2^{d-1}}\right)\zeta(d)}{\vol\, S^{d-1}},\nn\\
    c_1 &= \frac{2^{\lfloor \frac d2 \rfloor+1}\left(1-\frac{1}{2^{d-3}}\right)\zeta(d-2)}{24(d-2)\vol\, S^{d-1}},\nn\\
    c_2 &= - \frac{2^{\lfloor \frac d2 \rfloor+1}\left(1-\frac{1}{2^{d-3}}\right)\zeta(d-2)}{96(d-2)\vol\, S^{d-1}},
\label{eq:fc1c2freefieldfermion}
\end{align}
for the free fermion.

\subsection{Holographic theories}
\label{sec:holo}

Finally, let us consider CFTs dual to semiclassical Einstein gravity via AdS/CFT. We can estimate the partition functions of such theories by studying the thermodynamics of Kerr-AdS black holes (see e.g.\ \cite{Gibbons_2005}).
By the holographic principle, a holographic CFT has the same partition function as its dual in AdS space. 
We get the following high-temperature partition function for a holographic CFT$_d$, $d\geq 3$:
\begin{align}
    \log Z=
    \frac{\vol\, S^{d-1}(4\pi)^{d-1}}{4 d^d G_N}\frac{\ell_{\text{AdS}}^{d-1}T^{d-1}}{\prod_{i=1}^{\lfloor d/2 \rfloor}\left(1+\Omega_i^2\right)}
    \left(1-\frac{d^2\left((d-1)+\sum_{i=1}^{\lfloor d/2 \rfloor}\Omega_i^2\right)}{16\pi^2T^2}+\mathcal{O}\left(\frac{1}{T^4}\right)\right),
\end{align}
where $\ell_{\text{AdS}}$ is the characteristic length of the dual asymptotic $\text{AdS}_{d+1}$ spacetime and $G_N$ is the $d+1$-dimensional Newton constant. 

This is indeed consistent with the result \eqref{eq:Zformula} from the thermal effective action. Matching coefficients, we find\footnote{The coefficients in (\ref{eq:holographicfc1c2}) were also independently computed by Edgar Shaghoulian. We thank him for discussions related to these coefficients.}
\begin{align}
    f &= \frac{(4\pi)^{d-1}\ell_{\text{AdS}}^{d-1}}{4d^d G_N}, \nonumber \\
    c_1 &= \frac{(4\pi)^{d-3}\ell_{\text{AdS}}^{d-1}}{4(d-2)d^{d-2}G_N}, \nonumber \\
    c_2 &= -\frac{(4\pi)^{d-3}\ell_{\text{AdS}}^{d-1}}{32(d-2)d^{d-3}G_N},
    \label{eq:holographicfc1c2}
\end{align}
for the thermal Wilson coefficients of a holographic CFT.

\subsubsection{Extended regime of validity for holographic theories}
\label{sec:extendedholographic}

For holographic theories, the entropy of local operators with certain dimensions and spins can be approximated by the entropy of a black hole with the same quantum numbers, as long as the black hole is stable and has large area (in Planck units). In this section we examine where the entropy of Kerr black holes is trustworthy, and compare it to the range of validity of the EFT expansion in section~\ref{sec:regime}. We will find that for holographic theories, the universal formula for entropy has an extended regime of validity, compared to general CFTs. This is reminiscent of what happens in two-dimensional CFTs, where the Cardy formula has an extended regime of validity for holographic theories \cite{Hartman:2014oaa}. 

When Kerr black holes in AdS spin too quickly, they suffer from a phenomenon called superradiant instability \cite{Cardoso:2004hs}. In the case of the Kerr-AdS black hole, it has an instability when any of the angular velocities $\Omega_i$ become larger than $\ell_{\text{AdS}}^{-1}$. In particular, a stable Kerr-AdS black hole has a bound on quantity $E-\sum_i|J_i|\ell_{\text{AdS}}^{-1}$. For instance, in AdS$_4$, the condition for stability is (see e.g.\ \cite{Kim:2023sig})
\begin{equation}
E - |J| \ell_{\text{AdS}}^{-1} > \sqrt{\frac{E \ell_{\text{AdS}}}{2G_N}},
\end{equation}
when the black hole has large mass and spin.\footnote{Meaning $E G_N \ell_{\text{AdS}}^{-1}, |J| G_N \ell_{\text{AdS}}^{-2} \gg 1$.}
Translated to CFT data, the entropy for holographic theories is trustworthy when the twist obeys
\begin{equation}
    \Delta - |J| \gtrsim \sqrt{f \Delta},
    \label{eq:regimeextend1}
\end{equation}
where by $\gtrsim$ we allow for an $O(1)$ constant on the RHS that we do not compute.\footnote{The reason we allow this freedom is the possibility of the black holes being stable, but not yet dominating the canonical ensemble. This can occur for sufficiently light black holes, sometimes called ``enigmatic black holes" \cite{deBoer:2008fk, Hartman:2014oaa}.} A similar calculation shows that, for a holographic theory where $m$ of the spins are taken to be large compared to the twist (i.e.\ $\Delta, |J_1|, \ldots |J_m| \gg \Delta - \sum_{i=1}^{\lfloor \frac d2 \rfloor} |J_i|$), then the entropy is trustworthy when the twist obeys
\begin{equation}
\Delta-\sum_i|J_i|\gtrsim \left(f\prod_{i=1}^m |J_i|\right)^{\frac{1}{m+1}}.
\label{eq:regimeextend2}
\end{equation}

We see that the functional form of the stability bound for Kerr-AdS black holes is very similar to the regime of validity (\ref{eq:twistcondition}) and (\ref{eq:generaltwistcond}) for the general entropy formula.

This is reminiscent of the extended regime of the Cardy formula for the case of CFT$_2$: for theories holographically dual to large-radius gravity in AdS$_3$, it was shown that there is a further extension of the validity of the Cardy formula in \cite{Hartman:2014oaa}. For holographic CFTs, the Cardy formula matches the Bekenstein-Hawking entropy of BTZ black holes, which only requires 
\begin{equation}
\Delta - |J| \gtrsim c,
\end{equation}
with $c\rightarrow\infty$ (rather than the usual $\Delta - |J| \gg c$ condition). In particular, it was shown that theories with a sufficiently sparse light spectrum (which is a necessary, but not sufficient condition for the theory to have a semiclassical Einstein gravity dual) have an extended Cardy regime of entropy. It would be interesting if one could prove that a similar statement for higher dimensional CFTs.\footnote{Some works studying sparseness in higher-$d$ CFT include \cite{Belin:2016yll, Horowitz:2017ifu,Mefford:2017oxy}. In particular it would be interesting if the precise sparseness conditions in \cite{Mefford:2017oxy} implied the extended entropy formulas described in this section.}

\section{A ``genus-2" partition function}
\label{sec:partitionfunction}

While the density of states of a CFT is encoded in the partition function on $S^1_\beta\x S^{d-1}$, OPE coefficients are encoded in partition functions on other manifolds. 
In this work, we will be interested in ``heavy-heavy-heavy" OPE coefficients $c_{ijk}$ between operators with parametrically large scaling dimension. To study them, we can consider the partition function of the CFT on a manifold constructed by gluing a pair of three-punctured spheres along their punctures. In two dimensions, this produces a genus-2 Riemann surface. However, a similar construction works in higher dimensions.  We will  continue to refer to such a manifold as ``genus-2" in higher dimensions, by analogy with the 2d case. 

We pause to note that our final expression for ``heavy-heavy-heavy" OPE coefficients is given at the end of Sec. \ref{sec:opecoeffhhh}, in Eqn (\ref{eq:generalope}). Readers only interested in the final result can skip to this part.

\subsection{Conformal structures of a genus-2 manifold}
\label{sec:conformalstructures}

In higher dimensions we can build a genus-2 manifold $M_2$ by taking two copies of the plane $\R^d$ (more precisely its conformal compactification $S^d$), removing three balls from each plane, and gluing the boundaries of the balls with cylinders. In this construction, we can choose the positions and radii of the balls, as well as the lengths of the cylinders. We can additionally add angular twists by elements of $\SO(d)$ as we move along each cylinder. 
This is a large number of parameters, but many of them are related by conformal symmetry.

In addition, if the CFT has a global symmetry $\G$, we can introduce topological defects on $M_2$, or equivalently a flat $\G$-bundle over $M_2$. Such flat bundles  are parametrized by homomorphisms of the fundamental group of $M_2$ to $\G$. For $d > 2$, 
$\R^d$ with balls removed is simply connected, and the only homotopically non-trivial cycles on $M_2$ are those going through the cylinders between the two copies of $\R^d$. Therefore, the fundamental group is a free group with $2$ generators, and flat $\G$-bundles can be parametrized by decorating the cylinders by inserting topological defects transverse to these cycles. The situation is different for $d=2$ since $\R^2$ with balls removed is already not simply connected, and there are additional generators of the fundamental group which go around the boundaries of the balls. Inserting topological defects transverse to these additional cycles on $M_2$ is equivalent to considering twisted sectors along the cylinders.

To understand the implications of conformal symmetry, it is helpful to ignore the Weyl anomaly and focus on the conformal structure of the manifold $M_2$ --- i.e.\ properties of $M_2$ that are independent of the Weyl class of the metric.
First, we can associate an (orientation-reversing) conformal group element to each cylinder as follows. Let $x,x'$ be flat coordinates on the two copies of $\R^d$. A cylinder $C$ that connects the two planes is Weyl-equivalent to an annulus in each plane. Using this Weyl-equivalence, each coordinate $x$ and $x'$ can be extended to cover $C$. Inside $C$, the coordinates $x$ and $x'$ are identified by an orientation-reversing conformal group element:
\be
x = g x',\quad g\in G^- \qquad \textrm{(inside $C$).}
\ee
Here, we denote the conformal group as $G=\SO(d+1,1)$, and we write the orientation-reversing component of $O(d+1,1)$ as $G^-$.
For example, if a cylinder of length $\b$ connects the unit spheres in each copy of the plane, then we have
\be
g &= e^{-\b D} I,
\ee
where $I(x)=\frac{x}{x^2}$ is an inversion. More generally, suppose the cylinder is centered at $x=a$, has radius $r$ and length $\b r$, and includes an angular twist by $h\in \SO(d)$. Then\footnote{The appearance of the quantity $\b-2\log r$ reflects the fact that two planes glued by a cylinder with radius $r$ and length $\b r$ is Weyl-equivalent to two planes glued by a cylinder with radius $1$ and length $\b-2\log r$. The Weyl transformation breaks the cylinder into three pieces of lengths $r\log r,r(\b-2\log r),r\log r$, and flattens out the first and third piece into annuli.}
\be
g &= e^{a\.P} e^{-(\b-2\log r) D} h I e^{-a\.P}.
\ee

In the case of interest, we have three cylinders connecting two copies of the plane. This gives three group elements $(g_1,g_2,g_3)\in (G^-)^3$. However, the conformal structure of the resulting manifold is unchanged if we perform a conformal transformation $x\mto g x$ on the first plane or $x'\mto g' x'$ on the second. These conformal transformations become gauge redundancies acting on the $g_i$:
\be
\label{eq:groupaction}
(g_1,g_2,g_3) &\sim (g g_1 g'^{-1}, g g_2 g'^{-1}, g g_3 g'^{-1}).
\ee
Modding out by this gauge-redundancy, we obtain the moduli space of conformal structures as a double-quotient
\be
\cM &= G\backslash(G^-)^3/G,
\ee
where the left and right factors of $G$'s act on $(G^-)^3$ via (\ref{eq:groupaction}). The action of the $G\x G$ gauge redundancies on $(G^-)^3$ is almost free, so the dimension of $\cM$ is
\be
\dim \cM &= 3 \dim G^- - 2 \dim G = \dim G = \frac{(d+1)(d+2)}{2}.
\ee

For orientation, in 2d the parametrization of a genus-2 Riemann surface in terms of $(g_1,g_2,g_3)$ is called a Whittaker parametrization. (The closely-related Schottky parametrization can be obtained by forming the combinations $\g_{ij}=g_i g_j^{-1}$, which satisfy $\g_{12}\g_{23}\g_{31}=1$.) In 2d, the true moduli space of genus-2 surfaces is a quotient of $\cM$ by the mapping class group. The action of the mapping class group is unfortunately somewhat complicated in the Whittaker/Schottky parameterizations.

In higher dimensions, $\cM$ is again a covering space of the moduli space of conformal structures on $M_2$. Topologically, $M_2$ is equivalent to a connected sum of two copies of $S^1 \x S^{d-1}$:\footnote{We thank Yifan Wang for pointing this out and directing us to reference~\cite{Brendle_2023}.}
\be
M_2 &\cong (S^1 \x S^{d-1}) \,\#\, (S^1 \x S^{d-1}).
\ee
This can be seen by decomposing $M_2$ in the ``dumbbell" channel where we slice figure~\ref{eq:gluedcylindersfigure} down the middle into a left half and a right half. Each half is topologically a copy of $S^1 \x S^{d-1}$ with a ball removed, and the two halves are glued along an $S^{d-1}$. The mapping class group of this space was computed for $d=3$ in \cite{Brendle_2023}. This mapping class group will not play a further role in the present work. It will be interesting to explore its implications and other global aspects of higher-dimensional ``higher-genus" surfaces in future work.

An important set of functions on $\cM$ are eigenvalues of the group elements $g_i^{-1} g_j\in \SO(d+1,1)$:
\be
\label{eq:relativecoords}
\mathrm{eigenvalues}_{(d+2)\x(d+2)}(g_i^{-1} g_j) = \begin{cases}
(e^{\pm\b_{ij}},e^{\pm i\vec \th_{ij}}) & (\textrm{even $d$}), \\
(e^{\pm\b_{ij}},e^{\pm i\vec \th_{ij}},1) & (\textrm{odd $d$}) .
\end{cases}
\ee
These are indeed invariant under gauge-redundancies (\ref{eq:groupaction}). We refer to the $\b_{ij}$ and $\vec \th_{ij}$ as ``relative" inverse temperatures and angles, for reasons that will become clear shortly. Note that there are $3\lfloor\frac{d+2}{2}\rfloor$ relative temperatures/angles, which is not enough to parametrize the full moduli space when $d\geq 3$.

\subsection{Choice of geometry}
\label{sec:choiceofgeometry}

The partition function of a CFT on $M_2$ factors into a theory-independent part that depends on the precise metric and is determined by the Weyl anomaly, times a theory-dependent part that depends only on the conformal structure of $M_2$. To get nontrivial information about the theory, it suffices to study only a single representative geometry for each conformal structure.

\begin{figure}
\centering
\begin{tikzpicture}[scale=1.2, line width=1pt]
    \fill[even odd rule, fill=gray!15] (0, 0) circle (2) (-1, 0) circle (0.99) (1, 0) circle (0.99);
    
    \draw [green!40!black] (0, 0) circle (2);
    
    \draw [red!60!black] (-1, 0) circle (0.99);
    
    \draw [blue!70!black] (1, 0) circle (0.99);
    
    \node [green!40!black] at (1.1,2.05) {$\ptl B_3$};
    \node [red!60!black] at (-1,0.7) {$\ptl B_1$};
    \node [blue!70!black] at (1,0.7) {$\ptl B_2$};
    \node at (0,-1.3) {$Y$};
\end{tikzpicture}
\caption{The space $Y=B_3\backslash (B_1\cup B_2)$. The boundary of $Y$ has three $S^{d-1}$ components given by $\ptl B_1,\ptl B_2,\ptl B_3$ with radii $1,1,2$, respectively. Note that in $d\geq 3$, $Y$ has an $\SO(d-1)$ rotational symmetry around the horizontal axis, and here we are depicting only a 2-dimensional slice. \label{eq:figureforY}}
\end{figure}
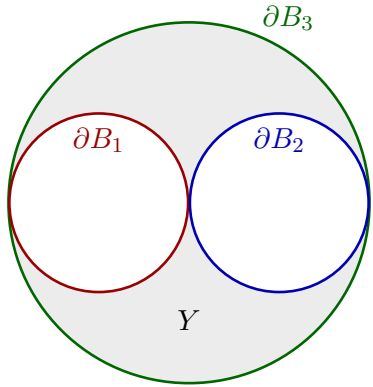

\begin{figure}
\centering
\begin{tikzpicture}
\begin{scope}[scale=1.2]
  \begin{scope}[yscale=0.3, xscale=1.3, line width=1pt]
      \draw [green!40!black] (2,0) arc (0:-180:2);
      \draw [dotted,green!40!black] (2,0) arc (0:180:2);
      
      \draw [dotted,red!60!black] (-1, 0) circle (1);
      
      \draw [dotted,blue!70!black] (1, 0) circle (1);
  \end{scope}
  
  \begin{scope}[yscale=0.3, xscale=1.3, shift={(0,8)}, line width=1pt]
      \draw [green!40!black] (0, 0) circle (2);
      
      \draw [red!60!black] (-1, 0) circle (1);
      
      \draw [blue!70!black] (1, 0) circle (1);
  \end{scope}
  
  \draw [dotted,thick,red!60!black] (-2.6,0) to[out=85,in=-85] (-2.6,2.4);
  \draw [thick,green!40!black] (-2.6,0) to[out=95,in=-95] (-2.6,2.4);
  \draw [dotted,thick,red!60!black] (0,0) to[out=95,in=-95] (0,2.4);
  \draw [dotted,thick,blue!70!black] (0,0) to[out=85,in=-85] (0,2.4);
  \draw [thick,green!40!black] (2.6,0) to[out=85,in=-85] (2.6,2.4);
  \draw [dotted,thick,blue!70!black] (2.6,0) to[out=95,in=-95] (2.6,2.4);
  
  \node [green!40!black] at (3.0,1.2) {$C_3$};
  \node [blue!70!black] at (2.23,1.2) {$C_2$};
  \node [red!60!black] at (-0.35,1.2) {$C_1$};
  \node at (-1,-0.8) {$Y'$};
  \node at (-1,3.2) {$Y$};
  \draw [thick,->,dotted] (-0.8,-0.8) to[out=0,in=-100] (0,-0.3);
  \draw [thick] (-0.8,-0.8) to[out=2,in=-143] (-0.15,-0.6);
  \draw [thick,->] (-0.8,3.15) to[out=0,in=100] (0,2.75);
\end{scope}
\end{tikzpicture}
\caption{The manifold $M_2$ is obtained by taking two copies of $Y$ and gluing corresponding boundary components with cylinders $C_1,C_2,C_3$ of inverse temperatures $\b_1,\b_2,\b_3$ and angular twists $h_1,h_2,h_3$, colored red, blue, and green, respectively. In the figure, we slightly bent the edges of the cylinders to help visualize them, but in the actual geometry the cylinders have constant radii. The figure also naively suggests that the lengths of the cylinders must be equal, but in reality they need not be related to each other.\label{eq:gluedcylindersfigure}}
\end{figure}
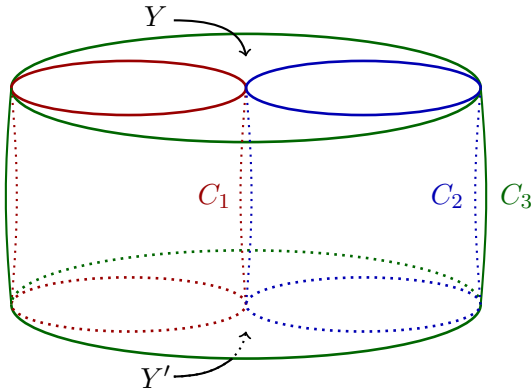

We will choose our geometry as follows. Let $B_1$ be the unit ball centered at $(-1,0,\dots,0)$, let $B_2$ be the unit ball centered at $(1,0,\dots,0)$, and let $B_3$ be the ball of radius $2$ centered at the origin. All three balls are mutually tangent. From $B_3$, we remove $B_1$ and $B_2$. The resulting space $Y=B_3\backslash (B_1\cup B_2)$ has three $S^{d-1}$ boundaries given by $\ptl B_1,\ptl B_2,\ptl B_3$, see figure~\ref{eq:figureforY}. We now take a second copy of $Y$, which we call $Y'$, containing boundaries $\ptl B_1',\ptl B_2',\ptl B_3'$. Finally, we glue each $\ptl B_i$ to $\ptl B_i'$ with cylinders $C_i$ whose ratios of length/radius are $\b_i$, and we include angular twists $h_i\in \SO(d)$ along each cylinder. See figure~\ref{eq:gluedcylindersfigure} for an illustration. 

This construction gives a particular choice of metric that is flat on $Y,Y'$ and is the usual cylinder metric on the $C_i$. Note that the metric has curvature localized at the junctions between $Y,Y'$ and the cylinders. However, it is everywhere conformally-flat because the plane, the cylinder, and a plane-cylinder junction are all conformally-flat. In terms of conformal structures, our geometry corresponds to 
\be
\label{eq:ourparametrization}
g_1 &= e^{-P^1}e^{-\b_1 D}h_1 I e^{P^1}, \nn\\
g_2 &= e^{P^1}e^{-\b_2 D}h_2 I e^{-P^1}, \nn\\
g_3 &= e^{D(2\log 2 + \b_3)}h_3 I,
\ee
which gives a point in $\cM$, parametrized by $\b_1,\b_2,\b_3,h_1,h_2,h_3$.

A slight disadvantage of the parametrization (\ref{eq:ourparametrization}) is that it is not permutation-symmetric among the three cylinders --- $C_3$ is treated differently. However, an advantage is that it makes manifest an important $\SO(d-1)$ symmetry that rotates all three balls around the $x^1$ axis, preserving their points of tangency.\footnote{By contrast, we could restore manifest permutation symmetry by taking the balls to all have the same radius and be mutually tangent, but then the $\SO(d-1)$ would act via a nontrivial conformal transformation.} We can act with an $\SO(d-1)$ rotation on either $Y$ or $Y'$, which means that the angular twists $(h_1,h_2,h_3)$ are subject to a residual gauge redundancy
\be
\label{eq:sodgaugeredudnancy}
(h_1,h_2,h_3) &\sim (k h_1 k'^{-1}, k h_2 k'^{-1}, k h_3k'^{-1}),\quad k,k'\in \SO(d-1).
\ee
Thus, overall, we can think of the parametrization (\ref{eq:ourparametrization}) as a map
\be
\label{eq:parametrization}
\SO(d-1)\backslash (\SO(1,1)\x \SO(d))^3 /\SO(d-1) &\to \cM,
\ee
where $\b_i$ parametrize the $\SO(1,1)$'s, the $h_i$ parametrize the $\SO(d)$'s, and the $\SO(d-1)$'s act on the $\SO(d)$'s via (\ref{eq:sodgaugeredudnancy}).

We claim that (\ref{eq:ourparametrization}) is injective and covers an open subset of $\cM$ --- in particular an open subset that contains the physical loci that will be important in what follows. These loci include the low temperature regime of large $\b_i$, and a high temperature limit that will control the asymptotics of OPE coefficients. A first important check is that the two spaces in (\ref{eq:parametrization}) have the same dimension. Indeed, they do, since
\be
3(1 + \dim \SO(d)) - 2 \dim \SO(d-1) = 3\p{1+\frac{d(d-1)}{2}}-2\frac{(d-1)(d-2)}{2} = \frac{(d+1)(d+2)}{2}.
\ee
In section~\ref{sec:measure}, we will show that the natural measure on $\cM$ is nonzero in the coordinates (\ref{eq:ourparametrization}) at both low and high temperatures. This establishes that $\b_1,\b_2,\b_3,h_1,h_2,h_3$ (modulo the gauge redundancy (\ref{eq:sodgaugeredudnancy}))  furnish good coordinates on $\cM$ for our purposes.

Consequently, it suffices to consider geometries of the form described above. These geometries contain all possible theory-dependent information about OPE coefficients of the CFT. In appendix~\ref{app:matchingofquantumnumbers}, we show that there is a matching between quantum numbers specifying OPE coefficients and the dimension of the genus-2 moduli space $\dim \cM$.

\subsection{The partition function as a sum over states}
\label{sec:cuttingandgluing}

The partition function on the above geometry is a weighted sum of squares of OPE coefficients. In this section, we derive this fact in detail, taking care with some of the subtleties of cutting and gluing in higher-dimensional CFTs.

Consider first the space $Y=B_3\backslash (B_1\cup B_2)$. This space has boundaries given by $\ptl Y = \ptl B_3 \sqcup -\ptl B_1 \sqcup -\ptl B_2$. Thus, the partition function $Z(Y)$ is an element of $\cH_2\otimes \cH_1^* \otimes \cH_1^*$, or equivalently a map $\cH_1\otimes \cH_1\to \cH_2$, where $\cH_r$ is the Hilbert space on a sphere of radius $r$.
A basis of states $|\cO(x)\>_r$ in $\cH_r$ is given by the insertion of an operator $\cO(x)$ inside a ball of radius $r$.  The defining property of $Z(Y)$ is that its pairing with three basis elements is a conformal three-point function. 

Let us state this more precisely. The Hermitian conjugate state to $|\cO(x)\>_r \in \cH_r$ can be obtained by inserting the following conjugate operator in a flat geometry outside the ball of radius $r$:
\be
\label{eq:modifiedbpz}
[\cO^a(x)]^{\dag_r} &\equiv r^{D}[r^{-D} \cO^a(x)]^{\dag} = \p{\frac{r}{|x|}}^{2\De} I^{-1}(\hat x)_a{}^{\bar a}\cO^\dag_{\bar a}\p{\frac{r^2 x}{x^2}}.
\ee
Note that $[\cdots]^\dag=[\cdots]^{\dag_1}$ is the usual BPZ conjugation.
The inversion tensor $I_{\bar a}{}^a$ is the solution to the conformal Ward identities for a two-point function of $\cO^a$ and $\cO_{\bar a}^\dag$:
\be
\< \cO^{\dag}_{\bar a}(x)\cO^a(0)\> &= \frac{I_{\bar a}{}^{a}(\hat x)}{x^{2\De}},
\ee
normalized so that $I I^\dag =1$.  With this notation, a projector onto the conformal multiplet of $\cO$ inside $\cH_r$ can be written
\be
\label{eq:canbewritten}
|\cO|_r &= |\cO^a(0)\>_r\,\<\cO^a(0)[\cO^{a'}(0)]^{\dag_r}\>^{-1}\,{}_r\<\cO^{a'}(0)| + \textrm{descendants},
\ee
where the inverse two-point function $\<\cO^a(0)[\cO^{a'}(0)]^{\dag_r}\>^{-1}$ should be understood as a matrix with indices $a,a'$. The indices $a,a'$ are implicitly summed over in (\ref{eq:canbewritten}). The form of the sum over descendants is determined by the conformal algebra. A resolution of the identity on $\cH_r$ is given by summing over projectors $1=\sum_\cO |\cO|_r$.

By composing $Z(Y)$ with resolutions of the identity on each of its three boundaries, we find an expression in terms of three-point functions:
\be
\label{eq:Zassumoverstates}
Z(Y) &= \sum_{\cO_1,\cO_2,\cO_3} |\cO_3^\dag|_2 Z(Y) (|\cO_1|_1 \otimes |\cO_2|_1) \nn\\
&= \sum_{\cO_1,\cO_2,\cO_3} \<\cO_1(-e) \cO_2(e)\cO_3(\oo e)\> \<\cO_1(0)[\cO_1(0)]^\dag\>^{-1} \<\cO_2(0)[\cO_2(0)]^\dag\>^{-1} \<\cO_3(\oo) [\cO_3(\oo)]^{\dag_2}\>^{-1} \nn\\
&\qquad\qquad \x
|[\cO_3(\oo)]^{\dag_2}\>_2\,{}_1\<\cO_1(0)|\otimes {}_1\<\cO_2(0)|.
\ee
For simplicity, we have omitted spin indices.
This is a sum over states with coefficients given by a three-point function, where $e=(1,0,\dots,0)$ is a unit vector along the $x^1$ direction. Note that we define a primary operator at infinity {\it without} an inversion tensor:
\be
\label{eq:defofprimaryatinfinity}
\cO_3^c(\oo e) &\equiv \lim_{L\to \oo} L^{2\De_3} \cO_3^c(L e).
\ee
To help restore symmetry among the three operators, we have chosen to insert the projector $|\cO_3^\dag|$, as opposed to $|\cO_3|$ --- this ensures that the three-point function contains $\cO_3(\oo)$ and not $\cO_3^\dag(\oo)$. 

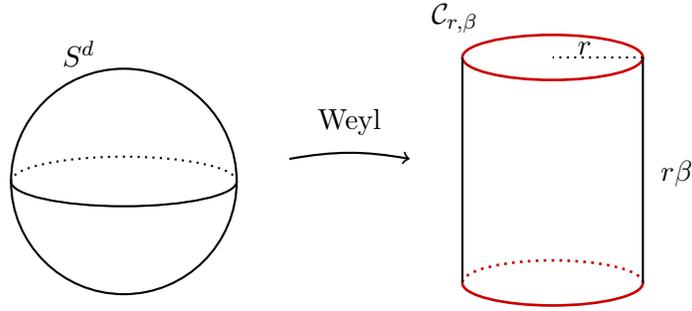
\begin{figure}
\centering
\begin{tikzpicture}
\begin{scope}[shift={(5.7,-1.35)},rotate=90]
\begin{scope}[xscale=0.3,yscale=1.2,line width=1pt]
    \draw  [red!80!black] (0,1) arc (90:270:1);
    \draw [red!80!black,dotted] (0,-1) arc (-90:90:1);
\end{scope}
\begin{scope}[xscale=0.3,yscale=1.2,shift={(10,0)},line width=1pt]
\draw [red!80!black] (0,0) circle (1);
\end{scope}
\draw [thick] (0,-1.2) -- (3,-1.2);
\draw [thick] (0,1.2) -- (3,1.2);
\draw [thick,dotted] (3,-1.2) -- (3,0);
\end{scope}
\node [right] at (5.9,1.77) {$r$};
\node [right] at (7,0.1) {$r\b$};
\draw [thick] (0,0) circle (1.5);
\begin{scope}[yscale=0.22]
    \draw [thick,dotted] (1.5,0) arc (0:180:1.5);
    
    \draw [thick] (1.5,0) arc (0:-180:1.5);
\end{scope}
\draw [thick,->] (2.2,0.3) to[out=10,in=170] (3.8,0.3);
\node [] at (3,0.8) {Weyl};
\node [above] at (4.4,1.85) {$\cC_{r,\b}$};
\node [above] at (-0.6,1.4) {$S^d$};
\end{tikzpicture}
\caption{The sphere $S^d$ is Weyl-equivalent to a ``capped cylinder" $\cC_{r,\b}$ with radius $r$ and length $r\b$. Each end cap is a ball (the interior of an $S^{d-1}$) of radius $r$. The ``closed junctions" where the cylinder meets the end caps are highlighted in red. \label{fig:weylcapped}}
\end{figure}

\begin{figure}
\centering
\begin{tikzpicture}
\begin{scope}[rotate=90]
\begin{scope}[xscale=0.35,yscale=1.2,line width=1pt]
    \draw  [blue!90!black] (0,1) arc (90:270:1);
    \draw [blue!90!black,dotted] (0,-1) arc (-90:90:1);
\end{scope}
\draw [thick] (0.65,-1.2) -- (0.65,-1.7) -- (-0.8,-2) -- (-0.8,2) -- (0.65,1.7) -- (0.65,1.2);
\draw [thick,dotted] (0.65,-1.2) -- (0.65,1.2);
\draw [thick] (0,1.2) -- (1.2,1.2);
\draw [thick] (0,-1.2) -- (1.2,-1.2);
\begin{scope}[xscale=0.35,yscale=1.2,line width=1pt,shift={(3.5,0)}]
\draw [] (0,0) circle (1);
\end{scope}
\end{scope}
\end{tikzpicture}
\caption{An ``open junction," where a cylinder meets the complement of a ball in a flat plane. The junction is highlighted in blue. \label{fig:openjunction}}
\end{figure}
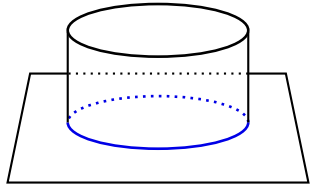

The partition function of each cylinder $C_i$ is simply $e^{-\b_i(D+\varepsilon_0)} h_i$, where $\varepsilon_0$ is the Casimir energy (\ref{eq:unambiguouscasimirenergy}), and $h_i\in \SO(d)$ is the angular twist along the cylinder. One subtle ingredient is that we must also associate nontrivial ``gluing" factors to junctions between cylinders and flat planes. To derive these gluing factors, let us start with the partition function on $S^{d}$ (with unit radius). This is Weyl-equivalent to a cylinder of radius $r$ and length $r\b$, capped off by flat balls, see figure~\ref{fig:weylcapped}. Let the capped cylinder be $\cC_{r,\b}$, and denote the Weyl factor going from $S^d$ to $\cC_{r,\b}$ by $e^{2\w_{r,\b}}$. The Weyl anomaly implies
\be
Z(\cC_{r,\b}) &= e^{-S_\textrm{anom}[g,\w_{r,\b}]} Z(S^d).
\ee
At the same time, $Z(\cC_{r,\b})$ can be computed by cutting and gluing. The end caps are simply identity operators in radial quantization, $|1\>_r$. The cylinder contributes $e^{-\varepsilon_0 \b}$, where $\varepsilon_0$ is the Casimir energy on $S^{d-1}$. Let us define $Z_\textrm{glue}(r)$ as the factor associated to a junction between a cylinder and a flat end-cap, which we call a ``closed junction". We find
\be
\label{eq:zglue}
Z(\cC_{r,\b}) = |Z_\textrm{glue}(r)|^2 e^{-\varepsilon_0 \b} \qquad\implies\qquad
|Z_\textrm{glue}(r)| = e^{\frac 1 2 \varepsilon_0 \b} e^{-\frac 1 2 S_\textrm{anom}[g,\w_{r,\b}]} Z(S^d)^{\frac 1 2}.
\ee
We calculate these gluing factors in various dimensions in appendix~\ref{app:gluing}. For example, we find\footnote{In $d=4$, we write $a_4$ as $a$.}
\be
\label{eq:thegluingfactors}
|Z_\textrm{glue}(r)| &= Z(S^d)^{\frac 1 2} \x \begin{cases} 
1 & \textrm{$d$ odd,} \\
e^{c/12}(r/2)^{c/6} & d=2, \\
e^{-7a/6}(r/2)^{-2a} & d= 4,\\
e^{37 a_6/10} (r/2)^{6 a_6} & d=6.
\end{cases}
\ee
Our geometry also contains four ``open junctions" with the opposite curvature, where a cylinder joins a flat region that locally looks like the complement of a ball, see figure~\ref{fig:openjunction}. The gluing factor associated to an open junction is the inverse $Z_\textrm{glue}(r)^{-1}$ of the one associated to a closed junction. The reason is that we can perform an infinitesimal Weyl transformation on a plane to create a closed junction infinitesimally-close to an open junction. The Weyl anomaly is infinitesimal, so the gluing factors must multiply to 1.

Putting everything together, the partition function on our genus-2 manifold is
\be
Z(M_2)&=\frac{|Z_\textrm{glue}(2)|^2}{|Z_\textrm{glue}(1)|^4} \Tr(Z(Y)^\dag e^{-\b_3(D+\varepsilon_0)}h_3^{-1} Z(Y) (e^{-\b_1(D+\varepsilon_0)} h_1\otimes e^{-\b_2(D+\varepsilon_0)} h_2)).
\ee
Each group element $e^{-\b_i D}h_i$ acts on the Hilbert space corresponding to the boundary component $\ptl B_i$.
Inserting our expression (\ref{eq:Zassumoverstates}) for $Z(Y)$, we obtain the partition function as a sum over a triplet of primary operators
\be
\label{eq:zassumoverstates}
Z(M_2) &= \frac{|Z_\textrm{glue}(2)|^2}{|Z_\textrm{glue}(1)|^4} e^{-\varepsilon_0(\b_1+\b_2+\b_3)} \nn\\
&\quad \x \sum_{\cO_1,\cO_2,\cO_3} \Big(e^{-\b_1 \De_1-\b_2 \De_2 -\b_3 \De_3}\nn\\
&\quad\qquad\qquad\quad \<\cO_1^{a'}(-e)\cO_2^{b'}(e) \cO_3^{c'}(\oo e)\>^*\<h_1\.\cO_1^a(-e)\, h_2\.\cO_2^b(e)\,h_3\.\cO_3^c(\oo e)\>\nn\\
&\quad\qquad\qquad\quad  \x  \<\cO_1^a(0) [\cO_1^{a'}(0)]^\dag\>^{-1} \<\cO_2^b(0) [\cO_2^{b'}(0)]^\dag\>^{-1} \<\cO_3^c(\oo) [\cO_3^{c'}(\oo)]^{\dag_2}\>^{-1} \nn\\
&\quad\qquad\qquad\quad   + \textrm{descendants}\Big).
\ee
Here, $h\cdot \cO$ denotes the action of a rotation $h\in \SO(d)$ on a local operator:
\be
h\cdot \cO^a &= h \cO^a h^{-1} = \lambda(h^{-1})^a{}_b \cO^b,
\ee
where $\lambda$ is the $\SO(d)$ representation of $\cO$.

Let us choose the $\cO_i$ to be an orthonormal basis of primaries with respect to the BPZ inner product. Using (\ref{eq:modifiedbpz}) and (\ref{eq:defofprimaryatinfinity}), we find
\be
\<\cO_3^c(\oo) [\cO_3^{c'}(\oo)]^{\dag_2}\> &= 2^{2\De_3} \de^c{}_{c'},
\ee
while the other two-point functions in (\ref{eq:zassumoverstates}), which involve standard BPZ conjugates $[\cdots]^{\dag}$, are identity matrices.

Let us furthermore expand the three-point functions in a basis of conformally-invariant three-point structures:
\be
\<\cO_1^a(-e) \cO_2^b(e) \cO_3^c(\oo e)\> &= \frac{1}{2^{\De_1+\De_2-\De_3}} \<\cO_1^a(0) \cO_2^b(e) \cO_3^c(\oo e)\> \nn\\
&= \frac{1}{2^{\De_1+\De_2-\De_3}} c_{123}^s V^{s;abc}(0,e,\oo).
\ee
Here, $s$ is a structure label, which runs over a finite-dimensional space of solutions $V^s$ to the 3-point conformal Ward identities. Meanwhile, $a,b,c$ are spin indices in the $\SO(d)$ representations associated to the three operators. Each three-point structure comes with an associated OPE coefficient $c_{123}^s$, and a sum over $s$ is implicit. We will discuss the space of three-point structures in more detail in section~\ref{sec:conformalblock}.

Plugging everything in,  we find an expression for the ``genus-2" partition function as a sum over conformal blocks
\be
\label{eq:genus2blockexpansion}
Z(M_2) &= \frac{|Z_\textrm{glue}(2)|^2}{|Z_\textrm{glue}(1)|^4} e^{-\varepsilon_0(\b_1+\b_2+\b_3)}  \sum_{\cO_1\cO_2\cO_3} (c_{123}^{s'})^* c_{123}^s B^{s's}_{123},
\ee
where we have introduced the ``genus-2" block $B_{123}^{s's}$ 
\be
\label{eq:blocklowtemperature}
B^{s's}_{123}(\b_i,h_i) &=2^{-2\De_1-2\De_2} e^{-\b_1 \De_1-\b_2\De_2-\b_3\De_3} (V^{s';abc}(0,e,\oo))^* (h_1h_2h_3\.V^s)^{abc}(0,e,\oo)
\nn\\
&\quad +\textrm{descendants}.
\ee
The first term in (\ref{eq:blocklowtemperature}) comes from primary states, and dominates in the ``low temperature" limit $\b_1,\b_2,\b_3\gg 1$. The descendent terms involve three-point functions of descendant operators, contracted using the inverse of the Gram matrix. Such terms are determined by the conformal algebra.

The block $B_{123}^{s's}$ is naturally a function on the moduli space $\cM$ of conformal structures, and doesn't depend on the Weyl class of the metric. This fact is already hinted at in (\ref{eq:blocklowtemperature}). Note that the factor $2^{-2\De_1-2\De_2}$ seems to violate permutation symmetry among the three operators. However, this is an artifact of our asymmetric conformal frame. We can restore manifest permutation symmetry by rewriting the block in terms of the ``relative" temperatures $\b_{ij}$, defined in (\ref{eq:relativecoords}), which are permutation-symmetric functions on $\cM$. 
At low temperatures, we find
\be
2^{-2\De_1-2\De_2} e^{-\b_1 \De_1-\b_2\De_2-\b_3\De_3} &= e^{-\frac{\b_{12}+\b_{31}-\b_{23}}{2} \De_1-\frac{\b_{12}+\b_{23}-\b_{31}}{2} \De_2-\frac{\b_{31}+\b_{23}-\b_{12}}{2} \De_3} + \dots,
\ee
which is manifestly symmetric under permuting $1,2,3$. This is a nontrivial check on (\ref{eq:blocklowtemperature}).

In appendix~\ref{app:matchingofquantumnumbers}, we point out that the number of unbounded quantum numbers needed to specify the block $B_{123}^{s's}$ matches the dimension of the moduli space $\cM$. This is analogous to the fact that the number of unbounded quantum numbers needed to specify a four-point conformal block (two: $\De$ and $J$) matches the number of cross-ratios for a four-point function (two: $z$ and $\bar z$).
 
\subsection{Hot spots and the thermodynamic limit}

Because the geometry described in section~\ref{sec:choiceofgeometry} is not a circle fibration, it is not immediately obvious how to compute the partition function using the thermal effective action. To make progress, we adopt the following assumption:
\begin{assumption}
The thermal effective action describes the contribution to the partition function from any region where the geometry locally looks like a circle fibration with a large local temperature.
\end{assumption}
\noindent We call such a region a ``hot spot."

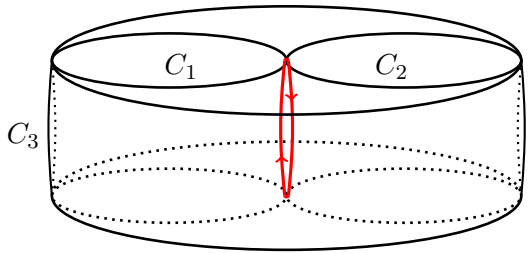
\begin{figure}
\centering
\begin{tikzpicture}
\begin{scope}[scale=1.2]
  \begin{scope}[yscale=0.3, xscale=1.3, line width=1pt]
      \draw [dotted] (-2,0) arc (180:0:2);
      \draw [] (2,0) arc (0:-180:2);
      
      \draw [dotted] (-1, 0) circle (1);
      
      \draw [dotted] (1, 0) circle (1);
  \end{scope}
  
  \begin{scope}[yscale=0.3, xscale=1.3, shift={(0,5)}, line width=1pt]
      \draw [] (-1, 0) circle (1);
      
      \draw [] (1, 0) circle (1);
  \end{scope}
  \draw [thick,dotted] (-2.6,0) to[out=85,in=-85] (-2.6,1.5);
  \draw [thick] (-2.6,0) to[out=95,in=-95] (-2.6,1.5);
  \draw [very thick,red,postaction={decorate,decoration={markings,mark=at position 0.3 with {\arrow[scale=0.8]{>}}}}] (-0.02,0) to[out=96,in=-96] (-0.02,1.5);
  \draw [very thick,red,postaction={decorate,decoration={markings,mark=at position 0.3 with {\arrow[scale=0.8]{>}}}}]  (0.02,1.5) to[out=-84,in=84] (0.02,0);
  \draw [very thick,red] (-0.02,0) to[out=-84,in=-96] (0.02,0);
  \draw [very thick,red] (-0.02,1.5) to[out=84,in=96] (0.02,1.5);
  \draw [thick] (2.6,0) to[out=85,in=-85] (2.6,1.5);
  \draw [thick,dotted] (2.6,0) to[out=95,in=-95] (2.6,1.5);
  \end{scope}
  \begin{scope}[scale=1.2]
  \begin{scope}[yscale=0.3, xscale=1.3, shift={(0,5)}, line width=1pt]
      \draw [] (0, 0) circle (2);
  \end{scope}
  \end{scope}
  \node [] at (-1.4,1.72) {$C_1$};
  \node [] at (1.4,1.72) {$C_2$};
  \node [] at (-3.5,0.8) {$C_3$};
\end{tikzpicture}
\caption{The thermal circle near the hot spot at the origin, highlighted in red. Starting from the top, we move along cylinder $C_2$ to the top, and then back along cylinder $C_1$ to the top. \label{eq:hotspotorigin}}
\end{figure}

For example, consider the origin in one of the copies of $\R^d$, where the balls $B_1$ and $B_2$ are tangent. Starting at the origin, there is a circular path of length $\b_1+\b_2$ that runs along one cylinder $C_2$, and then back along the other $C_1$, see figure~\ref{eq:hotspotorigin}. In the limit where $\b_1,\b_2$ are both small, this circular path shrinks and we have a hot spot.

\begin{figure}
\centering
\begin{tikzpicture}
\draw [very thick,red,postaction={decorate,decoration={markings,mark=at position 0.65 with {\arrow[scale=0.8]{>}}}}
      ] (0.38,0.45) -- (2.15,0.45);
\draw [very thick,red,postaction={decorate,decoration={markings,mark=at position 0.7 with {\arrow[scale=0.8]{>}}}}
      ] (2.15,0.45) -- (2.15,-1.35);
\draw [very thick,red,postaction={decorate,decoration={markings,mark=at position 0.65 with {\arrow[scale=0.8]{>}}}}
      ] (2.15,-1.35) -- (0.38,-1.35);
\draw [very thick,red,postaction={decorate,decoration={markings,mark=at position 0.65 with {\arrow[scale=0.8]{>}}}}
      ] (0.38,-1.35) -- (0.38,0.45);
\begin{scope}[yscale=0.3,xscale=1.3,line width=1pt]
\draw [white!70!black] (-0.3,0) arc (0:48:0.7);
\draw [thick,white!70!black] (-1,0) -- (0.3,1.5);
\draw [thick,white!70!black] (-1,0) -- (1,0);
\draw [] (-1,-2) arc (-90:90:2);
\draw [] (3,2) arc (90:270:2);
\end{scope}
\begin{scope}[yscale=0.3,xscale=1.3,line width=1pt,shift={(0,-6)}]
\draw [] (-1,-2) arc (-90:0:2);
\draw [dotted] (1,0) arc (0:90:2);
\draw [dotted] (3,2) arc (90:180:2);
\draw [] (1,0) arc (180:270:2);
\end{scope}
\draw [thick] (1.3,0) -- (1.3,-1.8);
\node [] at (-0.75,0.35) {$\theta$};
\node [] at (2.2,0.75) {$\ptl B_2$};
\node [] at (0.3,0.75) {$\ptl B_1$};
\node [] at (2.3,-1.65) {$\ptl B_2'$};
\node [] at (0.3,-1.65) {$\ptl B_1'$};
\end{tikzpicture}
\caption{An approximation to the thermal circle, at an angle $\theta$ away from the origin. Starting at the top left, we move horizontally from $\ptl B_1$ to $\ptl B_2$. Then we move down along $C_2$ to $\ptl B_2'$. Then horizontally to the left to $\ptl B_1'$, then up along $C_1$ to the starting point. The path has approximate length $\b_1+\b_2+2\theta^2$. \label{fig:approxthermcircle}}
\end{figure}
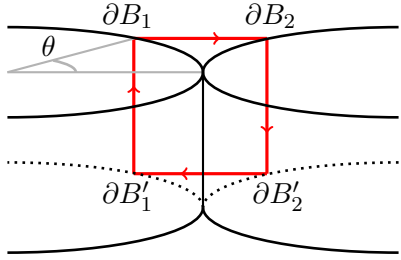

To build some intuition, let us compute the leading contribution to the thermal effective action near this hot spot. The local temperature is highest at the origin, and decays away from it. To determine the local temperature precisely, we should find a (locally defined) conformal Killing vector field that moves around the hot spot's thermal circle. At leading order near the origin, we can guess what it looks like without too much calculation. Consider a path that starts at the point $(-1+\cos \theta,\sin\theta,0,\dots)$ on $\ptl B_1$. Move horizontally to the point $(1-\cos\th,\sin\th,0,\dots,0)$ on $\ptl B_2$, through cylinder 2 to $\ptl B_2'$, horizontally from $\ptl B_2'$ to $\ptl B_1'$, and back through cylinder 1 to the initial point on $\ptl B_1$, see figure~\ref{fig:approxthermcircle}. This path has length 
\be
\beta_1+\beta_2 + 4(1-\cos \theta) \approx \beta_1 + \beta_2 + 2\theta^2 \qquad (\th\ll 1).
\ee
When $\theta$ is small, we expect this path to be close to the orbit of a local conformal Killing vector. Thus, the local temperature is approximately
\be
\beta(\theta) &\approx \b_1+\b_2 + 2\th^2.
\ee

The leading contribution to the thermal effective action near this hot spot is thus
\be
-S_{\textrm{hot}}(\b_1,\b_2) &\sim f \vol\, S^{d-2} \int d\th \frac{\sin^{d-2} \th}{(\b_1+\b_2+2\th^2)^{d-1}}.
\ee
Here, $\vol\, S^{d-2} \sin^{d-2} \th$ comes from an integral over azimuthal angles. When $\b_1+\b_2$ is small, the integral will be dominated by small $\th \sim \sqrt{\b_1+\b_2}$. To compute it, we can approximate $\sin^{d-2}\th\approx \th^{d-2}$ and extend the $\th$-integral from 0 to $\oo$:
\be
\label{eq:actionfromhotspot}
-S_{\textrm{hot}}(\b_1,\b_2) &\sim f \vol\, S^{d-2} \int_0^\oo \frac{d\th\,\th^{d-2}}{(\b_1+\b_2+2\th^2)^{d-1}} = \frac{f \vol\, S^{d-1}}{(8(\b_1+\b_2))^{\frac{d-1}{2}}}.
\ee

Note that the integral is dominated near the hot spot, i.e.\ in the neighborhood $\theta\sim \sqrt{\b_1+\b_2}$. This justifies our use of the thermal effective action everywhere inside the integrand. Furthermore, when $\b_1+\b_2$ is small, we find a large negative action from the hot spot, which translates into a large multiplicative contribution to the partition function $Z\sim e^{-S}$.

\subsection{A more precise formula for the action of a hot spot}

The fact that the integral (\ref{eq:actionfromhotspot}) is dominated near the hot spot suggests a more precise and illuminating way to derive it. Let us assume that the action of a hot spot doesn't depend on the geometry far outside the neighborhood $\theta\sim \sqrt{\b_1+\b_2}$. Thus, to compute it, it suffices to consider a ``genus-1" version of the geometry discussed in section~\ref{sec:choiceofgeometry}, where we have only two balls $B_1$ and $B_2$.

We claim that this ``genus-1" geometry is Weyl equivalent to $S^1_{\b_{12}} \x S^{d-1}$ with a special inverse temperature $\b_{12}$ that depends on $\b_1$ and $\b_2$. The Weyl transformation that implements this equivalence essentially spreads out the hot-spot over the entire $S^{d-1}$, resulting in a uniform inverse temperature $\b_{12}$. The result is
\be
\label{eq:hotspotformula}
-S_{\textrm{hot}}(\b_1,\b_2) &\sim \log Z_{S^1\x S^{d-1}}(\b_{12}) + \textrm{Weyl terms},
\ee
where ``Weyl terms" are possible contributions from the Weyl anomaly, and ``$\sim$" indicates that both sides have the same singular parts as $\b_1,\b_2\to 0$.

Now, $\b_{12}$ is determined by the conformal structure of our ``genus-1" manifold, so we can read it off from the gluing group elements $g_1$ and $g_2$ associated to the two cylinders, given in (\ref{eq:ourparametrization}). Gluing two copies of the plane with $g_1$ and $g_2$ is equivalent to gluing a single copy of the plane to itself with $g_1^{-1} g_2$. To read off $\b_{12}$, we must simply diagonalize $g_1^{-1} g_2$:
\be
\label{eq:comparingthetrace}
g_1^{-1} g_2 &= U e^{-\b_{12} D+i\vec\th_{12}\.\vec M} U^{-1},\quad U\in \SO(d+1,1).
\ee
In other words, $\b_{12}$ is precisely the ``relative" inverse temperature defined in (\ref{eq:relativecoords}).

There is a particularly nice expression for $\b_{12}$ when the angular fugacities are turned off.
In this case, the group elements $g_1,g_2$ are built from conformal generators $P^1,D,K^1$, that generate a $\mathrm{PSL}(2,\R)$ subgroup of the conformal group. Thus, we can obtain $\b_{12}$ by computing $g_1^{-1} g_2$ inside $\SL(2,\R)$ and comparing the trace of both sides of (\ref{eq:comparingthetrace}) as $2\x 2$ matrices. We should compare them up to a sign, since the 1d conformal group $\mathrm{PSL}(2,\R)$ is a quotient of $\SL(2,\R)$ modulo $\pm 1$. This gives
\be
\pm\Tr\begin{pmatrix} e^{-\b_{12}/2} & 0 \\
0 & e^{\b_{12}/2}
\end{pmatrix} &= \Tr \begin{pmatrix}
e^{\frac{\b_1-\b_2}{2}}(1-2 e^{\b_2}) & e^{\frac{\b_1-\b_2}{2}}+e^{\frac{\b_2-\b_1}{2}}-2 e^{\frac{\b_1+\b_2}{2}} \\
-2 e^{\frac{\b_1+\b_2}{2}} & e^{\frac{\b_2-\b_1}{2}}(1-2 e^{\b_1})
\end{pmatrix}.
\ee
To find a solution, we must choose the $-$ sign, which gives
\be
\label{eq:formulaforbetaonetwo}
\b_{12} &= 2 \cosh^{-1} \p{2 e^{\frac{\b_1+\b_2}{2}} - \cosh \p{\frac{\b_1-\b_2}{2}}}.
\ee
This is the inverse temperature at which we should evaluate (\ref{eq:hotspotformula}).

In the limit where $\b_1,\b_2$ become small, the relative inverse temperature $\b_{12}$ has the expansion
\be
\b_{12} &\sim \sqrt{8(\b_1+\b_2)} - \frac{\b_1^2  - 10 \b_1\b_2+ \b_2^2}{12\sqrt 2\sqrt{\b_1+\b_2}} + O(\beta_i^{5/2})
\quad (\b_1,\b_2 \ll 1).
\ee
Consequently, the leading contribution to the thermal effective action (\ref{eq:freeenergy}) is
\be
-\log Z_{S^1\x S^{d-1}}(\b_{12}) &\sim \frac{f \vol\, S^{d-1}}{(8(\b_1+\b_2))^{\frac {d-1}{2}}} + O(\beta_i^{-\frac{d-3}{2}}),\quad (\b_1,\b_2\ll 1),
\ee
in perfect agreement with (\ref{eq:actionfromhotspot})! We have recovered our earlier result for the leading action of a hot spot. However, an advantage of this more abstract derivation is that we expect  (\ref{eq:hotspotformula}) and (\ref{eq:formulaforbetaonetwo}) to encompass all singular terms in the small $\b_1,\b_2$ limit.

This derivation is also straightforward to generalize to the case with angular fugacities. Let us think of the $S^1\x S^{d-1}$ partition function as a class function  $Z_{S^1\x S^{d-1}}(g)$ of a conformal group element $g$. The old notation $Z_{S^1\x S^{d-1}}(\b,\vec \th)$ is obtained by setting $g=e^{-\b D+\vec \th\.\vec M}$. Then the above argument implies that the singular part of the action of a hot spot associated to two group elements $g_1,g_2$ is
\be
\label{eq:estimateforhotspot}
-S_{\textrm{hot},12} &\sim \log Z_{S^1\x S^{d-1}}(g_1^{-1} g_2) + \textrm{Weyl terms}.
\ee

Let us comment on the Weyl anomaly terms in (\ref{eq:estimateforhotspot}). In the genus-1 case, one can check that contributions from the Weyl anomaly to (\ref{eq:estimateforhotspot}) vanish in the limit $\b_1,\b_2\to 0$. In particular, they do not contribute to the singular part of the partition function in the high temperature limit. In what follows, we will assume that the same is true at higher genus, so that analogous Weyl terms can be ignored for our purposes. It would be nice to make these contributions more precise in an example theory.

\subsection{The hot spot action for the genus-2 case}
\label{sec:hotspotgenustwo}

Let us finally apply this result to our ``genus-2" partition function. We conjecture that the singular part of the log of the partition function as $\b_1,\b_2,\b_3\to 0$ is given by a sum of hot spot actions for each pair of tangent balls. Combined with (\ref{eq:estimateforhotspot}), this implies
\be
\label{eq:hotspotgenustwo}
\log Z(M_2) &\sim -S_{\textrm{hot},12}-S_{\textrm{hot},23}-S_{\textrm{hot},31} \nn\\
&\sim \log Z_{S^1\x S^{d-1}}(g_1^{-1} g_2)+\log Z_{S^1\x S^{d-1}}(g_2^{-1} g_3)+\log Z_{S^1\x S^{d-1}}(g_3^{-1} g_1).
\ee
Another way to state the conjecture is as follows. Consider the ratio
\be
R(\beta_i) = \frac{Z(M_2)}{Z_{S^1\x S^{d-1}}(g_1^{-1} g_2) Z_{S^1\x S^{d-1}}(g_2^{-1} g_3) Z_{S^1\x S^{d-1}}(g_3^{-1} g_1)}.
\ee
We conjecture that $R(\beta_i)$ has a finite limit as $\b_i\to 0$:
\be
R = \lim_{\b_i\to 0}R(\beta_i) < \oo.
\label{eq:Rdefinition}
\ee
Intuitively, we imagine that dividing by the hot-spot partition function $Z_{S^1\x S^{d-1}}(g_i^{-1} g_j)$ allows us to define a kind of renormalized ``hot-spot operator" in the limit $\b_i\to 0$ --- a CFT operator that lives at a location where a circle shrinks to zero size. The quantity $R$ is then a correlator of three such hot-spot operators. It would be very interesting to make this statement more precise and compute $R$ in some example theories.

In this work, we will mostly be concerned with the leading singularity of the partition function that follows from (\ref{eq:hotspotgenustwo}). Using (\ref{eq:freeenergy}), this is
\be
\label{eq:leadingpartition}
Z(M_2) &\sim \exp\p{\frac{f\vol\, S^{d-1}}{\b_{12}^{d-1}\prod_a(1+\Omega_{12,a}^2)} + \frac{f\vol\, S^{d-1}}{\b_{23}^{d-1}\prod_a(1+\Omega_{23,a}^2)} + \frac{f\vol\, S^{d-1}}{\b_{31}^{d-1}\prod_a(1+\Omega_{31,a}^2)}},
\ee
where the relative angular velocities are given by $\Omega_{ij,a}=\b_{ij}\theta_{ij,a}$. We define the ``high temperature" regime of the genus-2 partition function as $\b_{ij}\to 0$, with $\vec \Omega_{ij}$ held fixed. This is the physical regime where the thermal effective action can be applied to each hot spot. Here, ``$\sim$" means that the logs of both sides agree, up to subleading terms as $\b_{ij}\to 0$.

If the CFT has a global symmetry $\G$, we can decorate each cylinder by a topological defect associated to a group element $\g_i$ ($i=1, 2, 3$). 
If we do so, the coefficient $f$ 
in (\ref{eq:leadingpartition})
for the $(i, j)$ hot spot becomes a function of
the conjugacy class of $\g_i \g_j^{-1} \in \Gamma$. Thus, in general, we can have 
a different $f$ for each hot spot. 

Note that the angular parameters $\vec \theta_{ij}$ scale to zero at high temperature. In this work, we will be particularly interested in a limit of low spin, where the $\vec \Omega_{ij}$ will scale to zero as well at an appropriate saddle point (so that the $\vec \theta_{ij}$ are parametrically smaller than the $\b_{ij}$). Let us further expand the partition function in this regime. It will be convenient to parametrize the rotations $h_i$ as a product of a rotation away from the $x^1$ axis, times an $\SO(d-1)$ rotation:
\be
\label{eq:hdecomp}
h_i &= \exp\p{\sum_{b=2}^d i\a_{i,b} M_{1b}}\exp\p{\sum_{2\leq a < b \leq d} i\Phi_{i,ab} M_{ab}},
\ee
Here, $\vec \a_i=(\a_{i,2},\dots,\a_{i,d})$ transforms like a vector under $\SO(d-1)$, while the $\SO(d-1)$ parameters $\vec \Phi_i=(\Phi_{i,23},\dots,\Phi_{i,d-1,d})$ transform like an adjoint under $\SO(d-1)$.

Recall that the $h_i$ are subject to the gauge-redundancy (\ref{eq:sodgaugeredudnancy}). The right action of $\SO(d-1)$ simultaneously shifts the $\vec \Phi_i$ (to leading order). Thus, the partition function must be translation-invariant in the $\vec \Phi_i$. Under the left action by $\SO(d-1)$, the $\vec \a_i$ and $\vec \Phi_i$ transform linearly as $\SO(d-1)$ vectors and adjoints, respectively. So the partition function must also be invariant under $\SO(d-1)$ rotations of these variables.

Indeed, expanding (\ref{eq:leadingpartition}) in small angles, we find 
\be
\label{eq:quadraticexpansion}
\frac{1}{\b_{12}^{d-1}\prod_{a=1}^n(1+\Omega_{12,a}^2)} &= \frac{1}{\b_{12,0}^{d-1}} \p{1- \frac{(\vec \Phi_1 - \vec \Phi_2)^2}{\b_{12,0}^2} - 8(d+1)\frac{(\vec\a_{1}+\vec \a_2)^2}{ \b_{12,0}^4} +\dots},
\nn\\
\frac{1}{\b_{23}^{d-1}\prod_{a=1}^n(1+\Omega_{23,a}^2)} &= \frac{1}{\b_{23,0}^{d-1}} \p{1- \frac{(\vec \Phi_2 - \vec \Phi_3)^2}{\b_{23,0}^2} - 8(d+1)\frac{(\tfrac 1 4\vec \a_2 - \tfrac 1 2 \vec \a_3)^2}{ \b_{23,0}^4} +\dots},
\nn\\
\frac{1}{\b_{31}^{d-1}\prod_{a=1}^n(1+\Omega_{31,a}^2)} &= \frac{1}{\b_{31,0}^{d-1}} \p{1- \frac{(\vec \Phi_3 - \vec \Phi_1)^2}{\b_{31,0}^2} - 8(d+1)\frac{(\tfrac 1 4\vec \a_1 - \tfrac 1 2 \vec \a_3)^2}{ \b_{31,0}^4} +\dots},
\ee
This formula is valid when $\vec \a/\b\ll 1$, and $\vec \Phi/\b^{1/2}\ll 1$, and $\b\ll 1$.
Here, $\b_{ij,0}$ denote the relative temperatures {\it when the angles are set to zero}. They are given by
\be
\label{eq:hightemperaturebetas}
\b_{12,0} &= \sqrt{8(\b_1+\b_2)} + \dots, \nn\\
\b_{23,0} &= \sqrt{8(\tfrac 1 4 \b_2 +\tfrac 1 2 \b_3)} + \dots, \nn\\
\b_{31,0} &= \sqrt{8(\tfrac 1 4 \b_1 +\tfrac 1 2 \b_3)} + \dots.
\ee

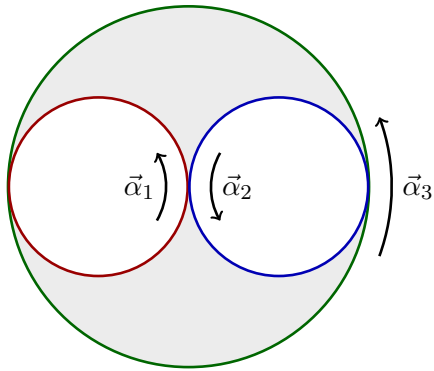
\begin{figure}
\centering
\begin{tikzpicture}[scale=1.2, line width=1pt]
    \fill[even odd rule, fill=gray!15] (0, 0) circle (2) (-1, 0) circle (0.99) (1, 0) circle (0.99);
    
    \draw [green!40!black] (0, 0) circle (2);
    
    \draw [red!60!black] (-1, 0) circle (0.99);
    
    \draw [blue!70!black] (1, 0) circle (0.99);
    \draw [->] (-0.350481, -0.375) arc (-30:30:0.75);
    \draw [->] (0.350481, 0.375) arc (150:210:0.75);
    \draw [->] (2.11431, -0.769545) arc (-20:20:2.25);
    \node [left] at (-0.25,0) {$\vec \a_1$};
    \node [right] at (0.25,0) {$\vec \a_2$};
    \node [right] at (2.25,0) {$\vec \a_3$};
    \end{tikzpicture}
\caption{The partition function (\ref{eq:quadraticexpansion}) penalizes rotations $\vec \a_i$ in such a way that the spheres behave like three interlocked gears. No matter what signs we choose for the $\vec \a_i$, there is no way to rotate the gears, since two of them will always be counter-rotating at their points of contact. \label{fig:rotatinggears}}
\end{figure}

Let us understand some physical implications of (\ref{eq:quadraticexpansion}). Terms like $(\vec \a_1+\vec \a_2)^2$ come from rotating the spheres so that they rub against each other (figure~\ref{fig:rotatinggears}). The thermal effective action penalizes such rotations --- the spheres behave like gears that are interlocked. It follows that there is no zero mode associated with moving the $\vec \a_i$'s: three mutually interlocked circular gears cannot be rotated. Meanwhile, terms like $(\vec \Phi_i-\vec \Phi_j)^2$ represent the effect of twisting the spheres by different amounts around their point of tangency. Such twists are also penalized by the effective action, but by a smaller power of $\b_{ij,0}^2$. To summarize, the only zero mode in the angular parameters is associated to the gauge symmetry of right multiplication by $\SO(d-1)$.

\section{Genus-2 global conformal blocks}
\label{sec:conformalblock}

To determine the asymptotics of CFT OPE coefficients, we must invert the conformal block expansion (\ref{eq:genus2blockexpansion}) of the genus-2 partition function $Z(M_2)$. In particular, we will need the large-$\De$ limit of the genus-2 conformal blocks $B_{123}^{s's}$ in the high-temperature regime discussed in section~\ref{sec:hotspotgenustwo}.

Our strategy will be to write an integral representation for the block using the ``shadow formalism." In the large-$\De$ regime, the integral can be evaluated by saddle point, yielding simple closed-form expressions in the regimes of interest. 

\subsection{Review: shadow integrals for four-point blocks}

Let us first review this strategy in the more familiar case of conformal blocks for four-point functions of local operators on $\R^d$ \cite{Ferrara1972,Dolan:2000ut,Simmons-Duffin:2012juh,Karateev:2018oml}. We will follow the notation and conventions of \cite{Karateev:2018oml}.

The central objects in the shadow formalism are principal series representations and their matrix elements. Let $\pi=(\De,\lambda)$ denote a conformal representation, where $\lambda$ is a representation of $\SO(d)$. The principal series corresponds to (unphysical) complex dimensions of the form $\De=\frac d 2 + i s$, where $s\in \R$. States in a principal series representation are given by functions $f^a(x)$ that transform like conformal primaries with dimension $\De$ and rotation representation $\lambda$. Here, $a$ is an index for $\lambda$. Such states admit a Hermitian inner product
\be
\label{eq:principalseriesinnerproduct}
(g|f) &\equiv \int d^d x (g^a(x))^* f^a(x),
\ee
where the index $a$ is summed over. Note that $(g^a(x))^*$ has scaling dimension $\frac d 2 - i s = d - \p{\frac d 2 + i s}$, so the integrand (including the measure $d^dx$) has scaling dimension 0. Furthermore, it transforms in the dual rotation representation $\lambda^*$, so the integrand is rotation-invariant. It follows that the pairing $(g|f)$ is conformally-invariant.

The principal series representation $\pi=(\frac d 2 + i s, \lambda)$ is isomorphic to the ``shadow" representation $\tl \pi=(\frac d 2 - i s, \lambda^R)$, where $\lambda^R$ denotes the reflection of $\lambda$. This isomorphism is implemented by the shadow transform:
\be
\label{eq:shadowtrans}
\mathbf{S}[f](x) &= \int d^d y \<\tl\cO(x)\tl\cO^\dag(y)\> f(y),
\ee
where $\<\tl \cO(x) \tl \cO^\dag(y)\>$ denotes the unique (up to scale) conformal two-point structure between operators in the representations $\tl \pi$ and $\tl\pi^\dag \equiv (\frac d 2 - i s, \lambda^*)$.

Conformal three-point functions can be thought of as Clebsch-Gordon coefficients for a tensor product of principal series representations. Such three-point functions carry a structure label $s$ that corresponds to different solutions of the conformal Ward identities:
\be
\label{eq:solutiontothreeptward}
V^{s;abc}(x_1,x_2,x_3) &= \<\cO^a_1(x_1)\cO^b_2(x_2) \cO^c_3(x_3)\>^{(s)}.
\ee
Here, $\<\cdots\>^{(s)}$ denotes a solution to the conformal Ward identities --- not a physical three-point function. (In particular, it does not include an OPE coefficient.) The space of three-point structures is given by $(\lambda_1\otimes \lambda_2 \otimes \lambda_3)^{\SO(d-1)}$, where $\SO(d-1)$ indicates the $\SO(d-1)$-invariant subspace \cite{Kravchuk:2016qvl}. We sometimes write $V^{s}$ for a three-point structure, and we sometimes use the notation on the right-hand side of (\ref{eq:solutiontothreeptward}).

With these ingredients, we are ready to build conformal blocks.
Four-point blocks are eigenfunctions of the conformal Casimir acting simultaneously on points $1$ and $2$, obeying certain boundary conditions. Using the inner product on principal series representations, we can instead easily build an eigenfunction called a ``conformal partial wave" from two three-point structures:
\be
\Psi^{s's}_{\pi}(x_1,\cdots,x_4) &= \int d^d x \<\cO_3(x_3)\cO_4(x_4)\tl\cO^\dag(x)\>^{(s')} \<\cO_1(x_1)\cO_2(x_2)\cO(x)\>^{(s)} .
\ee
Here, $\tl \cO^\dag$ has representation $\tl \pi^\dag=(\frac d 2 - i s,\lambda^*)$, so that it can be paired with $\cO(x)$ inside the integral. We omit spin indices for brevity.

The partial wave $\Psi_{\pi}^{ss'}$ satisfies the same Casimir differential equations as a conformal block, but obeys different boundary conditions. However, it gets us ``most of the way" to a block, and the block can be extracted from it with a small amount of extra work. The key point is that the space of solutions of the Casimir equations is 2-dimensional. It is spanned by the conformal block, and a so-called ``shadow" block for the representation $\tl \pi$.  It follows that the partial wave can be written as a linear combination of the block and its shadow:
\be
\label{eq:relationbetweenpartialwaveandblock}
\Psi^{s's}_\pi &= S(\pi_3\pi_4[\tl\pi^\dag])^{s'}{}_{t'} G_\pi^{t's} + S(\pi_1\pi_2[\pi])^{s}{}_{t} G^{s't}_{\tl \pi}.
\ee
Here, the ``shadow" coefficients $S(\pi_1\pi_2[\pi])^s{}_t$ are obtained by applying shadow transformations to a three-point structure. For example,
\be
\mathbf{S}_3V^s(x_1,x_2,x_3) &= S(\pi_1\pi_2[\pi])^s{}_tV^t(x_1,x_2,x_3),
\ee
where $\mathbf{S}_3$ denotes the shadow transform (\ref{eq:shadowtrans}) acting at $x_3$. The reason these coefficients appear in (\ref{eq:relationbetweenpartialwaveandblock}) is explained in \cite{Karateev:2018oml}. 
Starting from (\ref{eq:relationbetweenpartialwaveandblock}), we can isolate the block $G_\pi^{s's}$ using a ``monodromy projection" \cite{Simmons-Duffin:2012juh}, as we explain in more detail later.

To summarize, the shadow formalism gives a convenient integral representation of a partial wave, satisfying the same differential equations as a conformal block, and from which the block can be extracted. This approach will work for the genus-2 blocks $B_{123}^{s's}$ as well. Integral representations are particularly useful for studying large quantum number asymptotics, since we can use saddle point methods.

There exist alternative constructions of four-point conformal blocks via shadow-like integrals in Lorentzian signature \cite{Polyakov:1974gs}. These have the advantage of giving the block ``on the nose," eschewing the need for a monodromy projection. Finding a similar Lorentzian shadow representation for the genus-2 block is an interesting problem for the future.

\subsection{A genus-2 partial wave}
\label{sec:genustwopw}

The genus-2 conformal block $B_{123}^{s's}$ is a simultaneous eigenfunction of the conformal Casimir operators acting on each of the group elements $g_1,g_2,g_3$. In more detail, let $L^A$ $(A=1,\dots,\dim G)$ be the generators of the Lie algebra of $G$, realized as left-invariant vector fields on $G$. Then $\cD=L^A L_A$ is a differential operator on $G$ such that for any irrep $\pi$, we have
\be
\cD \pi(g) = \pi(g)  C_2(\pi),
\ee
where $C_2(\pi)$ is the Casimir eigenvalue for $\pi$. Any matrix element of $g$ in the representation $\pi$ is thus an eigenfunction of $\cD$. Viewing the conformal block as a matrix element of three group elements $g_1,g_2,g_3$ in the representations $\pi_i$, it follows that it must be a simultaneous eigenfunction of $\cD$, acting on each of the $g_i$:
\be
\label{eq:genustwocasimireqs}
\cD_i B_{123}^{s's} = C_2(\pi_i) B_{123}^{s's} \qquad i=1,2,3,
\ee
where $\cD_i$ indicates the action of $\cD$ on $g_i$. The block also diagonalizes the higher Casimirs of the conformal group, acting on each $g_i$.

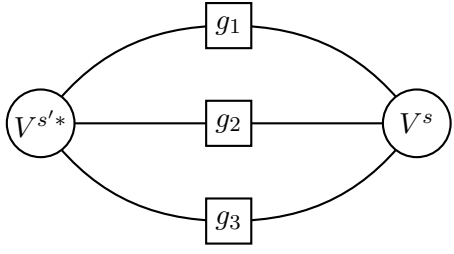
\begin{figure}
\centering
\begin{tikzpicture}
\draw [thick] (0,0) to[out=55,in=180] (2.5,1.3) to[out=0,in=125] (5,0);
\draw [thick] (0,0) -- (2.5,0) -- (5,0);
\draw [thick] (0,0) to[out=-55,in=180] (2.5,-1.3) to[out=0,in=-125] (5,0);
\begin{scope}[shift={(0,0)}]
\draw [thick,fill=white] (0,0) circle (0.45);
\node [] at (0,0) {$V^{s'*}$};
\end{scope}
\begin{scope}[shift={(5,0)}]
\draw [thick,fill=white] (0,0) circle (0.45);
\node [] at (0,0) {$V^{s}$};
\end{scope}
\begin{scope}[shift={(2.5,1.3)}]
\draw [thick,fill=white] (-0.3,-0.3) -- (0.3,-0.3) -- (0.3,0.3) -- (-0.3,0.3) -- cycle;
\node [] at (0,0) {$g_1$};
\end{scope}
\begin{scope}[shift={(2.5,0)}]
\draw [thick,fill=white] (-0.3,-0.3) -- (0.3,-0.3) -- (0.3,0.3) -- (-0.3,0.3) -- cycle;
\node [] at (0,0) {$g_2$};
\end{scope}
\begin{scope}[shift={(2.5,-1.3)}]
\draw [thick,fill=white] (-0.3,-0.3) -- (0.3,-0.3) -- (0.3,0.3) -- (-0.3,0.3) -- cycle;
\node [] at (0,0) {$g_3$};
\end{scope}
\end{tikzpicture}
\caption{A ``tensor diagram" for the genus 2 partial wave (\ref{eq:genustwopartialwave}). The three-point structures $V^s$ and $V^{s'*}$ are invariant tensors for a tensor product of three principal series representations, so they are represented as trivalent nodes. Each group element acts as a linear operator on a representation, so is a bivalent node. Lines connect together using the inner product (\ref{eq:principalseriesinnerproduct}). We act on each of the legs of $V^s$ with group elements $g_1,g_2,g_3$ before contracting with $V^{s'*}$. \label{fig:genustwodiag}}
\end{figure}

By analogy with the four-point case, we can define a genus-2 partial wave as a principal series matrix element of $g_i$'s between a pair of three-point structures:
\be
\Psi^{s's}_{123} &\equiv (V^{s'}|g_1\otimes g_2\otimes g_3|V^{s}) \nn\\
&= \int d^d x_1 d^d x_2 d^d x_3 V^{s'*}(x_1,x_2,x_3) g_1 g_2 g_3 \. V^{s}(x_1,x_2,x_3).
\label{eq:genustwopartialwave}
\ee
This is illustrated diagrammatically in figure~\ref{fig:genustwodiag}.
The action of a conformal group element on an operator is
\be
\label{eq:theformula}
g\cdot \cO^a(x) &= \Omega(x')^{\De} \lambda^a{}_b(R^{-1}(x')) \cO^b(x').
\ee
The notation $g_1 g_2  g_3 \cdot V^{s}(x_1,x_2,x_3)$ indicates the simultaneous actions of $g_1,g_2,g_3$ on the three-point structure $V^{s}$, using the formula (\ref{eq:theformula}) at each point. The spin indices of the operators are implicitly contracted in (\ref{eq:genustwopartialwave}). By construction, the partial wave also solves the Casimir equations (\ref{eq:genustwocasimireqs}).

Though we have not proven it, we expect that the space of solutions of the Casimir equations (\ref{eq:genustwocasimireqs}) is 8-dimensional, and is spanned by the block $B_{123}^{s's}$ and seven ``shadow" blocks obtained by replacing $\pi_i \to \tl \pi_i$ in various combinations: $\{B_{\tl 1 23}^{s's}, B_{1\tl 2 3}^{s's},\cdots, B_{\tl 1 \tl 2 \tl 3}^{s's}\}$.  The genus-2 partial wave is a linear combination of these 8 solutions. Applying similar logic to the derivation of (\ref{eq:relationbetweenpartialwaveandblock}), we expect it to have the form
\be
\label{eq:partialwaveintermsofblocks}
\Psi^{s's}_{123} &= 
(I^{-3} S_{\tl 1^\dag \tl 2^\dag \tl 3^\dag}^3)^{s'}{}_{t'} B^{t's}_{123} + (\textrm{7 shadow blocks}).
\ee
Here $(S^3_{\tl 1^\dag \tl 2^\dag \tl 3^\dag})^s{}_t$ denotes the product of three shadow coefficients coming from performing the shadow transform of $\tl V ^{\dag s'}$ on each of its external legs:
\be
\mathbf{S}_1\mathbf{S}_2\mathbf{S}_3 \tl V^{\dag s'} &= (S^3_{\tl 1^\dag \tl 2^\dag \tl 3^\dag})^{s'}{}_{v'} V^{v'}\nn\\
\label{eq:tripleshadowcoefficient}
&= S([\tl \pi_1]^\dag\tl\pi_2^\dag\tl\pi_3^\dag)^{s'}{}_{t'} S(\pi_1^\dag [\tl \pi_2^\dag] \tl \pi_3^\dag)^{t'}{}_{u'} S(\pi_1^\dag\pi_2^\dag[\tl \pi_3^\dag])^{u'}{}_{v'} V^{v'}.
\ee
(other expressions are possible, coming from doing the shadow transforms in other orders). Meanwhile, $I^{-3}$ indicates the action of inverse inversion tensors $(I(e)^{-1})_{a}{}^{\bar a}$ on each operator. These are needed in order for the resulting three-point structure to transform in the dual representations $\lambda_1^*,\lambda_2^*,\lambda_3^*$, so that it can be paired with $V^s$. The 7 shadow blocks in (\ref{eq:partialwaveintermsofblocks}) will have similar coefficients, though we have not written them explicitly for brevity. We have not attempted to give a rigorous proof of (\ref{eq:partialwaveintermsofblocks}), but instead have motivated it by analogy with the four-point case. We will verify (\ref{eq:partialwaveintermsofblocks}) explicitly in the large-$\De$ limit where it is needed in section~\ref{sec:lowtempsaddles}.

\subsection{Symmetries of the saddle point equations}
\label{sec:saddlesyms}

We will be interested in the blocks and partial waves in the large-$\De$ limit. In this limit, the shadow integral (\ref{eq:genustwopartialwave}) for the genus-2 partial wave can be evaluated by saddle point. The structure of the saddle point equations is complicated, and we will not attempt to solve them exactly for arbitrary $g_i$. However, there are some simple operations that permute the saddle points that will be helpful in exploring their structure. We derive them in this section.

Let us begin with the integral for a partial wave. For simplicity, we will work in $d=1$; the results of this section will generalize straightforwardly to any $d$. The integral takes the form
\be
\Psi_{\De_1,\De_2,\De_3} &= \int dz_1 dz_2 dz_3  V_{\tl \De_1,\tl\De_2,\tl \De_3}(z_1,z_2,z_3) g_1 g_2 g_3\cdot V_{\De_1,\De_2,\De_3}(z_1,z_2,z_3),
\label{eq:partialwaveintegral1d}
\ee
where 
\be
V_{\De_1,\De_2,\De_3}(z_1,z_2,z_3) &= \frac{1}{|z_{12}|^{\De_1+\De_2-\De_3}|z_{23}|^{\De_2+\De_3-\De_1}|z_{31}|^{\De_3+\De_1-\De_2}}
\ee
is a three-point structure for primaries with dimensions $\De_1,\De_2,\De_3$ in $1d$.  Thinking of each $g_i$ as an $\SL(2,\R)$ element, the action of $g_i$ on each operator is given by
\be
\label{eq:actionofgi}
g_i \cdot \cO(z_i) &= (c_i z_i + d_i)^{-2 \De_i} \cO\p{\frac{a_i z_i + b_i}{c_i z_i + d_i}}.
\ee

In the limit of large $\De_i$, the integral is dominated by saddle points. Let us split the integrand into a rapidly-varying part that depends exponentially on $\De_i$, and a part that is slowly-varying at large $\De_i$. We define the saddle point equations as stationarity equations for the rapidly-varying part of the integrand. Concretely they are
\be
\label{eq:thesaddlepointeqs}
\ptl_{z_i} \log \left[ V_{-\De_1,-\De_2,-\De_3}(z_1,z_2,z_3) g_1 g_2 g_3\cdot V_{\De_1,\De_2,\De_3}(z_1,z_2,z_3)\right] &= 0 \qquad (i=1,2,3).
\ee
This is a system of three coupled polynomial equations in the $z_i$, with coefficients that depend on the $g_i$. Note that the saddle-point equations are homogeneous in the $\De_i$ in our conventions. We denote the coordinates of the three points collectively as $\vec p=(z_1,z_2,z_3)$.

Suppose that we can find a saddle point $\vec p_*=(z_{1*},z_{2*},z_{3*})$, i.e.\ a solution of (\ref{eq:thesaddlepointeqs}), as a function of the $\De_i$. A simple operation that relates different saddle points is
\be
\tau \vec p_* &\equiv \left.(g_1^{-1} z_1, g_2^{-1} z_2,g_3^{-1} z_3)\right|_{\De_i\to \tl \De_i},
\ee
where we can approximate $\tl \De_i\approx -\De_i$ at large $\De_i$.
The fact that $\tau \vec p_*$ is a saddle point of (\ref{eq:partialwaveintegral1d}) follows from symmetry of the integrand under the change of variables $z_i\to g_i^{-1} z_i$ and $\De_i\to\tl \De_i$.

However there is another less-obvious operation that relates saddle points to each other, coming from the $\Z_2^3$ shadow symmetry of the partial wave. We start by rewriting (\ref{eq:partialwaveintegral1d}) by introducing a shadow transformation on $z_1$:
\be
&\Psi_{\De_1,\De_2,\De_3} 
\nn\\
&= \frac{1}{S([\De_1]\tl \De_2\tl \De_3)}\int dz_1 dz_1' dz_2 dz_3 \frac{1}{z_{11'}^{2\De_1}}V_{\De_1,\tl\De_2,\tl \De_3}(z_1',z_2,z_3) g_1 g_2 g_3\cdot V_{\De_1,\De_2,\De_3}(z_1,z_2,z_3) 
\nn\\
&= \frac{S([\De_1]\De_2\De_3)}{S([\De_1]\tl \De_2\tl \De_3)}
\int dz_1' dz_2 dz_3 V_{\De_1,\tl\De_2,\tl \De_3}(z_1',z_2,z_3) g_1 g_2 g_3\cdot V_{\tl\De_1,\De_2,\De_3}(z_1',z_2,z_3).
\label{eq:clevertrick}
\ee
In the second line, we performed the integral over $z_1$ and used that the two-point function $z_{11'}^{-2\De_1}$ is $G$-invariant.

The resulting integral (\ref{eq:clevertrick}) has the same form as (\ref{eq:partialwaveintegral1d}), except that $\De_1$ and $\tl \De_1$ have been swapped, and $z_1$ has been swapped with $z_1'$. Thus, a saddle point of (\ref{eq:clevertrick}) is given by
\be
(z_1',z_2,z_3) &= \vec p_*|_{\De_1\to \tl \De_1}.
\ee
So far, we have managed to find a saddle point for a different integral --- not the original integral we started with. 

However, in the large $\De_i$ limit, $z_1'$ can be related to $z_1$ using the integral on the first line of (\ref{eq:clevertrick}). The integral over $z_1$ takes the form of a shadow transform, and the shadow transform is dominated by its own saddle point at large $\De_i$, as we explain in appendix~\ref{sec:shadowintegralslargedelta}. Let us denote the saddle point obtained by shadow-transforming $V_{\De_1,\De_2,\De_3}(z_1,z_2,z_3)$ at site $z_1$ by
\be
 s_{[\De_1]\De_2\De_3}(\vec p) 
&\equiv-\frac{2 \De_1 z_2 z_3-(\De_1+\De_2-\De_3)z_1 z_3 - (\De_1-\De_2+\De_3)z_1 z_2}{2\De_1 z_1 - (\De_1+\De_2-\De_3)z_2 - (\De_1-\De_2+\De_3)z_3} \qquad (\De_i \gg 1).
\ee
Note that $s_{[\De_1]\De_2\De_3}$ satisfies the identity
\be
\label{eq:anidentity}
z_1 &= s_{[\tl\De_1]\De_2\De_3}(s_{[\De_1]\De_2\De_3}(z_1,z_2,z_3),z_2,z_3),
\ee
which is related to the fact that the square of the shadow transform is proportional to the identity.

In our case, we have
\be
z_1' &= s_{[\De_1]\tl\De_2\tl\De_3}(z_1,z_2,z_3).
\ee
Using (\ref{eq:anidentity}), we can solve this as
\be
z_1 &= s_{[\tl \De_1]\tl \De_2\tl \De_3}(z_1',z_2,z_3).
\ee
Putting everything together, we find a saddle point of the original integrand (\ref{eq:partialwaveintegral1d})
\be
\s_1 \vec p_* &\equiv (s_{[\tl \De_1]\tl \De_2\tl \De_3}(\vec p_*|_{\De_1\to \tl\De_1}),\,z_{2*}|_{\De_1\to \tl\De_1},\,z_{3*}|_{\De_1\to \tl\De_1}) \nn\\
&= (s_{[\De_1]\tl \De_2 \tl \De_3}(\vec p_*),z_{2*},z_{3*})|_{\De_1\to \tl\De_1}.
\label{eq:definitionofsigmaone}
\ee
Note that $\s_1 \vec p_*$ may or may not coincide with $\vec p_*$.
We can similarly define operations $\s_2$ and $\s_3$ by cyclic permutations of (\ref{eq:definitionofsigmaone}). Note that $\s_i^2=1$ and the $\s_i$ are mutually commuting. One can also show that $\tau=\s_1\s_2\s_3$.

The $\s_i$ operations give a homomorphism from $\Z_2^3$ into the group of permutations of the saddle solutions. The behavior of this homomorphism can jump when any of the $\De_i$ crosses $0$, since the saddle point analysis of the shadow integral for $z_1,z_1'$ becomes invalid if $\De_1$'s is small. Indeed, we will see that such jumps happen in practice. 

\subsection{Low temperature saddles}
\label{sec:lowtempsaddles}

We are now ready to explore the saddle points of the partial wave integral in different regimes. Let us begin by exploring low temperature $\b_i\to \oo$, where it will be easy to distinguish the block from shadow blocks. We study high-temperature saddles (which will be our main interest) in the next section.

As a reminder, we will use the parametrization of $\cM$ given in (\ref{eq:ourparametrization}). For simplicity, let us first turn off the angular fugacities by setting $h_i=1$. The shadow integral (\ref{eq:genustwopartialwave}) then has an $\SO(d-1)$ symmetry, so we can locate its saddle points by specializing the points to the $x^1$ axis: $x_i=(z_i,0,\dots,0)$.

The $\SO(d-1)$-symmetry also means that the local rotations $R_\mu{}^\nu(x')$ associated to each group element $g_i$ are trivial, since the centralizer of $\SO(d-1)\subset \SO(d)$ is trivial. Thus, each group element acts in a simple way on the $x$-axis:
\be
g_i \cdot \cO^a(x) &= \Omega_i(x')^{\De} \cO^a(g_i x).
\ee
The tensor structures in the numerators of the three-point structures are identical to what they would be in a standard configuration $(x_1,x_2,x_3)=(0,e,\oo)$. The remaining factors are the same as in the $\SL(2,\R)$ transformation of conformal 3-point structures in 1d. Thus, the integrand restricted to the $x^1$-axis becomes
\be
I(x_i)&= V^{s'*}(x_1,x_2,x_3) g_1g_2g_3 \cdot V^{s}(x_1,x_2,x_3)
\nn\\
&= V^{s' *}(0,e,\oo)V^{s}(0,e,\oo) V_{\tl \De_1,\tl\De_2,\tl \De_3}(z_1,z_2,z_3) g_1 g_2 g_3\cdot V_{\De_1,\De_2,\De_3}(z_1,z_2,z_3),
\ee 
where $V_{\De_1,\De_2,\De_3}(z_1,z_2,z_3)$ are 1d conformal three-point functions, and the $g_i$ act via (\ref{eq:actionofgi}).

In the small temperature limit $\b_i\to \oo$, it is straightforward to find at least one solution to the saddle point equations (\ref{eq:thesaddlepointeqs}). We naively expand the equations in the $\b_i\to \oo$ limit to obtain
\be
\label{eq:lowtempexpansionofeqns}
0 &= \frac{(\Delta _1-\Delta _2+\Delta _3) z_2+(\Delta _1+\Delta _2-\Delta _3) z_3-2 \Delta _1 z_1}{z_{12} z_{31}}-\frac{2 \Delta _1}{z_1+1} + O(e^{-\b_i}),\nn\\
0 &= \frac{(\Delta _2+\Delta _3-\Delta _1) z_1+(\Delta _1+\Delta _2-\Delta _3) z_3-2 \Delta _2 z_2}{z_{12} z_{23}}-\frac{2 \Delta _2}{z_2-1}  + O(e^{-\b_i}),\nn\\
0 &= \frac{(\Delta _2+\Delta _3-\Delta _1) z_1+(\Delta _1+\Delta _3-\Delta _2) z_2-2 \Delta _3 z_3}{z_{31} z_{23}} + O(e^{-\b_i}).
\ee
These have the solution 
\be
\vec p_{0,0,0} &\equiv
\p{\frac{3 \Delta _1-\Delta _2+\Delta _3}{\Delta _1+\Delta _2-\Delta _3}
,\frac{\Delta _1-3 \Delta _2-\Delta _3}{\Delta _1+\Delta _2-\Delta _3}
, \frac{\Delta _2-\Delta _1}{\Delta _3}} + O(e^{-\b_i}).
\ee

Then, using the operations defined in section~\ref{sec:saddlesyms}, we can generate 7 additional saddles:
\be
\vec p_{1,0,0} &\equiv \s_1 \vec p_{0,0,0}, \nn\\
\vec p_{0,1,0} &\equiv \s_2 \vec p_{0,0,0},\nn\\
\vec p_{0,0,1} &\equiv \s_3 \vec p_{0,0,0},\nn\\
\vec p_{1,1,0} &\equiv \s_1\s_2 \vec p_{0,0,0} = \tau \vec p_{0,0,1},\nn\\
\vec p_{1,0,1} &\equiv \s_1\s_3 \vec p_{0,0,0}\nn = \tau \vec p_{0,1,0},\\
\vec p_{1,0,1} &\equiv \s_1\s_3 \vec p_{0,0,0}\nn = \tau \vec p_{1,0,0},\\
\vec p_{1,1,1} &\equiv \s_1\s_2\s_3 \vec p_{0,0,0} = \tau \vec p_{0,0,0}.
\ee
As an aside, these additional saddles are more subtle to see directly from the saddle point equations because they involve points scaling towards singularities. For example, in the solution
\be
\label{eq:equationforx100}
\vec p_{1,0,0} &= \p{-1 + \frac{\De_1-\De_2+\De_3}{4\De_1} e^{-\b_1} + \dots,\frac{\De_1+3\De_2+\De_3}{\De_1-\De_2+\De_3} + \dots,\frac{\De_1+\De_2}{\De_3}+\dots},
\ee
the point $z_1$ approaches the center of the ball $B_1$ at $z_1=-1$, which is a singularity of the saddle point equations at low temperatures. To find the solution $\vec p_{1,0,0}$ directly, we cannot use (\ref{eq:lowtempexpansionofeqns}). Instead, we must re-expand the equations near the singularity and re-solve them in a small-temperature expansion, resulting in (\ref{eq:equationforx100}). 

Let us denote the saddle point integral along a steepest descent contour through $\vec p_{a,b,c}$ by $I_{a,b,c}$. Plugging in the different solutions, we find that the $I_{a,b,c}$ have the following behavior in the small temperature regime (as a function of the $\b_i$):
\be
I_{0,0,0} &\sim e^{-\b_1\De_1-\b_2\De_2-\b_3\De_3}, \nn\\
I_{1,0,0} &\sim e^{-\b_1\tl\De_1-\b_2\De_2-\b_3\De_3}, \nn\\
I_{0,1,0} &\sim e^{-\b_1\De_1-\b_2\tl\De_2-\b_3\De_3}, \nn\\
&\dots \nn\\
I_{1,1,1} &\sim e^{-\b_1 \tl \De_1 - \b_2 \tl \De_2 - \b_3 \tl \De_3}.
\ee
More formally, if we consider monodromies $M_i:\b_i \to \b_i + 2\pi i$, then each saddle point integral is an eigenfunction of the monodromies $M_{1,2,3}$ with different eigenvalues. The block is the solution to the Casimir equations with monodromies $M_i B_{123}^{s's}=e^{-2\pi i \De_i} B_{123}^{s's}$. It follows that $I_{0,0,0}$ is the block at low temperatures, while the other saddle point contours give shadow blocks.

Let us finally turn back on the angular fugacities $h_1,h_2,h_3$. In the low-temperature limit, they do not move the saddle point $\vec p_{0,0,0}$. Performing the gaussian integral around $\vec p_{0,0,0}$, and multiplying by the inverse of the triple shadow coefficient computed in (\ref{eq:exprfortripleshadow}), we find a nontrivial cancellation of $\De$-dependent factors, resulting in
\be
\label{eq:gettinglowtempblock}
\left.((I^{-3} S^3_{\tl 1^\dag \tl 2^\dag \tl 3^\dag})^{-1})^{s'}{}_{t'}\Psi_{123}^{t's}\right|_{\vec p_{0,0,0}}
 &= 2^{-2\De_1-2\De_2} e^{-\De_1\b_1-\De_2\b_2-\De_3\b_3} V^{s' *}(0,e,\oo) (h_1 h_2 h_3\. V^{s})(0,e,\oo)\nn\\
&\quad \x (1+O(\De_i^{-1},e^{-\b_i})),
\ee
where the $\vec p_{0,0,0}$ subscript means we evaluate the saddle point integral around $\vec p_{0,0,0}$.
This result agrees precisely with the formula for the block from summing over states (\ref{eq:blocklowtemperature}). This is a check on our assertion that $I_{0,0,0}$ computes the block, and also on our ansatz (\ref{eq:partialwaveintermsofblocks}) for the partial wave as a sum of blocks.

\subsection{High temperature saddles}
\label{sec:hightempsaddle}

We define the high temperature regime as $\b_{ij}\to 0$ with $\vec \Omega_{ij}=\b_{ij}^{-1} \vec \th_{ij}$ fixed. In terms of the coordinates $\b_i,h_i$, this means that $h_i\to 1$ at high temperatures. Our strategy will be to start with the infinite temperature case $\b_i=0$ and $h_i=1$, and then work in perturbation theory at small $\b_i$. At infinite temperature, we restore $\SO(d-1)$ symmetry, so we can again look for solutions along the $x^1$ axis.

It is not obvious a-priori that perturbation theory around infinite temperature makes sense --- what if the block had a singularity at infinite temperature? However, we find in practice that the block is nonsingular at infinite temperature, and this strategy works. Relatedly, we do not find any evidence of a nontrivial difference in the order of limits $\b_i\to 0$ and $\De_i\to \oo$. This situation is somewhat different from the ``$t$-channel" $z,\bar z\to 1$ limit of four-point conformal blocks, where the blocks have nontrivial log or power-law singularities, and one must be careful about orders of limits \cite{Pappadopulo:2012jk,Rychkov:2015lca}. It would be nice to understand these differences in more detail.

In the infinite temperature limit $\b_i=0$, one saddle point is relatively easy to find. We naively expand the saddle point equations and solve them to give:
\be
\vec q_0 &\equiv \p{\tfrac{2 \Delta _3}{2 \Delta _2-\Delta _3},-\tfrac{2 \Delta _3}{2 \Delta _1-\Delta _3},\tfrac{2 \left(\Delta _2-\Delta _1\right)}{\Delta _1+\Delta _2}} + O(\b_i). 
\ee
Interestingly, it turns out that $\s_1 \vec q_0=\s_2 \vec q_0=\s_3 \vec q_0=\tau \vec q_0$, so the operations defined in section~\ref{sec:saddlesyms} generate only one additional high temperature saddle, namely $\tau\vec q_0$.

However, it turns out that there are three additional high temperature saddles where the points $x_1,x_2,x_3$ scale towards each other in the high temperature limit. For example, we find a solution $\vec q_{12}$ given by
\be
\label{eq:equationforq12}
\vec q_{12} &\equiv \p{
-\tfrac{\beta _2 \Delta _1+\beta _1 \left(\Delta _1+\Delta _3\right)}{2 \Delta _3},
\tfrac{\beta _1 \Delta _2+\beta _2 \left(\Delta _2+\Delta _3\right)}{2 \Delta _3},
\tfrac{\beta _1 \left(\Delta _1^2-\Delta _2^2-\Delta _3^2\right)+\beta _2 \left(\Delta _1^2-\Delta _2^2+\Delta _3^2\right)}{4 \Delta _3^2}
} + O(\b_i^2).
\ee
Here, all three points scale toward $x=0$ (the point where balls $B_1$ and $B_2$ are tangent) as $\b_i\to 0$. Similarly, we find a solution $\vec q_{23}$ where the $x_i$ scale toward the point where balls $B_2$ and $B_3$ are tangent, and a solution $\vec q_{31}$ where the $x_i$ scale toward the point where balls $B_3$ and $B_1$ are tangent. The action of $\s_i$ and $\tau$ on these solutions is given by
\be
\s_2 \vec q_{12} = \s_1 \vec q_{12} = \vec q_{12},\qquad \s_3 \vec q_{12}=\tau \vec q_{12},
\ee
and cyclic permutations of these relations. The saddle points $\tau \vec q_{ij}$ are new --- they involve two points scaling towards a singularity, while one remains at a finite position. Thus, overall, we have $8$ high-temperature saddles given by $\vec q_0,\vec q_{12},\vec q_{23},\vec q_{31}$, and their images under $\tau$.

\begin{figure}[ht]
\centering
\begin{subfigure}{0.43\textwidth}
\includegraphics[width=\textwidth,trim={0 0 1.8cm 0},clip]{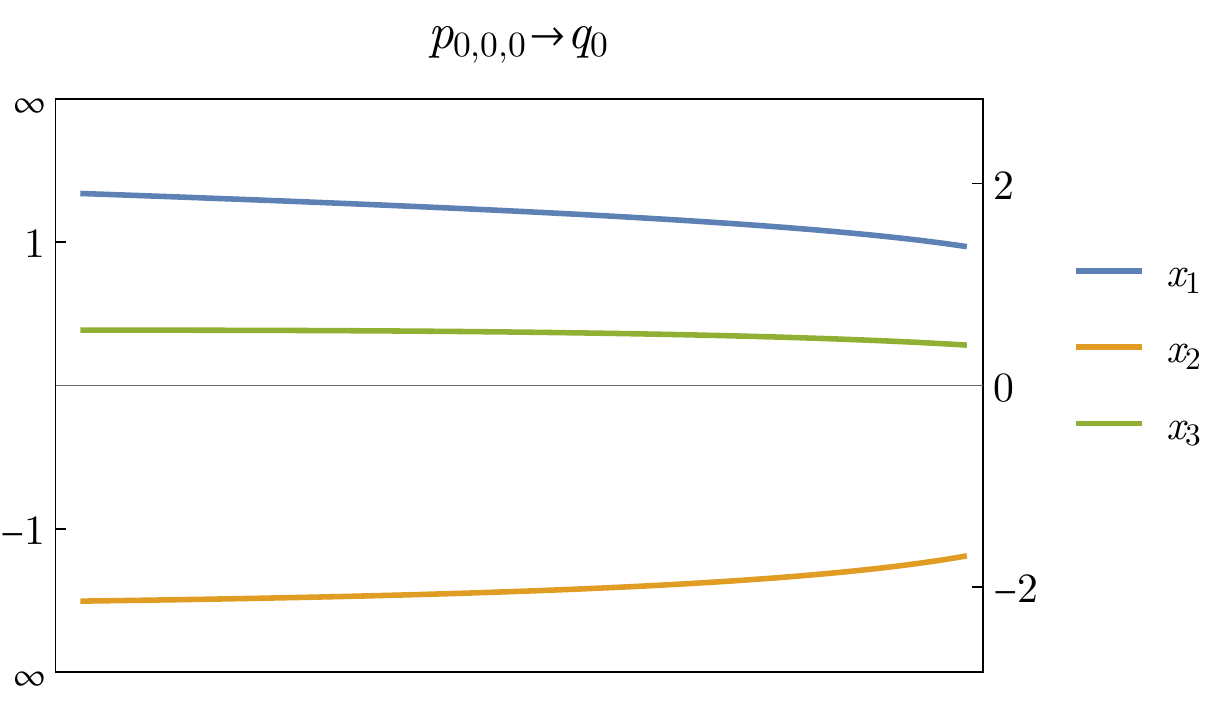}
\end{subfigure}
\begin{subfigure}{0.43\textwidth}
\includegraphics[width=\textwidth]{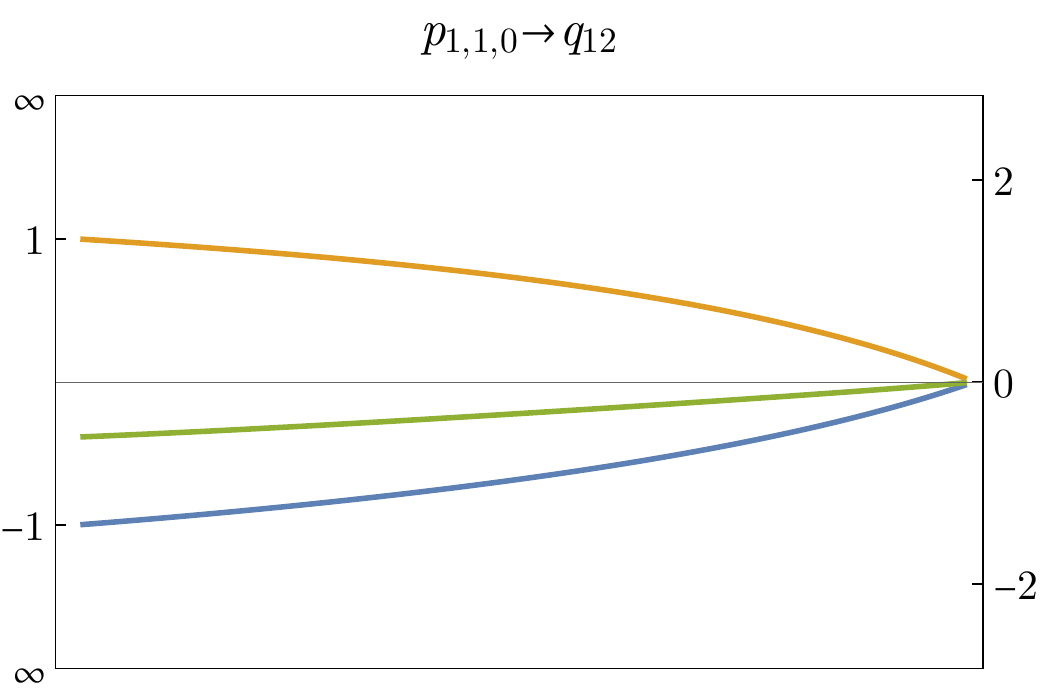}
\end{subfigure}
\begin{subfigure}{0.43\textwidth}
\includegraphics[width=\textwidth]{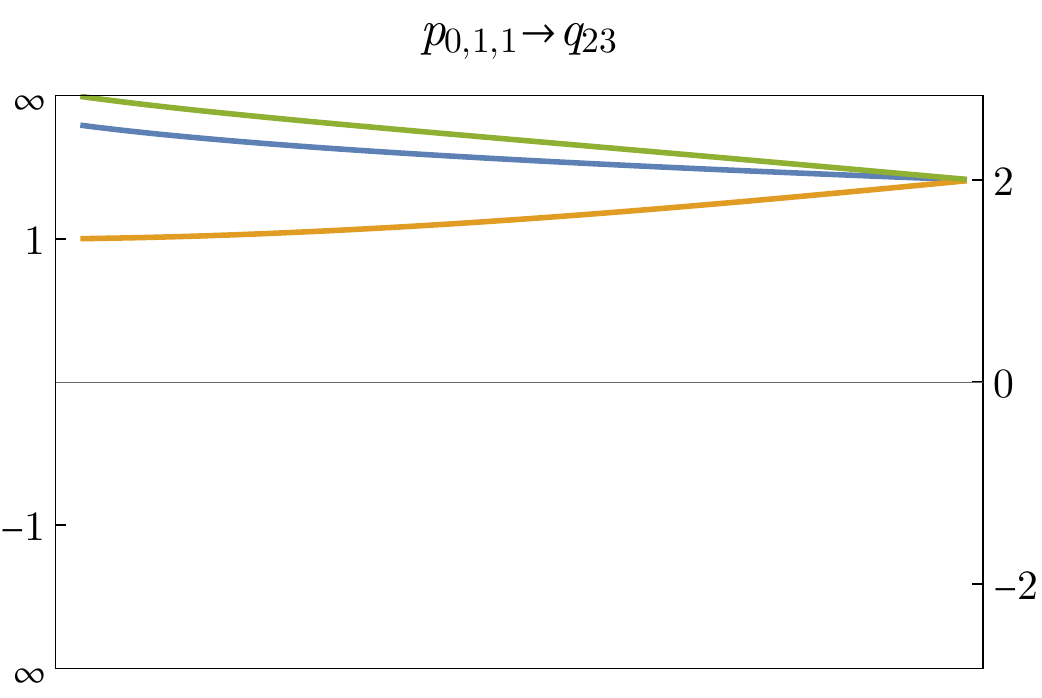}
\end{subfigure}
\begin{subfigure}{0.43\textwidth}
\includegraphics[width=\textwidth]{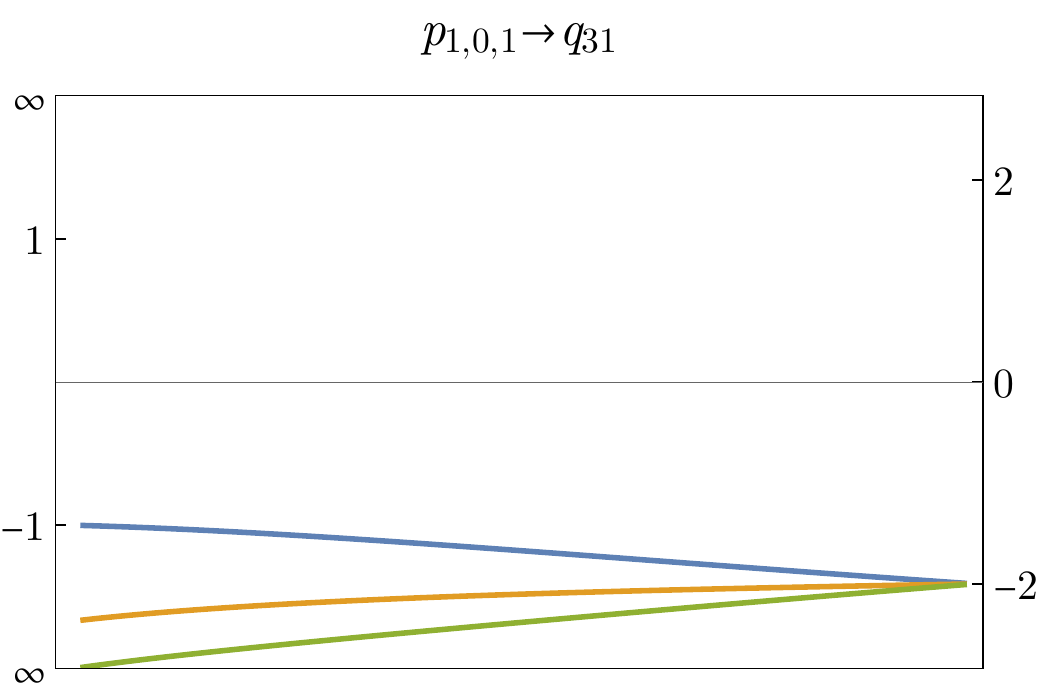}
\end{subfigure}
\caption{\label{fig:evolution}Evolution of saddle points from low to high temperature when $\De_i/\De_j>0$ for some representative values of $\De_i$. The blue curve is $x_1$, the orange curve is $x_2$, and the green curve is $x_3$. Here, we set all three temperatures equal $\b_i=\b$. The horizontal axis is $t=e^{-\b}$, with low temperatures near $t=0$ and high temperatures near $t=1$.}
\end{figure}

These saddle points yield 8 solutions of the conformal Casimir equations in the high temperature regime. But which one(s) corresponds to the block? To answer this, let us start at low temperature, where we know that the block corresponds to $\vec p_{0,0,0}$. As we dial from low to high temperature, we find that each low temperature saddle point transitions smoothly to a high temperature saddle. Although we have not proved analytically which saddle becomes which, we can track them numerically, see for example figure~\ref{fig:evolution}.

Interestingly, the matching between low- and high-temperature saddles depends on the signs of the $\De_i$.  The saddle point equations depend projectively on the $\De_i$, so more precisely only the signs of their ratios matter. We will be most interested in the case where all $\De_i/\De_j$ are positive, where we find
\be
\begin{tabular}{ccc}
low temperature & $\longrightarrow$ & high temperature \\
\hline
$\vec p_{0,0,0}$ &$\longrightarrow$ &$\vec q_0$ \\
$\vec p_{1,1,0}$ &$\longrightarrow$ &$\vec q_{12}$ \\
$\vec p_{0,1,1}$ &$\longrightarrow$ &$\vec q_{23}$ \\
$\vec p_{1,0,1}$ &$\longrightarrow$ &$\vec q_{31}$
\end{tabular}
\qquad (\De_i/\De_j>0).
\label{eq:ipluscontinuation}
\ee
The remaining mappings from low to high temperature are obtained by acting with $\tau$, for example $\tau \vec p_{0,0,0}\to \tau \vec q_0$.

How is this compatible with the fact that the $\s_i$ operations act differently on the high temperature and low temperature saddle points? The key is that the map from low to high temperature depends on the signs of the $\De_i$'s, and the $\s_i$ operations can flip these signs. We can capture these rules as follows. Let us define maps $H_{\pm,\pm,\pm}$ that take low temperature saddle points to high temperature saddle points by continuation in $\b_i$, with the signs of $(\De_1,\De_2,\De_3)$ corresponding to the signs in the subscript of $H_{\pm,\pm,\pm}$. For example, when all of the $\De_i$ are positive, the map $H_{+++}$ is given by (\ref{eq:ipluscontinuation}). We claim that the $\s_i$ interchange the maps $H_{\pm\pm\pm}$ in the following way:
\be
\label{eq:crazyintertwining}
H_{s_1 s_2 s_3}(\vec p) &= \s_1 H_{(-s_1) s_2 s_3}(\s_1 \vec p), \nn\\
H_{s_1 s_2 s_3}(\vec p) &= \s_2 H_{s_1 (-s_2) s_3}(\s_2 \vec p),\nn\\
H_{s_1 s_2 s_3}(\vec p) &= \s_3 H_{s_1 s_2 (-s_3)}(\s_3 \vec p).
\ee
In other words, conjugating by $\s_i$ flips the sign of the $i$-th subscript in $H_{\pm\pm\pm}$. Intuitively, this is because $\s_i$ flips the sign of $\De_i$ in (\ref{eq:definitionofsigmaone}).

With the rules (\ref{eq:ipluscontinuation}) defining $H_{+++}$, the relations (\ref{eq:crazyintertwining}), and the action of $\s_i,\tau$ on the low and high temperature saddles, we can predict how any low temperature saddle continues to a high temperature saddle, for various signs of the $\De_i$. We have verified these predictions numerically in examples. It would be nice to prove them analytically.

To summarize, as we continue from low to high temperature, with $\De_i/\De_j>0$, the saddle point $\vec p_{0,0,0}$ continues to $\vec q_0$. We will assume that no Stokes phenomena occur during this continuation, so that the saddle point integral through $\vec p_{0,0,0}$ continues to the saddle point integral through $\vec q_0$. Another way to think about this is that the saddle point integral automatically solves the conformal Casimir equation, in perturbation theory in $1/\De$. The assumption of no Stokes phenomena is the same as the assumption that perturbation theory in $1/\De$ can be used to solve the Casimir equations at large $\De$ for all temperatures.

Thus, let us focus on the saddle point $\vec q_0$. Away from infinite temperature, the positions of the points in the $\vec q_0$ saddle get corrected, and we can compute these corrections in a systematic expansion in $\b_i$ and the angles $\vec \a_i$ and $\vec \Phi_i$ defined in (\ref{eq:hdecomp}) (still in the large-$\De$ limit). For example, $x_1$ shifts by
\be
\de x_1^1 &=
\frac{(\Delta _1^2+(\Delta _2-\Delta _3)^2) (2 \Delta _2+\Delta _3)}{4 \Delta _2 (2 \Delta _2-\Delta _3) \Delta _3}\beta _1 
+\frac{\Delta _2 (4 \Delta _1^2-\Delta _3^2) (\Delta _1^2+\Delta _2^2-\Delta _3^2)}{4 \Delta _1^2 \Delta _3 (2 \Delta _2-\Delta _3)^2}\beta _2 \nn\\
&\quad-\frac{(\Delta _1^2-\Delta _2^2) \Delta _3 (\Delta _1^2-\Delta _2^2+\Delta _3^2)}{2 \Delta _1^2 \Delta _2 (2 \Delta _2-\Delta _3)^2}\beta _3  + O(\b^2,\vec\a^2), \nn\\
\de \vec x_1 &= 
\frac{(\Delta _1^2+(\Delta _2-\Delta _3)^2) (2\De_2+\De_3)}{4 \Delta _2 \Delta _3 (2\Delta _2-\Delta _3)} \vec \a _1
+ \frac{\Delta _2 (4 \Delta _1^2-\Delta _3^2) (\Delta _1^2+\Delta _2^2-\Delta _3^2)}{4 \Delta _1^2  \Delta _3 (2 \Delta _2-\De_3)^2} \vec \a_2
\nn\\
&\quad +\frac{(\Delta _1^2-\Delta _2^2) \Delta _3 (\Delta _1^2-\Delta _2^2+\Delta _3^2)}{2 \Delta _1^2 \Delta _2 (2 \Delta _2-\De_3)^2} \vec \a _3 + O(\b\vec \a,\vec\Phi \vec \a).
\ee
Here, $\de \vec x_1$ indicates the components of $x_1$ perpendicular to the $x_1^1$ axis, and $O(\b^2,\vec\a^2)$ and $O(\b \vec \a,\vec\Phi \vec\a)$ stand for quadratic corrections in the $\b_i,\vec \a_i,\vec\Phi_i$ of the indicated form. The $\vec\a_i$ appear at second order in $\de x_1^1$, as required by $\SO(d-1)$ invariance.

Plugging this corrected $\vec q_0$ into the saddle point integral, and taking into account the 1-loop determinant, we finally find the high temperature behavior of the block at large $\De$:
\be
\label{eq:hightemperatureblock}
B_{123}^{s's} &= \frac{(2\De_1)^{2\De_1-d}(2\De_2)^{2\De_2-d}(2\De_3)^{2\De_3-d}}{2^{3d/2}(\De_1+\De_2+\De_3)^{2\De_1+2\De_2+2\De_3-3d}} V^{s' *}(0,e,\oo) h_1 h_2 h_3 V^s(0,e,\oo)
\nn\\
&\quad \x \exp\p{-\frac{\Delta _1 \Delta _2}{\Delta _3} (\b_1+\b_2)-\frac{\Delta _2 \Delta _3}{ \Delta _1} \p{\frac {\b_2} 4 +\frac {\b_3} 2 }-\frac{\Delta _1 \Delta _3}{ \Delta _2} \p{\frac {\b_1} 4 +\frac {\b_3} 2 } + O(\De\b^2,\De\vec \a^2)}
\nn\\
&
 (\b_i,|\vec \a_i|,|\vec \Phi_i| \ll 1,\De_i\gg 1).
\ee
Recall that this formula only holds in the chamber $\De_i/\De_j>0$. When the signs of ratios $\De_i/\De_j$ are different, the high temperature behavior of the block is in general controlled by a different saddle. Our derivation so far has been for principal series representations $\De_i\in \frac d 2 + i\R_{\geq 0}$. However, we can now analytically continue the result to real $\De_i$ by simultaneously rotating the $\De_i$ clockwise in the complex plane.

In (\ref{eq:hightemperatureblock}), we only kept linear order terms in $\b_i$ in the exponent. Later, we will argue that the higher order terms in $\b,\vec\a,\vec \Phi$ do not contribute to the leading asymptotics of OPE coefficients, at large $\De$ and finite $J$. Such terms can potentially become important at large-$J$, but we leave the analysis of this case to future work. Note also that in (\ref{eq:hightemperatureblock}), we have $h_i=1+O(\a_i,\Phi_i)$. We study one consequence of the $\Phi_i^2$ terms in $h_i$ later in section~\ref{sec:subleadingcorrection}.

As was the case at low temperatures, the apparent breaking of 1-2-3 permutation symmetry in (\ref{eq:hightemperatureblock}) is due to using non-permutation symmetric coordinates on the moduli space $\cM$. Switching to the relative temperatures $\b_{ij}$, the exponent in (\ref{eq:hightemperatureblock}) becomes the beautifully permutation-symmetric
\be
 \exp\p{-\frac{\De_1\De_2}{8\De_3}\b_{12}^2  - \frac{\De_2\De_3}{8\De_1} \b_{23}^2 - \frac{\De_3\De_1}{8\De_2} \b_{31}^2 +  O(\De\b^2,\De\vec \a^2)}.
\ee

Let us make a few additional observations about the result (\ref{eq:hightemperatureblock}). We define a ``scaling block" as the primary term in~(\ref{eq:blocklowtemperature}). Because the full block is a sum of scaling blocks with nonnegative coefficients, we should have the inequality
\be
B_{123} &\geq 2^{-2\De_1-2\De_2} e^{-\b_1 \De_1-\b_2 \De_2-\b_3\De_3} \qquad (h_i=1).
\ee
Let us check this at high temperatures. Our calculation of the block is valid in the regime $\De \gg 1$ and $\b\ll 1$. Thus, we can ignore $\b\De$ compared to $\De$ and just compare the $\De$-dependent terms out front:
\be
\frac{(2\De_1)^{2\De_1-d}(2\De_2)^{2\De_2-d}(2\De_3)^{2\De_3-d}}{2^{3d/2}(\De_1+\De_2+\De_3)^{2\De_1+2\De_2+2\De_3-3d}} \geq 2^{-2\De_1-2\De_2}.
\ee
Indeed, we find that numerically, the above inequality holds when $\De_i>0$. It is saturated when $\De_1=\De_2=\De_3/2$, and in this case the high temperature block and the scaling block are exactly the same (up to the order we've computed them)! One speculative interpretation is that the full genus-2 block at large $\De$ may be a scaling block in an appropriate Weyl frame that depends on the $\De_i$. Our choice of Weyl frame in section~\ref{sec:choiceofgeometry} happens to be the appropriate frame for $\De_1=\De_2=\De_3/2$. Other Weyl frames would be best suited to other $\De_i$. The large-$\De$ limit of Virasoro blocks also simplifies in an appropriate Weyl frame \cite{Zamolodchikov1987,Maldacena:2015iua}.

\section{OPE coefficients of heavy operators}
\label{sec:opecoeffhhh}
``Heavy-heavy-heavy" OPE coefficients are encoded in the partition function of the CFT on the genus-2 manifold $M_2$ via \eqref{eq:genus2blockexpansion}.   
In section \ref{sec:partitionfunction}, using the thermal effective action and the ``hot spot" hypothesis, we calculated the leading expression \eqref{eq:leadingpartition} for the partition function in the high-temperature regime discussed in section \ref{sec:hotspotgenustwo}.
In section \ref{sec:conformalblock}, we obtained an expression \eqref{eq:hightemperatureblock} for a conformal block in the same regime. Finally, in this section, we will combine these ingredients to obtain an asymptotic formula for ``heavy-heavy-heavy" OPE coefficients by inverting the conformal block decomposition of the partition function.

\subsection{Review: inverting a genus-1 partition function}

Before discussing how to invert a genus-2 partition function, let us revisit the genus-1 case, phrasing it in language that will generalize to genus-2.
Conformal blocks for the genus-1 partition function on $S^1 \x S^{d-1}$ are just conformal characters $\chi_{\De,J}(\b,\Omega_i)$.  For simplicity, let us work in $d=1$, where the characters have the simple form
\be
\chi_\De(\b) &= \frac{e^{-\De \b}}{1-e^{-\b}} \qquad (d=1),
\ee
and the partition function has the decomposition
\be
Z(\b) &= \int d\De\,p(\De) \chi_\De(\b),
\ee
which is essentially a Laplace transform of the density of states $p(\De)$.
It is straightforward to decompose $Z(\b)$ into characters via an inverse Laplace transform, as we did in section~\ref{sec:laplacetransf}. However, let us pause to understand this transform in group-theoretic language.
 
The conformal characters can be viewed as functions on the group $\SL(2,\R)$ that are invariant under conjugation --- i.e.\ class functions. They are naturally eigenfunctions of the Casimir differential operator $\cD$ defined in section~\ref{sec:genustwopw}. In terms of $\b$, this leads to the eigenvalue equation
\be
\label{eq:diffeqforchi}
\cD \chi_\De(\b) &= \frac{1+e^{-\b}}{1-e^{-\b}}\chi_\De '(\beta )+\chi_\De ''(\beta )=\De(\De-1)\chi_\De(\b).
\ee
(Here, we abuse notation and write $\cD$ both for the differential operator $L^A L_A$ on the group $\SL(2,\R)$, and for the differential operator (\ref{eq:diffeqforchi}) acting on $\b$.)
Because the Casimir eigenvalue $\De(\De-1)$ is the same for $\De$ and $\tl \De=1-\De$, the shadow character $\chi_{\tl \De}(\b)$ satisfies the same differential equation as $\chi_\De(\b)$.

Because of its group-theoretic origin, $\cD$ is naturally self-adjoint in the Haar measure on $\SL(2,\R)$. When acting on class functions, this implies that $\cD$ defined in (\ref{eq:diffeqforchi}) is self-adjoint with respect to the quotient measure on the space of conjugacy classes of $\SL(2,\R)$. The quotient measure is given by the famous Weyl integration formula (and can be computed  using the Faddeev-Popov procedure):
\be
d\mu &= d\b (e^{\b/2}-e^{-\b/2})^2.
\ee

Self-adjointness of $\cD$ immediately implies an orthogonality relation
\be
\int d\mu \chi_\De(\b) \chi_{\tl \De'}(\b) &= 0 \qquad \textrm{unless $\De=\De'$ or $\De=\tl \De'$},
\ee
where the integral can follow any contour such that the boundary terms from integrating $\cD$ by parts vanish. For our applications, we can integrate over an infinite contour parallel to the imaginary axis $\b = \b_0+it$, which gives
\be
\oint d\mu \chi_\De(\b) \chi_{\tl \De'}(\b) &= \oint d\b\, e^{(\De'-\De)\b} = 2\pi i \de(\De-\De').
\ee
This allows us to invert the partition function by integrating against a shadow block:
\be
\label{eq:inverselap}
p(\De) &= \frac{1}{2\pi i}\oint d\mu \chi_{\tl \De}(\b) Z(\b) = \frac{1}{2\pi i}\oint d\b e^{\De \b} (1-e^{-\b}) Z(\b),
\ee
which is the usual inverse Laplace transform.

Before proceeding, let us make a comment about the choice of contour. By analogy with Euclidean inversion formulae for local correlation functions, we could have instead tried  to decompose $Z(\b)$ in characters for principal series representations, which take the form
\be
\chi'_s(\b) &= \chi_{\frac 1 2 + i s}(\b)+\chi_{\frac 1 2 - i s}(\b) = \frac{e^{is\b}+e^{-i s\b}}{e^{\b/2}-e^{-\b/2}}.
\ee
Principal series characters are naturally orthonormal with respect to $d\mu$, when integrated along a {\it real} contour $\b\in \R$. However, this kind of orthogonality is unsuitable for decomposing a physical partition function. The reason is that  $Z(\b)$ typically possesses a high-temperature singularity on the real axis of the form $Z(\b)\sim e^{1/\b^a}$ for some positive power $a$. This high-temperature singularity cannot be integrated against $\chi'_s(\b)$ along a real contour in a simple way.\footnote{Interestingly, there is usually no problem with decomposing correlators of local operators in principal series representations. Doing so leads to  Euclidean inversion formulas, which typically involve integrable power-like singularities. It would be nice to better understand the distinction between these cases.} Using a complex contour as in (\ref{eq:inverselap}) bypasses this issue by avoiding the singularity.

\subsection{An inverse Laplace transform for genus-2 blocks}

Let us now assemble analogous ingredients in the genus-2 case. Let $d\mu$ be the natural quotient measure on the moduli space $\cM=G\backslash (G^-)^3 /G$, descending from the product of Haar measures on $(G^-)^3$. Consider a contour integral of a block against a shadow block
\be
\oint d\mu\, B_{123}^{s's}\, B_{\tl 1^{'\dag}\tl 2^{'\dag}\tl 3^{'\dag}}^{t't}.
\ee
(We do not specify the precise contour for now.) The block and shadow block are both simultaneous eigenfunction of the Casimir operators $\cD_i$ acting on each group element $g_i$. Because of their group-theoretic origin, these operators are naturally self-adjoint in the measure $d\mu$. If the contour is such that there are no boundary terms from integrating the $\cD_i$ by parts, then the above integral must be proportional to $\de$-functions restricting $\pi_i$ and $\pi_i'$ to be the same
\be
\oint d\mu\, B_{123}^{s's}\, B_{\tl 1^{'\dag}\tl 2^{'\dag}\tl 3^{'\dag}}^{t't}
&\propto \de_{\pi_1\pi_1'}\de_{\pi_2\pi_2'}\de_{\pi_3\pi_3'},
\label{eq:orthogonalityofgenustwo}
\ee
where
\be
\de_{\pi\pi'} &= \de(\De-\De') \de_{\lambda\lambda'}.
\ee

To find the constant of proportionality in (\ref{eq:orthogonalityofgenustwo}), we will assume that the contour is a complex contour in $\b_i$ that can be deformed to low temperature and evaluated in that regime. This is analogous to the inverse Laplace transform (\ref{eq:inverselap}), where can choose any contour of the form $\b=\b_0+i t$. Moving the contour to low temperatures, we can evaluate the orthogonality relation (\ref{eq:orthogonalityofgenustwo}) using the low-temperature expansion of the blocks (\ref{eq:blocklowtemperature}). When inverting a partition function, we can instead deform the contour into the high temperature regime and look for a saddle point.

\subsection{Computing the measure}
\label{sec:measure}

The first step is to compute the quotient measure $d\mu$ via the Faddeev-Popov procedure. There are a few wrinkles in doing so, so let us work through the computation in full. Recall that the quotient space $\cM$ can be redundantly parametrized by $(g_1,g_2,g_3)\in G^-\x G^-\x G^-$. We would like to fix a gauge by writing $g_i$ in terms of $\b_i,h_i$ as in (\ref{eq:ourparametrization}). For the moment, let us also imagine that we have chosen a non-redundant parametrization of the $h_i$ in terms of angles, so that overall the $g_i$ are specified in terms of $n=\dim G$ parameters which we call $y$. Our gauge-fixing condition is $g_i=\bar g_{i}(y)$. An appropriate gauge-fixing function is
\be
Q(g_1,g_2,g_3) &= \int d y\, \de(g_1,\bar g_1) \de(g_2,\bar g_2) \de(g_3,\bar g_3),
\ee
where $d y$ is any measure on the coordinates $y$, and $\de(g,g')$ is a unit-normalized $\de$-function (in the Haar measure) on the group supported at $g=g'$. 

Consider now a gauge-invariant function $f(g_1,g_2,g_3)$. We formally define its integral over the moduli space as
\be
\int_\cM d\mu\,f &= \int \frac{dg_1\,dg_2\,dg_3}{(\vol G)^2} f(g_1,g_2,g_3),
\ee
where $dg_i$ are Haar measures. Following the usual FP procedure, we insert $1$ in the form of an integral over gauge orbits of $Q$, divided by its average over gauge orbits
\be
\int_\cM d\mu\,f &= \int \frac{dg_1\,dg_2\,dg_3}{(\vol G)^2} dg dg' \frac{Q(gg_1g^{'-1},gg_2g^{'-1},gg_3g^{'-1})}{\hat Q(g_1,g_2,g_3)} f(g_1,g_2,g_3),
\ee
where
\be
\hat Q(g_1,g_2,g_3) &\equiv \int dg\, dg' Q(gg_1g^{'-1},gg_2g^{'-1},gg_3g^{'-1}).
\ee
Now we change variables $g_i\to g^{-1} g_i g'$ and use gauge-invariance of $f$ and $\hat Q$ to obtain
\be
\int_\cM d\mu\,f &= \int dg_1\,dg_2\,dg_3 \frac{f(g_1,g_2,g_3)}{\hat Q(g_1,g_2,g_3)} Q(g_1,g_2,g_3) = \int d y\, \frac{f(\bar g_1,\bar g_2 ,\bar g_3)}{\hat Q(\bar g_1,\bar g_2 ,\bar g_3)}.
\ee
This is our gauge-fixed integral and $1/\hat Q$ is the FP determinant, which we now compute.

We have
\be
\hat Q(\bar g_1,\bar g_2 ,\bar g_3) &= 
\int dg\,dg'\,d y'\, \prod_{i=1}^3 \de(g \bar g_{i}( y)g'^{-1},\bar g_{i}( y')).
\ee
The $\de$-functions are supported for $g,g'$ near the identity and $ y'$ near $ y$. Thus, we can write
\be
g = 1+\xi,\quad g'=1+\xi',
\ee
where $\xi,\xi'$ are elements of the Lie algebra of $G$, and we can furthermore Taylor expand the $\bar g_i$ in $ y'= y+d  y$. We have
\be
\de(g \bar g_{i}( y) g'^{-1}, \bar g_{i}( y')) &= \de(\bar g_{i}( y)^{-1} g \bar g_{i}( y) g'^{-1}, \bar g_{i}( y)^{-1} \bar g_{i}( y')) \nn\\
&= \de(1+\bar g_{i}( y)^{-1} \xi \bar g_{i}( y) - \xi', 1 + d  y\.\bar g_{i}( y)^{-1} \ptl_{ y}\bar g_{i}( y))
\nn\\
&= \de(\textrm{Ad}_{\bar g_{i}^{-1}} \xi - \xi' - d  y\.\bar g_{i}^{-1} \ptl_{ y}\bar g_{i}),
\ee
where in the last line, we have a $\de$ function on the Lie algebra $\mathfrak{g}$, and $\textrm{Ad}_{g}$ denotes the adjoint action of $g$.
 The gauge-fixed measure is thus
\be
d\mu &= \frac{d y}{\hat Q} 
= \det\begin{pmatrix}
\textrm{Ad}_{\bar g_{1}^{-1}} & -\mathbf{1} & -\bar g_{1}^{-1} \ptl_{y^1} \bar g_{1} & \cdots &  -\bar g_{1}^{-1} \ptl_{y^n} \bar g_{1}\\
\textrm{Ad}_{\bar g_{2}^{-1}} & -\mathbf{1} & -\bar g_{2}^{-1} \ptl_{y^1} \bar g_{2} & \cdots &  -\bar g_{2}^{-1} \ptl_{y^n} \bar g_{2} \\
\textrm{Ad}_{\bar g_{3}^{-1}} & -\mathbf{1} & -\bar g_{3}^{-1} \ptl_{y^1} \bar g_{3} & \cdots & -\bar g_{3}^{-1} \ptl_{y^n} \bar g_{3}\\
\end{pmatrix} dy^1 \cdots dy^n,
\label{eq:quotientmeasure}
\ee
The object inside the determinant is a $3n \x 3n$ matrix. Choosing an orthonormal basis of $\mathfrak{g}$, $\textrm{Ad}_{\bar g_i^{-1}}$ becomes an $n\x n$ block, and each $\mathbf{1}$ becomes an $n\x n$ identity matrix. Finally, $-\bar g_{i}^{-1} \ptl_{y^j} \bar g_{i}$ is an element of $\mathfrak{g}$, which we can think of as a column vector of height $n$.

\subsubsection{Partial gauge fixing}

In our parametrization of $\cM$ in terms of temperatures $\b_1,\b_2,\b_3$ and rotations $h_1,h_2,h_3\in \SO(d)$, we write the moduli space as
\be
\cM &= G\backslash (G^-)^3 /G \cong \SO(d-1)\backslash (\SO(1,1)\x SO(d))^3 /\SO(d-1),
\ee
where the two copies of $\SO(d-1)$ act by left and right multiplication on the $h_i$.
Above, we obtained the measure from fully gauge-fixing both the left and right action of $G$. However, it will be more convenient to only partially fix the gauge, leaving the $\SO(d-1)\x\SO(d-1)$ gauge redundancy un-fixed. Let us determine how the above computation should be modified in this case.

Let $y=(\b_1,\b_2,\b_3,h_1,h_2,h_3)$ now be coordinates on $(\SO(1,1)\x SO(d))^3$, and let us write $K=\SO(d-1)$, with Lie algebra $\mathfrak k$. The $y$ have an action of $K\x K$ given by $ y\mto k y k'^{-1}$ and an invariant measure $dy$. Once again, we should consider the average over gauge orbits of the gauge-fixing function
\be
\label{eq:thingweshouldconsider}
\hat Q(g_1,g_2,g_3) &\equiv \int dg dg' d y \prod_i \de(g g_1 g'^{-1}, \bar g_i( y)).
\ee
We can factorize $g$ into
\be
g &= k\,\g,
\ee
where $k\in K$, and $\g$ is a representative of the quotient $K\backslash G$. The measure on $G$ similarly splits as
\be
dg = dk\, d\g,
\ee
where $dk$ is the measure on $K$, and $d\g$ is a right-$G$-invariant measure on $K\backslash G$. To be more precise, let $T^a$ be an orthonormal basis of generators of $\mathfrak{g}$, and let the generators of $\mathfrak{k}\subset \mathfrak{g}$ be the $T^a$ with $a=1,\dots,\dim K$. Then we can take the $\g$ to be the image under the exponential map of the remaining generators $T^a$ with $a=\dim K+1,\dots,\dim G$.

Inserting this decomposition into (\ref{eq:thingweshouldconsider}), we find
\be
\hat Q(g_1,g_2,g_3) &\equiv \int dk\,d\g\,dk'\, d\g'\, d y \prod_i \de(k\g g_1  \g'^{-1}k'^{-1}, \bar g_i( y))
\nn\\
&= \int dk\,d\g\,dk'\, d\g'\, d y \prod_i \de(\g g_1  \g'^{-1}, \bar g_i(k^{-1} yk'))\nn\\
&= (\vol K)^2\int d\g\, d\g'\, d y \prod_i \de(\g g_1  \g'^{-1}, \bar g_i( y)) \nn\\
&= (\vol K)^2\int d\xi\, d\xi'\, d y \prod_i \de(\Ad_{g_i^{-1}} \xi - \xi' - d y \. \bar g_i^{-1} \ptl_{ y} \bar g),
\ee
where we have written
\be
\g = \exp(\xi),\quad \g'=\exp(\xi').
\ee
The only differences from before are that now $\xi$ and $\xi'$ are restricted to generators of $\g$ --- i.e.\ $T^a$ with $a=\dim K+1,\dots,\dim G$, and furthermore we must divide by $(\vol K)^2$. The partially gauge-fixed measure is thus
\be
\label{eq:partiallygaugefixedmeasure}
d\mu &= \frac{1}{(\vol K)^2} \det\begin{pmatrix}
\textrm{Ad}_{\bar g_{1}^{-1}}\Pi_\g^\dag & -\Pi_\g^\dag & -\bar g_{1}^{-1} \ptl_{y^1} \bar g_{1} & \cdots &  -\bar g_{1}^{-1} \ptl_{y^m} \bar g_{1}\\
\textrm{Ad}_{\bar g_{2}^{-1}}\Pi_\g^\dag & -\Pi_\g^\dag & -\bar g_{2}^{-1} \ptl_{y^1} \bar g_{2} & \cdots &  -\bar g_{2}^{-1} \ptl_{y^m} \bar g_{2} \\
\textrm{Ad}_{\bar g_{3}^{-1}}\Pi_\g^\dag & -\Pi_\g^\dag & -\bar g_{3}^{-1} \ptl_{y^1} \bar g_{3} & \cdots & -\bar g_{3}^{-1} \ptl_{y^m} \bar g_{3}\\
\end{pmatrix} dy^1 \cdots dy^m,
\ee
where $\Pi_\g$ is an $(n-\dim K)\x n$ matrix implementing the orthogonal projection onto the generators of $\g$, and $\Pi_\g^\dag$ is its adjoint. Finally, $m=n+2\dim K$. Overall, this again gives a $3n\x3n$ matrix.

Let us finally plug in (\ref{eq:ourparametrization}) to write the measure (\ref{eq:partiallygaugefixedmeasure}) in terms of the parameters $\b_i,h_i$. We found it difficult to compute the measure exactly for generic parameters.\footnote{The reason is that (\ref{eq:partiallygaugefixedmeasure}) is the determinant of a large symbolic matrix, which is extremely difficult for {\it Mathematica} to handle.} However, we can compute it in various limits. At low temperatures, we find
\be
\label{eq:lowtempmeasure}
d\mu &= \frac{2^{5d}}{(\vol\,\SO(d-1))^2} \prod_{i=1}^3 e^{d\,\b_i} d\b_i dh_i,\qquad (\textrm{low temperature}),
\ee
where $dh_i$ denotes Haar measures on $\SO(d)$. At high temperature, we find
\be
\label{eq:hightempmeasure}
d\mu &= \frac{2^{4d}}{(\vol\,\SO(d-1))^2}\prod_{i=1}^3 d\b_i d\vec \a_i d\vec \Phi_i, \qquad (\textrm{high temperature}),
\ee
where $\vec \a_i$ and $\vec \Phi_i$ are the angles defined in (\ref{eq:hdecomp}).\footnote{We can find an interpolating result between high and low temperatures by setting the angles to zero, $\vec\a_i=0,\vec \Phi_i=0$. In this case, {\it Mathematica} is able to compute the determinant for general $\b_i$, giving
\be
d\mu &= \p{-8 e^{\beta _1}-8 e^{\beta _2}+\frac{1}{2} e^{\beta _1-\beta _2-\beta _3}+\frac{1}{2} e^{-\beta _1+\beta _2-\beta _3}-2 e^{\beta _1+\beta _2-\beta _3}+e^{-\beta _3}+32 e^{\beta _1+\beta _2+\beta _3}}^d \nn\\
&\quad \x \frac{1}{(\vol\,\SO(d-1))^2}\prod_{i=1}^3 d\b_i d\vec \a_i d\vec \Phi_i \qquad\qquad (\vec \a_i=0,\vec\Phi_i=0).
\ee
This indeed agrees with both (\ref{eq:lowtempmeasure}) and (\ref{eq:hightempmeasure}) in the appropriate limits.}

\subsection{Orthogonality relation for genus-2 blocks}

With the measure in hand, let us determine the correct orthogonality relation for genus-2 blocks. We will take the contour to be a real contour for the group elements $h_i\in \SO(d)$, and a complex contour running parallel to the imaginary axis for the $\b_i$.  The key idea will be to deform the $\b_i$ contour into the low-temperature region, where the blocks are given by the simple formula
\be
B_{123}^{s' s} &= 2^{-2\De_1-2\De_2} e^{-\De_1\b_1-\De_2\b_2-\De_3 \b_3} V^{s'}(0,e,\oo)^* h_1h_2h_3\,V^s(0,e,\oo)
\nn\\
&\quad \x(1+ O(e^{-\b_i})).
\ee

Consider a shadow block with complex-conjugated three-point structures
\be
B_{\tl 1^{'\dag}\tl 2^{'\dag}\tl 3^{'\dag}}^{t'^{*} t^*} 
&=2^{-2(d-\De_1')-2(d-\De_2')} e^{-(d-\De_1')\b_1-(d-\De_2')\b_2-(d-\De_3') \b_3} V^{t'}(0,e,\oo) h_1h_2h_3\,V^t(0,e,\oo)^*\nn\\
&\quad \x(1+ O(e^{-\b_i})).
\ee
Integrating the block and the shadow block against each other using the low-temperature measure (\ref{eq:lowtempmeasure}), we find
\be
\oint d\mu B_{123}^{s' s} B_{\tl 1^{'\dag}\tl 2^{'\dag}\tl 3^{'\dag}}^{t'^{*} t^*}
&= \frac{2^{5d}}{(\vol\,\SO(d-1))^2} \int \p{\prod_i d\b_i dh_i} 2^{-4d} e^{-\b_1(\De_1-\De_1') - \b_2(\De_2-\De_2') -\b_3 (\De_3-\De_3')} \nn\\
&\quad \x V^{s'}(0,e,\oo)^* h_1h_2h_3\,V^s(0,e,\oo) \x V^{t'}(0,e,\oo) h_1h_2h_3\,V^{t}(0,e,\oo)^*.
\ee

To perform the integral over the $h_i$, we can use the Schur orthogonality formula, which states that for any compact group $G$ with unitary representations $\lambda,\lambda'$
\be
\int dg \<a|\lambda(g)|b\>\<c|\lambda'(g^{-1}) |d\> = \<a|d\>\<c |b\> \frac{\vol G}{\dim \lambda}\de_{\lambda\lambda'}.
\ee
The integral over $\b_i$ gives $\de$-functions of the form $\de(\De_i-\De_i')$, as in the inverse Laplace transform (\ref{eq:inverselap}). Overall, we find the orthogonality relation
\be
&=\frac{V^t(0,e,\oo)^*V^s(0,e,\oo)}{2^d\vol\,\SO(d-1)} \frac{V^{s'}(0,e,\oo)^*V^{t'}(0,e,\oo)}{2^d\vol\,\SO(d-1)} \prod_{i=1}^3 2\pi\de(\De_i-\De_i')\frac{2^d\vol\,\SO(d)}{\dim \lambda_i}\de_{\lambda_i\lambda_i'}
\nn\\
&= T^{ts} T^{s' t'} \prod_{i=1}^3 2\pi\de(\De_i-\De_i')\frac{2^d\vol\,\SO(d)}{\dim \lambda_i}\de_{\lambda_i\lambda_i'},
\ee
where we have introduced the 3-point pairing matrix
\be
\label{eq:thptpair}
T^{ts} &= \frac{V^t(0,e,\oo)^*V^s(0,e,\oo)}{2^d\vol\,\SO(d-1)}.
\ee
Recall that the structure $V^s$ is a tensor with an index for each of the representations $\lambda_1,\lambda_2,\lambda_3$, and $V^{t*}$ carries indices for the dual representations $\l_1^*,\l_2^*,\l_3^*$. The indices of $V^s$ and $V^{t*}$ are implicitly contracted in the three point pairing (\ref{eq:thptpair}).
The pairing matrix $T^{ts}$ shows up in other contexts related to harmonic analysis on the conformal group, and is discussed more extensively in \cite{Liu:2018jhs,Karateev:2018oml}.

To summarize, if the partition function has an expansion in conformal blocks
\be
\label{eq:zblocksum}
Z &= \sum_{\lambda_1,\lambda_2\lambda_3} \int d\De_1 d\De_2 d\De_3 P_{123}^{ss'} B^{s's}_{123},
\ee
then the conformal block coefficients $P_{123}^{s's}$ are given by the inversion formula
\be
\label{eq:inversionformula}
P^{ss'}_{123}  &= (T^{-1})^{st}(T^{-1})^{t's'}\frac 1 {(2\pi)^3}\prod_{i=1}^3 \p{\frac{\dim \lambda_i}{2^d\vol\,\SO(d)}} \oint d\mu \, Z B^{t'^{*}t^*}_{\tl 1^\dag \tl 2^\dag \tl 3^\dag}.
\ee
In both (\ref{eq:zblocksum}) and (\ref{eq:inversionformula}), a sum over repeated three point structure indices $s,t,s',t'$ is implicit. 

\subsection{Putting everything together}

We are finally ready to put everything together and perform the genus-2 Laplace transform at high temperature. Let us recall the important formulas. The shadow block at high temperature is given by
\be
B^{t'^*t^*}_{\tl \pi_1 \tl \pi_2 \tl \pi_3} &= \frac{1}{2^{3d/2}} \p{\prod_{i=1}^3 \p{\frac{2\De_i}{\De_1+\De_2+\De_3}}^{d-2\De_i}} e^{\frac{\De_1\De_2}{8\De_3}\b_{12,0}^2 + \frac{\De_2\De_3}{8\De_1}\b_{23,0}^2 + \frac{\De_1\De_3}{8\De_2}\b_{31,0}^2 +
O(\De\b_i^2,\De\vec \a^2)}\nn\\
&\quad \x V^{t'}(0,e,\oo)\prod_{i=1}^3 e^{i\a_i\.M_{i,\a}}e^{i\Phi_i\.M_{i,\Phi}} V^t(0,e,\oo)^*,
\label{eq:blockathightempresult}
\ee
where we use the shorthand notation
\be
\a_i\.M_{i,\a} &= \sum_{b=2}^d\a_{i,b}M_{i,1b}, \nn\\
\Phi_i\.M_{i,\Phi} &=\sum_{2\leq a<b\leq d}\Phi_{i,ab}M_{i,ab},
\ee
where $M_{i,ab}$ denotes a rotation generator with indices $ab$ acting on the $i$-th point in the representation $\lambda_i$.
The partition function at high temperature is given by
\be
Z &= \exp \Bigg(
\frac{f_{12}\vol\, S^{d-1}}{\b_{12,0}^{d-1}} \p{1- \frac{(\vec \Phi_1 - \vec \Phi_2)^2}{\b_{12,0}^2} - 8(d+1)\frac{(\vec\a_{1}+\vec \a_2)^2}{ \b_{12,0}^4} +\dots} \nn\\
&\qquad\qquad + \frac{f_{23}\vol\, S^{d-1}}{\b_{23,0}^{d-1}} \p{1- \frac{(\vec \Phi_2 - \vec \Phi_3)^2}{\b_{23,0}^2} - 8(d+1)\frac{(\tfrac 1 4\vec \a_2 - \tfrac 1 2 \vec \a_3)^2}{ \b_{23,0}^4} +\dots}
\nn\\
&\qquad\qquad + \frac{f_{31}\vol\, S^{d-1}}{\b_{31,0}^{d-1}} \p{1- \frac{(\vec \Phi_3 - \vec \Phi_1)^2}{\b_{31,0}^2} - 8(d+1)\frac{(\tfrac 1 4\vec \a_1 - \tfrac 1 2 \vec \a_3)^2}{ \b_{31,0}^4} +\dots}
\Bigg).
\ee
Here, we have allowed for different free energy densities $f_{ij}$ at each hot spot. This would arise if we inserted topological defects into the partition function, for example symmetry operators, as discussed in section~\ref{sec:hotspotgenustwo}. Our main case of interest is where $f_{ij}=f$ (the thermal free energy density), but it is just as straightforward to do the computation for general $f_{ij}$. Finally, the measure at high temperature is
\be
d\mu &= \frac{2^{4d}}{(\vol\,\SO(d-1))^2}\prod_{i=1}^3 d\b_i\, d^{d-1}\vec\a_i\, d^{\frac{(d-1)(d-2)}{2}}\vec\Phi_i.
\ee


We would now like to integrate (\ref{eq:inversionformula}) to extract the density of OPE coefficients. We deform the contour so that it passes through the regime of high temperature.
We will organize the calculation as follows. We split the integrand into the form
\be
(\textrm{quickly varying}) \x (\textrm{slowly varying}).
\ee
Here ``quickly varying" includes terms that are exponential in large parameters like $\De$ or $1/\b^\#$, while ``slowly varying" includes everything else. We look for a saddle point of the ``quickly varying" terms, writing them as a gaussian centered at this saddle point, times perturbative corrections. Meanwhile, we expand the $(\textrm{slowly varying})$ part perturbatively around the saddle point.

One simplification of this way of organizing the calculation is that, because we are working in the regime $J \ll \De$, terms in the conformal block of the form $V^* h_1 h_2 h_3 V$ will be included among the slowly-varying terms, and will not affect the location of the saddle point. By symmetry, the saddle point will be located at $\vec \a_i=\vec \Phi_i=0$.

Let us analyze the size of fluctuations around the saddle point. In particular, we would like to determine which terms must be kept in our approximation to the conformal block and the partition function. The quickly-varying (i.e.\ exponential in $\De$) part of the block has the schematic form
\be
\label{eq:rapidblock}
B &\sim e^{-\De\beta  - \De\b^2 -\De\vec \a^2 + \dots},
\ee
where ``$\dots$" are higher-order corrections in $\b,\a,\Phi$.
Meanwhile, the quickly-varying part of the partition function has the form
\be
\label{eq:rapidZ}
Z &\sim \exp\p{\frac{1}{\b^{\frac{d-1}{2}}}\p{1 - \frac{\vec\Phi^2}{\b} - \frac{\vec \a^2}{\b^2}}}.
\ee
Here, $\b$ schematically denotes the individual $\b_i$, not the relative $\b_{ij,0}$.

The saddle point equation for $\b$ will set $\De\sim \b^{-(d+1)/2}$. Plugging this in, we find
\be
Z B &\sim \exp\p{\frac{1}{\b^{\frac{d-1}{2}}} - \frac{\vec\Phi^2}{\b^{\frac{d+1}{2}}} - \frac{\vec \a^2}{\b^{\frac{d+3}{2}}} - \frac{\vec \a^2}{\b^{\frac{d+1}{2}}}}.
\ee
There are two $\vec \a^2$ terms: one coming from the block (\ref{eq:rapidblock}) and one coming from the partition function (\ref{eq:rapidZ}). 
We see that the $\vec\a^2$ term coming from the partition function is more important --- it is enhanced by an additional power of $1/\beta$. Thus, we can ignore the quadratic $\vec\a$-dependence of the conformal block. In other words, the terms written explicitly in (\ref{eq:blockathightempresult}) are sufficient for our purposes. Overall, the characteristic size of fluctuations in the angular variables coming from (\ref{eq:quadraticexpansion}) is
\be
\vec \a \sim \b^{\frac{d+3}{4}} \sim \De^{-\frac 1 2 - \frac{1}{d+1}},\qquad
\vec \Phi\sim \b^{\frac{d+1}{4}} \sim \De^{-\frac 1 2}.
\label{eqn:anglefluctuationsize}
\ee


The saddle point for the quickly-varying terms is located at $\vec \a_i=0, \vec\Phi_i=0$, and
\be
\label{eq:saddlepointbetavalue}
\b_{12,0} &= \p{\frac{4(d-1)f_{12}\vol\, S^{d-1}\De_3}{\De_1\De_2}}^{\frac{1}{d+1}},
\ee
together with cyclic permutations of (\ref{eq:saddlepointbetavalue}). The Hessian matrix at the saddle point splits into three separate blocks: a block for the $\b_i$, a block for the $\vec\a_i$, and a block for the $\vec\Phi_i$. Thus, we can calculate the 1-loop determinant separately in each of these sets of variables. The determinant for the $\b_i$ and $\vec \a_i$ variables is straightforward:
\be
\textrm{1-loop $\b_i$ factor} &= 
 \p{\frac{(\beta _{12,0} \beta _{23,0} \beta _{31,0})^{d+3}}{4 f_{12} f_{23} f_{31} } \p{\frac{\pi}{2(d^2-1)\vol\,S^{d-1}}}^3}^{\frac 1 2}, \nn\\
\textrm{1-loop $\vec \a_i$ factor} &= \p{\frac{(\beta _{12,0} \beta _{23,0} \beta _{31,0})^{d+3}}{4 f_{12} f_{23} f_{31} } \p{\frac{\pi}{2(d+1)\vol\,S^{d-1}}}^3}^{\frac{d-1}{2}}.
\ee

To compute the $\vec\Phi_i$ determinant, we must fix the $\SO(d-1)$ gauge redundancy that simultaneously shifts the $\vec\Phi_i$. For example, we can set $\vec\Phi_3=0$ and compute the 1-loop determinant in $\vec\Phi_1,\vec\Phi_2$, multiplying the result by $\vol\,\SO(d-1)$ to account for the volume of the $\SO(d-1)$ orbit. This gives
\be
&\textrm{1-loop $\vec\Phi_i$ factor} \nn\\
&= \vol\,\SO(d-1) \left(\p{\frac{\vol\, S^{d-1}}{\pi}}^2 \left(\frac{f_{12} f_{23}}{\left(\beta _{12,0} \beta _{23,0}\right)^{d+1}}+\frac{f_{12} f_{31}}{ \left(\beta _{12,0} \beta _{31,0}\right)^{d+1}}+\frac{f_{23} f_{31}}{\left(\beta _{23,0} \beta _{31,0}\right)^{d+1}}\right)\right)^{-\frac{(d-2) (d-1)}{4} }.
\ee

Putting everything together, plugging in the saddle point values (\ref{eq:saddlepointbetavalue}), we find the asymptotic conformal block coefficients 
\be
\label{eq:generalope}
P^{ss'}_{123} &\sim (T^{-1})^{ss'} \p{\prod_{i=1}^3\frac{\dim \lambda_i}{\vol\,\SO(d)}} \frac{\pi ^{\frac{(d+2) (d-2)}{2} } (4 (d-1))^{\frac{d^2}{2}-\frac{3}{d+1}+\frac{5}{2}}}{2^{\frac{d}{2}} (d+1)^{\frac{3 d}{2}}}
\left(8 f_{12} f_{23} f_{31} (\vol\,S^{d-1})^3\right)^{\frac{d}{d+1}}
\nn\\
&\quad \x \frac{(\De_1+\De_2+\De_3)^{2 (\De_1+\De_2+\De_3)-3 d}}{(\De_1^2+\De_2^2+\De_3^2)^{\frac{(d-2) (d-1)}{4} } \prod_{i=1}^3(2 \De_i)^{2 \De_i-\frac{d (d-1)}{2 (d+1)}} } \nn\\
&\quad \x 
\exp \left[(d+1)  \left(\frac{ \vol\,S^{d-1}}{2}\right)^{\frac{2}{d+1}}\left(
f_{12} ^{\frac{2}{d+1}}\p{\frac{ \De_1 \De_2}{8 (d-1) \De_3}}^{\frac{d-1}{d+1}}
+\textrm{cycl.}
\right)\right],
\ee
where ``$+\textrm{cycl.}$" denotes a sum over cyclic permutations of $123$.
This result is valid for large $\De_i$ with the spin-representations $\lambda_i$ fixed, up to subleading corrections at large $\De_i$. Our approximation for the asymptotic squared OPE coefficients is then
\be
(c_{123}^{s'})^* c_{123}^s &\sim \frac{P^{ss'}_{1 2 3}}{\rho_1\rho_2\rho_3},
\ee
where $\rho_i$ are the densities of states of the CFT computed in section~\ref{sec:density} for the representations $\pi_i=(\De_i,\l_i)$. (In the case, where we refine the partition function with topological defects, the density $\rho_i$ should be the appropriate density of states with that defect inserted.)

As an example, let us study $P^{ss'}_{123}$ for identical-dimension scalars in various dimensions. In 2d, the rotation representations are one dimensional, there is a unique conformal three-point structure, and the corresponding $T$ matrix is simply $T=1/2$. Plugging this in, we find the high energy density of (global primary) OPE coefficients in 2d:
\be
2d &: P_{\De\De\De} \sim \p{\frac 3 2}^{6\De-9}\frac{f^2 e^{\frac{9}{2} (\pi^2 f^2 \Delta)^{1/3}}}{\pi\Delta ^5}.
\ee

In 3d, the $T$ matrix is diagonalized in the $q$-basis of \cite{Kravchuk:2016qvl}. Specifically, it is given by \cite{Karateev:2018oml}
\be
T^{[q_1 q_2 q_3],[q_1' q_2' q_3']} &= \frac{1}{2^3 2\pi} \prod_{i=1}^3{2J_i \choose J_i+q_i}^{-1} \de_{q_i q_i'}.
\ee
Plugging this into (\ref{eq:generalope}), we find the high energy density of OPE coefficients in 3d: 
\be
3d &: P^{[q_1q_2q_3][q_1'q_2'q_3']}_{(\De,J_1)(\De,J_2)(\De,J_3)} \sim 
\p{\frac 3 2}^{6\De}\frac{2^{\frac{49}{4}}   f^{\frac 9 4} e^{3 \sqrt{2 \pi f\Delta } }}{3^{\frac{19}{2}}\pi^{\frac 1 4}\Delta ^{\frac{31}{4}} }
\prod_{i=1}^3(2J_i+1){2J_i \choose J_i+q_i} \de_{q_i q_i'}.
\ee

\subsubsection{A subleading correction}
\label{sec:subleadingcorrection}

It is straightforward to take into account perturbative corrections around the saddle point. For example, let us highlight the leading correction that is not proportional to the $T^{ss'}$ three-point structure matrix. It comes from the $\Phi^2$ term in the expansion of
\be
V^{t'}(0,e,\oo) \prod_i e^{i\Phi_i \. M_{i,\Phi}} V^t(0,e,\oo)^*
\ee
in the genus-2 block.
Performing the Gaussian integral and using the fact that $M_{1,\Phi}+M_{2,\Phi}+M_{3,\Phi}=0$ (because the three-point structures are $\SO(d-1)$-invariant), we find a multiplicative correction of the form
\be
\textrm{$\vec\Phi_i^2$-correction}
&= \p{1-(d-1) \frac{\De_1^2 \De_2^2 M_{3,\Phi}^2 + \De_2^2 \De_3^2 M_{1,\Phi}^2 + \De_3^2 \De_1^2 M_{2,\Phi}^2}{\De_1\De_2\De_3(\De_1^2+\De_2^2+\De_3^2)}}.
\ee
Here, $M_{i,\Phi}^2$ are the Casimirs of the $\SO(d-1)$ subgroup of $\SO(d)$. As discussed in \cite{Karateev:2018oml}, the three-point pairing matrix $T^{s's}$ can be simultaneously diagonalized together with the $M_{i,\Phi}^2$. For example, in 3 dimensions, we have $M_{i,\Phi}^2 = q_i^2$ in the $q$-basis. It will be interesting in the future to compute the full spin-dependence of the asymptotic OPE coefficients by computing the genus-2 blocks in the regime of finite $J/\De$.

\section{Asymptotics of thermal 1-point functions}
\label{sec:thermalonept}

We can also use the techniques developed in this work to determine asymptotics of thermal 1-point functions. (In fact, one can think of high-energy thermal 1-point functions as a particular limit of heavy-heavy-heavy OPE coefficients.) One nice thing about this exercise is that, because the blocks are so simple, we can easily invert the partition function for arbitrary $J/\De$. For brevity, we will only determine the leading exponential form of the thermal 1-point coefficients, leaving 1-loop determinants and subleading corrections for later work.

Recall that the 1-point function of a primary operator $\cO$ at inverse temperature $\b_0$ is fixed by symmetries to be \cite{Iliesiu:2018fao}
\be
\<\cO^{\mu_1\cdots\mu_J}\>_{\b_0} &= \frac{b_\cO}{\b_0^\De} (e^{\mu_1}\cdots e^{\mu_J}-\textrm{traces}), 
\ee
where $e=(1,0,\dots,0)$ is a unit vector in the Euclidean time direction, and $b_\cO$ is an operator-dependent thermal 1-point coefficient. Only even-spin traceless symmetric tensors have nonvanishing thermal 1-point functions.

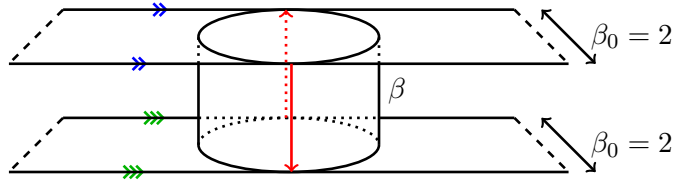
\begin{figure}
\centering
\begin{tikzpicture}[scale=1.2, line width=1pt]
\draw [blue] (1,0.07) -- (1.07,0) -- (1,-0.07);
\draw [blue] (1.07,0.07) -- (1.14,0) -- (1.07,-0.07);
\begin{scope}[shift={(-0.23,-0.6)}]
\draw [blue] (1,0.07) -- (1.07,0) -- (1,-0.07);
\draw [blue] (1.07,0.07) -- (1.14,0) -- (1.07,-0.07);
\end{scope}
\begin{scope}[shift={(-0.1,-1.2)}]
\draw [green!70!black] (1,0.07) -- (1.07,0) -- (1,-0.07);
\draw [green!70!black] (1.07,0.07) -- (1.14,0) -- (1.07,-0.07);
\draw [green!70!black] (1.14,0.07) -- (1.21,0) -- (1.14,-0.07);
\begin{scope}[shift={(-0.23,-0.6)}]
\draw [green!70!black] (1,0.07) -- (1.07,0) -- (1,-0.07);
\draw [green!70!black] (1.07,0.07) -- (1.14,0) -- (1.07,-0.07);
\draw [green!70!black] (1.14,0.07) -- (1.21,0) -- (1.14,-0.07);
\end{scope}
\end{scope}
\draw [] (0,0) -- (5,0);
\draw [] (-0.6,-0.6) -- (5.6,-0.6);
\draw [dashed] (0,0) -- (-0.6,-0.6);
\draw [dashed] (5,0) -- (5.6,-0.6);
\begin{scope}[yscale=0.3,shift={(2.5,-1)}]
\draw [] (0,0) circle (1);
\end{scope}
\begin{scope}[shift={(0,-1.2)}]
\draw [] (0,0) -- (1.5,0);
\draw [] (3.5,0) -- (5,0);
\draw [dotted] (1.5,0) -- (3.5,0);
\draw [] (-0.6,-0.6) -- (5.6,-0.6);
\draw [dashed] (0,0) -- (-0.6,-0.6);
\draw [dashed] (5,0) -- (5.6,-0.6);
\begin{scope}[yscale=0.3,shift={(2.5,-1)}]
\draw [] (-1,0) arc (-180:0:1);
\draw [dotted] (-1,0) arc (180:0:1);
\end{scope}
\end{scope}
\draw [dotted] (1.5,-0.6) -- (1.5,-0.3);
\draw [dotted] (3.5,-0.6) -- (3.5,-0.3);
\draw [] (1.5,-0.6) -- (1.5,-1.5);
\draw [] (3.5,-0.6) -- (3.5,-1.5);
\draw [<->]  (5.3,0) -- (5.9,-0.6);
\node at (5-1.3,-0.9) {$\b$};
\node at (6.3,-0.3) {$\b_0=2$};
\begin{scope}[shift={(0,-1.2)}]
\draw [<->]  (5.3,0) -- (5.9,-0.6);
\node at (6.3,-0.3) {$\b_0=2$};
\end{scope}
\draw [red,->] (2.5+0.03,-0.6) -- (2.5+0.03,-1.8);
\draw [dotted,red,->] (2.5-0.03,-1.2) -- (2.5-0.03,0);
\end{tikzpicture}
\caption{A geometry that encodes a sum of squares of thermal one-point functions. The top surface is a copy of thermal flat space $S_{\b_0=2}^1 \x \R^{d-1}$, with a unit ball removed. The ball is tangent to itself because it wraps completely around the thermal circle of length $\b_0=2$. The bottom is the same as the top. The top and bottom are connected by a cylinder of length $\b$ and angular twist $h\in \SO(d)$. The periodicity of each copy of thermal flat space is illustrated via arrow marks, indicating loci that should be identified. The hot spot thermal circle (red) runs down the cylinder in the front of the figure, and back up the cylinder in the back of the figure. \label{fig:thermalonepointgluing}}
\end{figure}

We can build a geometry that measures squares of the $1$-point coefficients $b_\cO$ as follows.  We start with two copies of thermal flat space $S^1_{2} \x \R^{d-1}$ at inverse temperature $\b_0=2$. From each copy, we drill out unit balls, so that the balls wrap around the thermal circle and are self-tangent. We then glue the boundaries of the two balls together with a cylinder of length $\b$ (not equal to $\b_0$!) and angular twist $h\in \SO(d)$, see figure~\ref{fig:thermalonepointgluing}. By the cutting and gluing arguments of section~\ref{sec:cuttingandgluing}, this geometry computes
\be
Z &= \frac{1}{|Z_\textrm{glue}(1)|^2}\sum_\cO \<\cO\>_{\b_0=2}\<h\.\cO\>_{\b_0=2} e^{-\b (\De+\varepsilon_0)}
\nn\\
&= \frac{1}{|Z_\textrm{glue}(1)|^2} \sum_\cO \frac{b_\cO^2}{2^{2\De}} q_J \cP_J(\cos \th) e^{-\b(\De+\varepsilon_0)}.
\label{eq:zthermalsumstates}
\ee
Here, we used the fact that thermal 1-point functions are $\SO(d-1)$-invariant to write $h$ as a rotation away from the $e$-axis by an angle $\theta$. The function $\cP_J(\cos\th)$ is a Gegenbauer polynomial, given by
\be
\cP_J(x) &= {}_2F_1(-J,J+d-2,\tfrac{d-1}{2},\tfrac{1-x}{2}),
\ee
and $q_J$ is given by
\be
q_J &= \frac{\G(\frac{d-2}{2})\G(J+d-2)}{2^J \G(d-2)\G(J+\frac{d-2}{2})}.
\ee

In the limit $\b\to 0$, this geometry develops a hot spot. The emergent thermal circle goes down the cylinder starting at a point of tangency, and then back up the diametrically opposite side of the cylinder to the starting point. To determine the hot-spot partition function, we should find the conformal group element that glues the plane to itself near this hot spot. On the each copy of the plane, we have a thermal periodicity
\be
x\sim e^{2P^1} x,\quad x'\sim e^{2P^1} x',
\ee
where $P^1$ generates translations in the Euclidean time direction. Meanwhile, the cylinder induces an identification between the two coordinates
\be
x &= e^{-\b D} h I x'.
\ee
The hot-spot thermal circle thus corresponds to the group element
\be
g_\textrm{hot} &= e^{-2P^1}(e^{-\b D} h I)^{-1}e^{2P^1}(e^{-\b D} h I).
\ee

We can define a relative temperature and angle from the eigenvalues of $g_\textrm{hot}$:
\be
(e^{\pm \b_\textrm{rel}},e^{\pm i \th_\textrm{rel}},1,\dots,1) = \textrm{eigenvalues}(g_\textrm{hot}).
\ee
Note that $h$ can be brought to the form of a rotation between the $1$ and $2$ axes. Then $\SO(d-2)$ symmetry guarantees that the eigenvalues of $g_\textrm{hot}$ take the above form. The leading contribution to the partition function from the hot spot is 
\be
Z &\sim \exp\p{\frac{f \vol S^{d-1}}{\b_\textrm{rel}^{d-1}(1 + \Omega_\textrm{rel}^2)}} = \exp\p{\frac{f\vol S^{d-1}}{\b_{\textrm{rel},0}^{d-1}}\p{1-32(d+1)\frac{\th^2}{\b_{\textrm{rel},0}^4} +\dots}},
\label{eq:thermaloneptZ}
\ee
where
\be
\b_{\textrm{rel},0} = \cosh ^{-1}\left(1-8 e^{\beta }+8 e^{2 \beta }\right) = 4\sqrt \b + \dots
\ee
is the relative inverse temperature when $\th=0$, and this formula is valid for $\th^2/\b^2\ll 1$ and $\b\ll 1$. Formula (\ref{eq:thermaloneptZ}) is the analog of (\ref{eq:quadraticexpansion}) from the genus-2 calculation.

All that remains is to invert (\ref{eq:thermaloneptZ}) to determine the asymptotics of the coefficients $b_\cO$. In doing so, we can use the orthogonality relation for Gegenbauer polynomials
\be
\int_0^\pi d\th \sin^{d-2}\th\, \cP_J(\cos\th) \cP_{J'}(\cos\th) &= n_J \de_{JJ'},
\ee
where
\be
n_J &= \frac{\pi \G(J+1)}{\G(\frac{d-2}{2})^2 (J+\frac{d-2}{2})\G(J+d-2)}.
\ee
When we integrate $\cP_J(\cos \th)$ in $\th$ against the Gaussian in (\ref{eq:thermaloneptZ}), the integral will be dominated by small $\th$ with fixed $\th J$. In this regime, we can use an approximation for the Gegenbauer function in terms of a Bessel function (which plays an important role in dispersive bounds on scattering amplitudes \cite{Caron-Huot:2021rmr}):
\be
\lim_{J\to \oo,\, \th J\textrm{ fixed}}\cP_J(\cos \th) &= \frac{\G(\frac{d-1}{2})}{(\th J/2)^{\frac{d-3}{2}}}J_{\frac{d-3}{2}}(\th J) = \int \frac{d\vec n}{\vol\, S^{d-2}} e^{i J \vec n \.\vec \th}.
\ee
Here, $J_\a(x)$ is a Bessel function. In the right-hand formula, $\vec n$ is a point on the unit $S^{d-2}$, and we think of $\vec \th$ as a vector in $\R^{d-1}$ with norm $|\vec \th|=\th$. The idea is that $\cP_J(\cos\th)$ satisfies a wave equation on $S^{d-1}$. In the limit of large $J$ with small $\th$, we can zoom in near the locus $\th=0$, where the $S^{d-1}$ becomes flat space $\R^{d-1}$. We are left with a linear combination of solutions to the wave equation in flat space --- i.e.\ plane waves. In this limit, the measure $d\th \sin^{d-2} \th$ becomes equivalent to the usual measure $d^{d-1}\vec \th$ on $\R^{d-1}$.

Thus, overall, the angular integral from inverting the partition function takes the form
\be
&\int \frac{d\vec n}{\vol\, S^{d-2}} \int d^{d-1} \vec \th\, e^{i J \vec n \.\vec \th} \exp\p{-\frac{32(d+1)f\vol S^{d-1}}{\b_{\textrm{rel},0}^{d+3}}\vec \th^2}\nn\\
&= \exp\p{-\frac{\b_{\textrm{rel},0}^{d+3}}{128 (d+1) f \vol\,S^{d-1}}J^2 } \x \textrm{1-loop determinant}.
\ee
At the same time, we must perform an inverse Laplace transform in $\b$. This integral can be done by saddle point, with the overall result
\be
\label{eq:thermalonepointresult}
\frac{b_\cO^2 \rho(\De,J)}{2^\De} q_J n_J &\sim \exp \left(\frac{1}{\Delta }\left(\frac{d+1}{d-1} \Delta ^2-\frac{d-1}{d+1} J^2\right) \left(\frac{(d-1) f\vol\,S^{d-1}}{2^{2 d-1} \Delta }\right)^{\frac{2}{d+1}}\right),
\ee
where $\rho(\De,J)$ is the density of states for traceless symmetric tensors.

\section{Discussion}
\label{sec:discuss}

In this paper, we studied the asymptotic behavior of CFT data at large energy. Using the thermal effective action, we looked at both the density of states and the three-point-functions of heavy operators as a function of $\Delta, J$. There are a number of interesting future directions to study.

\subsection{Density of states}

The formula (\ref{eq:rhogenerald}) for the density of states is valid in a specific region of $\Delta, J$. For example, in CFT$_3$, it is valid when 
\begin{equation}
\Delta - |J| \gg \sqrt{f \Delta}.
\end{equation}
This notably has no overlap with the regime of large spin with fixed twist  described by the lightcone bootstrap \cite{Fitzpatrick:2012yx, Komargodski:2012ek}. Naively, the lightcone bootstrap suggests that the spectrum of interacting CFT should look like Mean Field Theory in this regime \cite{Alday:2007mf,Fitzpatrick:2015qma,Simmons-Duffin:2016wlq}. It would be interesting if one could prove this statement using some kind of effective action, perhaps by compactifying the CFT on a null circle. 
It would be also interesting to study how the spectrum of operators can behave \emph{between} these regions. For instance, for interacting 3d CFTs, is the density of states at large spin with twist obeying
\begin{equation}
\Delta^0 \ll \Delta - |J| \ll \sqrt{f \Delta}
\end{equation}
universal or theory-dependent?

One could also ask to further refine our general entropy formulas. In \cite{Kang:2022orq}, a universal formula for CFT$_d$ with global symmetry was found. It would be nice to combine them and obtain universal formulas as a function of energy, spin, and global charges.

In section~\ref{sec:examplesdensityofstates}, we compared the predictions of the thermal effective action to exact results in free theories and Einstein gravity, finding excellent agreement. In appendix~\ref{sec:3dising}, we give a preliminary comparison between the thermal effective action for the $S^1\x S^2$ partition function and numerical bootstrap data for the 3d Ising CFT. One could also consider other theories where a large number of operators are known from numerics, e.g.\ the $O(2)$ model \cite{Chester:2019ifh,Liu:2020tpf}. Obtaining accurate information about large-twist operators is a challenge for the numerical bootstrap, which seems to be most sensitive to the lowest-twist Regge trajectories \cite{Simmons-Duffin:2016wlq}. (Computing a large number of heavy-heavy-heavy OPE coefficients with the numerical bootstrap is likely even more challenging.)

In 2d CFTs, it is possible to make very precise statements about the spectrum of high energy states using more sophisticated tools than Laplace transforms and saddle point approximations, see e.g.~\cite{Mukhametzhanov:2019pzy,Benjamin:2019stq,Ganguly:2019ksp,Pal:2019zzr,Mukhametzhanov:2020swe,Pal:2020wwd,Das:2020uax}. Such techniques typically rely on nonperturbative input coming from modular invariance. Is it possible to derive similarly precise statements in higher dimensions? What additional information about the partition function is needed?

\subsection{Effective actions}

We parametrized our ignorance of the $d{-}1$-dimensional gapped theory upon compactifying on a thermal circle via an effective action, with an infinite set of Wilson coefficients. Can we place bounds on these Wilson coefficients? For instance, as discussed in Section \ref{sec:cosmoconst}, we know that $f > 0$. Are there similar bounds (in either direction) on $c_1$, $c_2$, or other higher-derivative Wilson coefficients? One possible approach is to consider Weinberg-like sum rules relating two-point functions in the IR (described by the thermal effective action) and the UV (described by the CFT), as recently done in \cite{Creminelli:2022onn}. Another approach is to consider the compactified theory in $(d-2,1)$ (Lorentzian) signature and study dispersive bounds on scattering, following e.g.\ \cite{Adams:2006sv,Komargodski:2011vj,Caron-Huot:2020cmc}. 

It may also be interesting to study perturbative examples. For example, the value of $f$ in the 3d $O(N)$ models at large $N$ was computed long ago by Sachdev \cite{Sachdev:1993pr}. To our knowledge, higher Wilson coefficients in the thermal effective action, like the coefficients of $F^2$ and $\hat R$, have not yet been computed for the $O(N)$ models.

In this work, we obtained all of our results purely using equilibrium hydrodynamical information. Recently, there has been a surge of progress in non-equilibrium hydrodynamics for CFTs, see  \cite{Liu:2018kfw,Brauner:2022rvf} for reviews. What additional CFT data can be predicted using this more sophisticated machinery? See \cite{Delacretaz:2020nit,Karlsson:2022osn} for recent work in this direction. Can one study non-equilibrium dynamics at higher genus?

There has also been tremendous recent progress applying other effective actions to characterize asymptotic CFT data, for example the effective theory of large charge \cite{Hellerman:2015nra,Monin:2016jmo,Jafferis:2017zna}. One way of summarizing our ``hot spot" analysis of the genus-2 partition function is the idea of using an EFT in the part of a geometry where it is valid (the hot spots), and factoring out the part where the EFT is not valid (the region away from the hot spots). Can this ``hot spot" idea be useful in other contexts like large charge?

\subsection{Three-point functions and genus-2 blocks}

So far, we calculated asymptotic OPE coefficients to leading order at large $\Delta$, with fixed spin. It would interesting to allow the spin to grow large with $\De$, as we did for the density of states. In particular, this would require a more general expression for the genus-2 conformal block at large quantum numbers.

Genus-2 blocks are interesting objects in their own right, and it would be interesting to study their properties more systematically, both at large and non-large quantum numbers. For example, can we find recursion relations for genus-2 blocks similar to those in \cite{Zamolodchikov1987,Kos:2013tga,Kos:2014bka,Penedones:2015aga,Erramilli:2019njx}? Can we explore genus-2 blocks from the perspective of integrability \cite{Isachenkov:2017qgn}? Is there a clearer understanding of the interesting saddle-point dynamics uncovered in section~\ref{sec:hightempsaddle}? Do there exist Lorentzian shadow representations \cite{Polyakov:1974gs} or holographic representations \cite{Hijano:2015zsa} for higher-genus blocks, and do they admit any interesting kinematic limits? The literature on global conformal blocks for correlation functions of local operators is vast, but global genus-2 blocks are essentially unexplored.

We would also like to understand how to systematically improve our three-point function result. For the density of states, we understand how to systematically improve the result by keeping further terms in the thermal effective action. However, for the three-point function, corrections come in two types: higher derivative terms in the effective action (which are easy to include), and corrections to the hot spot assumption, as discussed in Sec. \ref{sec:hotspotgenustwo}. In order to understand how to systematically improve the estimate for the HHH three-point functions, we need to understand contributions to the partition function outside of the hot spot regions, namely understanding the quantity $R$ defined in (\ref{eq:Rdefinition}). Explicitly computing examples of ``genus-two" partition functions, either for free or holographic theories, could be instructive.

It is also worthwhile to compare our result to known results CFT$_2$. In \cite{Collier:2019weq}, it was shown that HHH, HHL, and HLL asymptotic density of states for Virasoro primary operators are all related to analytic continuations of the DOZZ formula --- the structure constants of Liouville theory. In higher dimensions, asymptotic formulas for HLL OPE coefficients have been studied using Tauberian techniques and inversion formulae \cite{Pappadopulo:2012jk,Rychkov:2015lca,Mukhametzhanov:2018zja}. Furthermore, it is well-known that HHL OPE coefficients are related to thermal one-point functions. These computations, together with our genus-2 computation seem to involve different physics. It would be extremely interesting if there were a unifying perspective or formula similarly to 2d.

\subsection{Bootstrap axioms and crossing equations}

To what extent are our results for the density of states and OPE coefficients encoded in the usual bootstrap conditions  --- namely unitarity and crossing symmetry of local correlation functions? In 2d, modular invariance is known to be independent from crossing symmetry of local correlators. By analogy, this suggests that perhaps the formulas we derived from the thermal effective action are {\it independent} from the usual bootstrap axioms.\footnote{We thank Dalimil Maz\'{a}\v{c} for pointing this out.}  If so, should we enlarge the axioms to include them? What is the minimal set of extra axioms that we need? In 2d, modular invariance can be interpreted as crossing symmetry of twist operators. Can our results in higher dimensions be interpreted in terms of traditional bootstrap axioms applied to appropriate twist operators?

As we mentioned briefly in section~\ref{sec:conformalstructures}, there exists another decomposition of the genus-2 partition function $Z(M_2)$ into a sum over states: the ``dummbell" channel, which expresses $Z(M_2)$ as a sum of squares of 1-point functions on $S^1 \x S^{d-1}$. The dumbell channel has its own conformal blocks, which as far as we know have not been studied in detail. (The blocks discussed in section~\ref{sec:thermalonept} can be thought of as a limiting case of these dumbbell blocks.) Furthermore, one can formulate a crossing equation relating the dumbbell channel to the channel considered in this work. As pointed out in \cite{Cho:2017fzo} for $d=2$, this crossing equation enjoys manifest positivity properties needed for numerical bootstrap applications. It would be very interesting to explore it in both 2-dimensional and higher-dimensional theories.

\subsection{Ensembles and holographic theories}

There is an important difference between our higher-dimensional result for asymptotic OPE coefficients and the 2d results of \cite{Cardy:2017qhl,Collier:2019weq}. The results of \cite{Cardy:2017qhl,Collier:2019weq} were for OPE coefficients of {\it Virasoro} primaries, while our results are for OPE coefficients of {\it global} primaries. In the case of the density of states, there isn't a huge difference between Virasoro and global primaries. But the story is different for OPE coefficients, where descendant states play an important role. We can see this by comparing the leading exponential behavior of Virasoro and global OPE coefficients in 2d:
\be
\label{eq:globalvsvirasoro}
P^\textrm{global}_{\De\De\De} \sim\p{\frac 3 2}^{6 \De} \quad\gg\quad P^\textrm{Virasoro}_{\De\De\De} \sim \p{\frac{27}{16}}^{3\De}.
\ee
We see that typical global primaries have much larger OPE coefficients than typical Virasoro primaries.\footnote{Similarly, by comparing scaling blocks and full genus-2 blocks in the high temperature regime, we conclude that typical states have much larger OPE coefficients than typical global primaries.} In other words, the statistics of CFT data in a theory with Virasoro symmetry has more structure than is captured by $P_{\De\De\De}^\textrm{global}$.

These statements are interesting to consider in a holographic CFT. In a holographic 2d CFT, a high energy Virasoro primary is interpreted as a black hole microstate, while a Virasoro descendant is a black hole orbited by boundary gravitons. We have found that states with boundary gravitons typically have much larger OPE coefficients than pure black hole microstates. While we don't have an analog of Virasoro symmetry in higher dimensions, we can conjecture an analogous statement for higher-dimensional CFTs: we expect that typical states of black holes with orbiting matter have much larger OPE coefficients than pure black hole microstates. It would be very interesting to make this more precise, for example by performing a holographic computation of OPE coefficients of pure black hole microstates via an appropriate wormhole geometry.

The authors of \cite{Chandra:2022bqq} used $P^\textrm{Virasoro}_{\De\De\De}$ to define an interesting ``ensemble" of CFT data. In their ensemble, OPE coefficients are (almost) gaussian random variables whose variance is set by $P^\textrm{Virasoro}_{\De\De\De}/\rho(\De)^3$. Remarkably, the predictions of this ensemble turn out to agree with bulk 3d gravity. The result (\ref{eq:globalvsvirasoro}) indicates that an analogous ensemble based on $P^\textrm{global}_{\De\De\De}$ would not have refined-enough information to recover bulk gravity in 2d (presumably also in higher $d$). However, it is interesting to ask whether any interesting physics {\it would} be captured by an ensemble built from $P^\textrm{global}_{\De\De\De}$. In the spirit of \cite{Belin:2021ibv,Belin:2021ryy,Chandra:2022bqq,Jafferis:2022uhu}, we can imagine starting with a completely general ensemble of CFT data. We can refine this ensemble with knowledge of the partition functions on $S^1_\b \x S^{d-1}$ and the ``genus-2" geometry $M_2$. We could additionally refine the ensemble with other information like local correlation functions and thermal one-point functions. At what point does the refined ensemble begin to make nontrivial predictions that can be tested in additional observables, and what are those predictions? 

\subsection{Bulk locality for thermal observables}

HPPS famously conjectured that any unitary CFT with large $c_T$ and a large gap $\De_\textrm{gap}$ in the spectrum of higher-spin single-trace operators should agree with a local gravitational EFT in AdS \cite{Heemskerk:2009pn}. Recently, there has been significant progress proving this statement for correlators of local operators in the CFT vacuum \cite{Hartman:2015lfa,Afkhami-Jeddi:2016ntf,Kologlu:2019bco,Belin:2019mnx,Caron-Huot:2021enk}. However, holography implies  analogous statements in nontrivial backgrounds as well --- in particular a thermal background. For example, the Wilson coefficients $f,c_1,c_2$, etc.\ in the thermal effective action of a theory satisfying HPPS conditions should agree with those of Einstein gravity, up to small corrections suppressed by $1/\De_\textrm{gap}$. How can we prove the emergence of black hole physics using field-theoretic methods? Can we formulate dispersion relations in a black hole background? For recent work in these direction, see \cite{Caron-Huot:2022lff}.

\subsection{Completing the square in the thermal bootstrap}

An interesting feature of our formulas (\ref{eq:zthermalsumstates}) and (\ref{eq:thermaloneptZ}) is that they provide a kind of sum rule for squares of thermal 1-point coefficients $b_\cO^2$. Such a sum rule could in principle be used to ``complete the square" in the bootstrap equations studied in \cite{Iliesiu:2018fao,Iliesiu:2018zlz}.

 The works \cite{Iliesiu:2018fao,Iliesiu:2018zlz} studied crossing symmetry of thermal two-point functions, which have an expansion in products $c_{\f\f\cO}b_\cO$, where $c_{\f\f\cO}$ are bulk OPE coefficients. Unfortunately, we do not know the sign of $c_{\f\f\cO}b_\cO$, and this prevents one from applying traditional numerical bootstrap techniques \cite{Rattazzi:2008pe}. (The same issue appears in the study of boundary and defect two-point functions \cite{Liendo:2012hy}.) Ideally, one would like to complete the square by finding other crossing equations in which $c_{\f\f\cO}$ and $b_\cO$ appear quadratically. Then, one can treat $c_{\f\f\cO}b_\cO$ as an off-diagonal element in a positive-definite $2\x 2$ matrix and apply numerical bootstrap techniques for mixed correlators \cite{Kos:2014bka}.

A crossing equation where $c_{\f\f\cO}$ appears quadratically is easy to find: it is just the usual crossing equation for vacuum four-point functions! Tantalizingly, the formulas (\ref{eq:zthermalsumstates}) and (\ref{eq:thermaloneptZ}) have $b_\cO$ appearing quadratically, but unfortunately they are not as precise as the usual four-point crossing equation, due to our use of the hot-spot ansatz. It would be interesting to go beyond the hot-spot ansatz and find a sum rule precise enough to be used in the numerical bootstrap.

\subsection{``Sphere packing" and other hot-spot geometries}

In addition to the ``genus-2" manifold $M_2$ studied in this work, there are many additional geometries that encode statistics of CFT data and can be studied using the hot-spot idea. Partition functions on these geometries are examples of ``generalized spectral form factors" \cite{Belin:2021ibv}.

As a simple example, consider a higher-genus generalization of $M_2$, where we take two copies of $\R^{d}$, drill out $n>3$ balls from each copy, and connect the boundaries of the balls with cylinders. The partition function on this geometry encodes a sum of squared $n$-point correlation functions of the CFT, schematically
\be
\sum_{\cO_1,\dots,\cO_n} |\<\cO_1\cdots\cO_n\>|^2 e^{-\sum_i \b_i \De_i}.
\ee
 If two balls are tangent in (both copies of) $\R^d$, we obtain a hot spot when the corresponding cylinders shrink to zero length.

\begin{figure}
\centering
\begin{tikzpicture}[scale=1.2, line width=1pt,rotate=30]
    \fill[even odd rule, fill=gray!15] (0, 0) circle (2) (0, 0.8) circle (1.2) (0.2,-1.18) circle (0.8) (-1.17,-0.65) circle (0.65);
    
    \draw [] (0, 0) circle (2);
    
    \draw [] (0, 0.8) circle (1.19);
    \draw [] (0.2,-1.18) circle (0.8);
    \draw [] (-1.17,-0.64) circle (0.655);
        
\end{tikzpicture}
\caption{A ball with three mutually-tangent balls removed. If we take two copies of this space and glue the boundaries of the balls together with cylinders, analogously to figure~\ref{eq:gluedcylindersfigure}, we obtain a geometry that computes a sum of squares of CFT four-point functions. When the cylinders shrink, this ``sphere packing" geometry contains hot spots at each of the six points of tangency. \label{eq:hotspotfour}}
\end{figure}
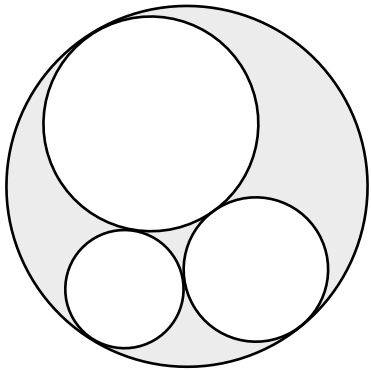

The hot spot ansatz is most useful when there are a maximal number of hot spots, and all other moduli of the geometry are frozen. Thus, we should consider configurations where most of the balls are mutually tangent --- i.e.\ sphere packings!\footnote{See \cite{Dias:2017coo,Hartman:2019pcd} for other connections between sphere packing and conformal field theory.} For example, the packing shown in figure~\ref{eq:hotspotfour} encodes interesting asymptotics of CFT four-point functions.

We can construct an even more general ``higher genus" manifold as follows. We take $m$ copies of $\R^d$ and drill out various numbers of balls from each copy, such that there are an even number of balls in total. We then connect pairs of balls with cylinders. This computes a sum of products of $m$ correlation functions
\be
\sum_{\cO_1,\cdots,\cO_n} \underbrace{\<\cdots\>\cdots \<\cdots\>}_m  e^{-\sum_i \b_i \De_i}.
\ee
Again, hot spots can emerge when cylinders shrink to zero. 

Of course, CFT $n$-point functions for $n>3$ are determined in terms of 2- and 3-point functions. In this work we have determined asymptotics of 2- and 3-point functions, and it is interesting to ask whether our results can be used to predict partition functions on higher genus geometries, and whether the results agree with the hot-spot ansatz at higher genus. In 2 dimensions, it is known that crossing symmetry of local four-point functions and torus one-point functions implies crossing symmetry on arbitrary Riemann surfaces. However, we expect that in higher dimensions, hot spot results for higher genus manifolds provide a nontrivial refinement of the statistics computed in this work.

\section*{Acknowledgements}	

We thank Alex Belin, Alejandra Castro, Stuart Dowker, Liam Fitzpatrick, Tom Hartman, Zohar Komargodski, Petr Kravchuk, Alex Maloney, Henry Maxfield, Dalimil Maz\'{a}\v{c}, Jake McNamara, Shiraz Minwalla, Sridip Pal, Julio Parra-Martinez, Mukund Rangamani, Edgar Shaghoulian, Douglas Stanford, Herman Verlinde, Pedro Vieira, Yifan Wang, and Sasha Zhiboedov for helpful discussions. 
We thank Zohar Komargodski, Edgar Shaghoulian, and Yifan Wang for comments on the draft.
This material is based upon work supported by the U.S. Department of Energy, Office of Science, Office of High Energy Physics, under Award Number DE-SC0011632. In addition, 
NB is supported in part by the Sherman Fairchild Foundation. 
HO is supported in part by the World Premier International Research Center Initiative, MEXT, Japan, and by
JSPS Grants-in-Aid for Scientific Research 20K03965 and 23K03379. 
DSD is supported in part by Simons Foundation grant 488657 (Simons Collaboration on the Nonperturbative Bootstrap) and a DOE Early Career Award under grant No.\ DE-SC0019085.

\pagebreak
	
\appendix

\section{Thermal two-point function of momentum generators}
\label{sec:warmup}

In section~\ref{sec:density}, we study the response of a CFT on $S^1_\beta \x S^{d-1}$ when we twist by a rotation of $S^{d-1}$. In this appendix, as a warmup, we study the leading term in the twisting parameter in the high temperature limit. This computation reduces to a two-point function of momenta in thermal flat space. We determine this two-point function directly from Ward identities, and then show how the same result can be understood using the thermal effective action.

At high temperature, the radius of curvature of the sphere becomes unimportant, and we can approximate $S^1_\beta \x S^{d-1}$ as thermal flat space $S^1_\beta \x \R^{d-1}$. A rotation generator on the sphere locally looks like a translation on $\R^{d-1}$. Thus, it suffices to study the thermal expectation value of a translation group element
\be
\label{eq:exponentiatedtranslation}
\<e^{\vec a\.\vec P}\>_\beta &= \<1\>_\beta + \frac 1 2 a_i a_j \<P^i P^j\>_\beta + \dots,
\ee
The momentum $P^i$ is the integral of $T^{0i}$ on a spatial slice at fixed Euclidean time $\tau$ (which can take any value, by conservation):
\be
P^i &= -\int d\vec x\, T^{0i}(\tau,\vec x).
\ee
The $O(a)$ term in (\ref{eq:exponentiatedtranslation}) vanishes by rotation-invariance. For simplicity, in this section we set $\beta=1$, restoring it when needed by dimensional analysis.

\subsection{Using Ward identities}

Let us focus on the quadratic term in (\ref{eq:exponentiatedtranslation}), given by an integrated two-point function of stress tensors:
\be
\label{eq:integratedcorrelator}
\frac{1}{\vol \R^{d-1}}\frac 1 2 a_i a_j \<P^i P^j\>_\beta &= \frac 1 2 a_i a_j \int d\vec x_1 \<T^{0i}(0,\vec x_1) T^{0j}(x_2)\>_\beta.
\ee
Here, we separated the momentum generators in Euclidean time, placing the first at time $\tau_1=0$ and the second at time $\tau_2$. We furthermore divided by $\vol\, \R^{d-1}$ to obtain a finite result, and used translation-invariance in $\R^{d-1}$ to fix the second stress tensor at $x_2=(\tau_2,\vec x_2)$.

We claim that the integrated two-point correlator (\ref{eq:integratedcorrelator}) is determined by the {\it one-point\/} function of the stress tensor at finite temperature. To understand why, we must express it in terms of operators in the dimensionally-reduced $d{-}1$-dimensional theory. The first step is to average over Euclidean times $\tau_1$ and $\tau_2$. However, this averaging is subtle because $T^{0i}(x_1)$ and $T^{0j}(x_2)$ become coincident, and contact terms can contribute. Such contact terms are actually crucial to the calculation, so let us take a moment to define them carefully.

We define (un-normalized) one- and two-point functions of the stress tensor by
\be
\sqrt{G(x)}  \<T^{\mu\nu}(x) \> &= 2 \frac{\de Z}{\de G_{\mu\nu}(x)},\\
\sqrt{G(x)} \sqrt{G(y)} \<T^{\mu\nu}(x) T^{\rho\s}(y)\> &= 4 \frac{\de^2 Z}{\de G_{\mu\nu}(x) \de G_{\rho\s}(y)},\label{eq:twopointfunction}
\ee
where $Z[G]$ is the partition function. Our definition of the two-point function (\ref{eq:twopointfunction}) applies at both coincident and non-coincident points, and thus suffices to specify all contact terms. Diffeomorphism invariance of $Z[G]$ implies that
\be
0=\int d^d x \cL_\xi G_{\mu\nu} \frac{\de Z}{\de G_{\mu\nu}},
\ee
where $\cL_\xi$ denotes the Lie derivative with respect to a vector field $\xi^\mu$. Taking an additional derivative with respect to $G_{\rho\s}(y)$, and evaluating in a locally flat metric $G_{\mu\nu}=\de_{\mu\nu}$, we obtain the following Ward identity for conservation of the stress tensor inside a two-point correlator
\be
&\int d^d x\, (\ptl_\mu \xi_\nu(x)) \<T^{\mu\nu}(x)T^{\rho\s}(y)\> \nn\\
&= (\xi\.\ptl) \<T^{\rho\s}(y)\> + (\ptl\.\xi)\<T^{\rho\s}(y)\> - \ptl_\mu \xi^\rho \<T^{\mu\s}(y)\>- \ptl_\mu \xi^\s \<T^{\mu\rho}(y)\>.
\label{eq:wardidentitythingy}
\ee

We can use this identity to average the correlator (\ref{eq:integratedcorrelator}) over Euclidean time. Consider the vector field
\be
\label{eq:specificvectorfield}
\xi^i(\tau,\vec x) = -a^i \tau,\quad \xi^0(\tau,\vec x) = 0,
\ee
where $\tau\in [0,1]$. Since $\tau$ is periodic, $\xi^i$ has the shape of a ``sawtooth" function, with a discontinuity at $\tau=0$.  In particular, we have
\be
\ptl_\tau \xi^i &= a^i (\de(\tau)-1).
\ee
Applying (\ref{eq:wardidentitythingy}), we find
\be
\label{eq:euclideanaveraged}
\int d\vec x_1 a_i a_j \<T^{0i}(0,\vec x_1) T^{0j}(y)\>_\beta
&=\int d\tau_1 d\vec x_1 a_i a_j \<T^{0i}(\tau_1,\vec x_1) T^{0j}(y)\>_\beta+a^2\<T^{00}(y)\>_\beta,
\ee
where we used $\ptl\.\xi=0$ and translation invariance $\xi\.\ptl\<T^{\rho\s}\>_\beta=0$.
The right-hand side of (\ref{eq:euclideanaveraged}) is the two-point correlator averaged over Euclidean time, plus a nontrivial contact term $a^2 \<T^{00}(y)\>$ that is a consequence of diffeomorphism invariance.

It is natural to define the $d{-}1$ dimensional stress tensor $t^{ij}(\vec x)$ and KK current $j^i(\vec x)$ via derivatives with respect to $g_{ij}$ and $A_i$ in the Kaluza-Klein parametrization (\ref{eq:kkgeometry}). For example, we have
\be
\label{eq:lowerdimstressandcurrent}
\sqrt{g(\vec x)}\<t^{ij}(\vec x)\> &\equiv \frac{\de Z}{\de g_{ij}(\vec x)}, \nn\\
\<j^{i}(\vec x)j^j(\vec y)\> &\equiv \frac{\de^2 Z}{\de A_i(\vec x) \de A_j(\vec y)}.
\ee
A key property of the KK parametrization (\ref{eq:kkgeometry}) is that gauge transformations $A_i(\vec x)\to A_i(\vec x) + \ptl_i \l(\vec x)$ are diffeomorphisms of the $d$-dimensional metric. Consequently, diff-invariance implies that $\<j^{i}(\vec x)j^j(\vec y)\>$ is exactly conserved, {\it even at coincident points\/}. This will be important in a moment.

At separated points $t^{ij}(\vec x)$ and $j^i(\vec x)$ are equivalent to Euclidean time averages of $T^{ij}(\tau,\vec x)$ and $T^{0i}(\vec x)$, respectively. However, at coincident points, they differ from na\"ive averages by contact terms. In particular, the definitions (\ref{eq:lowerdimstressandcurrent}) and (\ref{eq:twopointfunction}) imply (on a flat geometry)
\be
\label{eq:jjcorrelator}
\<j^i(\vec x_1) j^j(\vec x_2)\> &= \int d\tau_1 d\tau_2 \<T^{0i}(x_1) T^{0j}(x_2)\> + \de(\vec x_1 - \vec x_2) \int d\tau_1 \<T^{ij}(x_1)\>.
\ee
The contact term on the right-hand side arises from the quadratic term in $A_i$ in the KK metric: $G_{ij}=g_{ij}+e^{2\f} A_i A_j$.

Integrating (\ref{eq:jjcorrelator}) over $\vec x_1$, and combining it with the average of (\ref{eq:euclideanaveraged}) over $\tau_2$, we find
\be
\label{eq:rewrittenthing}
\int d\vec x_1 a_i a_j \<T^{0i}(0,\vec x_1) T^{0j}(y)\>_\beta
&=a_i a_j\int d \vec x_1  \<j^i(\vec x_1)j^j(\vec x_2)\> - a_i a_j \<T^{ij}\>_\beta + \vec a^2 \<T^{00}\>_\beta.
\ee
Finally, we will argue that the integrated correlator $\int d\vec x_1 \<j^i(\vec x_1)j^j(\vec x_2)\>$ vanishes.
We can think of it as the momentum-space two-point function $\<j^i(\vec p) j^j(-\vec p)\>$, evaluated at zero momentum. Rotation-invariance and conservation imply that the momentum-space two-point function takes the form
\be
\label{eq:jjansatz}
\<j^i(\vec p) j^j(-\vec p)\> &= (\vec p^i \vec p^j - \de^{ij} \vec p^2)G(|\vec p|).
\ee
If the finite-temperature theory has a nonzero mass gap, then $G(|\vec p|)$ must be regular near zero momentum. (Otherwise its Fourier transform would have support at long distances.) Thus, at low momenta, we have
\be
\<j^i(\vec p) j^j(-\vec p)\> &= c(\vec p^i \vec p^j - \de^{ij} \vec p^2) + O(\vec p^4).
\ee
In particular, $\<j^i(\vec p) j^j(-\vec p)\>$ vanishes at $\vec p=0$. Note that conservation of $j^i(\vec x)$ at coincident points is crucial here. Without it, the momentum-space correlator could have an $O(\vec p^0)$ contact term of the form $\de^{ij}$.

It is instructive to understand this vanishing result in position space as well. In the position-space integral $\int d\vec x_1 a_i a_j \<j^i(\vec x_1)j^j(\vec x_2)\>$, we can write $a_i=\ptl_i (\vec a\.\vec x_1)$. Integrating by parts and using conservation, we obtain a boundary term at infinity:
\be
a_i a_j\int d \vec x_1  \<j^i(\vec x_1)j^j(\vec x_2)\> &= \lim_{R\to \oo} a_j\int_{|\vec x|=R} dS_i (\vec a\.\vec x_1) \<j^i(\vec x_1) j^j(\vec x_2)\>,
\label{eq:boundarytermatinfinity}
\ee
where $dS_i$ is the surface normal for the sphere $|\vec x|=R$.
This boundary term (\ref{eq:boundarytermatinfinity}) vanishes provided the two-point function decays sufficiently quickly at large $|\vec x|$. In other words,
\be
\label{eq:decaycondition}
a_i a_j\int d \vec x_1  \<j^i(\vec x_1)j^j(\vec x_2)\> = 0\qquad \textrm{if}\qquad \lim_{|\vec x|\to \oo} |\vec x|^{d-1} \<j^i(\vec x) j^j(0)\> &= 0.
\ee
This condition certainly holds when the finite-temperature theory has a mass gap (since the correlator decays exponentially). However, it also holds more generally. For example, if the finite-temperature theory possesses a massless sector, we expect the current two-point function to decay no slower than a correlator of conserved currents in a $d{-}1$ dimensional CFT: $\<j(\vec x)j(0)\>\sim |\vec x|^{-2(d-2)}$. In that case, (\ref{eq:decaycondition}) will be satisfied as long as $d>3$.

Finally, using (\ref{eq:decaycondition}) in (\ref{eq:rewrittenthing}), we find
\be
\int d\vec x_1 a_i a_j \<T^{0i}(0,\vec x_1) T^{0j}(y)\>_\beta
&=- a_i a_j \<T^{ij}\>_\beta + \vec a^2 \<T^{00}\>_\beta 
= -(f d) \vec a^2,
\ee
where we used (\ref{eq:stressonept}) for the one-point functions $\<T^{\mu\nu}\>_\beta$.
We conclude 
\be
\<e^{\vec a\.\vec P}\>_\beta &= 1 - \frac {f d} 2 T^{d+1} \vec a^2 \vol\, \R^{d-1} + \dots,
\ee
where we have restored factors of $T$ by dimensional analysis. To apply this result to $S^1_\beta \x S^{d-1}$ with a twist by a rotation of $S^{d-1}$, we can make the replacement
\be
\vec a^2 \vol\,\R^{d-1} &\to \int_{S^{d-1}} d\Omega_{d-1} |v|^2,
\ee
where $v$ is the Killing vector on $S^{d-1}$ implementing the rotation.

\subsection{Using the thermal effective action}

The thermal effective action gives an efficient way to package the Ward identity calculations above and extend them to arbitrary nonlinear order in $\vec a$. Let us see how it recovers the result (\ref{eq:exponentiatedtranslation}). The correlator $\<e^{\vec a \. \vec P}\>$ is captured by the geometry $S^1_\beta \x \R^{d-1}$ with a twist of $\vec a$ around the thermal circle, i.e.\ an identification
\be
(\tau,\vec x) &\sim (\tau+1,x-\vec a).
\ee
To use the thermal effective action, we must put the metric into Kaluza Klein form. We undo the twist with a coordinate transformation
\be
\vec x' = \vec x - \tau \vec a.
\ee
This essentially implements averaging over Euclidean time (\ref{eq:euclideanaveraged}) by spreading out the twist over the thermal circle. The metric changes to 
\be
d\vec x^2 + d\tau^2
&= (d\vec x'+\vec a d\tau)^2 + d\tau^2  \nn\\
&= d\vec x'^2 - \frac{(\vec a\.d\vec x')^2}{1+\vec a^2} + (1+\vec a^2)\p{d\tau + \frac{\vec a\.d\vec x'}{1+\vec a^2}}^2.
\ee
The effective metric is thus
\be
\hat g_{ij} &= \frac{1}{1+\vec a^2}\p{\de_{ij} - \frac{a_i a_j}{1+ \vec a^2}},
\ee
and the thermal effective action is
\be
S[\hat g,A] &= -f\vol\, \R^{d-1} \sqrt{\hat g}  = -f\vol\, \R^{d-1} (1+\vec a^2)^{-d/2}.
\ee
Finally, the partition function is
\be
e^{-S[\hat g,A]} = e^{f\vol\, \R^{d-1} (1+\vec a^2)^{-d/2}} = e^{f \vol\, \R^{d-1}}\p{1-\frac{f d}{2} \vec a^2 \vol \,\R^{d-1} + \dots},
\ee
in agreement with (\ref{eq:exponentiatedtranslation}).

These manipulations are clearly easier and more efficient than those in the previous section. However, detailed manipulations of correlators are instructive as well. For instance, they tell us that (\ref{eq:exponentiatedtranslation}) holds even when the thermal theory is not gapped, as long as $d>3$. It would be interesting to determine which other results from the thermal effective action continue to hold in non-gapped thermal theories. We leave this problem for future work.

\section{Scheme independence}
\label{app:scheme}

In section~\ref{sec:cancelscheme}, we derived (\ref{eq:theresultfirst}) by working in a scheme where $b$-type terms $S_\textrm{ct}$ are absent from the Weyl anomaly. In this appendix, we describe how (\ref{eq:theresultfirst}) comes about in a general scheme. The point is that the scheme dependence of the Casimir energy and the thermal effective action cancel each other. For concreteness, let us work in 4d. The partition function on $S^1_\b \x S^3$ is
\be
\Tr\left[e^{-\b(D+\frac{3a}{4}-\frac{3b}{8})}\right] &\sim e^{-S_\textrm{th}} = e^{-S[\hat g,A]-S_\textrm{Euler}-\mathrm{DR}[S_\textrm{ct}[G]]+\mathrm{DR}[S_\textrm{ct}[\hat G]]},
\label{eq:ourequality}
\ee
where we have used that $S^1_\b \x S^{d-1}$ is conformally-flat to drop Weyl-invariants.
Meanwhile, we have
\be
\mathrm{DR}[S_\textrm{ct}[G]] = \mathrm{DR}[S_\textrm{ct}[\hat G]] &= \int_{S^3} d^3 x \sqrt g \int_0^1 \beta d\tau \p{-\frac{b}{12(4\pi)^2} R^2} = -\frac{3b}{8} \beta,
\ee
where we used that $R=6$ on $S^3$. Thus, we can cancel the $b$-dependence on the left-hand side with $\mathrm{DR}[S_\textrm{ct}[G]]$ on the right-hand side, leaving
\be
\Tr\left[e^{-\b(D+\frac{3a}{4})}\right] \sim e^{-S[\hat g,A]+\mathrm{DR}[S_\textrm{ct}[\hat G]]-S_\textrm{Euler}}.
\ee
The combination $-S[\hat g,A]+\mathrm{DR}[S_\textrm{ct}[\hat G]]$ is then scheme-independent, and equal to $S[\hat g,A]$ in the scheme of section~\ref{sec:cancelscheme}.

Of course, this cancellation was not an accident. The $b$-term contribution to the Casimir energy is
\be
E_0|_\textrm{$b$-type} &= \frac{1}{\vol\, \R} \left.S_\textrm{ct}\right|_{\R\x S^{d-1}}.
\ee
Because $S_\textrm{ct}$ is local, it follows that
\be
\beta E_0|_\textrm{$b$-type} &= \left.\textrm{DR}[S_\textrm{ct}]\right|_{S^1_\b\x S^{d-1}}.
\ee
Said another way, the $b$-term in the Casimir energy is precisely what matches the $b$-term contribution to the Weyl anomaly on the cylinder. Since the $b$-term in the thermal effective action was determined by Weyl anomaly matching, it must cancel with the Casimir energy. This argument generalizes to arbitrary $d$.

\section{More on free theories}
\label{app:freebosonconstants}

\subsection{Partition function of free scalar theories}
\label{app:partitionfunctionfreebosonderivation}

In this section we review the partition function of a free scalar theory on $\mathbb{R}\times S_R^{d-1}$. This space is conformally equivalent to the Euclidean space $\mathbb{R}^d$. The energy $E$ of the state on $\mathbb{R}\times S_R^{d-1}$ is related to the scaling dimension $\Delta$ of the corresponding field on $\mathbb{R}^d$ via:
\begin{equation}
    E=\Delta/R. 
\end{equation}
The equation of motion of the free scalar field on $\mathbb{R}\times S_R^{d-1}$ is 
\begin{equation}
    \left[-\frac{\partial^2}{\partial t^2}+\nabla_{S_R^{d-1}}^2-\xi\mathcal{R}\right]\phi=0,
\label{eq:EOM}
\end{equation}
where $\xi$ is the conformal coupling in $d$ dimensions, $\xi=\frac{d-2}{4(d-1)}$, and $\mathcal{R}$ is the Ricci scalar of $S_R^{d-1}$, $\mathcal{R}=\frac{(d-1)(d-2)}{R^2}$. The spherical harmonics in $d$ dimension, $Y^{(d)}_l$, are eigenfunctions of $\nabla^2_{S^{d-1}_R}$, with eigenvalue $-l(l+d-2)$, where $l$ is a non-negative integer.
We can then construct an orthonormal set of solutions as
\begin{equation}
    \phi_l \propto e^{-iEt}Y^{(d)}_l, \;\;\;\; E=\frac{l+\frac{d-2}{2}}{R},
\end{equation}
whose elements become each mode after quantization. Note that $Y^{(d)}_l$ is the representation of $Spin(d)$ with the highest weight $\left(l,0,\dots,0\right)$ in the standard Cartan-Weyl labeling scheme. 

We will write down the case of even dimension and odd dimension separately because the group structure of $SO(d)$ is slightly different in these two cases.

\emph{Even dimensions}

The spherical harmonics $Y_{l}^{(d)}$ are also eigenfunctions of $\partial_{\theta_a}$ in the coordinate system \eqref{eq:themetricintwisted}. The eigenvalues $m_a$ $(a=1,\dots,n)$ of $\partial_{\theta_a}$ are integers, and they obey the following relation:
\begin{equation}
    l = 2m_0 + |m_1| + \dots + |m_n|,
\end{equation}
where $m_0$ is a non-negative integer.
The multiplicity of this specific eigenvalue is 
$\binom{n+m_0-2}{m_0}$.
Therefore, in even $d$ ($d>2$), the partition function of a free scalar field is:
\begin{equation}
Z(T, \Omega_i) = \prod_{m_0=0}^\infty \left(\prod_{m_1=-\infty}^\infty \cdots \prod_{m_{d/2}=-\infty}^\infty \frac{1}{1-e^{-\frac1T(2m_0 + \sum_i |m_i| + d/2 - 1 + i \sum_i m_i \Omega_i)}}\right)^{\binom{d/2+m_0-2}{m_0}},
\label{eq:freefieldeven}
\end{equation}
where the sums over $i$ in (\ref{eq:freefieldeven}) run from $i=1, \cdots, \frac d2$.
From this we can read off $\log Z$. The first two terms in the high-temperature expansion are as in (\ref{eq:freefieldlogeven}).

The higher order terms in $\log Z$ come with a factor proportional to $\sim \zeta(d-2k)T^{d-2k-1}$ for integer $k$. Because the zeta function vanishes at negative even integers, this means the high-temperature perturbative expansion for $\log Z$ of a free scalar field in even dimensions truncates after the $O(1/T)$ term \cite{Kutasov:2000td}. The further corrections after the perturbative expansion in $1/T$ are non-perturbatively suppressed in $T$. In fact, using techniques similar to \cite{Cardy:1991kr}, we can get an exact expression for $\log Z$ for free scalar field theories in even $d$. For completeness, we write these expressions in Section \ref{app:nonpert}.

\emph{Odd dimensions}

In odd $d$, the relation between the eigenvalues $l$ and $m_a$ is:
\begin{equation}
    l = m_0' + |m_1| + \dots + |m_{\frac{d-1}2}|,
\end{equation}
where $m_0'$ is a non-negative integer.
The multiplicity of this specific eigenvalue state is 
   $\binom{\frac{d-1}2+m_0-1}{m_0}$,
where $m_0=\lfloor\frac{m_0'}{2}\rfloor$. Therefore, the partition function of a free scalar field is:
\begin{align}
Z(T, \Omega_i) &= \prod_{m_0=0}^\infty \left(\prod_{m_1=-\infty}^\infty \cdots \prod_{m_{\frac{d-1}2}=-\infty}^\infty \frac{1}{1-e^{-\frac1T(2m_0 + \sum_i |m_i| + d/2 - 1 + i \sum_i m_i \Omega_i)}}\right)^{\binom{\frac{d-3}2+m_0}{m_0}} \nonumber \\
&\quad\times \prod_{m_0=0}^\infty \left(\prod_{m_1=-\infty}^\infty \cdots \prod_{m_{\frac{d-1}2}=-\infty}^\infty \frac{1}{1-e^{-\frac1T(2m_0 + \sum_i |m_i| + d/2 + i \sum_i m_i \Omega_i)}}\right)^{\binom{\frac{d-3}2+m_0}{m_0}},
\label{eq:freefieldodd}
\end{align}
 where the sums over $i$ in (\ref{eq:freefieldodd}) run from $i=1, \cdots, \frac{d-1}2$. When we take the $\log$ and expand at high temperature, we arrive at the same result as in (\ref{eq:freefieldlogeven}).

In even $d$, the expansion in inverse powers of $T$ truncates after the $O(T^{-1})$ term. In odd $d$, however, the expansion never truncates. This is because the higher order terms have a factor proportional to $\sim \zeta(d-2k)T^{d-2k-1}$. For odd $d$, this never vanishes. Moreover, due to the factorial growth of the zeta function at large, negative, odd values of the argument, the expansion in inverse powers of $T$ is in fact asymptotic rather than convergent. Finally (due to a pole of the zeta function at argument $1$), there is a $\log T$ term in the high-temperature expansion as well. (See (4.13) of \cite{Kang:2022orq} for this $\log T$ term in the case of $d=3$.) We write an explicit expression for all perturbative terms in odd dimensions in Section \ref{app:nonpert}.

\subsection{Gapless sector in the free scalar}
\label{app:gapless}

As noted in Sec \ref{sec:scalarsfree}, the free scalar in $d$ dimensions is a somewhat pathological example for our purposes, due to the presence of a gapless sector upon compactification on $S^1$, namely the $d{-}1$-dimensional free scalar CFT. As a result, the partition function at high temperature contains terms proportional to $O(T^0)$ and $O(\log T)$ in even-$d$ and odd-$d$ respectively. These terms cannot be produced by the thermal effective action, and must come from the gapless sector. In this appendix we understand them explicitly (see also \cite{Chang:1992fu} for earlier discussion of such terms). 

An important subtlety is that the $d{-}1$ dimensional gapless sector is {\it not\/} conformally coupled to curvature in $d{-}1$ dimensions. To see why, we start with a conformally-coupled scalar in $d$-dimensions. This contains a term in the Lagrangian $\frac 1 2 \xi_d \cR \f^2$ with coefficient
\be
\xi_d &= \frac{d-2}{4(d-1)} \qquad (\textrm{conformal coupling in $d$-dimensions}).
\ee
When we dimensionally reduce on $S^1$, we do not obtain a conformally-coupled scalar in $d{-}1$ dimensions, because $\xi_d\neq \xi_{d-1}$. Instead, the $d{-}1$-dimensional scalar has a particular mass deformation turned on. To be more precise, it satisfies the equation of motion
\be
\label{eq:ourequationofmotion}
\cD \f &= \p{-\nabla^2_{d-1} + \xi_d \cR}\f=0,
\ee
where $\nabla^2_{d-1} $ is the $d{-}1$ dimensional Laplacian. By contrast, a conformally-coupled scalar would satisfy
\be
\tl \cD \f &= \p{-\nabla^2_{d-1}  + \xi_{d-1} \cR}\f=0 \qquad (\textrm{$d{-}1$-dimensional conformal coupling}).
\ee

The partition function of the gapless sector in our case is $(\det \cD)^{-1/2}$. By contrast, reference \cite{Klebanov:2011gs} computed the sphere partition function of a conformally-coupled free scalar, i.e.\ $(\det \tl \cD)^{-1/2}$. For our purposes, we can follow the methods of \cite{Klebanov:2011gs}, but we will obtain different results because we have a different equation of motion (\ref{eq:ourequationofmotion}).\footnote{We are grateful to Yifan Wang for discussions related to this point.}

Following \cite{Klebanov:2011gs}, the contribution to $-\log Z_{S^1 \x S^{d-1}}$ from the gapless sector is
\be
F = \sum_{n=0}^{\infty} m_n \left[-\log(\mu R) + \log\left(n + \frac{d-2}2\right)\right],
\label{eq:divergentsum}
\ee
where
\be
m_n \coloneqq \frac{(2n+d-2)(n+d-3)!}{(d-2)!n!}
\ee
is the dimension of the $n$-th traceless symmetric tensor representation of $\SO(d)$. Here, $R$ is the radius of the $S^{d-1}$, and $\mu$ is a mass scale coming from the regulator. In our case, the temperature sets the regulator scale, so we have $\mu=T$. 

The sum (\ref{eq:divergentsum}) diverges, but we can use $\z$-function regularization to make it finite. In even $d$, the $R$-dependence of (\ref{eq:divergentsum}) formally drops out. However, in odd $d$ it remains, giving a nontrivial $\log T$ term in $\log Z$. The $\z$-function regulated sum is
\begin{align}
F_s \coloneqq \sum_{n=0}^{\infty} \frac{m_n}{\left(n+\frac{d-2}2\right)^s}.
\label{eq:fsdef}
\end{align}
The $\log T$ term is given by $F_{s=0}$ and the $T^0$ term is given by $\partial_s F_s|_{s=0}$. These values for the first few $d$ are given in Table \ref{tab:fsstuff}. They indeed agree with the corresponding terms in the high-temperature expansion of the partition function of the free scalar on $S^1\x S^{d-1}$, as we explicitly write in Section \ref{app:nonpert}.

\begin{table}[h!]
\begin{center}
    \begin{tabular}{| c | c | c |}
    \hline
    $d$ & $F_{s=0}$ & $\partial_s F_s|_{s=0}$  \\ \hline
    3 & $\frac1{12}$ & $-\frac{\log 2}{12} - \zeta'(-1)$  \\ \hline
    4 & 0 &  $-\frac{\zeta(3)}{4\pi^2}$ \\ \hline
    5 & $-\frac{17}{2880}$ & $\frac{11\log2}{2880} + \frac{\zeta'(-1)}{24} - \frac{7\zeta'(-3)}{24}$ \\ \hline
    6 & 0 & $\frac{\pi^2 \zeta(3) + 3\zeta(5)}{48\pi^4}$\\ \hline
    7 & $\frac{367}{483840}$ & $-\frac{211\log2}{483840} - \frac{3\zeta'(-1)}{640}  + \frac{7\zeta'(-3)}{192} - \frac{31\zeta'(-5)}{1920}$ \\
    \hline
    8 & 0 & $-\frac{8\pi^4 \zeta(3) + 30\pi^2 \zeta(5) + 45\zeta(7)}{2880\pi^6}$ \\ \hline
    \end{tabular}
\caption{Values for $F_{s=0}$ and $\partial_s F_s|_{s=0}$ for various dimensions, with $F_s$ defined in (\ref{eq:fsdef}). These provide the coefficients of the $O(\log T)$ and $O(T^0)$ terms respectively in the free energy of a free scalar field in $d$ dimensions.}
\label{tab:fsstuff}
\end{center} 
\end{table}

\subsection{The $a$-anomaly of the free scalar}

As an aside, we can use similar techniques to compute the $a$-anomaly of a free scalar theory in $d$ dimensions. The value of the $a$ anomaly is well-known in $d=2,4,6$ \cite{Duff:1993wm}, see e.g.\ \cite{Bastianelli:2000hi} for a calculation in 6d. In general $d$ it was computed in \cite{Giombi:2013yva, Giombi:2014xxa}, which we review here. 

Here we study the free scalar field in $d$ dimensions conformally coupled to $S^d$, as was precisely done in \cite{Klebanov:2011gs}. As discussed in appendix~\ref{app:gluing}, the $a$-anomaly is related to the sphere partition function by\footnote{References \cite{Giombi:2013yva,Giombi:2014xxa} use a different convention for the $a$ anomaly where $\frac{(-1)^{d/2} a_d^{\text{here}} d! \vol\, S^d}{(4\pi)^{d/2}} = a_d^{\text{there}}$.}
\be
\log Z(S^d_r) &= -\frac{(-1)^{d/2}a_d d!}{(4\pi)^{d/2}} \vol\, S^d \log (\mu r),
\ee
where $\mu$ is a regulator scale. Thus, we can read off $a_d$ from the $\log r$ term in the sphere free energy. We find the general answer
\be
a_d &= \frac{(-1)^{\frac d 2 + 1}}{2 \G(\frac{d+2}{2})\G(d+2)}\int_0^1 dt (d+4 t^2) \p{t-\frac d 2 + 1}_{d-1} \qquad (\textrm{$d$ even}, d>2),
\label{eq:aanomaly}
\ee
where $(x)_n$ is the Pochhammer symbol $(x)_n \coloneqq \frac{\Gamma(x+n)}{\Gamma(x)}$. The integrand is of course a polynomial in $t$ for positive integer $d$, so the integral is trivial in practice. This formula requires $d>2$ because the sphere partition function of the (noncompact) free boson in $d=2$ is ill-defined. The first few values of $a_d$ are listed in table~\ref{tab:anomaly}.
\begin{table}[h!]
\begin{center}
    \begin{tabular}{| c | c | c | c | c | c |}
    \hline
    $d$ & 4 & 6 & 8 & 10 & 12 \\
    \hline
    $a_d$ & $\frac{1}{360}$ & $\frac{1}{9072}$ & $\frac{23}{5443200}$
    & $\frac{263}{1796256000}$ & $\frac{133787}{29422673280000}$
    \\ \hline
    \end{tabular}
\caption{Values for the conformal $a$-anomaly of a free scalar field in $d$ dimensions, with $d$ even. For general $d$, see (\ref{eq:aanomaly}).}
\label{tab:anomaly}
\end{center} 
\end{table}

\subsection{Non-perturbative corrections for free scalars}
\label{app:nonpert}

From section \ref{app:gapless}, we have the perturbative corrections in $1/T$ for the free scalar field in $d$ dimensions. For even $d$, they truncate after the $O(1/T)$ term (see e.g. \cite{Dowker:1978md,Candelas:1978gf}). From the techniques in \cite{Cardy:1991kr} we can in fact compute the exact high-temperature partition function. It is given by the following.

Define the auxiliary function for even $d$: 
\begin{align}
    f(d,T) \coloneqq \zeta(d) T^{d-1} - \frac{(-1)^{d/2}}{(2\pi)^{d-2}}\left[ \frac{\zeta(d-1)}2 - \frac{(d-1)\zeta(d)}{4\pi^2 T} + \sum_{n=1}^\infty \frac{e^{-4\pi^2 T n} \sigma_{d-1}(n)}{n^{d-1}} \sum_{i=0}^{d-2} \frac{(4\pi^2 T n)^i}{\Gamma(i+1)} \right],
    \label{eq:auxhighT}
\end{align}
where $\sigma$ is the divisor sigma function: $\sigma_{d-1}(n) \coloneqq \sum_{\ell | n} \ell^{d-1}$.
Then the general even $d$ free scalar is
\begin{align}
\log Z_{d}(T) = \sum_{i=0}^{d/2-2} c_{2i-(d-1)}(d) f(d-2i,T)
\label{eq:nonpertfreebosoneven}
\end{align}
where $c_{2i}(d)$ is the coefficient of the $\beta^{2i}$ term in the expansion of $\frac{\sinh(\beta)}{2^{d-1}\sinh^{d}(\beta/2)}$ about $\beta=0$.\footnote{This function comes about from writing the logarithm of the free boson partition function as \begin{equation}\log Z(T) = -\sum_{j=0}^\infty \frac{j+d/2-1}{d/2-1} \binom{j+d-3}{d-3} \log(1-e^{-\frac1T(j+\frac{d-2}{2})}),\end{equation} doing the Taylor expansion of the logarithm, and finally resumming over $j$, giving (see e.g. \cite{Chang:1992fu}) \begin{equation}\log Z(T) = \sum_{n=1}^\infty \frac{\sinh(\frac n T)}{n 2^{d-1} \sinh^d(\frac{n}{2T})}.\end{equation}}

For example at $d=4$ and $d=6$, (\ref{eq:nonpertfreebosoneven}) reduces to the following two equations\footnote{These perturbative terms here reproduce e.g. \cite{Melia:2020pzd}.}:
\begin{align}
\log Z_{d=4}(T) = \frac{\pi^4}{45}T^3 - \frac{\zeta(3)}{4\pi^2} + \frac{1}{240T} - \sum_{n=1}^\infty \frac{4 \sigma_3(n) e^{-4\pi^2 T n}}n \left(\pi^2 T^2 + \frac{T}{2n} + \frac{1}{8\pi^2 n^2}\right),
\end{align}
\begin{align}
\log Z_{d=6}(T) &= \frac{2\pi^6}{945}T^5 - \frac{\pi^4}{540}T^3 + \frac{\pi^2\zeta(3)+3\zeta(5)}{48\pi^4} - \frac{31}{60480T} \nonumber \\ &~~~~+ \sum_{n=1}^{\infty} e^{-4\pi^2 n T}\Bigg[\frac{4\sigma_5(n)}{3n}\left(\pi^4 T^4 + \frac{\pi^2 T^3}{n} + \frac{3T^2}{4n^2} + \frac{3T}{8n^3\pi^2} + \frac{3}{32n^4\pi^4}\right) \nonumber \\ &~~~~~~~~~~~~~~~~~~~~~~~~+ \frac{\sigma_3(n)}{3n}\left(\pi^2 T^2 + \frac{T}{2n} + \frac{1}{8n^2\pi^2}\right)\Bigg].
\end{align}

For a free scalar field in odd dimensions, the perturbative expansion in $1/T$ no longer truncates. The perturbative expansion is
\begin{align}
\log Z_d(T) \sim \left(\sum_{n=1}^{\frac{d-1}{2}} {c}_{-2n}(d) \zeta(2n+1) T^{2n}\right) + F_{s=0} \log T + \partial_s F_s|_{s=0} + \left(\sum_{n=1}^\infty c_{2n}(d) \zeta(-2n+1) T^{-2n} \right),
\label{eq:odddpert}
\end{align}
where in (\ref{eq:odddpert}), $F_{s=0}$ and $\partial_s F_s|_{s=0}$ are defined in (\ref{eq:fsdef}), and $c_{2i}(d)$ is defined again as the coefficient of the $\beta^{2i}$ term in the expansion of $\frac{\sinh(\beta)}{2^{d-1}\sinh^{d}(\beta/2)}$ about $\beta=0$. 

At large $n$, $|c_{2n}(d)| \sim (2\pi)^{-2n}$. On the other hand, $|\zeta(-2n+1)| \sim \frac{(2n)!}{(2\pi)^{2n}}$. Thus the expression (\ref{eq:odddpert}) is an asymptotic series with divergent piece growing like 
\be
\log Z_d(T) \sim \sum_n \frac{(2n)!}{(4\pi^2 T)^{2n}}.
\ee
From techniques in resurgence, this implies the first non-perturbative correction to (\ref{eq:odddpert}) scales as $e^{-4\pi^2 T}$, just as in even dimensions.

These results are consistent with the worldline instanton corrections discussed in section~\ref{sec:nonpertcorrections} (even though in this example there is also a gapless sector upon compactification). When the free boson is compactified on a circle with thermal boundary conditions, the mass of the lightest KK mode is $m_{KK}=2\pi T$. Therefore, (\ref{eq:firstnonpert}) predicts a correction to $\log Z$ of the form $e^{-4\pi^2 T}$, which is precisely consistent with what we found in both even and odd $d$.

We can also study the free scalar with a $\Z_2$ twist around the thermal circle. For example, in $d=4$, we find  
\be
\log Z_{d=4}^\textrm{$\Z_2$-twisted}(T) &= -\frac{7\pi^4}{360} T^3 + \frac{1}{240T} + O(e^{-2\pi^2 T}).
\ee
In this case, the lightest KK mode has mass $m_{KK}=\pi T$, and the nonperturbative corrections are indeed of the form $e^{-2\pi m_{KK}}=e^{-2\pi^2 T}$.

\subsection{Partition function of free fermion theories}

In this section, we review the partition function of a free massless Dirac fermion in $d$ dimensions on $\mathbb{R}\times S_R^{d-1}$. We can construct the partition function from the solution of the Dirac equation:
 \begin{equation}
    \left(\Gamma^0\frac{\partial}{\partial t} 
    +
    \Gamma^i\nabla_i\right)\psi=0,
 \end{equation}
 where $\Gamma^\mu$ $(\mu=0,1,\dots,d-1)$ are gamma matrices, and $\nabla_i$ $(i=1,\dots,d-1)$ is the covariant derivative on the sphere.

The spectrum of the Dirac operator on $S^{d-1}$ has been considered in e.g. \cite{Camporesi_1996}. The Dirac operator on $S^{d-1}$, $\cancel{\nabla}\equiv \Gamma^i\nabla_i$, has the following eigenvalues:
\begin{equation}
    \cancel{\nabla}\psi_{\pm\rho}=\pm i \left(\rho+\frac{d-1}{2}\right)\psi_{\pm \rho},
    \;\;\;\;
    \rho=0,1,2,\dots.
\label{eq:diracspectrum}
\end{equation}

Because $\Gamma^{0^2}=-\mathbf{1}$ and $\{\Gamma^0,\cancel{\nabla}\}=0$, we find the solution of the Dirac equation as
\begin{equation}
    \psi=e^{ i Et}\left(\psi_{\pm\rho}\pm\Gamma^0\psi_{\pm\rho}\right),
    \;\;\;\;
    E=\rho+\frac{d-1}{2}.
\end{equation}

From \cite{Camporesi_1996} we see the solutions are representations of $\mathrm{Spin}(d)$ with highest weight $\left(\rho+\frac{1}{2},\frac{1}{2},\dots,\frac{1}{2}\right)$ for odd $d$ and $\left(\rho+\frac{1}{2},\frac{1}{2},\dots,\frac{1}{2},\pm\frac{1}{2}\right)$ for even $d$ where $\rho$ is the eigenvalue of the Dirac operator in \eqref{eq:diracspectrum}. 
 
In odd dimensions,
we have a complete set of solution of the Dirac equation with eigenvalues as follows:
\begin{align}
    E &= \rho+\frac{d-1}{2},\nn \\
    \rho &= m_0+m_1'+\dots+m_{\frac{d-1}{2}}',\nn \\
    m_a &= \pm \left(m_a'+\frac{1}{2}\right),   
    \;\;\;\;
    \left(a=1,\dots,\frac{d-1}{2}\right),
\end{align}
where $m_0,m_1',\dots,m_n'$ are non-negative integers. Here $E$ is the energy of the state and $m_a$ is the eigenvalue of the rotation generator $\partial_{\theta_a}$. The multiplicity of states with eigenvalues $(E,m_1,\dots,m_\frac{d-1}{2})$ is $\binom{\frac{d-3}2+m_0}{m_0}$. Finally we get an additional tower of states from quantizing the field $\bar\psi$.
Therefore, the partition function of a free fermion in odd dimensions is:
\begin{equation}
    Z(T,\Omega_i)
    =
    \prod_{m_0=0}^\infty\prod_{m_1\in \mathbb Z + \frac12}\dots\prod_{m_\frac{d-1}{2}\in\mathbb Z + \frac12}\left(
    1+e^{-\frac 1T (m_0 + \sum_i |m_i|+\frac{d-1}{4} + i\sum_i m_i \Omega_i)}
    \right)^{2\binom{\frac{d-3}2+m_0}{m_0}}, 
\label{eq:freefermionpf}
\end{equation}
where the sums over $i$ run from $1, \dots, \frac{d-1}2$.
Taking a log and expanding at high temperature recovers (\ref{eq:fermionresult}).

In even dimensions, we can do a very similar calculation. We have a complete set of solutions to the Dirac equation with eigenvalues as follows:
\begin{align}
    E &= \rho+\frac{d}{2},\nn \\
    \rho &= m_0+m_1'+\dots+m_{\frac{d}{2}}',\nn \\
    m_a &= \pm \left(m_a'+\frac{1}{2}\right),   
    \;\;\;\;
    \left(a=1,\dots,\frac{d}{2}\right),
\end{align}
where $m_0,m_1',\dots,m_n'$ are nonnegative integers. The multiplicity of the states with eigenvalues $\left(E,m_1,\dots,m_\frac{d}{2}\right)$ is $\binom{\frac{d}{2}+m_0-2}{m_0}$.
Therefore, the partition function of a free fermion in even dimensions is:
\begin{equation}
    Z(T,\vec{\Omega})
    =
    \prod_{m_0=0}^\infty\prod_{m_1\in \mathbb Z + \frac12}\dots\prod_{m_\frac{d}{2}\in\mathbb Z + \frac12}\left(
    1+e^{-\frac 1T (m_0 + \sum_i |m_i|+\frac{d-2}{4} + i\sum_i m_i \Omega_i)}
    \right)^{2\binom{\frac{d}{2}+m_0-2}{m_0}}, 
\label{eq:freefermionpfeven}
\end{equation}
where the sums over $i$ in (\ref{eq:freefermionpfeven}) run from $1,\dots, \frac{d}{2}$. This gives the same answer as \eqref{eq:fermionresult}.

\subsection{Non-perturbative corrections for free fermions}
\label{app:nonpertferm}

We can repeat the same analysis in Section \ref{app:nonpert} to find the non-perturbative corrections for the free energy of a free Dirac fermion in $d$ dimensions. When we turn off the spin fugacities, we can rewrite (\ref{eq:freefermionpf}), (\ref{eq:freefermionpfeven}), as:
\begin{align}
    \log Z^f_d(T) &= \sum_{n=1}^\infty \frac{2^{\lfloor\frac d2\rfloor + 1}(-1)^{n+1} e^{-\frac{(d-1)n}{2T}}}{n(1-e^{-\frac nT})^{d-1}}.
    \label{eq:freefermsumform}
\end{align}
In even dimensions, this admits the following (exact) high-temperature expansion. First, we define the auxiliary function for even $d$ as:
\begin{align}
    g(d,T) &\coloneqq 2^{d/2}\left(1-\frac1{2^{d-1}}\right)\zeta(d) T^{d-1} + \frac{(-1)^{d/2}(d-1)(2^{d}-2)\zeta(d)}{\pi^d 2^{3d/2} T} \nonumber \\ &+ \frac{(-1)^{d/2}}{2^{\frac d2 -2}\pi^{d-2}}\sum_{n=1}^\infty \frac{e^{-2\pi^2 T n}(-1)^n \sigma^{\text{odd}}_{d-1}(n)}{n^{d-1}} \sum_{i=0}^{d-2} \frac{(2\pi^2 T n)^i}{\Gamma(i+1)},
\end{align}
where
\be
\sigma^{\text{odd}}_{d-1}(n) \coloneqq \sum_{\substack{\ell | n, \\ \ell ~\text{odd}}} \ell^{d-1}.
\ee
We also define the function $\tilde c_i(d)$ as the $\beta^i$ term in the expansion of $\frac{\beta^d e^{-\frac{\beta(d-1)}{2}}}{(1-e^{-\beta})^{d-1}}$ about $\beta=0$.\footnote{Like in the free scalar case, up to an overall constant, this comes from (\ref{eq:freefermsumform}) upon setting $n=1$.} Then the partition function of a free Dirac fermion in even $d$ dimensions at temperature $T$ is
\begin{equation}
    \log Z^f_d(T) = \sum_{i=1}^{d-1} 2^{\frac{i+1}2} \tilde c_i(d) g(d+1-i,T).
\label{eq:genansfermevend}
\end{equation}
For instance, (\ref{eq:genansfermevend}) for $d=4, 6$ reduces to:
\begin{align}
\log Z^f_{d=4}(T) &= \frac{7\pi^4}{90}T^3 - \frac{\pi^2}{12}T + \frac{17}{480T}  \nonumber \\ &~~~~~~~ + \sum_{n=1}^{\infty} (-1)^n e^{-2\pi^2 T n} \left(\frac{4\pi^2}n \sigma^{\text{odd}}_3(n) T^2 + \frac{4}{n^2} \sigma^{\text{odd}}_3(n) T + \frac{\sigma^{\text{odd}}_1(n)}{n} + \frac{2\sigma^{\text{odd}}_3(n)}{n^3\pi^2}\right),
\end{align}
and
\begin{align}
    &\log Z^f_{d=6}(T) = \frac{31\pi^6}{1890} T^5 - \frac{7\pi^4}{216} T^3 + \frac{\pi^2}{32} T - \frac{367}{24192T} \nonumber \\ & +\sum_{n=1}^\infty (-1)^{n+1} e^{-2\pi^2 T n} \Bigg[\frac{2\pi^4}{3n} \sigma_5^{\text{odd}}(n) T^4 + \frac{4\pi^2}{3n^2} \sigma_5^{\text{odd}}(n) T^3 +\left(\frac{2\sigma_5^{\text{odd}}(n)}{n^3} + \frac{5\pi^2 \sigma_3^{\text{odd}}}{3n} \right) T^2 \nonumber \\ &~~~~~~~~~~~~~~~~~~~~~~~~~~~~~~ + \left(\frac{2\sigma_5^{\text{odd}}(n)}{\pi^2n^4} + \frac{5 \sigma_3^{\text{odd}}(n)}{3n^2} \right) T + \left(\frac{\sigma_5^{\text{odd}}(n)}{\pi^4 n^5} + \frac{5\sigma_3^{\text{odd}}(n)}{6\pi^2 n^3} + \frac{3\sigma_1^{\text{odd}}(n)}{8n}\right)\Bigg].
\end{align}

In odd dimensions, the perturbation theory in $1/T$ no longer truncates. Rather, it looks like
\begin{align}
    \log Z^f_{d}(T) &\sim \sum_{n=1}^{\infty} 2^{\frac{d+1}2}\left(1-2^{n-d}\right)\zeta(d+1-n) \tilde c_n(d) T^{d-n},
\label{eq:pertfermoddasympt}
\end{align}
where the $n=d$ term in (\ref{eq:pertfermoddasympt}) is $\tilde c_d(d) 2^{\frac{d+1}2} \log 2$.
The sum in (\ref{eq:pertfermoddasympt}) is an asymptotic expansion. At large odd $n$, $|\tilde c_n(d)| \sim (2\pi)^{-n}$ and $|\zeta(d+1-n)| \sim \frac{n!}{(2\pi)^n}$. The sum then diverges, growing as
\be
\log Z^f_d(T) \sim\sum_n \frac{n!}{(2\pi^2 T)^n}.
\ee
From the techniques in resurgence, this implies the first non-perturbative correction of (\ref{eq:pertfermoddasympt}) scales as $e^{-2\pi^2 T}$, just as in even dimensions.

In a free fermion theory compactified on a circle with thermal boundary conditions, we have $m_{KK} = \pi T$, so that (\ref{eq:firstnonpert}) predicts a non-perturbative correction of the form $e^{-2\pi^2 T}$, consistent with what we found in both even and odd dimensions.

\section{Wilson coefficients for the 3d Ising model}
\label{sec:3dising}

In this appendix, we discuss estimates of the high-temperature partition function of the 3d Ising model. In Appendix A of \cite{Iliesiu:2018fao}, the coefficient $f$ for the 3d Ising model was estimated by constructing the partition function with $\Omega=0$ as a sum over the spectrum of known operators, which has numerically been computed up to about dimension 8 \cite{Simmons-Duffin:2016wlq}. In this appendix, we perform a similar analysis, but including spin-dependence. Note that there is a balance in choosing the temperature -- if the temperature is too low, then truncating the thermal effective action becomes a poor approximation; if the temperature is too high, then truncating the partition function to a finite sum of characters becomes inaccurate.\footnote{Much like the porridge in \emph{Goldilocks and the Three Bears} \cite{goldilocks}, it is important to pick a temperature that is just right.}

\begin{figure}[ht]
\centering
\begin{subfigure}{0.49\textwidth}
\includegraphics[width=\textwidth,trim={0.5cm 0.7cm 0.3cm 0.7cm},clip]{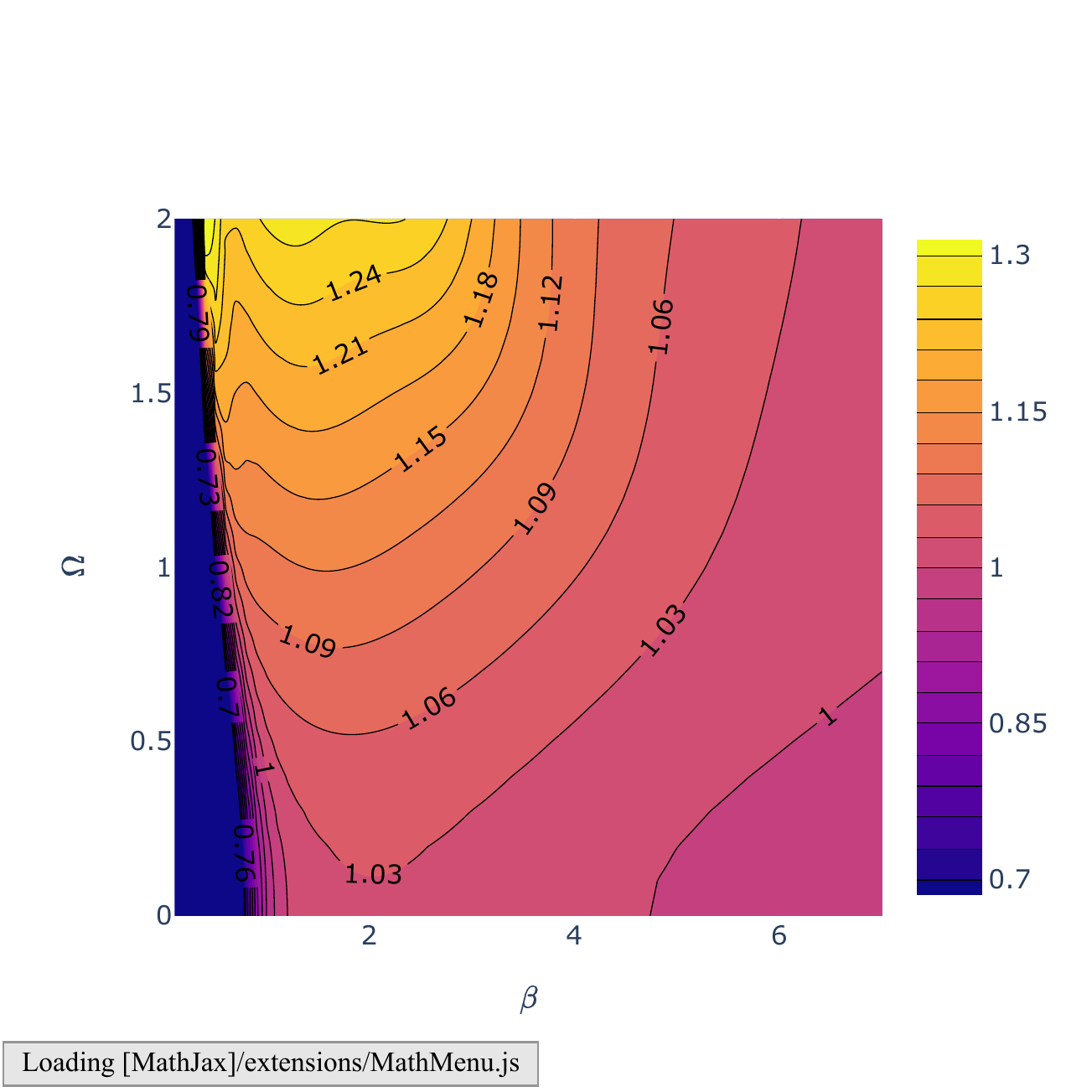}
\end{subfigure}
\begin{subfigure}{0.49\textwidth}
\includegraphics[width=\textwidth,trim={0.3cm 0.7cm 0.5cm 0.7cm},clip]{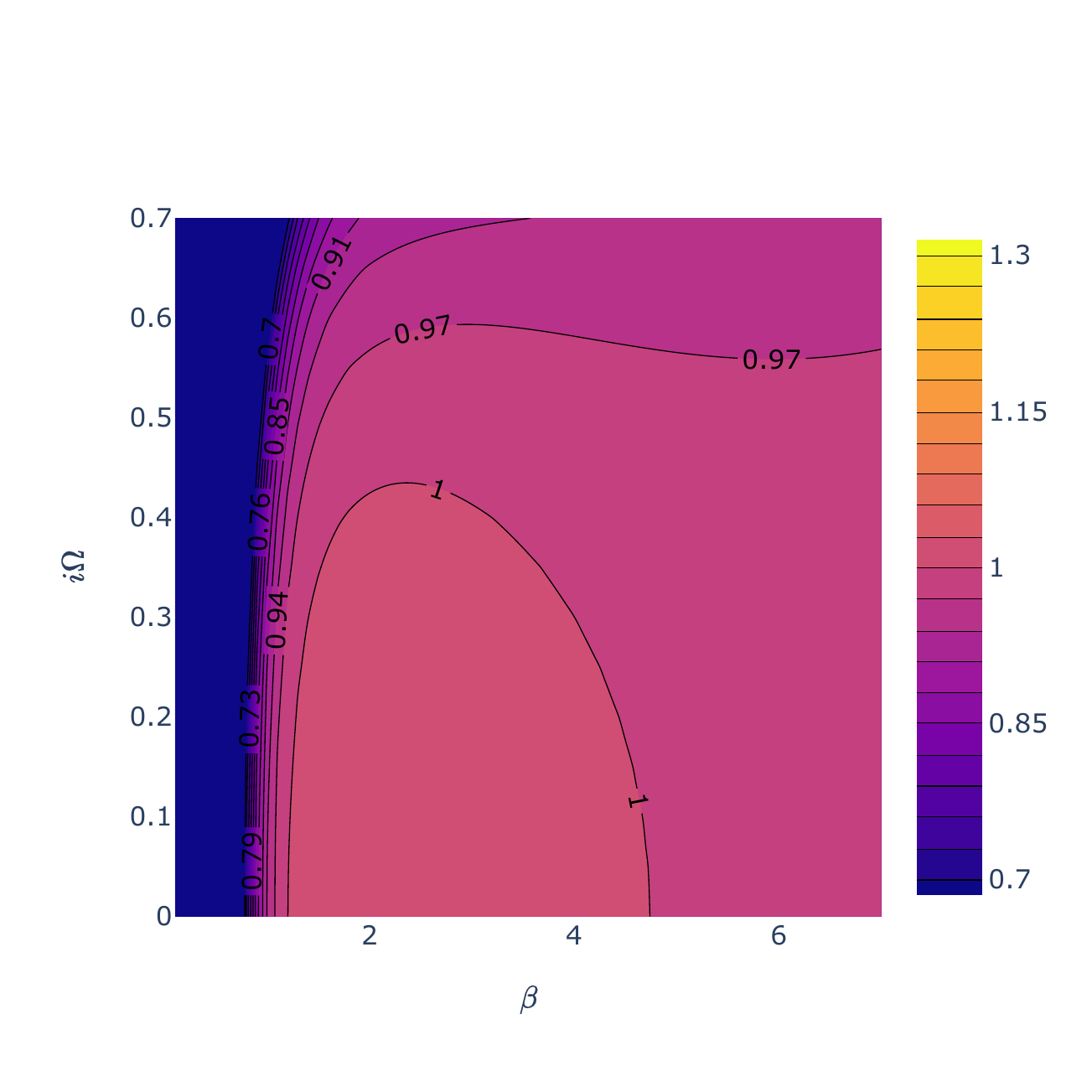}
\end{subfigure}
\caption{\label{fig:3dIsing} Contour plots of the ratio of the estimated partition function to the leading term in the thermal effective action with $f = 0.153$, i.e.\ $Z^\textrm{3d Ising}(\beta,\Omega) / \exp\p{4\pi \frac{0.153}{\beta^2(1+\Omega^2)}}$, as a function of $\beta, \Omega$ for real $\Omega$ (left) and imaginary $\Omega$ (right). The ratio is very close to 1 for intermediate temperatures $1\lesssim \b \lesssim 5$ and small angles $|\Omega|\lesssim 0.5$.}
\end{figure}

From Monte-Carlo techniques, it has been estimated that $f_{\text{3d Ising}} \sim 0.153$ \cite{PhysRevE.79.041142, casimir2, casimir3}. In figure~\ref{fig:3dIsing}, we plot the ratio of the computed partiton function to the estimate from the first term in the thermal effective action (\ref{eq:freeenergy}) with $f=0.153$. For the values of $\beta$ and $\Omega$ shown in the figure, the ratio is quite close to $1$. We can also independently fit $f$ from the partition function. By studying temperatures with $1.5 < \beta < 3$ and chemical potentials $|\Omega| < 0.6$, we estimate that $f \sim 0.15$, consistent with  \cite{PhysRevE.79.041142, casimir2, casimir3}. 

We can also try to estimate higher-derivative Wilson coefficients in the effective action of the 3d Ising model, such as $c_1, c_2$. Unfortunately, we do not have a clear enough picture of the high-dimension spectrum to estimate these coefficients with reliable precision. However, our best fits suggest that $c_2 < 0$ (we are not yet able to reliably estimate the sign of $c_1$).

In general, accessing large twist operators is challenging for the numerical bootstrap, which seems most sensitive to low-twist operators --- particularly ``double twist" operators \cite{Simmons-Duffin:2016wlq}. Furthermore, the numerical bootstrap studies done so far are blind to certain parts of the operator spectrum of the 3d Ising model, such as odd-spin $\Z_2$-even operators, or parity-odd operators.\footnote{Though there are preliminary results for some of these sectors from the stress tensor bootstrap \cite{Dymarsky:2017yzx}.} Such sectors would need to be included to reliably compute the partition function at higher temperatures. See \cite{Zhu:2022gjc,Hu:2023xak} for recent work accessing these sectors with other methods.

\section{The shadow transform of a three-point function at large $\De$}
\label{sec:shadowintegralslargedelta}

The formula for OPE coefficients depends on the triple shadow coefficient $S^3_{\tl 1^\dag \tl 2^\dag \tl 3^\dag}$ given in (\ref{eq:tripleshadowcoefficient}). In this appendix, we compute this coefficient at large $\De_i$. First consider a single shadow transform applied to a three-point function with large $\De$'s. The shadow transform is
\be
\<\cO_1^a(x_1)\cO_2^b(x_2) \mathbf{S}[\cO_3]^{\bar c}(x_3)\>^{(s)}  &\equiv \int d^dx_0 \<\tl \cO^{\bar c}(x_3) \tl \cO^\dag_c(x_0)\>\<\cO_1^a(x_1)\cO_2^b(x_2)\cO_3^c(x_0)\>^{(s)} .
\ee
Here, $a,b,c$ are spin indices for the representations $\lambda_1,\lambda_2,\lambda_3$. The operator $\tl \cO^\dag$ has Lorentz representation $\lambda_3^*$, so we write its index as a lowered $c$ index. The operator $\tl \cO$ has Lorentz representation $\lambda_3^R$ (the reflected representation), and we indicate this with a barred index.

The three-point function is
\be
\<\cO_1^a(x_1)\cO_2^b(x_2)\cO_3^c(x_3)\>^{(s)} &= V^{s;abc}(x_1,x_2,x_3).
\ee
We will be interested in restricting this three-point function to a single axis $x_i=z_i e$ with unit vector $e\in S^{d-1}\subset \R^d$. We get two different answers, depending on the cyclic ordering of the points
\be
V^{s;abc}(z_1 e,z_2e, z_3 e) =
\begin{cases}
\frac{V^{s;abc}(0,e,\oo)}{|z_{12}|^{\De_1+\De_2-\De_3} |z_{23}|^{\De_2+\De_3-\De_1} |z_{31}|^{\De_3+\De_1-\De_2}} & (z_1 < z_2 < z_3,\textrm{ or cycl.}), \\
\frac{V^{s;abc}(e,0,\oo)}{|z_{12}|^{\De_1+\De_2-\De_3} |z_{23}|^{\De_2+\De_3-\De_1} |z_{31}|^{\De_3+\De_1-\De_2}} & (z_2 < z_1 < z_3,\textrm{ or cycl.}).
\end{cases}
\ee
The tensors $V^{s;abc}(0,e,\oo)$ and $V^{s;abc}(e,0,\oo)$ are related to each other by a rotation by $\pi$ in the 1-2 direction applied simultaneously to all three indices. The operator at $\oo$ is defined by (\ref{eq:defofprimaryatinfinity}).

To compute the shadow transform, we can use conformal symmetry to choose a simple configuration of the points. We pick $(x_1,x_2,x_3)=(0,e,\oo)$, where $e$ is a unit vector in the $x^1$ direction. The two-point function becomes a tensor depending on the unit vector $e$ that maps $\lambda_3\to \lambda_3^R$:
\be
\label{eq:inversiontensors}
\<\tl \cO^{\bar c}(\oo e) \tl \cO^\dag_c(0)\> &= I^{\bar c}{}_c(e).
\ee
For example, in the case of a spinor representation in 4d, we have $I^{\dot \a}{}_\a(e)\propto(e\.\bar\s)^{\dot\a}{}_\a$. 

The shadow transform will be dominated by a saddle point on the $x^1$-axis, by $\SO(d-1)$ invariance. Its location depends only on the $z_0$-dependent factors in the three-point function
\be
V^{s;abc}(0,e, z_0 e) \propto |z_0|^{\De_1-\De_2-\De_3}|1-z_0|^{\De_2-\De_1-\De_3}.
\ee
This has the saddle solution
\be
z_{0*} &= \frac{\De_2+\De_3-\De_1}{2\De_3}.
\ee
The tensor structure that multiplies the answer depends on the location of the saddle. Taking into account the gaussian fluctuations around the saddle, we find
\be
&\<\cO_1^a(0)\cO_2^b(e) \mathbf{S}[\cO_3]^{\bar c}(\oo)\>^{(s)} \nn\\
&= i\p{\frac{\pi}{\De_3}}^{d/2}\p{\p{\frac{\De_1+\De_3-\De_2}{2\De_3}}^2}^{\frac{\De_2-\De_1-\De_3+\frac d 2}{2}}
\p{\p{\frac{\De_2+\De_3-\De_1}{2\De_3}}^2}^{\frac{\De_1-\De_2-\De_3+\frac d 2}{2}} \nn\\
&\quad\x I^{\bar c}{}_c(e) \x
\begin{cases}
 V^{s;abc}(0,e,\oo) & 0<z_{0*}<1, \\
V^{s;abc}(e,0,\oo) & \textrm{otherwise}.
\end{cases}
\ee

If we perform the shadow transform 3 times, we find that $z_{0*}\in(0,1)$ twice, and once it lies outside of this range.  Thus, the cyclic ordering of the arguments to $V^s$ get swapped twice, resulting in the same ordering after three transforms. The overall effect is to multiply by $\De$-dependent factors and act on each index with $I^{\bar c}{}_c(e)$.
Overall, we find
\be
&(I(e)^{-1})_a{}^{\bar a} (I(e)^{-1})_b{}^{\bar b} (I(e)^{-1})_c{}^{\bar c} \<\mathbf{S}[\tl\cO_1^\dag ]_{\bar a}(0)\mathbf{S}[\tl\cO_2^\dag ]_{\bar b}(e) \mathbf{S}[\tl\cO_3^\dag]_{\bar c}(\oo)\>^{(s'*)}
\nn\\
&= V^{s'*}_{abc}(0,e,\oo) \x e^{\frac{i \pi  (d-2)}{4}}
\left(\frac{\pi i}{\Delta _1}\right)^{d/2}
\left(\frac{\pi i}{\Delta _2}\right)^{d/2}
\left(\frac{\pi i}{\Delta _3}\right)^{d/2}
\nn\\
&\quad \x
\left(\tfrac{(\Delta _1+\Delta _2-\Delta _3) (\Delta _1+\Delta _2+\Delta _3)}{4\Delta _1 \Delta _2}\right)^{\Delta _1+\Delta _2-\Delta _3-\frac{d}{2}}
\left(\tfrac{(\Delta _3+\Delta _1-\Delta _2) (\Delta _1+\Delta _2+\Delta _3)}{4\Delta _3 \Delta _1}\right)^{\Delta _3+\Delta _1-\Delta _2-\frac{d}{2}}
\nn\\
&\quad \x
 \left(\tfrac{(\Delta _2+\Delta _3-\Delta _1) (\Delta _1+\Delta _2+\Delta _3)}{4\Delta _2 \Delta _3}\right)^{\Delta _2+\Delta _3-\De_1-\frac{d}{2}}.
\label{eq:exprfortripleshadow}
\ee
Here, $V^{s'*}$ indicates the complex conjugate of the three-point structure $V^{s'}$. For the purposes of this calculation, we should think of it simply as a three-point structure for operators in the representations $\tl \pi_i^\dag$.
We have written (\ref{eq:exprfortripleshadow}) so that its phase is manifest when $\De_i$ is on the principal series $\De_i\in\frac d 2+i\R_{\geq 0}$ (with positive imaginary part). Finally, we included inverse two-point structures $I(e)^{-1}$, since they are needed in (\ref{eq:gettinglowtempblock}).

\section{Gluing factors}
\label{app:gluing}

In this appendix, we determine the gluing factor $|Z_\textrm{glue}(r)|$ coming from a junction between a $d$-dimensional cylinder of radius $r$ and a flat end-cap given by a $d$-dimensional ball. Our strategy is to start with the partition function on $S^d$ (with radius $1$) and perform a Weyl transformation to a cylinder $\cC_{r,\b}$ of radius $r$ and length $\b r$ with two flat end-caps. We will integrate the Weyl anomaly to compute $S_\textrm{anom}$ and deduce $|Z_\textrm{glue}(r)|$ via (\ref{eq:zglue}).

Recall that on a conformally-flat geometry, with the scheme $S_\textrm{ct}=0$ discussed in section~\ref{sec:cancelscheme}, the finite form of the Weyl anomaly is
\be
\label{eq:integratedweyl}
\log Z[e^{2\w}g] - \log Z[g] &= -S_\textrm{anom}[g,\w]\nn\\
&=-\frac{(-1)^{d/2} a_d}{(4\pi)^{d/2}} \int_0^1 dt \int d^d x  \w  \sqrt{g}e^{dt\w}\,E_d[e^{2t\w} g].
\ee
To compute the Weyl anomaly between the sphere and the capped cylinder, we first need the Riemann tensor for a Weyl rescaling of $S^d$. Let us write the metric on $S^d$ as
\be
ds^2_{S^d} &= d\f^2 + \sin^2 \f\,ds^2_{S^{d-1}}.
\ee
We will be interested in Weyl rescalings $\tl g=e^{2\w} g_{S^d}$, where $\w$ is a function of $\f$ alone. In this case, the Riemann tensor simplifies:
\be
\tl R_{\mu\nu}{}^{\rho\s} &= 
e^{-2 \omega } \Big[
(1-2 \omega ' \cot \phi-\omega '^2) 
(\delta _{\mu}^{\rho } \delta _{\nu}^{\sigma }-\delta _{\mu}^{\sigma } \delta _{\nu}^{\rho })
\nn\\
&\qquad\qquad+
(\omega ' \cot \phi+\omega '^2 -\omega '')
(\delta _{\mu}^{1} \delta _{1}^\rho  \delta _{\nu}^{\sigma }
-\delta _{\nu}^{1} \delta ^{\rho}_{1} \delta _{\mu}^{\sigma }
-\delta _{\mu}^{1} \delta _{1}^{\sigma} \delta _{\nu}^{ \rho }
+\delta _{\nu}^{1} \delta ^{\sigma}_{1} \delta _{\mu}^{\rho })
\Big],
\ee
where the index $1$ represents the $\f$ coordinate, and $\w',\w''$ denote derivatives of $\w$ with respect to $\f$. The Euler density is
\be
\label{eq:oureuler}
\tl E_d &= \frac{1}{2^{d/2}} \e^{\mu_1\cdots \mu_d}\e_{\nu_1\cdots\nu_d} \tl R_{\mu_1\mu_2}{}^{\nu_1\nu_2}\cdots \tl R_{\mu_{d-1}\mu_d}{}^{\nu_{d-1}\nu_d} \nn\\
&= d!e^{-d \w}(1-2\w'\cot \f -\w'^2)^{\frac{d-2}{2}}(1- \w'\cot \f-\w'').
\ee
Plugging this result into (\ref{eq:integratedweyl}), we find
\be
\label{eq:partitionshift}
&\log Z[e^{2\w}g] - \log Z[g]\nn\\
 &= -\frac{(-1)^{d/2} d! a_d}{2^{d-1}\Gamma (\tfrac{d}{2})} \int_0^1 dt \int_0^\pi d\f\,
\omega(\sin \phi)^{d-1} (1-2 t \omega ' \cot \phi-t^2 \omega '^2)^{\frac{d-2}{2}} \left(1-t \omega ' \cot \phi - t \omega ''\right).
\ee

Now let us examine the Weyl factor that relates the sphere to $\cC_{r,\b}$. We will impose a symmetry under $\f\to \pi-\f$, so that it suffices to consider the range $0\leq \f \leq \frac\pi 2$. The Weyl factor is
\be
\label{eq:ourweylfactor}
\w(\f) &= \begin{cases}
\log \frac{e^{\b/2}r \tan(\f/2)}{\sin \f} & 0\leq \f \leq \f_0 \\
\log \frac{r}{\sin \f} & \f_0 \leq \f \leq \pi/2,
\end{cases}
\ee
where $\f_0=2\tan^{-1}(e^{-\b/2})$.  As a check, consider first the range $0\leq \f<\f_0$. There, we have
\be
e^{2\w}(d\f^2+\sin^2 \f\,ds_{S^{d-1}}^2) &= e^{\b}r^2\p{\frac{\tan^2(\f/2)}{\sin^2 \f} d\f^2 +\tan^2(\f/2)ds_{S^{d-1}}^2} \nn\\
&= d\rho^2 +\rho^2 ds_{S^{d-1}}^2,
\ee
where $\rho=e^{\b/2} r \tan(\f/2)$. This is the metric on the flat ball, i.e.\ the first end cap. Similarly, for $\f_0 \leq \f \le \pi/2$, we have
\be
e^{2\w}(d\f^2+\sin^2 \f\,ds_{S^{d-1}}^2) &= \p{\frac{r d\f}{\sin\f}}^2 + r^2 ds_{S^{d-1}}^2 = d\tau^2 + r^2 ds^2_{S^{d-1}},
\ee
where $\tau = r\log \tan(\f/2)$, which is the metric on the cylinder. Thus, (\ref{eq:ourweylfactor}) describes how the first hemisphere $0\leq \f \leq \pi/2$ maps to half of the capped cylinder. The remaining hemisphere should be treated symmetrically under $\f\to \pi-\f$. In practice, this means integrating the anomaly over $\f\in [0,\pi/2]$ and including a factor of 2.

Plugging the Weyl factor (\ref{eq:ourweylfactor}) into (\ref{eq:partitionshift}) is subtle because $\w''$ has a $\de$-function singularity at $\f=\f_0$. In (\ref{eq:partitionshift}), this $\de$-function gets multiplied by a function of $\w'$, which is discontinuous at the support of the $\de$-function! To get the correct result, we must regularize $\w$ by smoothing out the discontinuities in its derivatives. For example, we can model $\w'$ near $\f_0$ as
\be
\w'(\f) &= \frac 1 2\p{\w'_++\w'_-+(\w'_+-\w'_-)\mathrm{erf}(\tfrac{\f-\f_0}{\e})},
\ee
where $\e$ is a small regulator, and $\w'_\pm$ are the values of $\w'$ to the left and the right of the discontinuity.
Plugging this into (\ref{eq:partitionshift}), expanding to leading order in $\e$, and writing $\f=\f_0+\e x$, we obtain integrals of the form
\be
\int dx\, e^{-x^2} \mathrm{erf}(x)^n &= \frac{1+(-1)^n}{2}\frac{\sqrt \pi}{n+1},
\ee
which give finite, calculable contributions. Applying this procedure, we can obtain the contribution to (\ref{eq:partitionshift}) from an infinitesimal neighborhood of $\f_0$:
\be
\label{eq:deltacontrb}
(\textrm{contribution near $\f_0$}) &=
\begin{cases}
\frac{a_2}{2}\log (r \cosh\frac \b 2) & d=2, \\
\frac{a_4}{4}\log (r \cosh \frac{\beta }{2})\frac{\sinh \beta -4 \cosh \beta -6 }{4 (\cosh \beta +1)} & d=4, \\
&\cdots
\end{cases}
\ee
The detailed expressions here will not matter for our purposes. The important observation is that the contributions (\ref{eq:deltacontrb}) all vanish when $r=1$ and $\b=0$. We will take advantage of this fact shortly.

Before computing the full result from plugging (\ref{eq:ourweylfactor}) into (\ref{eq:partitionshift}), let us use a shortcut to determine its $r$-dependence.
From cutting and gluing, we expect the capped cylinder partition function to take the form
\be
\label{eq:expectedcapped}
\log Z(\cC_{r,\b}) &= \log |Z_\textrm{glue}(r)|^2 - \varepsilon_0 \b,
\ee
where $\varepsilon_0$ is the Casimir energy on a unit $S^{d-1}$,  given in (\ref{eq:unambiguouscasimirenergy}).
We can determine the $r$-dependence of the right-hand side by starting with a cylinder $\cC_{1,\b}$ of radius $1$ and performing a Weyl rescaling $g\to r^2 g$ to get $\cC_{r,\b}$. Because the integral of the Euler density is topological, a constant Weyl rescaling gives the same anomaly on the capped cylinder as on the sphere. In other words, we have
\be
\label{eq:equalweyl}
\log Z(\cC_{r,\b})-\log Z(\cC_{1,\b}) &= \log Z(S^d_r) - \log Z(S^d),
\ee
where $S^d_r$ is the sphere with radius $r$. On the sphere, we can easily integrate the Weyl anomaly using $E_d[g_{S^d}]=d!$ to give
\be
\label{eq:weylsphereresult}
\log Z(S^d_r) - \log Z(S^d) &= -\frac{(-1)^{d/2} a_d d!}{(4\pi)^{d/2}} \vol\, S^d \log r = -2(-1)^{d/2}(\tfrac d 2)! a_d \log r.
\ee
Combining (\ref{eq:expectedcapped}), (\ref{eq:equalweyl}), and (\ref{eq:weylsphereresult}), we conclude
\be
\log | Z_\textrm{glue}(r)|^2 &= \log |Z_\textrm{glue}(1)|^2 -2(-1)^{d/2}(\tfrac d 2)! a_d \log r.
\ee

Thus, we have completely fixed the $\b$ and $r$ dependence of $\log Z(\cC_{r,\b})$, and the only remaining unknown is $\log Z(\cC_{1,0})=\log |Z_\textrm{glue}(1)|^2$. As noted above, when $r=1$ and $\b=0$, the contribution to the Weyl anomaly near $\f_0$ vanishes. Furthermore, the contribution from the cylinder region vanishes as well since $\f_0=\frac \pi 2$. We are left with an integral over the end cap $\f\in [0,\pi/2]$ alone:
\be
\label{eq:zglueconst}
&\log |Z_\textrm{glue}(1)|^2 - \log Z(S^d) \nn\\
&= \frac{(-1)^{d/2} a_d d!}{2^{d-2}\G(d/2)} \int_0^1 dt \int_0^1 dx \left[(1-t) (1-x) (1+t+x-xt)\right]^{\frac{d -2}{2}}(1-t) \log (x+1),
\ee
where we made the change of variables $x=\cos \f$. We have not found a simple closed-form formula for $\log |Z_\textrm{glue}(1)|^2$ in general $d$. However, it is straightforward to plug different values of $d$ into (\ref{eq:zglueconst}) and perform the resulting elementary integrals. We find that
\be
\label{eq:finalzglueanswer}
\log |Z_\textrm{glue}(r)|^2 &= \log Z(S_{r/2}^d) - (-1)^{d/2}  f(d) a_d,
\ee
where $Z(S_{r/2}^d)$ is the partition function on a sphere of radius $r/2$ (determined by (\ref{eq:weylsphereresult})), and $f(d)$ takes rational values for even $d$, see table~\ref{tab:zglue}.
\begin{table}
\centering
\begin{tabular}{|c|c|c|c|c|c|c|c|}
\hline
$d$ & 2 & 4 & 6 & 8 & 10 & 12 & 14 \\
\hline
$f(d)$
& 1
& $\frac{7}{3}$
& $\frac{37}{5}$
& $\frac{1066}{35}$
& $\frac{3254}{21}$
& $\frac{72428}{77}$
& $\frac{949484}{143}$\\
\hline
\end{tabular}
\caption{Values of $f(d)$ for the first few even $d$. \label{tab:zglue}}
\end{table}

\section{Counting quantum numbers in the genus-2 block}
\label{app:matchingofquantumnumbers}

A four point function of local operators depends on two independent cross ratios $z$ and $\bar z$. These cross ratios are roughly conjugate to the quantum numbers $\De$ and $J$ labeling the internal representation. By varying $z,\bar z$ independently, we can extract information about $\De$ and $J$ independently (for example, using Caron-Huot's Lorentzian inversion formula \cite{Caron-Huot:2017vep}).

Note that the internal operator in a four-point function may transform in a complicated $\SO(d)$ representation whose Young diagram has multiple rows with lengths $(m_1,m_2,\dots,m_n)$, where $n=\lfloor\frac d 2 \rfloor$ and $m_1=J$. However, in a fixed four-point function, only $m_1=J$ is unbounded. The remaining $m_i$ are bounded.

In this appendix, we point out a similar match between unbounded quantum numbers in the genus-2 block $B_{123}^{s's}$ and the dimension of the moduli space of genus-2 conformal structures $\dim\cM=\dim \SO(d+1,1)$. Before explaining the general case, let us describe the matching in $d=2$ and $d=3$.

In $d=2$, there is a unique three-point structure, so the labels $s,s'$ take only one value. The only unbounded quantum numbers in the block are the dimensions and spins of the three exchanged operators. This gives $6$ quantum numbers, which matches $\dim\cM=6$ in $d=2$.

In $d=3$, we again have $6$ unbounded quantum numbers from the dimensions and spins of the exchanged operators. However, we must also take into account the 3-point structure labels $s,s'$. One way to count them is to work in the embedding space. (The counting is even easier in the $q$-basis, but our embedding-space discussion will be useful later.) In the embedding space formalism, a spin-$J$ operator becomes a homogeneous function $\cO(X,Z)$ of an embedding coordinate $X\in \R^{d+1,1}$ and a polarization vector $Z\in \C^{d+1,1}$, subject to orthogonality conditions $X^2=Z^2=X\.Z=0$, and a gauge redundancy $Z\sim Z+\l X$. The operator $\cO(X,Z)$ has homogeneity $-\De$ in $X$ and $J$ in $Z$. A general three-point structure for such operators is given by
\be
\<\cO_1(X_1,Z_1)\cO_2(X_2,Z_2)\cO_3(X_3,Z_3)\> &\ni \frac{V_1^{J_1-\ell_2-\ell_3}V_{2}^{J_2-\ell_3-\ell_1}V_3^{J_3-\ell_1-\ell_2}H_{23}^{\ell_1} H_{31}^{\ell_2} H_{12}^{\ell_3}}{X_{12}^{\De_1+\De_2-\De_3}X_{23}^{\De_2+\De_3-\De_2} X_{31}^{\De_3+\De_1-\De_2}}\nn\\
&  (\ell_2+\ell_3\leq J_1, \ell_3+\ell_1\leq J_2, \ell_1+\ell_2\leq J_3),
\ee
where $H_{ij}$ and $V_i$ are standard polynomials in the polarization vectors \cite{Costa:2011mg}. The three point structure is labeled by integers $\ell_1,\ell_2,\ell_3$, which are constrained by the requirement that the correlator should be a polynomial in the $Z_i$. 

The $\ell_i$ are unbounded in the limit of large spin $J_i$.  However, there is a relation between the $H_{ij}$ and $V_i$:
\be
(V_1H_{23}+V_2H_{13}+V_3 H_{12}+2 V_1 V_2 V_3)^2 &= 2 H_{12} H_{13} H_{23}.
\ee
Using this relation, we can always reduce one of the $\l_i$ to zero, so the number of unbounded quantum numbers labeling the three-point structures in 3d is $3-1=2$. The conformal block is labeled by two three-point structures, so this gives $2\x 2=4$ additional unbounded quantum numbers for the block. Overall, we have $6+4=10=\dim\cM$ in $3d$.

We are now ready to tackle the $d$-dimensional case. In the $d$-dimensional embedding formalism, an operator $\cO$ becomes a homogeneous function of an embedding coordinate $X \in \R^{d+1,1}$ and polarization vectors $W_1,\dots,W_n\in \C^{d+1,1}$, where $n=\lfloor \frac d 2 \rfloor$. (We conventionally write $W_1=Z$.) More formally, $\cO$ is a locally holomorphic section of a line bundle over the flag variety $\cV_{d+1,1}$ of $\SO(d+1,1)$, which has $(X,W_1,\cdots,W_n)$ as projective coordinates. The number of these coordinates, subject to orthogonality relations and modulo gauge redundancies, is the same as the complex dimension of the flag variety, which is
\be
\dim_\C \cV_{d+1,1} &= \frac 1 2 \p{\dim \SO(d+1,1) - \dim T} 
= \frac 1 2 \p{\frac{(d+1)(d+2)}{2}-\left\lfloor\frac{d+2}{2}\right\rfloor},
\ee
where $T$ is the maximal torus of $\SO(d+1,1)$. See \cite{Costa:2014rya,Kologlu:2019mfz} for more discussion on the embedding formalism for general tensors.

For example, in $d=3$, this gives $\dim_\C \cV_{4,1}=4$, which is the correct number of degrees of freedom in the vectors $X,Z\in \R^{4,1}$. We can see this explicitly by restricting to the Poincare section $X=(1,x^2,x)$ and $Z=(1,2x\.z,z)$. Here $x\in \R^3$ is unconstrained, and $z$ is a null vector in 3d, modulo rescaling, which corresponds to a single angle on the celestial circle.

A three-point function $\<\cO_1 \cO_2 \cO_3\>$ is a section of a line bundle over three copies of $\cV_{d+1,1}$, satisfying invariance under $\SO(d+1,1)$. In the large quantum number limit, the number of quantum numbers labeling such sections is
\be
\#(\textrm{$\Z$-valued 3-pt structure labels}) &= 3\dim_\C \cV_{d+1,1} - \dim \SO(d+1,1).
\ee
Finally, a genus-2 block has two three-point structure labels $s,s'$, together with $\dim T = \lfloor \frac{d+2}{2}\rfloor$ quantum numbers for each of the three internal operators. Overall, the number of unbounded quantum numbers is
\be
2\p{3\dim \cV_{d+1,1} - \dim \SO(d+1,1)} + 3 \dim T = \dim \SO(d+1,1) = \dim \cM.
\ee

\bibliographystyle{JHEP}
\bibliography{refs}

\providecommand{\href}[2]{#2}\begingroup\raggedright\begin{thebibliography}{100}

\bibitem{Cardy:1986ie}
J.~L. Cardy, \emph{{Operator Content of Two-Dimensional Conformally Invariant
  Theories}}, \href{http://dx.doi.org/10.1016/0550-3213(86)90552-3}{\emph{Nucl.
  Phys. B} {\bf 270} (1986) 186--204}.

\bibitem{Strominger:1997eq}
A.~Strominger, \emph{{Black hole entropy from near horizon microstates}},
  \href{http://dx.doi.org/10.1088/1126-6708/1998/02/009}{\emph{JHEP} {\bf 02}
  (1998) 009}, [\href{https://arxiv.org/abs/hep-th/9712251}{{\tt
  hep-th/9712251}}].

\bibitem{Brown:1986nw}
J.~D. Brown and M.~Henneaux, \emph{{Central Charges in the Canonical
  Realization of Asymptotic Symmetries: An Example from Three-Dimensional
  Gravity}}, \href{http://dx.doi.org/10.1007/BF01211590}{\emph{Commun. Math.
  Phys.} {\bf 104} (1986) 207--226}.

\bibitem{Cardy:2017qhl}
J.~Cardy, A.~Maloney and H.~Maxfield, \emph{{A new handle on three-point
  coefficients: OPE asymptotics from genus two modular invariance}},
  \href{http://dx.doi.org/10.1007/JHEP10(2017)136}{\emph{JHEP} {\bf 10} (2017)
  136}, [\href{https://arxiv.org/abs/1705.05855}{{\tt 1705.05855}}].

\bibitem{Kraus:2016nwo}
P.~Kraus and A.~Maloney, \emph{{A cardy formula for three-point coefficients or
  how the black hole got its spots}},
  \href{http://dx.doi.org/10.1007/JHEP05(2017)160}{\emph{JHEP} {\bf 05} (2017)
  160}, [\href{https://arxiv.org/abs/1608.03284}{{\tt 1608.03284}}].

\bibitem{Collier:2019weq}
S.~Collier, A.~Maloney, H.~Maxfield and I.~Tsiares, \emph{{Universal dynamics
  of heavy operators in CFT$_{2}$}},
  \href{http://dx.doi.org/10.1007/JHEP07(2020)074}{\emph{JHEP} {\bf 07} (2020)
  074}, [\href{https://arxiv.org/abs/1912.00222}{{\tt 1912.00222}}].

\bibitem{Chandra:2022bqq}
J.~Chandra, S.~Collier, T.~Hartman and A.~Maloney, \emph{{Semiclassical 3D
  gravity as an average of large-c CFTs}},
  \href{http://dx.doi.org/10.1007/JHEP12(2022)069}{\emph{JHEP} {\bf 12} (2022)
  069}, [\href{https://arxiv.org/abs/2203.06511}{{\tt 2203.06511}}].

\bibitem{Shaghoulian:2015kta}
E.~Shaghoulian, \emph{{Modular forms and a generalized Cardy formula in higher
  dimensions}}, \href{http://dx.doi.org/10.1103/PhysRevD.93.126005}{\emph{Phys.
  Rev. D} {\bf 93} (2016) 126005},
  [\href{https://arxiv.org/abs/1508.02728}{{\tt 1508.02728}}].

\bibitem{Belin:2016yll}
A.~Belin, J.~de~Boer, J.~Kruthoff, B.~Michel, E.~Shaghoulian and M.~Shyani,
  \emph{{Universality of sparse $d > 2$ conformal field theory at large $N$}},
  \href{http://dx.doi.org/10.1007/JHEP03(2017)067}{\emph{JHEP} {\bf 03} (2017)
  067}, [\href{https://arxiv.org/abs/1610.06186}{{\tt 1610.06186}}].

\bibitem{Shaghoulian:2016gol}
E.~Shaghoulian, \emph{{Modular Invariance of Conformal Field Theory on
  $S^1×S^3$ and Circle Fibrations}},
  \href{http://dx.doi.org/10.1103/PhysRevLett.119.131601}{\emph{Phys. Rev.
  Lett.} {\bf 119} (2017) 131601},
  [\href{https://arxiv.org/abs/1612.05257}{{\tt 1612.05257}}].

\bibitem{Horowitz:2017ifu}
G.~T. Horowitz and E.~Shaghoulian, \emph{{Detachable circles and
  temperature-inversion dualities for CFT$_{d}$}},
  \href{http://dx.doi.org/10.1007/JHEP01(2018)135}{\emph{JHEP} {\bf 01} (2018)
  135}, [\href{https://arxiv.org/abs/1709.06084}{{\tt 1709.06084}}].

\bibitem{Belin:2018jtf}
A.~Belin, J.~De~Boer and J.~Kruthoff, \emph{{Comments on a state-operator
  correspondence for the torus}},
  \href{http://dx.doi.org/10.21468/SciPostPhys.5.6.060}{\emph{SciPost Phys.}
  {\bf 5} (2018) 060}, [\href{https://arxiv.org/abs/1802.00006}{{\tt
  1802.00006}}].

\bibitem{Luo:2022tqy}
C.~Luo and Y.~Wang, \emph{{Casimir Energy and Modularity in Higher-dimensional
  Conformal Field Theories}},  \href{https://arxiv.org/abs/2212.14866}{{\tt
  2212.14866}}.

\bibitem{Bhattacharyya:2007vs}
S.~Bhattacharyya, S.~Lahiri, R.~Loganayagam and S.~Minwalla, \emph{{Large
  rotating AdS black holes from fluid mechanics}},
  \href{http://dx.doi.org/10.1088/1126-6708/2008/09/054}{\emph{JHEP} {\bf 09}
  (2008) 054}, [\href{https://arxiv.org/abs/0708.1770}{{\tt 0708.1770}}].

\bibitem{Jensen:2012jh}
K.~Jensen, M.~Kaminski, P.~Kovtun, R.~Meyer, A.~Ritz and A.~Yarom,
  \emph{{Towards hydrodynamics without an entropy current}},
  \href{http://dx.doi.org/10.1103/PhysRevLett.109.101601}{\emph{Phys. Rev.
  Lett.} {\bf 109} (2012) 101601}, [\href{https://arxiv.org/abs/1203.3556}{{\tt
  1203.3556}}].

\bibitem{Banerjee:2012iz}
N.~Banerjee, J.~Bhattacharya, S.~Bhattacharyya, S.~Jain, S.~Minwalla and
  T.~Sharma, \emph{{Constraints on Fluid Dynamics from Equilibrium Partition
  Functions}}, \href{http://dx.doi.org/10.1007/JHEP09(2012)046}{\emph{JHEP}
  {\bf 09} (2012) 046}, [\href{https://arxiv.org/abs/1203.3544}{{\tt
  1203.3544}}].

\bibitem{Shaghoulian:2015lcn}
E.~Shaghoulian, \emph{{Black hole microstates in AdS}},
  \href{http://dx.doi.org/10.1103/PhysRevD.94.104044}{\emph{Phys. Rev. D} {\bf
  94} (2016) 104044}, [\href{https://arxiv.org/abs/1512.06855}{{\tt
  1512.06855}}].

\bibitem{DiPietro:2014bca}
L.~Di~Pietro and Z.~Komargodski, \emph{{Cardy formulae for SUSY theories in $d
  =$ 4 and $d =$ 6}},
  \href{http://dx.doi.org/10.1007/JHEP12(2014)031}{\emph{JHEP} {\bf 12} (2014)
  031}, [\href{https://arxiv.org/abs/1407.6061}{{\tt 1407.6061}}].

\bibitem{Jensen:2012kj}
K.~Jensen, R.~Loganayagam and A.~Yarom, \emph{{Thermodynamics, gravitational
  anomalies and cones}},
  \href{http://dx.doi.org/10.1007/JHEP02(2013)088}{\emph{JHEP} {\bf 02} (2013)
  088}, [\href{https://arxiv.org/abs/1207.5824}{{\tt 1207.5824}}].

\bibitem{Bhattacharyya:2013lha}
S.~Bhattacharyya, \emph{{Entropy current and equilibrium partition function in
  fluid dynamics}},
  \href{http://dx.doi.org/10.1007/JHEP08(2014)165}{\emph{JHEP} {\bf 08} (2014)
  165}, [\href{https://arxiv.org/abs/1312.0220}{{\tt 1312.0220}}].

\bibitem{Crossley:2015evo}
M.~Crossley, P.~Glorioso and H.~Liu, \emph{{Effective field theory of
  dissipative fluids}},
  \href{http://dx.doi.org/10.1007/JHEP09(2017)095}{\emph{JHEP} {\bf 09} (2017)
  095}, [\href{https://arxiv.org/abs/1511.03646}{{\tt 1511.03646}}].

\bibitem{Jensen:2013kka}
K.~Jensen, R.~Loganayagam and A.~Yarom, \emph{{Anomaly inflow and thermal
  equilibrium}}, \href{http://dx.doi.org/10.1007/JHEP05(2014)134}{\emph{JHEP}
  {\bf 05} (2014) 134}, [\href{https://arxiv.org/abs/1310.7024}{{\tt
  1310.7024}}].

\bibitem{Jensen:2012jy}
K.~Jensen, \emph{{Triangle Anomalies, Thermodynamics, and Hydrodynamics}},
  \href{http://dx.doi.org/10.1103/PhysRevD.85.125017}{\emph{Phys. Rev. D} {\bf
  85} (2012) 125017}, [\href{https://arxiv.org/abs/1203.3599}{{\tt
  1203.3599}}].

\bibitem{Verlinde:2000wg}
E.~P. Verlinde, \emph{{On the holographic principle in a radiation dominated
  universe}},  \href{https://arxiv.org/abs/hep-th/0008140}{{\tt
  hep-th/0008140}}.

\bibitem{Kutasov:2000td}
D.~Kutasov and F.~Larsen, \emph{{Partition sums and entropy bounds in weakly
  coupled CFT}},
  \href{http://dx.doi.org/10.1088/1126-6708/2001/01/001}{\emph{JHEP} {\bf 01}
  (2001) 001}, [\href{https://arxiv.org/abs/hep-th/0009244}{{\tt
  hep-th/0009244}}].

\bibitem{Anous:2021caj}
T.~Anous, A.~Belin, J.~de~Boer and D.~Liska, \emph{{OPE statistics from
  higher-point crossing}},
  \href{http://dx.doi.org/10.1007/JHEP06(2022)102}{\emph{JHEP} {\bf 06} (2022)
  102}, [\href{https://arxiv.org/abs/2112.09143}{{\tt 2112.09143}}].

\bibitem{Belin:2021ibv}
A.~Belin, J.~de~Boer, P.~Nayak and J.~Sonner, \emph{{Generalized spectral form
  factors and the statistics of heavy operators}},
  \href{http://dx.doi.org/10.1007/JHEP11(2022)145}{\emph{JHEP} {\bf 11} (2022)
  145}, [\href{https://arxiv.org/abs/2111.06373}{{\tt 2111.06373}}].

\bibitem{Fukushima:2010bq}
K.~Fukushima and T.~Hatsuda, \emph{{The phase diagram of dense QCD}},
  \href{http://dx.doi.org/10.1088/0034-4885/74/1/014001}{\emph{Rept. Prog.
  Phys.} {\bf 74} (2011) 014001}, [\href{https://arxiv.org/abs/1005.4814}{{\tt
  1005.4814}}].

\bibitem{Chai:2020zgq}
N.~Chai, S.~Chaudhuri, C.~Choi, Z.~Komargodski, E.~Rabinovici and M.~Smolkin,
  \emph{{Thermal Order in Conformal Theories}},
  \href{http://dx.doi.org/10.1103/PhysRevD.102.065014}{\emph{Phys. Rev. D} {\bf
  102} (2020) 065014}, [\href{https://arxiv.org/abs/2005.03676}{{\tt
  2005.03676}}].

\bibitem{Halperin_2018}
B.~I. Halperin, \emph{{On the Hohenberg{\textendash}Mermin{\textendash}Wagner
  Theorem and Its Limitations}},
  \href{http://dx.doi.org/10.1007/s10955-018-2202-y}{\emph{Journal of
  Statistical Physics} {\bf 175} (2018) 521--529},
  [\href{https://arxiv.org/abs/1812.00220}{{\tt 1812.00220}}].

\bibitem{Jensen:2013rga}
K.~Jensen, R.~Loganayagam and A.~Yarom, \emph{{Chern-Simons terms from thermal
  circles and anomalies}},
  \href{http://dx.doi.org/10.1007/JHEP05(2014)110}{\emph{JHEP} {\bf 05} (2014)
  110}, [\href{https://arxiv.org/abs/1311.2935}{{\tt 1311.2935}}].

\bibitem{Golkar:2015oxw}
S.~Golkar and S.~Sethi, \emph{{Global Anomalies and Effective Field Theory}},
  \href{http://dx.doi.org/10.1007/JHEP05(2016)105}{\emph{JHEP} {\bf 05} (2016)
  105}, [\href{https://arxiv.org/abs/1512.02607}{{\tt 1512.02607}}].

\bibitem{Chang:2019uag}
C.-M. Chang, M.~Fluder, Y.-H. Lin and Y.~Wang, \emph{{Proving the 6d Cardy
  Formula and Matching Global Gravitational Anomalies}},
  \href{http://dx.doi.org/10.21468/SciPostPhys.11.2.036}{\emph{SciPost Phys.}
  {\bf 11} (2021) 036}, [\href{https://arxiv.org/abs/1910.10151}{{\tt
  1910.10151}}].

\bibitem{Eling:2013bj}
C.~Eling, Y.~Oz, S.~Theisen and S.~Yankielowicz, \emph{{Conformal Anomalies in
  Hydrodynamics}}, \href{http://dx.doi.org/10.1007/JHEP05(2013)037}{\emph{JHEP}
  {\bf 05} (2013) 037}, [\href{https://arxiv.org/abs/1301.3170}{{\tt
  1301.3170}}].

\bibitem{Komargodski:2011vj}
Z.~Komargodski and A.~Schwimmer, \emph{{On Renormalization Group Flows in Four
  Dimensions}}, \href{http://dx.doi.org/10.1007/JHEP12(2011)099}{\emph{JHEP}
  {\bf 12} (2011) 099}, [\href{https://arxiv.org/abs/1107.3987}{{\tt
  1107.3987}}].

\bibitem{Iliesiu:2018fao}
L.~Iliesiu, M.~Kolo\u{g}lu, R.~Mahajan, E.~Perlmutter and D.~Simmons-Duffin,
  \emph{{The Conformal Bootstrap at Finite Temperature}},
  \href{http://dx.doi.org/10.1007/JHEP10(2018)070}{\emph{JHEP} {\bf 10} (2018)
  070}, [\href{https://arxiv.org/abs/1802.10266}{{\tt 1802.10266}}].

\bibitem{Schwimmer:2010za}
A.~Schwimmer and S.~Theisen, \emph{{Spontaneous Breaking of Conformal
  Invariance and Trace Anomaly Matching}},
  \href{http://dx.doi.org/10.1016/j.nuclphysb.2011.02.003}{\emph{Nucl. Phys. B}
  {\bf 847} (2011) 590--611}, [\href{https://arxiv.org/abs/1011.0696}{{\tt
  1011.0696}}].

\bibitem{LBonora1986}
L.~Bonora, P.~Pasti and M.~Bregola, \emph{Weyl cocycles},
  \href{http://dx.doi.org/10.1088/0264-9381/3/4/018}{\emph{Classical and
  Quantum Gravity} {\bf 3} (1986) 635}.

\bibitem{Boulanger:2007st}
N.~Boulanger, \emph{{General solutions of the Wess-Zumino consistency condition
  for the Weyl anomalies}},
  \href{http://dx.doi.org/10.1088/1126-6708/2007/07/069}{\emph{JHEP} {\bf 07}
  (2007) 069}, [\href{https://arxiv.org/abs/0704.2472}{{\tt 0704.2472}}].

\bibitem{Assel:2015nca}
B.~Assel, D.~Cassani, L.~Di~Pietro, Z.~Komargodski, J.~Lorenzen and
  D.~Martelli, \emph{{The Casimir Energy in Curved Space and its Supersymmetric
  Counterpart}}, \href{http://dx.doi.org/10.1007/JHEP07(2015)043}{\emph{JHEP}
  {\bf 07} (2015) 043}, [\href{https://arxiv.org/abs/1503.05537}{{\tt
  1503.05537}}].

\bibitem{Herzog:2013ed}
C.~P. Herzog and K.-W. Huang, \emph{{Stress Tensors from Trace Anomalies in
  Conformal Field Theories}},
  \href{http://dx.doi.org/10.1103/PhysRevD.87.081901}{\emph{Phys. Rev. D} {\bf
  87} (2013) 081901}, [\href{https://arxiv.org/abs/1301.5002}{{\tt
  1301.5002}}].

\bibitem{Carlip:2000nv}
S.~Carlip, \emph{{Logarithmic corrections to black hole entropy from the Cardy
  formula}}, \href{http://dx.doi.org/10.1088/0264-9381/17/20/302}{\emph{Class.
  Quant. Grav.} {\bf 17} (2000) 4175--4186},
  [\href{https://arxiv.org/abs/gr-qc/0005017}{{\tt gr-qc/0005017}}].

\bibitem{Harlow:2021trr}
D.~Harlow and H.~Ooguri, \emph{{A universal formula for the density of states
  in theories with finite-group symmetry}},
  \href{http://dx.doi.org/10.1088/1361-6382/ac5db2}{\emph{Class. Quant. Grav.}
  {\bf 39} (2022) 134003}, [\href{https://arxiv.org/abs/2109.03838}{{\tt
  2109.03838}}].

\bibitem{Kang:2022orq}
M.~J. Kang, J.~Lee and H.~Ooguri, \emph{{Universal formula for the density of
  states with continuous symmetry}},
  \href{http://dx.doi.org/10.1103/PhysRevD.107.026021}{\emph{Phys. Rev. D} {\bf
  107} (2023) 026021}, [\href{https://arxiv.org/abs/2206.14814}{{\tt
  2206.14814}}].

\bibitem{Dondi:2021buw}
N.~Dondi, I.~Kalogerakis, D.~Orlando and S.~Reffert, \emph{{Resurgence of the
  large-charge expansion}},
  \href{http://dx.doi.org/10.1007/JHEP05(2021)035}{\emph{JHEP} {\bf 05} (2021)
  035}, [\href{https://arxiv.org/abs/2102.12488}{{\tt 2102.12488}}].

\bibitem{Grassi:2019txd}
A.~Grassi, Z.~Komargodski and L.~Tizzano, \emph{{Extremal correlators and
  random matrix theory}},
  \href{http://dx.doi.org/10.1007/JHEP04(2021)214}{\emph{JHEP} {\bf 04} (2021)
  214}, [\href{https://arxiv.org/abs/1908.10306}{{\tt 1908.10306}}].

\bibitem{Hellerman:2021yqz}
S.~Hellerman and D.~Orlando, \emph{{Large R-charge EFT correlators in N=2
  SQCD}},  \href{https://arxiv.org/abs/2103.05642}{{\tt 2103.05642}}.

\bibitem{Hellerman:2021duh}
S.~Hellerman, \emph{{On the exponentially small corrections to ${\cal N} = 2$
  superconformal correlators at large R-charge}},
  \href{https://arxiv.org/abs/2103.09312}{{\tt 2103.09312}}.

\bibitem{Caetano:2023zwe}
J.~Caetano, S.~Komatsu and Y.~Wang, \emph{{Large Charge 't Hooft Limit of
  $\mathcal{N}=4$ Super-Yang-Mills}},
  \href{https://arxiv.org/abs/2306.00929}{{\tt 2306.00929}}.

\bibitem{Luscher1986}
M.~L{\"u}scher, \emph{Volume dependence of the energy spectrum in massive
  quantum field theories},
  \href{http://dx.doi.org/10.1007/BF01211589}{\emph{Communications in
  Mathematical Physics} {\bf 104} (Jun, 1986) 177--206}.

\bibitem{Dowker:2021gqj}
J.~S. Dowker, \emph{{Remarks on spherical monodromy defects for free scalar
  fields}},  \href{https://arxiv.org/abs/2104.09419}{{\tt 2104.09419}}.

\bibitem{Gibbons_2005}
G.~W. Gibbons, M.~J. Perry and C.~N. Pope, \emph{{The first law of
  thermodynamics for Kerr-anti-de Sitter black holes}},
  \href{http://dx.doi.org/10.1088/0264-9381/22/9/002}{\emph{Class. Quant.
  Grav.} {\bf 22} (2005) 1503--1526},
  [\href{https://arxiv.org/abs/hep-th/0408217}{{\tt hep-th/0408217}}].

\bibitem{Hartman:2014oaa}
T.~Hartman, C.~A. Keller and B.~Stoica, \emph{{Universal Spectrum of 2d
  Conformal Field Theory in the Large c Limit}},
  \href{http://dx.doi.org/10.1007/JHEP09(2014)118}{\emph{JHEP} {\bf 09} (2014)
  118}, [\href{https://arxiv.org/abs/1405.5137}{{\tt 1405.5137}}].

\bibitem{Cardoso:2004hs}
V.~Cardoso and O.~J.~C. Dias, \emph{{Small Kerr-anti-de Sitter black holes are
  unstable}}, \href{http://dx.doi.org/10.1103/PhysRevD.70.084011}{\emph{Phys.
  Rev. D} {\bf 70} (2004) 084011},
  [\href{https://arxiv.org/abs/hep-th/0405006}{{\tt hep-th/0405006}}].

\bibitem{Kim:2023sig}
S.~Kim, S.~Kundu, E.~Lee, J.~Lee, S.~Minwalla and C.~Patel, \emph{{`Grey
  Galaxies' as an endpoint of the Kerr-AdS superradiant instability}},
  \href{https://arxiv.org/abs/2305.08922}{{\tt 2305.08922}}.

\bibitem{deBoer:2008fk}
J.~de~Boer, F.~Denef, S.~El-Showk, I.~Messamah and D.~Van~den Bleeken,
  \emph{{Black hole bound states in AdS(3) x S**2}},
  \href{http://dx.doi.org/10.1088/1126-6708/2008/11/050}{\emph{JHEP} {\bf 11}
  (2008) 050}, [\href{https://arxiv.org/abs/0802.2257}{{\tt 0802.2257}}].

\bibitem{Mefford:2017oxy}
E.~Mefford, E.~Shaghoulian and M.~Shyani, \emph{{Sparseness bounds on local
  operators in holographic CFT$_{d}$}},
  \href{http://dx.doi.org/10.1007/JHEP07(2018)051}{\emph{JHEP} {\bf 07} (2018)
  051}, [\href{https://arxiv.org/abs/1711.03122}{{\tt 1711.03122}}].

\bibitem{Brendle_2023}
T.~Brendle, N.~Broaddus and A.~Putman, \emph{{The mapping class group of
  connect sums of $S^2 \times S^1$}},
  \href{http://dx.doi.org/10.1090/tran/8758}{\emph{Trans. Amer. Math. Soc.}
  {\bf 376} (2023) 2557--2572}, [\href{https://arxiv.org/abs/2012.01529}{{\tt
  2012.01529}}].

\bibitem{Ferrara1972}
S.~Ferrara, A.~F. Grillo, G.~Parisi and R.~Gatto, \emph{The shadow operator
  formalism for conformal algebra. vacuum expectation values and operator
  products}, \href{http://dx.doi.org/10.1007/BF02907130}{\emph{Lettere al Nuovo
  Cimento (1971-1985)} {\bf 4} (May, 1972) 115--120}.

\bibitem{Dolan:2000ut}
F.~A. Dolan and H.~Osborn, \emph{{Conformal four point functions and the
  operator product expansion}},
  \href{http://dx.doi.org/10.1016/S0550-3213(01)00013-X}{\emph{Nucl. Phys. B}
  {\bf 599} (2001) 459--496}, [\href{https://arxiv.org/abs/hep-th/0011040}{{\tt
  hep-th/0011040}}].

\bibitem{Simmons-Duffin:2012juh}
D.~Simmons-Duffin, \emph{{Projectors, Shadows, and Conformal Blocks}},
  \href{http://dx.doi.org/10.1007/JHEP04(2014)146}{\emph{JHEP} {\bf 04} (2014)
  146}, [\href{https://arxiv.org/abs/1204.3894}{{\tt 1204.3894}}].

\bibitem{Karateev:2018oml}
D.~Karateev, P.~Kravchuk and D.~Simmons-Duffin, \emph{{Harmonic Analysis and
  Mean Field Theory}},
  \href{http://dx.doi.org/10.1007/JHEP10(2019)217}{\emph{JHEP} {\bf 10} (2019)
  217}, [\href{https://arxiv.org/abs/1809.05111}{{\tt 1809.05111}}].

\bibitem{Kravchuk:2016qvl}
P.~Kravchuk and D.~Simmons-Duffin, \emph{{Counting Conformal Correlators}},
  \href{http://dx.doi.org/10.1007/JHEP02(2018)096}{\emph{JHEP} {\bf 02} (2018)
  096}, [\href{https://arxiv.org/abs/1612.08987}{{\tt 1612.08987}}].

\bibitem{Polyakov:1974gs}
A.~M. Polyakov, \emph{{Nonhamiltonian approach to conformal quantum field
  theory}}, {\emph{Zh. Eksp. Teor. Fiz.} {\bf 66} (1974) 23--42}.

\bibitem{Pappadopulo:2012jk}
D.~Pappadopulo, S.~Rychkov, J.~Espin and R.~Rattazzi, \emph{{OPE Convergence in
  Conformal Field Theory}},
  \href{http://dx.doi.org/10.1103/PhysRevD.86.105043}{\emph{Phys. Rev. D} {\bf
  86} (2012) 105043}, [\href{https://arxiv.org/abs/1208.6449}{{\tt
  1208.6449}}].

\bibitem{Rychkov:2015lca}
S.~Rychkov and P.~Yvernay, \emph{{Remarks on the Convergence Properties of the
  Conformal Block Expansion}},
  \href{http://dx.doi.org/10.1016/j.physletb.2016.01.004}{\emph{Phys. Lett. B}
  {\bf 753} (2016) 682--686}, [\href{https://arxiv.org/abs/1510.08486}{{\tt
  1510.08486}}].

\bibitem{Zamolodchikov1987}
A.~B. Zamolodchikov, \emph{Conformal symmetry in two-dimensional space:
  Recursion representation of conformal block},
  \href{http://dx.doi.org/10.1007/BF01022967}{\emph{Theoretical and
  Mathematical Physics} {\bf 73} (Oct, 1987) 1088--1093}.

\bibitem{Maldacena:2015iua}
J.~Maldacena, D.~Simmons-Duffin and A.~Zhiboedov, \emph{{Looking for a bulk
  point}}, \href{http://dx.doi.org/10.1007/JHEP01(2017)013}{\emph{JHEP} {\bf
  01} (2017) 013}, [\href{https://arxiv.org/abs/1509.03612}{{\tt 1509.03612}}].

\bibitem{Liu:2018jhs}
J.~Liu, E.~Perlmutter, V.~Rosenhaus and D.~Simmons-Duffin,
  \emph{{$d$-dimensional SYK, AdS Loops, and $6j$ Symbols}},
  \href{http://dx.doi.org/10.1007/JHEP03(2019)052}{\emph{JHEP} {\bf 03} (2019)
  052}, [\href{https://arxiv.org/abs/1808.00612}{{\tt 1808.00612}}].

\bibitem{Caron-Huot:2021rmr}
S.~Caron-Huot, D.~Mazac, L.~Rastelli and D.~Simmons-Duffin, \emph{{Sharp
  boundaries for the swampland}},
  \href{http://dx.doi.org/10.1007/JHEP07(2021)110}{\emph{JHEP} {\bf 07} (2021)
  110}, [\href{https://arxiv.org/abs/2102.08951}{{\tt 2102.08951}}].

\bibitem{Fitzpatrick:2012yx}
A.~L. Fitzpatrick, J.~Kaplan, D.~Poland and D.~Simmons-Duffin, \emph{{The
  Analytic Bootstrap and AdS Superhorizon Locality}},
  \href{http://dx.doi.org/10.1007/JHEP12(2013)004}{\emph{JHEP} {\bf 12} (2013)
  004}, [\href{https://arxiv.org/abs/1212.3616}{{\tt 1212.3616}}].

\bibitem{Komargodski:2012ek}
Z.~Komargodski and A.~Zhiboedov, \emph{{Convexity and Liberation at Large
  Spin}}, \href{http://dx.doi.org/10.1007/JHEP11(2013)140}{\emph{JHEP} {\bf 11}
  (2013) 140}, [\href{https://arxiv.org/abs/1212.4103}{{\tt 1212.4103}}].

\bibitem{Alday:2007mf}
L.~F. Alday and J.~M. Maldacena, \emph{{Comments on operators with large
  spin}}, \href{http://dx.doi.org/10.1088/1126-6708/2007/11/019}{\emph{JHEP}
  {\bf 11} (2007) 019}, [\href{https://arxiv.org/abs/0708.0672}{{\tt
  0708.0672}}].

\bibitem{Fitzpatrick:2015qma}
A.~L. Fitzpatrick, J.~Kaplan, M.~T. Walters and J.~Wang, \emph{{Eikonalization
  of Conformal Blocks}},
  \href{http://dx.doi.org/10.1007/JHEP09(2015)019}{\emph{JHEP} {\bf 09} (2015)
  019}, [\href{https://arxiv.org/abs/1504.01737}{{\tt 1504.01737}}].

\bibitem{Simmons-Duffin:2016wlq}
D.~Simmons-Duffin, \emph{{The Lightcone Bootstrap and the Spectrum of the 3d
  Ising CFT}}, \href{http://dx.doi.org/10.1007/JHEP03(2017)086}{\emph{JHEP}
  {\bf 03} (2017) 086}, [\href{https://arxiv.org/abs/1612.08471}{{\tt
  1612.08471}}].

\bibitem{Chester:2019ifh}
S.~M. Chester, W.~Landry, J.~Liu, D.~Poland, D.~Simmons-Duffin, N.~Su et~al.,
  \emph{{Carving out OPE space and precise $O(2)$ model critical exponents}},
  \href{http://dx.doi.org/10.1007/JHEP06(2020)142}{\emph{JHEP} {\bf 06} (2020)
  142}, [\href{https://arxiv.org/abs/1912.03324}{{\tt 1912.03324}}].

\bibitem{Liu:2020tpf}
J.~Liu, D.~Meltzer, D.~Poland and D.~Simmons-Duffin, \emph{{The Lorentzian
  inversion formula and the spectrum of the 3d O(2) CFT}},
  \href{http://dx.doi.org/10.1007/JHEP09(2020)115}{\emph{JHEP} {\bf 09} (2020)
  115}, [\href{https://arxiv.org/abs/2007.07914}{{\tt 2007.07914}}].

\bibitem{Mukhametzhanov:2019pzy}
B.~Mukhametzhanov and A.~Zhiboedov, \emph{{Modular invariance, tauberian
  theorems and microcanonical entropy}},
  \href{http://dx.doi.org/10.1007/JHEP10(2019)261}{\emph{JHEP} {\bf 10} (2019)
  261}, [\href{https://arxiv.org/abs/1904.06359}{{\tt 1904.06359}}].

\bibitem{Benjamin:2019stq}
N.~Benjamin, H.~Ooguri, S.-H. Shao and Y.~Wang, \emph{{Light-cone modular
  bootstrap and pure gravity}},
  \href{http://dx.doi.org/10.1103/PhysRevD.100.066029}{\emph{Phys. Rev. D} {\bf
  100} (2019) 066029}, [\href{https://arxiv.org/abs/1906.04184}{{\tt
  1906.04184}}].

\bibitem{Ganguly:2019ksp}
S.~Ganguly and S.~Pal, \emph{{Bounds on the density of states and the spectral
  gap in CFT$_{2}$}},
  \href{http://dx.doi.org/10.1103/PhysRevD.101.106022}{\emph{Phys. Rev. D} {\bf
  101} (2020) 106022}, [\href{https://arxiv.org/abs/1905.12636}{{\tt
  1905.12636}}].

\bibitem{Pal:2019zzr}
S.~Pal and Z.~Sun, \emph{{Tauberian-Cardy formula with spin}},
  \href{http://dx.doi.org/10.1007/JHEP01(2020)135}{\emph{JHEP} {\bf 01} (2020)
  135}, [\href{https://arxiv.org/abs/1910.07727}{{\tt 1910.07727}}].

\bibitem{Mukhametzhanov:2020swe}
B.~Mukhametzhanov and S.~Pal, \emph{{Beurling-Selberg Extremization and Modular
  Bootstrap at High Energies}},
  \href{http://dx.doi.org/10.21468/SciPostPhys.8.6.088}{\emph{SciPost Phys.}
  {\bf 8} (2020) 088}, [\href{https://arxiv.org/abs/2003.14316}{{\tt
  2003.14316}}].

\bibitem{Pal:2020wwd}
S.~Pal and Z.~Sun, \emph{{High Energy Modular Bootstrap, Global Symmetries and
  Defects}}, \href{http://dx.doi.org/10.1007/JHEP08(2020)064}{\emph{JHEP} {\bf
  08} (2020) 064}, [\href{https://arxiv.org/abs/2004.12557}{{\tt 2004.12557}}].

\bibitem{Das:2020uax}
D.~Das, Y.~Kusuki and S.~Pal, \emph{{Universality in asymptotic bounds and its
  saturation in $2$D CFT}},
  \href{http://dx.doi.org/10.1007/JHEP04(2021)288}{\emph{JHEP} {\bf 04} (2021)
  288}, [\href{https://arxiv.org/abs/2011.02482}{{\tt 2011.02482}}].

\bibitem{Creminelli:2022onn}
P.~Creminelli, O.~Janssen and L.~Senatore, \emph{{Positivity bounds on
  effective field theories with spontaneously broken Lorentz invariance}},
  \href{http://dx.doi.org/10.1007/JHEP09(2022)201}{\emph{JHEP} {\bf 09} (2022)
  201}, [\href{https://arxiv.org/abs/2207.14224}{{\tt 2207.14224}}].

\bibitem{Adams:2006sv}
A.~Adams, N.~Arkani-Hamed, S.~Dubovsky, A.~Nicolis and R.~Rattazzi,
  \emph{{Causality, analyticity and an IR obstruction to UV completion}},
  \href{http://dx.doi.org/10.1088/1126-6708/2006/10/014}{\emph{JHEP} {\bf 10}
  (2006) 014}, [\href{https://arxiv.org/abs/hep-th/0602178}{{\tt
  hep-th/0602178}}].

\bibitem{Caron-Huot:2020cmc}
S.~Caron-Huot and V.~Van~Duong, \emph{{Extremal Effective Field Theories}},
  \href{http://dx.doi.org/10.1007/JHEP05(2021)280}{\emph{JHEP} {\bf 05} (2021)
  280}, [\href{https://arxiv.org/abs/2011.02957}{{\tt 2011.02957}}].

\bibitem{Sachdev:1993pr}
S.~Sachdev, \emph{{Polylogarithm identities in a conformal field theory in
  three-dimensions}},
  \href{http://dx.doi.org/10.1016/0370-2693(93)90935-B}{\emph{Phys. Lett. B}
  {\bf 309} (1993) 285--288}, [\href{https://arxiv.org/abs/hep-th/9305131}{{\tt
  hep-th/9305131}}].

\bibitem{Liu:2018kfw}
H.~Liu and P.~Glorioso, \emph{{Lectures on non-equilibrium effective field
  theories and fluctuating hydrodynamics}},
  \href{http://dx.doi.org/10.22323/1.305.0008}{\emph{PoS} {\bf TASI2017} (2018)
  008}, [\href{https://arxiv.org/abs/1805.09331}{{\tt 1805.09331}}].

\bibitem{Brauner:2022rvf}
T.~Brauner, S.~A. Hartnoll, P.~Kovtun, H.~Liu, M.~Mezei, A.~Nicolis et~al.,
  \emph{{Snowmass White Paper: Effective Field Theories for Condensed Matter
  Systems}},  in \emph{{Snowmass 2021}}, 3, 2022.
\newblock \href{https://arxiv.org/abs/2203.10110}{{\tt 2203.10110}}.

\bibitem{Delacretaz:2020nit}
L.~V. Delacretaz, \emph{{Heavy Operators and Hydrodynamic Tails}},
  \href{http://dx.doi.org/10.21468/SciPostPhys.9.3.034}{\emph{SciPost Phys.}
  {\bf 9} (2020) 034}, [\href{https://arxiv.org/abs/2006.01139}{{\tt
  2006.01139}}].

\bibitem{Karlsson:2022osn}
R.~Karlsson, A.~Parnachev, V.~Prilepina and S.~Valach, \emph{{Thermal stress
  tensor correlators, OPE and holography}},
  \href{http://dx.doi.org/10.1007/JHEP09(2022)234}{\emph{JHEP} {\bf 09} (2022)
  234}, [\href{https://arxiv.org/abs/2206.05544}{{\tt 2206.05544}}].

\bibitem{Hellerman:2015nra}
S.~Hellerman, D.~Orlando, S.~Reffert and M.~Watanabe, \emph{{On the CFT
  Operator Spectrum at Large Global Charge}},
  \href{http://dx.doi.org/10.1007/JHEP12(2015)071}{\emph{JHEP} {\bf 12} (2015)
  071}, [\href{https://arxiv.org/abs/1505.01537}{{\tt 1505.01537}}].

\bibitem{Monin:2016jmo}
A.~Monin, D.~Pirtskhalava, R.~Rattazzi and F.~K. Seibold, \emph{{Semiclassics,
  Goldstone Bosons and CFT data}},
  \href{http://dx.doi.org/10.1007/JHEP06(2017)011}{\emph{JHEP} {\bf 06} (2017)
  011}, [\href{https://arxiv.org/abs/1611.02912}{{\tt 1611.02912}}].

\bibitem{Jafferis:2017zna}
D.~Jafferis, B.~Mukhametzhanov and A.~Zhiboedov, \emph{{Conformal Bootstrap At
  Large Charge}}, \href{http://dx.doi.org/10.1007/JHEP05(2018)043}{\emph{JHEP}
  {\bf 05} (2018) 043}, [\href{https://arxiv.org/abs/1710.11161}{{\tt
  1710.11161}}].

\bibitem{Kos:2013tga}
F.~Kos, D.~Poland and D.~Simmons-Duffin, \emph{{Bootstrapping the $O(N)$ vector
  models}}, \href{http://dx.doi.org/10.1007/JHEP06(2014)091}{\emph{JHEP} {\bf
  06} (2014) 091}, [\href{https://arxiv.org/abs/1307.6856}{{\tt 1307.6856}}].

\bibitem{Kos:2014bka}
F.~Kos, D.~Poland and D.~Simmons-Duffin, \emph{{Bootstrapping Mixed Correlators
  in the 3D Ising Model}},
  \href{http://dx.doi.org/10.1007/JHEP11(2014)109}{\emph{JHEP} {\bf 11} (2014)
  109}, [\href{https://arxiv.org/abs/1406.4858}{{\tt 1406.4858}}].

\bibitem{Penedones:2015aga}
J.~Penedones, E.~Trevisani and M.~Yamazaki, \emph{{Recursion Relations for
  Conformal Blocks}},
  \href{http://dx.doi.org/10.1007/JHEP09(2016)070}{\emph{JHEP} {\bf 09} (2016)
  070}, [\href{https://arxiv.org/abs/1509.00428}{{\tt 1509.00428}}].

\bibitem{Erramilli:2019njx}
R.~S. Erramilli, L.~V. Iliesiu and P.~Kravchuk, \emph{{Recursion relation for
  general 3d blocks}},
  \href{http://dx.doi.org/10.1007/JHEP12(2019)116}{\emph{JHEP} {\bf 12} (2019)
  116}, [\href{https://arxiv.org/abs/1907.11247}{{\tt 1907.11247}}].

\bibitem{Isachenkov:2017qgn}
M.~Isachenkov and V.~Schomerus, \emph{{Integrability of conformal blocks. Part
  I. Calogero-Sutherland scattering theory}},
  \href{http://dx.doi.org/10.1007/JHEP07(2018)180}{\emph{JHEP} {\bf 07} (2018)
  180}, [\href{https://arxiv.org/abs/1711.06609}{{\tt 1711.06609}}].

\bibitem{Hijano:2015zsa}
E.~Hijano, P.~Kraus, E.~Perlmutter and R.~Snively, \emph{{Witten Diagrams
  Revisited: The AdS Geometry of Conformal Blocks}},
  \href{http://dx.doi.org/10.1007/JHEP01(2016)146}{\emph{JHEP} {\bf 01} (2016)
  146}, [\href{https://arxiv.org/abs/1508.00501}{{\tt 1508.00501}}].

\bibitem{Mukhametzhanov:2018zja}
B.~Mukhametzhanov and A.~Zhiboedov, \emph{{Analytic Euclidean Bootstrap}},
  \href{http://dx.doi.org/10.1007/JHEP10(2019)270}{\emph{JHEP} {\bf 10} (2019)
  270}, [\href{https://arxiv.org/abs/1808.03212}{{\tt 1808.03212}}].

\bibitem{Cho:2017fzo}
M.~Cho, S.~Collier and X.~Yin, \emph{{Genus Two Modular Bootstrap}},
  \href{http://dx.doi.org/10.1007/JHEP04(2019)022}{\emph{JHEP} {\bf 04} (2019)
  022}, [\href{https://arxiv.org/abs/1705.05865}{{\tt 1705.05865}}].

\bibitem{Belin:2021ryy}
A.~Belin, J.~de~Boer and D.~Liska, \emph{{Non-Gaussianities in the statistical
  distribution of heavy OPE coefficients and wormholes}},
  \href{http://dx.doi.org/10.1007/JHEP06(2022)116}{\emph{JHEP} {\bf 06} (2022)
  116}, [\href{https://arxiv.org/abs/2110.14649}{{\tt 2110.14649}}].

\bibitem{Jafferis:2022uhu}
D.~L. Jafferis, D.~K. Kolchmeyer, B.~Mukhametzhanov and J.~Sonner,
  \emph{{Matrix models for eigenstate thermalization}},
  \href{https://arxiv.org/abs/2209.02130}{{\tt 2209.02130}}.

\bibitem{Heemskerk:2009pn}
I.~Heemskerk, J.~Penedones, J.~Polchinski and J.~Sully, \emph{{Holography from
  Conformal Field Theory}},
  \href{http://dx.doi.org/10.1088/1126-6708/2009/10/079}{\emph{JHEP} {\bf 10}
  (2009) 079}, [\href{https://arxiv.org/abs/0907.0151}{{\tt 0907.0151}}].

\bibitem{Hartman:2015lfa}
T.~Hartman, S.~Jain and S.~Kundu, \emph{{Causality Constraints in Conformal
  Field Theory}}, \href{http://dx.doi.org/10.1007/JHEP05(2016)099}{\emph{JHEP}
  {\bf 05} (2016) 099}, [\href{https://arxiv.org/abs/1509.00014}{{\tt
  1509.00014}}].

\bibitem{Afkhami-Jeddi:2016ntf}
N.~Afkhami-Jeddi, T.~Hartman, S.~Kundu and A.~Tajdini, \emph{{Einstein gravity
  3-point functions from conformal field theory}},
  \href{http://dx.doi.org/10.1007/JHEP12(2017)049}{\emph{JHEP} {\bf 12} (2017)
  049}, [\href{https://arxiv.org/abs/1610.09378}{{\tt 1610.09378}}].

\bibitem{Kologlu:2019bco}
M.~Kologlu, P.~Kravchuk, D.~Simmons-Duffin and A.~Zhiboedov, \emph{{Shocks,
  Superconvergence, and a Stringy Equivalence Principle}},
  \href{http://dx.doi.org/10.1007/JHEP11(2020)096}{\emph{JHEP} {\bf 11} (2020)
  096}, [\href{https://arxiv.org/abs/1904.05905}{{\tt 1904.05905}}].

\bibitem{Belin:2019mnx}
A.~Belin, D.~M. Hofman and G.~Mathys, \emph{{Einstein gravity from ANEC
  correlators}}, \href{http://dx.doi.org/10.1007/JHEP08(2019)032}{\emph{JHEP}
  {\bf 08} (2019) 032}, [\href{https://arxiv.org/abs/1904.05892}{{\tt
  1904.05892}}].

\bibitem{Caron-Huot:2021enk}
S.~Caron-Huot, D.~Mazac, L.~Rastelli and D.~Simmons-Duffin, \emph{{AdS bulk
  locality from sharp CFT bounds}},
  \href{http://dx.doi.org/10.1007/JHEP11(2021)164}{\emph{JHEP} {\bf 11} (2021)
  164}, [\href{https://arxiv.org/abs/2106.10274}{{\tt 2106.10274}}].

\bibitem{Caron-Huot:2022lff}
S.~Caron-Huot, \emph{{Holographic cameras: an eye for the bulk}},
  \href{http://dx.doi.org/10.1007/JHEP03(2023)047}{\emph{JHEP} {\bf 03} (2023)
  047}, [\href{https://arxiv.org/abs/2211.11791}{{\tt 2211.11791}}].

\bibitem{Iliesiu:2018zlz}
L.~Iliesiu, M.~Kolo\u{g}lu and D.~Simmons-Duffin, \emph{{Bootstrapping the 3d
  Ising model at finite temperature}},
  \href{http://dx.doi.org/10.1007/JHEP12(2019)072}{\emph{JHEP} {\bf 12} (2019)
  072}, [\href{https://arxiv.org/abs/1811.05451}{{\tt 1811.05451}}].

\bibitem{Rattazzi:2008pe}
R.~Rattazzi, V.~S. Rychkov, E.~Tonni and A.~Vichi, \emph{{Bounding scalar
  operator dimensions in 4D CFT}},
  \href{http://dx.doi.org/10.1088/1126-6708/2008/12/031}{\emph{JHEP} {\bf 12}
  (2008) 031}, [\href{https://arxiv.org/abs/0807.0004}{{\tt 0807.0004}}].

\bibitem{Liendo:2012hy}
P.~Liendo, L.~Rastelli and B.~C. van Rees, \emph{{The Bootstrap Program for
  Boundary CFT$_d$}},
  \href{http://dx.doi.org/10.1007/JHEP07(2013)113}{\emph{JHEP} {\bf 07} (2013)
  113}, [\href{https://arxiv.org/abs/1210.4258}{{\tt 1210.4258}}].

\bibitem{Dias:2017coo}
O.~J.~C. Dias, J.~E. Santos and B.~Way, \emph{{Lattice Black Branes: Sphere
  Packing in General Relativity}},
  \href{http://dx.doi.org/10.1007/JHEP05(2018)111}{\emph{JHEP} {\bf 05} (2018)
  111}, [\href{https://arxiv.org/abs/1712.07663}{{\tt 1712.07663}}].

\bibitem{Hartman:2019pcd}
T.~Hartman, D.~Maz\'a\v{c} and L.~Rastelli, \emph{{Sphere Packing and Quantum
  Gravity}}, \href{http://dx.doi.org/10.1007/JHEP12(2019)048}{\emph{JHEP} {\bf
  12} (2019) 048}, [\href{https://arxiv.org/abs/1905.01319}{{\tt 1905.01319}}].

\bibitem{Cardy:1991kr}
J.~L. Cardy, \emph{{Operator content and modular properties of higher
  dimensional conformal field theories}},
  \href{http://dx.doi.org/10.1016/0550-3213(91)90024-R}{\emph{Nucl. Phys. B}
  {\bf 366} (1991) 403--419}.

\bibitem{Chang:1992fu}
P.~Chang and J.~S. Dowker, \emph{{Vacuum energy on orbifold factors of
  spheres}}, \href{http://dx.doi.org/10.1016/0550-3213(93)90223-C}{\emph{Nucl.
  Phys. B} {\bf 395} (1993) 407--432},
  [\href{https://arxiv.org/abs/hep-th/9210013}{{\tt hep-th/9210013}}].

\bibitem{Klebanov:2011gs}
I.~R. Klebanov, S.~S. Pufu and B.~R. Safdi, \emph{{F-Theorem without
  Supersymmetry}}, \href{http://dx.doi.org/10.1007/JHEP10(2011)038}{\emph{JHEP}
  {\bf 10} (2011) 038}, [\href{https://arxiv.org/abs/1105.4598}{{\tt
  1105.4598}}].

\bibitem{Duff:1993wm}
M.~J. Duff, \emph{{Twenty years of the Weyl anomaly}},
  \href{http://dx.doi.org/10.1088/0264-9381/11/6/004}{\emph{Class. Quant.
  Grav.} {\bf 11} (1994) 1387--1404},
  [\href{https://arxiv.org/abs/hep-th/9308075}{{\tt hep-th/9308075}}].

\bibitem{Bastianelli:2000hi}
F.~Bastianelli, S.~Frolov and A.~A. Tseytlin, \emph{{Conformal anomaly of (2,0)
  tensor multiplet in six-dimensions and AdS / CFT correspondence}},
  \href{http://dx.doi.org/10.1088/1126-6708/2000/02/013}{\emph{JHEP} {\bf 02}
  (2000) 013}, [\href{https://arxiv.org/abs/hep-th/0001041}{{\tt
  hep-th/0001041}}].

\bibitem{Giombi:2013yva}
S.~Giombi, I.~R. Klebanov, S.~S. Pufu, B.~R. Safdi and G.~Tarnopolsky,
  \emph{{AdS Description of Induced Higher-Spin Gauge Theory}},
  \href{http://dx.doi.org/10.1007/JHEP10(2013)016}{\emph{JHEP} {\bf 10} (2013)
  016}, [\href{https://arxiv.org/abs/1306.5242}{{\tt 1306.5242}}].

\bibitem{Giombi:2014xxa}
S.~Giombi and I.~R. Klebanov, \emph{{Interpolating between $a$ and $F$}},
  \href{http://dx.doi.org/10.1007/JHEP03(2015)117}{\emph{JHEP} {\bf 03} (2015)
  117}, [\href{https://arxiv.org/abs/1409.1937}{{\tt 1409.1937}}].

\bibitem{Dowker:1978md}
J.~S. Dowker and G.~Kennedy, \emph{{Finite Temperature and Boundary Effects in
  Static Space-Times}},
  \href{http://dx.doi.org/10.1088/0305-4470/11/5/020}{\emph{J. Phys. A} {\bf
  11} (1978) 895}.

\bibitem{Candelas:1978gf}
P.~Candelas and J.~S. Dowker, \emph{{Field theories on conformally related
  space-times: Some global considerations}},
  \href{http://dx.doi.org/10.1103/PhysRevD.19.2902}{\emph{Phys. Rev. D} {\bf
  19} (1979) 2902}.

\bibitem{Melia:2020pzd}
T.~Melia and S.~Pal, \emph{{EFT Asymptotics: the Growth of Operator
  Degeneracy}},
  \href{http://dx.doi.org/10.21468/SciPostPhys.10.5.104}{\emph{SciPost Phys.}
  {\bf 10} (2021) 104}, [\href{https://arxiv.org/abs/2010.08560}{{\tt
  2010.08560}}].

\bibitem{Camporesi_1996}
R.~Camporesi and A.~Higuchi, \emph{{On the eigenfunctions of the Dirac operator
  on spheres and real hyperbolic spaces}},
  \href{http://dx.doi.org/10.1016/0393-0440(95)00042-9}{\emph{J. Geom. Phys.}
  {\bf 20} (1996) 1--18}, [\href{https://arxiv.org/abs/gr-qc/9505009}{{\tt
  gr-qc/9505009}}].

\bibitem{goldilocks}
R.~Southey, \emph{The story of the three bears},  in \emph{The Doctor \& C.},
  pp.~318--326.
\newblock Longman, Rees, Orme, Brown, Green and Longman, 1837.

\bibitem{PhysRevE.79.041142}
O.~Vasilyev, A.~Gambassi, A.~Macioek and S.~Dietrich, \emph{{Universal scaling
  functions of critical Casimir forces obtained by Monte Carlo simulations}},
  \href{http://dx.doi.org/10.1103/PhysRevE.79.041142}{\emph{Phys. Rev. E} {\bf
  79} (2009) 041142}, [\href{https://arxiv.org/abs/0812.0750}{{\tt
  0812.0750}}].

\bibitem{casimir2}
M.~Krech and D.~P. Landau, \emph{{Casimir effect in critical systems: A Monte
  Carlo simulation}},
  \href{http://dx.doi.org/10.1103/PhysRevE.53.4414}{\emph{Phys. Rev. E} {\bf
  53} (1996) 4414--4423}.

\bibitem{casimir3}
M.~Krech, \emph{Casimir forces in binary liquid mixtures},
  \href{http://dx.doi.org/10.1103/PhysRevE.56.1642}{\emph{Phys. Rev. E} {\bf
  56} (1997) 1642--1659}, [\href{https://arxiv.org/abs/cond-mat/9703093}{{\tt
  cond-mat/9703093}}].

\bibitem{Dymarsky:2017yzx}
A.~Dymarsky, F.~Kos, P.~Kravchuk, D.~Poland and D.~Simmons-Duffin, \emph{{The
  3d Stress-Tensor Bootstrap}},
  \href{http://dx.doi.org/10.1007/JHEP02(2018)164}{\emph{JHEP} {\bf 02} (2018)
  164}, [\href{https://arxiv.org/abs/1708.05718}{{\tt 1708.05718}}].

\bibitem{Zhu:2022gjc}
W.~Zhu, C.~Han, E.~Huffman, J.~S. Hofmann and Y.-C. He, \emph{{Uncovering
  Conformal Symmetry in the 3D Ising Transition: State-Operator Correspondence
  from a Quantum Fuzzy Sphere Regularization}},
  \href{http://dx.doi.org/10.1103/PhysRevX.13.021009}{\emph{Phys. Rev. X} {\bf
  13} (2023) 021009}, [\href{https://arxiv.org/abs/2210.13482}{{\tt
  2210.13482}}].

\bibitem{Hu:2023xak}
L.~Hu, Y.-C. He and W.~Zhu, \emph{{Operator Product Expansion Coefficients of
  the 3D Ising Criticality via Quantum Fuzzy Sphere}},
  \href{https://arxiv.org/abs/2303.08844}{{\tt 2303.08844}}.

\bibitem{Caron-Huot:2017vep}
S.~Caron-Huot, \emph{{Analyticity in Spin in Conformal Theories}},
  \href{http://dx.doi.org/10.1007/JHEP09(2017)078}{\emph{JHEP} {\bf 09} (2017)
  078}, [\href{https://arxiv.org/abs/1703.00278}{{\tt 1703.00278}}].

\bibitem{Costa:2011mg}
M.~S. Costa, J.~Penedones, D.~Poland and S.~Rychkov, \emph{{Spinning Conformal
  Correlators}}, \href{http://dx.doi.org/10.1007/JHEP11(2011)071}{\emph{JHEP}
  {\bf 11} (2011) 071}, [\href{https://arxiv.org/abs/1107.3554}{{\tt
  1107.3554}}].

\bibitem{Costa:2014rya}
M.~S. Costa and T.~Hansen, \emph{{Conformal correlators of mixed-symmetry
  tensors}}, \href{http://dx.doi.org/10.1007/JHEP02(2015)151}{\emph{JHEP} {\bf
  02} (2015) 151}, [\href{https://arxiv.org/abs/1411.7351}{{\tt 1411.7351}}].

\bibitem{Kologlu:2019mfz}
M.~Kologlu, P.~Kravchuk, D.~Simmons-Duffin and A.~Zhiboedov, \emph{{The
  light-ray OPE and conformal colliders}},
  \href{http://dx.doi.org/10.1007/JHEP01(2021)128}{\emph{JHEP} {\bf 01} (2021)
  128}, [\href{https://arxiv.org/abs/1905.01311}{{\tt 1905.01311}}].

\end{thebibliography}\endgroup

\end{document}